\documentclass{aa}

\usepackage{graphicx}
\usepackage{txfonts}
\usepackage{amsmath}
\usepackage[colorlinks=true,linkcolor=blue,citecolor=blue,urlcolor=blue]{hyperref}

\newcommand{\GA}{\mathrm{GA}}
\newcommand{\HHO}{\mathrm{H_2O}}
\newcommand{\COO}{\mathrm{CO_2}}
\newcommand{\kb}{k_\mathrm{B}}
\newcommand{\nstar}{N^{*}}
\newcommand{\natas}{Nat. Astron.}   %
\newcommand{\percm}{\mathrm{cm^{-2}}}
\newcommand{\wmk}{\mathrm{W\,m^{-1}\,K^{-1}}}
\newcommand{\Ncom}{2\,940}
\newcommand{\Ncomsd}{230}
\newcommand{\tcom}{65}
\newcommand{\zetacom}{9.3\times10^{-14}}
\newcommand{\zetacomten}{6.5\times10^{-14}}
\newcommand{\dhacom}{15}
\newcommand{\gldcom}{2.0}
\newcommand{\glycom}{0.029}
\newcommand{\ketcom}{76\%}
\newcommand{\cotwocom}{11\%}
\newcommand{\hcoohcom}{15\%}
\newcommand{\egcom}{0.45}
\newcommand{\meancom}{3.7}
\newcommand{\cthreesum}{1.7}
\newcommand{\dhameohoff}{14}
\newcommand{\dhadhmoff}{15}
\newcommand{\cthreeald}{17}
\newcommand{\ketcomwin}{100\%}
\newcommand{\Nec}{180}
\newcommand{\tec}{4.1}
\newcommand{\dhaec}{10}
\newcommand{\gldec}{9.0}

\newcommand{\ecfacinv}{16}
\newcommand{\zetaec}{5.8\times10^{-15}}
\newcommand{\ecrad}{4.1\times10^{5}}

\newcommand{\egcolda}{3.0}

\newcommand{\Ncoldc}{2\,170}
\newcommand{\dhacoldc}{16}

\newcommand{\ketcoldc}{79\%}
\newcommand{\tcoldc}{48}

\newcommand{\egcoldc}{0.94}
\newcommand{\Ndhmgld}{2\,890}
\newcommand{\dhadhmgld}{15}
\newcommand{\glddhmgld}{2.2}

\newcommand{\Nhfone}{1\,810}
\newcommand{\dhahfone}{19}
\newcommand{\gldhfone}{2.6}
\newcommand{\kethfone}{66\%}
\newcommand{\thfone}{40}

\newcommand{\zetahfone}{5.7\times10^{-14}}
\newcommand{\Nhfhalf}{1\,410}
\newcommand{\dhahfhalf}{17}
\newcommand{\gldhfhalf}{2.5}
\newcommand{\kethfhalf}{77\%}
\newcommand{\thfhalf}{31}

\newcommand{\zetahf}{4.5\times10^{-14}}
\newcommand{\Nhfonek}{1\,340}
\newcommand{\dhahfonek}{16}

\newcommand{\kethfonek}{100\%}

\newcommand{\Nhfhalfk}{1\,250}

\newcommand{\Ngemtwo}{2\,980}

\newcommand{\ketgemtwo}{75\%}

\newcommand{\Ngemfive}{3\,120}

\newcommand{\ketgemfive}{70\%}

\newcommand{\Nkfree}{2\,500}

\newcommand{\ketkfree}{77\%}

\newcommand{\depthreefac}{2.3}
\newcommand{\hnonecom}{39\%}
\newcommand{\hnonecomq}{the observed ratio is reached only after 3\,770 events, with DHA at 16 times its limit}
\newcommand{\hdhafac}{a factor of 11.6}
\newcommand{\Ncomdx}{2\,600--3\,250}
\newcommand{\nstarlo}{1\,250}
\newcommand{\nstarhi}{12\,000}
\newcommand{\tlo}{28}
\newcommand{\thi}{260}

\newcommand{\Nres}{6\,470}

\newcommand{\comfacinv}{2.2}

\newcommand{\Rwge}{45}

\newcommand{\zetares}{2.0\times10^{-13}}

\newcommand{\Nressd}{300}
\newcommand{\nresseq}{16}
\newcommand{\Nreslo}{5\,980}
\newcommand{\zetareslo}{1.9\times10^{-13}}
\newcommand{\Nreshi}{7\,160}
\newcommand{\zetareshi}{2.3\times10^{-13}}
\newcommand{\Nrev}{23\,700}

\newcommand{\sdrange}{1--15\%}
\newcommand{\capfac}{$+62$\% and $-22$\%}
\newcommand{\Ncomfour}{2\,890}
\newcommand{\gldcomfour}{2.1}
\newcommand{\dhacomfour}{15}
\newcommand{\glycomfour}{0.030}

\newcommand{\facgins}{9.3}
\newcommand{\facdense}{130--230}

\newcommand{\facecdense}{8--14}
\newcommand{\powerfac}{9}
\newcommand{\powerfrac}{1/30}
\newcommand{\gacolcom}{$4.1\times10^{16}$}
\newcommand{\erucolcom}{$3.9\times10^{16}\,\percm$}
\newcommand{\desfrac}{$2.3\times10^{-3}$}
\newcommand{\taubarps}{19}
\newcommand{\liqtot}{56}
\newcommand{\gaevap}{1.1\%}
\newcommand{\eruevap}{$2.8\times10^{-6}$}
\newcommand{\cyclmax}{$5.6\times10^{-4}$}
\newcommand{\tkcom}{42}
\newcommand{\tecom}{160}
\newcommand{\tlcom}{63}
\newcommand{\glynoohcom}{0.29}
\newcommand{\Nrevcom}{10\,400}

\newcommand{\tecl}{150}
\newcommand{\trcentre}{240}
\newcommand{\trninety}{100}
\newcommand{\trniner}{37}
\newcommand{\trsurf}{9.3}
\newcommand{\trlcentre}{160}

\newcommand{\tNab}{51}

\newcommand{\Nflat}{1\,100}
\newcommand{\tflatfar}{61--150}
\newcommand{\flatfarfac}{0.9--1.0}
\newcommand{\depfac}{3.2}
\newcommand{\ldafac}{1.2}
\newcommand{\corefac}{$-7$\%}
\newcommand{\vitfac}{$-54$\%}
\newcommand{\hdafac}{$+14$\%}
\newcommand{\tkthree}{55}
\newcommand{\kfac}{1.5}
\newcommand{\Ntw}{1\,040}
\newcommand{\Ntwmob}{240}
\newcommand{\zetakcom}{6.0\times10^{-14}}
\newcommand{\erucomtwo}{1.9}
\newcommand{\hexcomtwo}{4.6}
\newcommand{\bigcomtwo}{34\%}
\newcommand{\dhacomtwo}{21}
\newcommand{\hmargin}{43}
\newcommand{\gldescdrop}{24\%}

\newcommand{\Ngemdehyda}{2\,890}
\newcommand{\ketgemdehyda}{76\%}
\newcommand{\Ngemdehydb}{2\,900}
\newcommand{\ketgemdehydb}{76\%}
\newcommand{\coldcfac}{0.75}
\newcommand{\dhashare}{94\%}
\newcommand{\tepmol}{2.2}
\newcommand{\tepthirteen}{6.5}
\newcommand{\Ncomet}{324}
\newcommand{\erucomet}{$2\times10^{-2}$}
\newcommand{\dhacomet}{0.22}
\newcommand{\ohfracfa}{34\%}
\newcommand{\ohfrachcho}{30\%}
\newcommand{\ohfracmeoh}{10\%}
\newcommand{\ohfracga}{4\%}
\newcommand{\cratio}{9.5}
\newcommand{\cfourdhm}{82\%}
\newcommand{\cfourdhmd}{61\%}
\newcommand{\cfourmeoh}{21\%}
\newcommand{\cfourgem}{13\%}
\newcommand{\cfourgemx}{3\%}
\newcommand{\cfourald}{0.1\%}
\newcommand{\ndhalim}{200}

\newcommand{\Ndosetwo}{1\,290}
\newcommand{\Ndhmoff}{10\,800}
\newcommand{\Nhdha}{8\,890}
\newcommand{\Nescape}{4\,930}
\newcommand{\escfac}{1.7}
\newcommand{\dosehalffac}{2.4}
\newcommand{\Nsmalloff}{850}
\newcommand{\zetasmalloff}{2.7\times10^{-14}}

\newcommand{\Nhnone}{3\,770}
\newcommand{\dxpctlo}{$-10$\%}
\newcommand{\dxpcthi}{$+12$\%}
\newcommand{\restpct}{50\%}

\newcommand{\dharange}{9.6--19}
\newcommand{\dhahdha}{177}
\newcommand{\gldcthreeald}{51}
\newcommand{\cthreesumrange}{9--17}
\newcommand{\cthreesumhdha}{118}
\newcommand{\gldhgld}{10}
\newcommand{\gldch}{17}

\newcommand{\gldmostrange}{0.2--3.0}
\newcommand{\gldlit}{12}
\newcommand{\glymax}{0.09}
\newcommand{\glych}{0.50}
\newcommand{\glyec}{0.23}
\newcommand{\ketmostrange}{70--85\%}
\newcommand{\ketsmalloff}{90\%}
\newcommand{\ketclosed}{58\%}
\newcommand{\ketmeohoff}{99\%}
\newcommand{\ketdhmoff}{37\%}
\newcommand{\kethnone}{61\%}
\newcommand{\keterysym}{8\%}
\newcommand{\ketrange}{37--100\%}
\newcommand{\pkerysym}{0.1}
\newcommand{\dhacomhalf}{8}
\newcommand{\gafrac}{0.69}
\newcommand{\sugfracrange}{0.4--0.6\%}
\newcommand{\sugfractot}{2.1\%}
\newcommand{\gagrain}{$4.8\times10^{5}$}
\newcommand{\cthreegrain}{$8\times10^{5}$}
\newcommand{\gaproton}{35}

\newcommand{\evlong}{5\%}
\newcommand{\cfourlong}{83\%}
\newcommand{\cfourverylong}{34\%}
\newcommand{\evshort}{88\%}
\newcommand{\cfourshort}{0.80\%}
\newcommand{\ohgashort}{38\%}
\newcommand{\cthreeborn}{$2.1\times10^{6}$}
\newcommand{\cfourborn}{$1.1\times10^{6}$}
\newcommand{\cthreedhm}{76\%}
\newcommand{\cfourgemall}{16\%}
\newcommand{\cfourglyc}{2.1\%}
\newcommand{\aldolgld}{$7\times10^{-8}$}
\newcommand{\aldolery}{$4\times10^{-5}$}
\newcommand{\nionsstar}{$3.0\times10^{7}$}
\newcommand{\gldres}{0.037}
\newcommand{\dhares}{0.39}
\newcommand{\ketres}{99.6\%}
\newcommand{\hexres}{0.57}
\newcommand{\glyres}{0.020}
\newcommand{\hcoohres}{17\%}
\newcommand{\cotwores}{12\%}
\newcommand{\aldres}{0.004}
\newcommand{\dhabelow}{2.5}
\newcommand{\dhagares}{0.04}
\newcommand{\gldgares}{0.002}
\newcommand{\gldgareshyd}{0.017}
\newcommand{\gldreshyd}{0.32}
\newcommand{\dhareshyd}{0.71}
\newcommand{\gldoverdha}{21}
\newcommand{\Nresdxmid}{6\,710}
\newcommand{\Nresgrid}{5\,980 and 7\,750}
\newcommand{\dhagrid}{0.23--0.40}
\newcommand{\gldgrid}{0.030--0.36}
\newcommand{\sumgrid}{0.27--0.28}
\newcommand{\dhasys}{0.004--0.71}
\newcommand{\gldsys}{0.005--0.79}
\newcommand{\Nsys}{820--25\,200}
\newcommand{\dhafix}{0.07--0.50}
\newcommand{\gldfix}{0.005--0.29}
\newcommand{\dhagafix}{0.007--0.05}
\newcommand{\glyresdx}{0.010--0.025}
\newcommand{\glyresnooh}{0.68}
\newcommand{\glyresnoohhi}{1.03}
\newcommand{\glyressys}{0.006--0.08}
\newcommand{\glyresch}{0.89}
\newcommand{\hexresdx}{0.53--0.60}
\newcommand{\bigres}{22\%}
\newcommand{\hexrestwo}{3.1}
\newcommand{\bigrestwo}{45\%}
\newcommand{\erurestwo}{2.4}
\newcommand{\meohres}{13}
\newcommand{\meohresdx}{11--16}
\newcommand{\meohressys}{8--26}
\newcommand{\hcoohrange}{15--17\%}
\newcommand{\hcoohcohalf}{6.0\%}
\newcommand{\Ncohalfpct}{6\%}
\newcommand{\Nrescaplo}{9\,440}
\newcommand{\Nrescaphi}{5\,210}
\newcommand{\Nresdfix}{8\,570}
\newcommand{\Nrestw}{3\,770}
\newcommand{\Nresmob}{23\,300}
\newcommand{\Nresch}{820}

\newcommand{\resdosefac}{2.4}
\newcommand{\resescfac}{1.7}
\newcommand{\Nresfour}{6\,450}
\newcommand{\gldresfour}{0.044}
\newcommand{\dharesfour}{0.39}
\newcommand{\glyresfour}{0.023}
\newcommand{\resdxpctlo}{$-7$\%}
\newcommand{\resdxpcthi}{$+11$\%}
\newcommand{\rescappctlo}{$-19$\%}
\newcommand{\rescappcthi}{$+46$\%}
\newcommand{\hgareket}{98.5\%}
\newcommand{\ketresrange}{98.5--99.9\%}

\newcommand{\hnoneresdha}{16}
\newcommand{\hnoneresald}{96\%}
\newcommand{\gldrescthreeald}{0.79}
\newcommand{\dharescthreeald}{0.004}
\newcommand{\keterysymres}{10\%}
\newcommand{\pkerysymres}{0.03}
\newcommand{\kettwhnone}{87\%}

\newcommand{\tresecom}{370}
\newcommand{\treslcom}{150}
\newcommand{\treskcom}{61}
\newcommand{\resagefac}{1.5--2.3}

\newcommand{\Nus}{330}
\newcommand{\tusrange}{66--153}
\newcommand{\zetausrange}{(0.9$--$2.2)\times10^{-13}}
\newcommand{\usfacrange}{9--22}
\newcommand{\tuscoll}{0.7}
\newcommand{\hexus}{11}
\newcommand{\gacolres}{$4.5\times10^{16}$}
\newcommand{\erucolres}{$4.2\times10^{16}$}
\newcommand{\gldcolres}{$8.9\times10^{13}$}
\newcommand{\dhacolres}{$1.9\times10^{15}$}
\newcommand{\desfracres}{$2.1\times10^{-3}$}
\newcommand{\aldfacres}{7\%}
\newcommand{\acylfullN}{1\%}
\newcommand{\acylfullgld}{4\%}
\newcommand{\audox}{78\%}
\newcommand{\audsug}{21\%}
\newcommand{\audpol}{1.1\%}
\newcommand{\audacid}{less than 0.1\%}
\newcommand{\audoxres}{90\%}
\newcommand{\audsugres}{10\%}
\newcommand{\chistar}{0.4}
\newcommand{\nchi}{200}
\newcommand{\omN}{9--16\%}
\newcommand{\omgld}{39\%}
\newcommand{\omother}{35\%}
\newcommand{\gldfixfac}{18}
\newcommand{\dhaescres}{0.30}
\newcommand{\dhachres}{0.07}
\newcommand{\dhahchores}{0.50}
\newcommand{\dhasmallres}{0.39}
\newcommand{\sdres}{5\%}
\newcommand{\hexcom}{0.74}
\newcommand{\hcoohover}{2.5}

\newcommand{\kgins}{6}
\newcommand{\cohalfgld}{19\%}
\newcommand{\cohalfgly}{16\%}
\newcommand{\cohalfdha}{5\%}
\newcommand{\Nresbasezero}{4\,790}
\newcommand{\pctresbasezero}{$-26$\%}
\newcommand{\hcoohresbasezero}{30\%}
\newcommand{\Nresbasetwo}{7\,340}
\newcommand{\pctresbasetwo}{$+14$\%}
\newcommand{\hcoohresbasetwo}{8\%}

\newcommand{\pctbasezero}{$-24$\%}

\newcommand{\pctbasetwo}{$+14$\%}

\newcommand{\gillcthree}{98$\pm$10}
\newcommand{\gillcthreemf}{98}
\newcommand{\gillcfour}{95$\pm$10}
\newcommand{\gillcfourmf}{92}
\newcommand{\gillcsix}{66$\pm$8}
\newcommand{\gillcsixmf}{67}
\newcommand{\gillnev}{$7.6\times10^{4}$}
\newcommand{\gillnrep}{40}
\newcommand{\gillcone}{1\%}
\newcommand{\erudiffuse}{1.0}
\newcommand{\dhadiffuse}{16}

\newcommand{\doseinmean}{2.5}
\newcommand{\doseinlong}{2.9}
\newcommand{\doseinbins}{0.9, 1.4, 2.5, 2.7, 2.7, 2.9, 2.9 and 2.9}
\newcommand{\doseout}{3.6}
\newcommand{\dosetotfac}{6.1}

\newcommand{\dosenonmelt}{1.9}
\newcommand{\doseoutmeltions}{1.7}

\newcommand{\evout}{0.41}
\newcommand{\evtot}{0.70}
\newcommand{\evmyr}{31}

\newcommand{\zwratioH}{2.9}
\newcommand{\zwratioL}{4.0}
\newcommand{\zwratioHfar}{3.7--4}
\newcommand{\doseoutrange}{1.9--79}
\newcommand{\ohevent}{$2.0\times10^{6}$}
\newcommand{\oheventmelt}{$8.1\times10^{5}$}
\newcommand{\ohlongM}{0.93}
\newcommand{\ohlongpct}{1.8\%}
\newcommand{\ohmeltM}{0.32}
\newcommand{\hinsitu}{$3.0\times10^{6}$}

\newcommand{\doseouttot}{1\,200}

\newcommand{\evoutgly}{0.15}
\newcommand{\evoutga}{2.0}

\newcommand{\radreached}{With $D_{1/2}$ of 6\,900 and 2\,300~eV per molecule the count rises by 5\% and 20\%}
\newcommand{\radnot}{With 1\,200~eV three of the four sequences reach the observed value and the mean peak is 0.93; with 690, 230, 69, 23, 6.9 and 2.3~eV it peaks at 0.67, 0.27, 0.063, 0.016, 0.0038 and 0.0018}
\newcommand{\dhalfthr}{1\,000}
\newcommand{\dhalffac}{220}
\newcommand{\dhalfgafac}{5\,000}
\newcommand{\dhalfgap}{a factor of 33--100}
\newcommand{\radreachedres}{With $D_{1/2}$ of 6\,900 and 2\,300~eV per molecule the count rises by 12\% and 72\%}
\newcommand{\radnotres}{With 1\,200~eV the ratio peaks at 0.53 and does not reach the observed value; with 690, 230, 69, 23, 6.9 and 2.3~eV it peaks at 0.31, 0.085, 0.013, 0.0023, 0.0004 and 0.0001}

\newcommand{\Nrevd}{10\,400}
\newcommand{\revfacd}{3.6}
\newcommand{\zetarevd}{3.3\times10^{-13}}
\newcommand{\doserevd}{4\,250}
\newcommand{\dosefaclo}{70}
\newcommand{\dosefachi}{250}

\newcommand{\zetarevresd}{7.5\times10^{-13}}

\newcommand{\Ndosehalf}{7\,070}
\newcommand{\Nchzero}{510}
\newcommand{\pkthirty}{0.0018}
\newcommand{\pkten}{0.0038}

\newcommand{\Nreshnone}{252\,800}
\newcommand{\zownH}{$4.9\times10^{-15}$}
\newcommand{\zownHp}{72\%}
\newcommand{\zfacH}{2.1}
\newcommand{\zownL}{$4.1\times10^{-17}$}
\newcommand{\zownLp}{61\%}
\newcommand{\zfacL}{250}
\newcommand{\sputrate}{$6.7\times10^{5}$}
\newcommand{\sputlo}{0.3}
\newcommand{\sputhi}{0.7}
\newcommand{\sputlight}{30\%}
\newcommand{\sputfeband}{15\%}
\newcommand{\Ndmelt}{12\,000}
\newcommand{\tdmelt}{260}
\newcommand{\zetadmelt}{3.8\times10^{-13}}
\newcommand{\facdmelt}{4.1}

\newcommand{\zetaesc}{1.6\times10^{-13}}
\newcommand{\facesc}{1.7}
\newcommand{\Ndmeltres}{25\,200}

\newcommand{\obsglypk}{0.16--0.57}
\newcommand{\obsdhapk}{0.25--0.78}
\newcommand{\obsglyrefseven}{1.2--2.1}
\newcommand{\obsglyrefthree}{0.8--2.9}
\newcommand{\obsglyweak}{5.1}
\newcommand{\obsglytime}{26}
\newcommand{\obsglyn}{16--19}
\newcommand{\obsdhan}{13--30}
\newcommand{\obsglystack}{0.98--2.9}
\newcommand{\obsdhastack}{1.1--3.2}
\newcommand{\obsglystackhi}{13--37}
\newcommand{\obsdhastackhi}{15--41}
\newcommand{\obssinglepk}{0.25}
\newcommand{\obssinglerms}{0.05}
\newcommand{\obssingle}{25}

\begin{document}

\title{Cosmic-ray melt windows in interstellar ice make a four-carbon ketose but not
       the sugar pattern of G+0.693$-$0.027}
\subtitle{Three quantitative obstacles: the three-carbon sugars, radiolysis between events, and the exposure}
\titlerunning{Melt windows and the sugars of G+0.693}
\authorrunning{S. Saito et al.}

\author{S. Saito\inst{1}
   \and M. Yajima\inst{2}
   \and T. Kenmotsu\inst{3}
   \and M. Wada\inst{4}
   \and K. Kobayashi\inst{5,6}
   \and H. Nakamura\inst{2}
   \and G. Chiaki\inst{7}}

\institute{Graduate School of Science and Engineering, Yamagata University,
           Yonezawa, Yamagata 992-8510, Japan\\ \email{saitos@yz.yamagata-u.ac.jp}
      \and National Institute for Fusion Science, Toki, Gifu 509-5292, Japan
      \and Faculty of Life and Medical Sciences, Doshisha University,
           Kyotanabe, Kyoto 610-0394, Japan
      \and Organization for Research Initiatives and Development,
           Doshisha University, Kyotanabe, Kyoto 610-0394, Japan
      \and Department of Chemistry and Life Science, Graduate School of
           Engineering Science, Yokohama National University, Hodogaya-ku,
           Yokohama 240-8501, Japan
      \and Department of Earth and Planetary Sciences, Institute of Science
           Tokyo, Meguro-ku, Tokyo 152-8551, Japan
      \and Department of Social Design Engineering, National Institute of
           Technology, Kochi College, Nankoku, Kochi 783-8508, Japan}

\date{Received ---; accepted ---}

\abstract
{Erythrulose (ERU), a four-carbon (C4) ketose, is reported towards the
Galactic-centre cloud G+0.693$-$0.027 (hereafter G+0.693) at a column density comparable to that of 
glycolaldehyde (GA), while the three-carbon (C3) sugars are not detected.}
{We test whether melt windows opened in interstellar ice by cosmic-ray heavy ions 
can reproduce this pattern, with each rate fixed to a published value where available.}
{An ion-track model for eleven cosmic-ray nuclei determines the melting-event rate, the liquid-window durations, and the energy deposited inside and outside the melted regions. 
A radical reaction network including C1 radicals derived from formaldehyde and methanol follows the chemical inventory through repeated events. 
Between melting events, radiolytic destruction of the sugars is estimated from measured half-life doses of organic molecules in cold ice.}
{Hydrated GA opens a reaction channel that preferentially forms the C4 ketose, and thus the melt-window chemistry can account for the ketose identity of ERU, although cold surface chemistry also yields it.
The overall abundance pattern, however, is not reproduced.
The C1 radicals add one carbon at a time: dihydroxyacetone is at \dhacom\ times its limit when ERU/GA reaches the observed value. The ions deposit \evout~eV per molecule per event outside the melts, which destroys the sugars between events; the ratio is reached only for half-life doses above \dhalfthr~eV, against measured values below 5~eV. The exposure is less certain: \Ncom\ whole-grain melting events, \tcom~Myr at an ionisation rate $\zeta=10^{-14}$~s$^{-1}$. Within 7~Myr it needs $\zeta\gtrsim\zetacom$~s$^{-1}$, \facgins\ times the limit on the dense gas of the Central Molecular Zone, or $\zetaec$~s$^{-1}$ if the radicals trapped at the end of a window recombine.}
{With standard assumptions the sugars of G+0.693 are not made in cosmic-ray melt
windows. The pattern would need C1 radicals that do not add to the sugar radicals, sugars that survive the solid-phase dose and a local source of MeV ions with $\zeta\gtrsim\zetares$~s$^{-1}$.
The ketose selection by hydrated GA and the reconstructed melting rate stand independently.}

\keywords{astrochemistry -- ISM: molecules -- ISM: individual objects: G+0.693$-$0.027 --
          cosmic rays -- molecular processes -- methods: numerical}

\maketitle
\nolinenumbers %%% DELETE THIS WHEN SUBMITTING THIS TO A&A %%%

\section{Introduction}
\label{sec:intro}

Sugars are among the least explained molecules in the interstellar inventory.
Glycolaldehyde (GA, $\mathrm{HOCH_2CHO}$), the two-carbon sugar-related aldehyde, has
been known towards the Galactic centre for a quarter of a century \citep{Hollis2000}.
Erythrulose (ERU, $\mathrm{C_4H_8O_4}$) has now been reported towards the
Galactic-centre cloud G+0.693$-$0.027 (hereafter G+0.693), which has no star formation but is shocked by a
cloud--cloud collision \citep{Zeng2020}; the detection is that of \citet{JimenezSerra2026}. It is the first
four-carbon sugar reported in the interstellar medium. This report changes the problem
for two reasons.

First, the abundance pattern is non-monotonic in carbon number. The reported columns
are $N(\GA)=(9.3\pm0.3)\times10^{13}\,\percm$ \citep[GA parameters
from][]{Rivilla2022} and $N(\mathrm{ERU})=(8.7\pm0.8)\times10^{13}\,\percm$: equal to within
errors. The intervening C3 sugars are \emph{not} detected, with
$N\leq0.5\times10^{13}\,\percm$ (glyceraldehyde, GLD) and
$N\leq1.0\times10^{13}\,\percm$ (dihydroxyacetone, DHA), so that ERU exceeds the C3 sugars
by a factor $\geq$8--17 \citep{JimenezSerra2026}. Other chemical families in this
cloud lose roughly an order of magnitude per added carbon \citep{JimenezSerra2026}. The sugars do the opposite
between C2 and C4, and collapse at C3 (C$n$ denotes a species with $n$ carbon atoms). Any growth scheme that adds one carbon at a
time cannot skip C3 and overproduces it.

Second, sugar growth by aldol condensation (the formose reaction;
\citealt{Butlerov1861,Breslow1959}) carries an activation energy of the order of
100~kJ\,mol$^{-1}$: the apparent barriers measured for the consumption of
GA and GLD are 64--69~kJ\,mol$^{-1}$ at pH 10.4 and
90--109~kJ\,mol$^{-1}$ at pH 7.3 \citep{Delidovich2009}. We adopt
$E_a=87$~kJ\,mol$^{-1}$, the middle of that range, for every aldol step. At the $\sim$10\,K of quiescent
ice mantles the rate is smaller than its 273\,K value by more than four hundred orders
of magnitude. Cold ice cannot run the thermal (formose) route. Energetic processing
of cold ice does make sugars: ultraviolet or electron irradiation of methanol-bearing
ices, analysed in the residues after warm-up, yields glyceraldehyde and larger sugars
\citep{deMarcellus2015,Meinert2016,Zhang2024}; DHA and the tetroses, erythrulose among them, are reported by \citet{Zhang2024}. The chemistry is a radical chemistry of the kind
that this paper follows in the melt (Sect.~\ref{sec:c3disc}).

Two ways out are available, and they are physically distinct.
\citet{JimenezSerra2026} propose the first: radical recombination on the cold ice surface, which needs no thermal activation. Radicals derived from GA and ethylene glycol recombine on amorphous solid water to the C4 ketose. Both parent molecules are abundant in this cloud and both form by CO
hydrogenation on grains \citep{Fedoseev2015}. Their quantum chemical calculations and
kinetic Monte Carlo model of ice build-up, run at 20\,K, reproduce the observed ERU/GA ratio of
$\sim$1, but the same model overproduces the two C3 sugars by factors of 25--70
relative to their observed upper limits (their Sect.~2.3).

Here we explore the second: the ice is not always cold. In the diffuse gas of the
Central Molecular Zone the cosmic-ray ionisation rate $\zeta$, the number of ionisations
per hydrogen molecule per second, is enhanced by three to
four orders of magnitude over the canonical value of local dense clouds, $\zeta_0\simeq10^{-17}$~s$^{-1}$,
\citep{LePetit2016,Padovani2020}; whether the dense gas of G+0.693 shares this
enhancement, and whether the ionising particles are cosmic rays at all, are open
questions that we return to in Sect.~\ref{sec:origin}. A single cosmic-ray impact can
melt the ice around its track for a short time. Our earlier work \citep{Saito2024}
modelled the melting and refreezing that one ion impact causes in an ice grain, with a
binary-collision cascade and a phase field. The results of the present paper use neither code. As a check, rerun with the properties of this paper and a 10~K background, the phase-field code, with the corrections listed in Appendix~\ref{app:trackcaveats}, cools a 1~MeV~u$^{-1}$ iron track below 200~K within 0.6~ns, as the simpler track model does, and closes the melt within 7~ns. The energy
deposited by an ion is taken from stopping-power tables, the liquid window from a track
model (Sect.~\ref{sec:chain}), and the chemistry of each window is solved in zero
dimensions (Appendices~\ref{app:network} and~\ref{app:numerics}). The window is the
time that the track spends as a liquid above the mobility limit of the molecules,
because below that limit the chemistry stops. During such a window the ice is briefly a warm liquid.
Barrier-carrying condensations and diffusion-limited radical chemistry both switch on.
Cosmic-ray heating of grains is an old idea \citep{Leger1985,HasegawaHerbst1993}, and
the localised heating by an iron ion can even trigger a thermal explosion of a reactive
mantle \citep{Ivlev2015spot}. What
is new here is to run a full sugar network inside the melt and to follow the ice
through many such events.

Two facts sharpen the problem. The C$_3$H$_6$O$_2$ isomers hydroxyacetone and
lactaldehyde, the deoxy trioses, and the ether methoxyacetaldehyde are detected in this
cloud, at a combined $3.5\times10^{13}\,\percm$, or 0.38$\times$GA (0.35$\times$GA for the
two deoxy trioses alone; \citealt{SanzNovo2026}). Carbon is therefore not missing at C3;
sugars are. In addition, the grain-surface model reproduces C2$\approx$C4 but not
the C3 limits, as noted above. The suppression of the C3 sugars is an open problem for
both routes.

Our aims are threefold. First, to establish which molecules the melt window can and
cannot make, with every rate that controls the result held to an independently
published value where one exists (laboratory aldol kinetics, quantum-chemical saddle
points, pulse-radiolysis branching ratios, measured hydration equilibria, measured
radiolytic half-life doses). Second, to reconstruct the rate of melting events from the
cosmic-ray spectrum itself. Third, to test the route against the observed pattern in
every way the model allows, and, where it fails, to state what would have to be
different for it to succeed.

The answer is negative on two firm counts and conditionally on a third, and positive on one mechanism. The mechanism: radical chemistry in a melt can make the C4 as the ketose. Hydrated GA is attacked at its gem-diol carbon, and the C4 that two such radicals form is born with its carbonyl on
that carbon. In the fiducial model \ketcom\ of the C4 is the ketose; the share depends on how the C3 radical and the trapped radicals are booked (Sect.~\ref{sec:ketose}). The three obstacles are these. The C1 radicals of the melt add one carbon at a time and overproduce the C3 sugars. The dose that the same particles deposit outside the melts destroys the sugars between events. The C2$\approx$C4 equality needs several thousand whole-grain melting events, more than the cosmic rays of the dense gas deliver in a cloud lifetime at the ionisation rates inferred for such gas, unless the radicals trapped at the end of a window recombine instead of being reduced. We
quantify the case in which all three obstacles are removed (Sect.~\ref{sec:hypothetical}):
the pattern is then reproduced, with testable predictions, and we state what
observation or experiment would decide each condition.

Section~\ref{sec:model} summarises the model chain. Section~\ref{sec:results}
presents the reconstructed event rate, the need for melting, the accumulation over
many events, the ketose mechanism, the C3 excess of the complete network, the exposure
that the C2$\approx$C4 equality needs, and the radiolysis between events. It closes
with the hypothetical case in which the obstacles are removed.
Section~\ref{sec:discussion} states the conditions on the ionising particles,
compares the two routes, lists the predictions of the hypothetical case and the
caveats. Section~\ref{sec:conclusions} concludes. Details of the network and of the
numerical scheme are given in Appendices~\ref{app:network} and \ref{app:numerics}.

\section{Model}
\label{sec:model}

\subsection{From a cosmic-ray impact to a chemistry window}
\label{sec:chain}

\begin{figure*}
\centering
\includegraphics[width=0.85\textwidth]{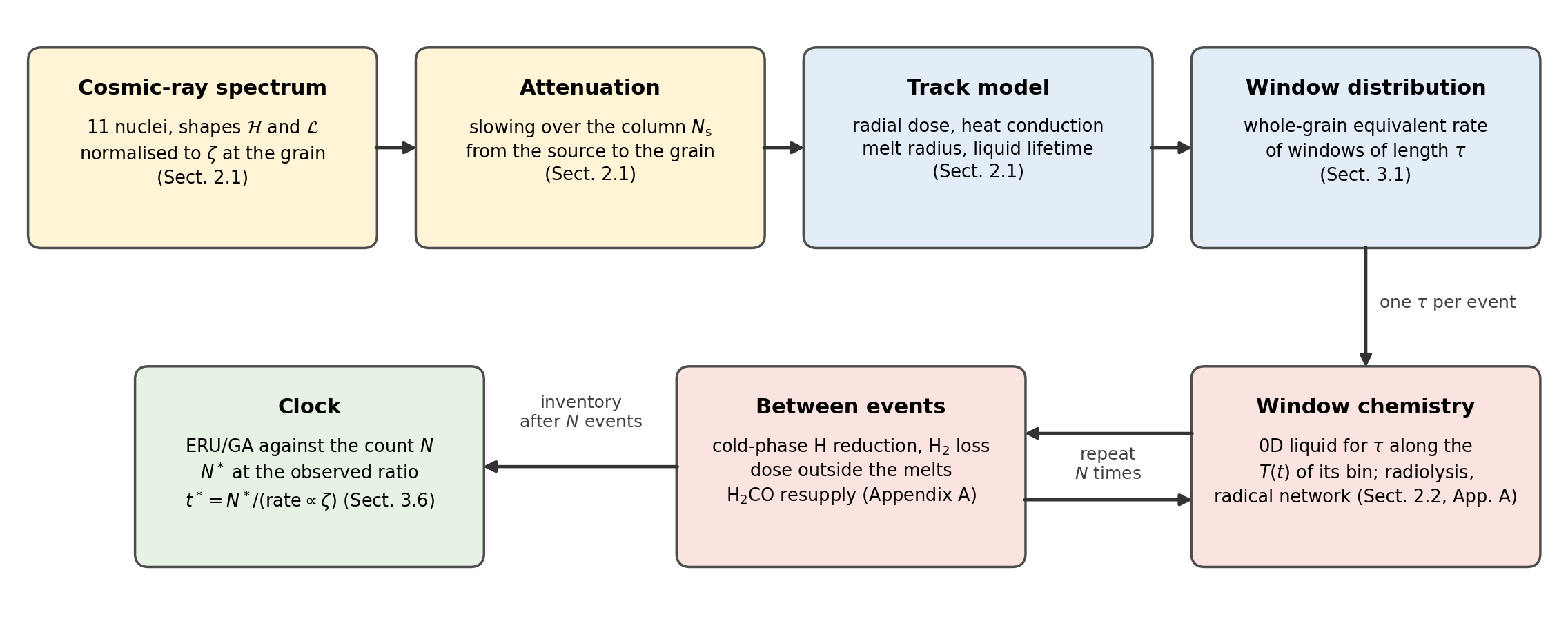}
\caption{The model chain. The cosmic-ray spectrum, attenuated over the column between
source and grain, is folded with the track model into a distribution of liquid windows.
Each melting event draws one window from it; the chemistry of the window is solved in
zero dimensions, and the inventory is then processed between events. The count $N^*$ at
which ERU/GA reaches the observed value, divided by the event rate, which scales with
$\zeta$, gives the ice age.}
\label{fig:chain}
\end{figure*}

The model treats one cosmic-ray impact on one grain in three steps: the energy
deposited in the grain, the liquid window that this energy creates along the ion
track, and the chemistry in that window. The three steps are computed separately and
feed forward only. Figure~\ref{fig:chain} shows how they connect to the repeated
melting that sets the clock (Sect.~\ref{sec:clock}).

\begin{figure*}
\centering
\includegraphics[width=\textwidth]{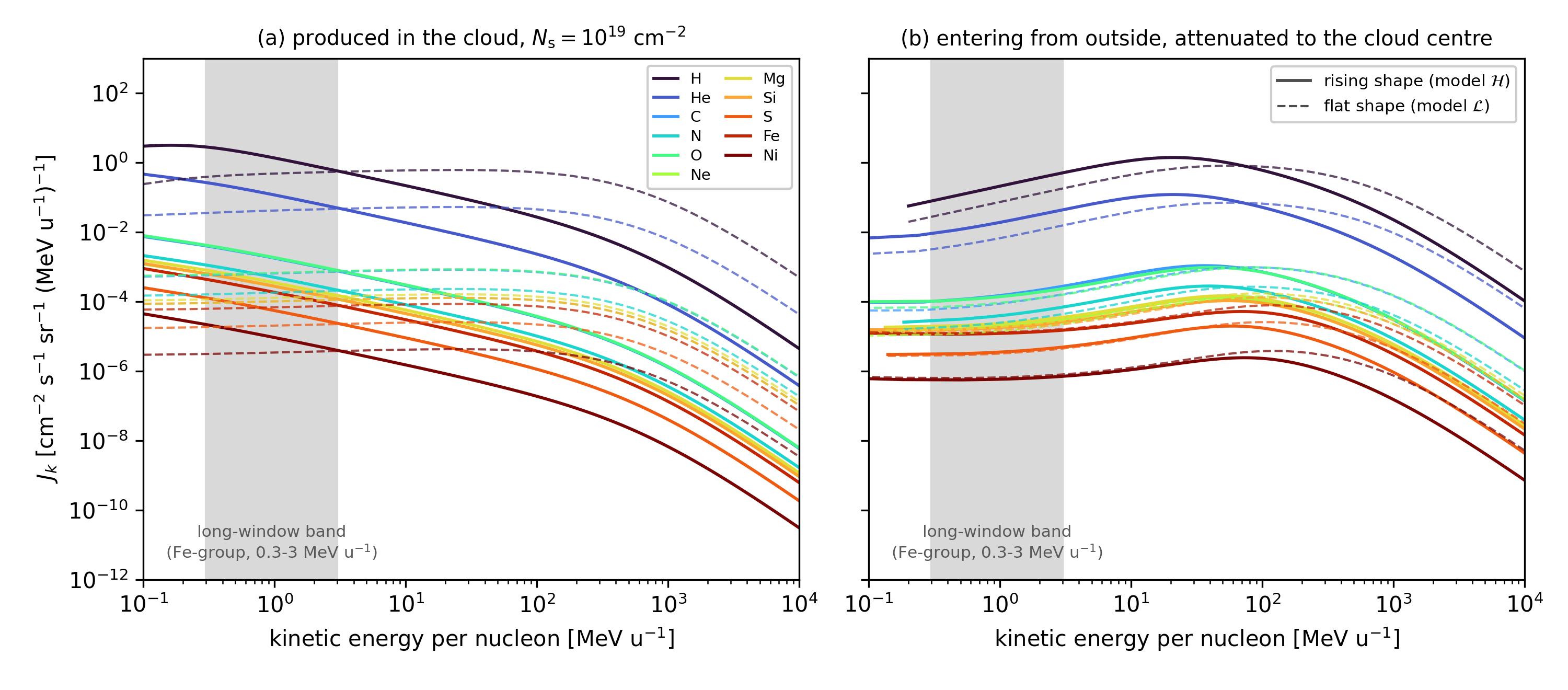}
\caption{The eleven per-nucleon spectra used in this work, each normalised to
$\zeta=10^{-14}$~s$^{-1}$ at the grain. Solid curves: the low-energy shape that rises
towards low energy (model $\mathcal{H}$). Dashed curves: the flat shape (model
$\mathcal{L}$), in the same colour for each nucleus. (a) Particles produced inside the
cloud, for a source-to-grain column $N_\mathrm{s}=10^{19}\,\percm$. (b) The same
spectra entering from outside and attenuated to the centre of a cloud of
$N(\mathrm{H_2})=1.35\times10^{23}\,\percm$. Only particles that enter above a threshold
cross the column (22~MeV for protons, 83~MeV~u$^{-1}$ for iron). Those that enter just
above it arrive with almost no energy left, so each curve extends down to the lower
limit of the range table. The intensity falls there, because a band of source
energies less than 1\% wide is spread over a decade of arrival energies. The shaded band marks the 0.3--3~MeV~u$^{-1}$
iron-group ions that make the long windows: for the rising shape their intensity in (b) is 3--18\% of that in (a) at the same arrival energy, and it is supplied by ions that entered near
84~MeV~u$^{-1}$. Inside the band in (a) the flat shape lies a factor of 1.1--7 below the rising one (1.1 at 3~MeV~u$^{-1}$, 7 at 0.3~MeV~u$^{-1}$), by the same factor for every nucleus, because the shape multiplies all eleven spectra alike. In (b) the two shapes differ by less than 10\%. The
abundances of He, C and O relative to protons are those of the local interstellar
spectra of \citet{Bisschoff2019} at 100~MeV~u$^{-1}$. The heavier nuclei are scaled to oxygen (Eq.~(\ref{eq:spec})).}
\label{fig:spectra}
\end{figure*}

\paragraph{Dose and event rate.} The energy deposited by an ion of mass number $A$ and
kinetic energy $E$ per nucleon on entering a grain of radius $a$ is
$\min[AE, S(E)\,4a/3]$. We take $a=0.1\,\mu$m (Appendix~\ref{app:network}), the single grain radius of gas--grain chemical models \citep{Hasegawa1992}. The first term, $AE$, is the total kinetic energy of the ion,
all of which is deposited if the ion stops inside the grain. The second is the energy
lost by an ion that crosses the grain: $S(E)$ is the electronic stopping power in
water, the energy lost per unit path length, and $4a/3$ is the mean chord of the grain
under isotropic irradiation. $S$ is the electronic stopping power of protons in liquid water from the PSTAR tables \citep{Berger2017} at the same velocity, scaled to the density of the ice and multiplied, for the nuclei heavier than hydrogen, by the square of the effective charge $Z_\mathrm{eff}=Z[1-\exp(-125\beta Z^{-2/3})]$ of \citet{Barkas1963}, with $Z$
the nuclear charge and $\beta$ the velocity in units of the speed of light. A heavy ion at a few MeV~u$^{-1}$
carries bound electrons, and $Z_\mathrm{eff}^{2}/Z^{2}$ is 0.23 for iron at
1~MeV~u$^{-1}$, 0.76 at 10~MeV~u$^{-1}$ and 0.58 for oxygen at 1~MeV~u$^{-1}$. For protons the tables already contain the charge exchange, and no factor is applied. The same effective charge is used for the attenuation in the gas below. We have not
compared this scaling with tabulated heavy-ion stopping powers. The long windows come
from iron-group ions at 0.3--3~MeV~u$^{-1}$, so an error of $S$ in that band would
shift the rate of the long windows in proportion. Evaluating $S$ at the entrance energy over the whole chord neglects the
slowing of the ion inside the grain. The neglected term is small: over the 133~nm mean
chord of the modelled grain, iron loses 0.9\% of its energy at 1~MeV~u$^{-1}$ and
0.05\% at 10~MeV~u$^{-1}$, and oxygen 0.7\% at 1~MeV~u$^{-1}$. No ion in the range of
the calculation, the eleven nuclei listed below at 0.1--1000~MeV~u$^{-1}$, stops inside
the grain; even iron at the lowest energy, 0.1~MeV~u$^{-1}$, has a range of several
micrometres in ice. The total dose of this bookkeeping and the ionisation rate rest on the same particles and the same stopping tables (Appendix~\ref{app:numerics}). The rate of events is obtained by folding the track model below
with a cosmic-ray spectrum. The spectrum is the sum over eleven nuclei (H, He, C, N, O,
Ne, Mg, Si, S, Fe, Ni). All eleven have the same shape in the kinetic energy per
nucleon $E$. This shape is the proton spectrum of the models $\mathcal{L}$ and
$\mathcal{H}$, introduced by \citet{Ivlev2015} and updated by \citet{Padovani2018}. At
the source, nucleus $k$ has the differential intensity (particles per unit area, time,
solid angle and energy per nucleon)
\begin{equation}
J_{0,k}(E)=f_k\,C\,\frac{E^{p}}{(E+E_\mathrm{c})^{q}},
\label{eq:spec}
\end{equation}
with the crossover energy $E_\mathrm{c}=650$~MeV~u$^{-1}$. The exponents are $p=0.1$ and
$q=2.8$ for model $\mathcal{L}$, and $p=-0.8$ and $q=1.9$ for model $\mathcal{H}$.
Below $E_\mathrm{c}$ the intensity follows $J_{0,k}\propto E^{p}$. It is nearly flat in
model $\mathcal{L}$ and rises towards low energy in model $\mathcal{H}$. Above
$E_\mathrm{c}$ both models fall as $E^{p-q}=E^{-2.7}$. The source spectrum is evaluated
from 0.1~MeV~u$^{-1}$ to 32~GeV~u$^{-1}$. The low-energy shape therefore sets the
spectrum over the whole range 0.1--650~MeV~u$^{-1}$. This range contains all the ions
that melt the grain: in the track model below, no nucleus melts ice above
100~MeV~u$^{-1}$. The constant $C$ is fixed by the normalisation to $\zeta$ described
below. The factor $f_k$ is the abundance of nucleus $k$ relative to protons at
$E=100$~MeV~u$^{-1}$, and it does not depend on energy. For He, C and O it is the
ratio of the local interstellar spectra of \citet{Bisschoff2019} at that energy. The
remaining nuclei are scaled to oxygen with the number ratios N 0.27, Ne 0.15, Mg 0.19,
Si 0.15, S 0.03, Fe 0.10 and Ni 0.005, within 35\% of the abundances measured by the Cosmic Ray Isotope Spectrometer on the
Advanced Composition Explorer at
160~MeV~u$^{-1}$ \citep{George2009}. The local interstellar spectra thus enter only
through these ratios, not through their shape. The spectrum is attenuated over
a column $N_\mathrm{s}$ of gas between the acceleration site and the grain in the
continuous-slowing-down approximation. A nucleus of charge $Z$ and mass number $A$
loses energy per nucleon $E$ along the column $N$ at the rate
\begin{equation}
\frac{\mathrm{d}E}{\mathrm{d}N}=-\frac{Z_\mathrm{eff}^{2}(E)}{A}\,L(E),
\label{eq:csda}
\end{equation}
with $L(E)$ the energy loss of a proton per H$_2$ molecule, taken from the PSTAR table for hydrogen \citep{Berger2017}, and the effective charge given above (unity for protons), evaluated along the slowing-down path. Between 0.5 and 100~MeV this loss function lies within 8\% of the power law $1.77\times10^{-10}E^{-0.82}$~eV~cm$^{2}$ of \citet{Padovani2018}; it is 17\% below it at 0.1~MeV and twice as large at 1~GeV, where the power law no longer applies. We integrate Eq.~(\ref{eq:csda}) into a range table
$\mathcal{R}(E)=\int_0^{E}A\,\mathrm{d}E'/[Z_\mathrm{eff}^{2}L]$ for each nucleus. A
particle that starts with energy $E_0$ arrives with the energy $E$ for which
$\mathcal{R}(E)=\mathcal{R}(E_0)-N_\mathrm{s}$, and does not arrive if
$\mathcal{R}(E_0)<N_\mathrm{s}$. The number of particles is conserved along the path,
$J_k(E)\,\mathrm{d}E=J_{0,k}(E_0)\,\mathrm{d}E_0$, where $J_k$ is the differential
intensity at the grain (particles per unit area, time, solid angle and energy per
nucleon). The
fall of the effective charge as a heavy ion slows lengthens its range, by a factor of 2.1 for iron at 1~MeV~u$^{-1}$ and 1.3 at 10~MeV~u$^{-1}$ relative to a charge held at
its entrance value. The particles travel in straight lines. We neglect nuclear
interactions, whose mean free path ($\gtrsim5\times10^{23}\,\percm$ of H$_2$ for iron) exceeds every
column used here, the straggling of the energy loss, and the lengthening of the path
by the magnetic field, which would raise the effective column. The spectrum is normalised so that the ionisation rate
at the grain is $\zeta$, the total ionisation rate per H$_2$ molecule including
secondary ionisation \citep{Padovani2018}. It is linear in the intensities,
\begin{equation}
\zeta=4\pi\sum_k\int_{I_\mathrm{H_2}}^{\infty}J_k(E)\,\big[1+\Phi_k(E)\big]\,\sigma^\mathrm{ion}_k(E)\,\mathrm{d}E ,
\label{eq:zeta}
\end{equation}
where the sum runs over the primary and secondary species $k$, $\sigma^\mathrm{ion}_k$ is
the ionisation cross section of H$_2$ by species $k$, $\Phi_k$ is the mean number of
further ionisations by the secondary electrons that it releases, and
$I_\mathrm{H_2}=15.44$~eV is the ionisation potential of H$_2$ \citep[their
Eq.~43]{Padovani2018}. For a fixed spectral shape $\zeta$ is therefore proportional to
the constant $C$ of Eq.~(\ref{eq:spec}), and fixing $\zeta$ fixes the absolute
intensity. We evaluate $\zeta$ for the nuclei of the model themselves, in the continuous-slowing-down picture of Eq.~(\ref{eq:csda}): the energy that the eleven propagated nuclei lose per H$_2$ molecule per unit time, divided by the mean energy per ion pair, $W_\mathrm{ion}=35$~eV, which includes the secondary ionisation,
\begin{equation}
\zeta=\frac{4\pi}{W_\mathrm{ion}}\sum_k\int J_k(E)\,Z_{\mathrm{eff},k}^{2}(E)\,L(E)\,\mathrm{d}E ,
\label{eq:zetaown}
\end{equation}
with $L$ the proton loss function and $Z_\mathrm{eff}$ the effective charge defined above. The constant $C$ is scaled so that Eq.~(\ref{eq:zetaown}) gives $\zeta$ at the column $N_\mathrm{s}$. The normalisation thus attributes all of $\zeta$ to the nuclei. Primary cosmic-ray electrons, which dominate the ionisation of the flat model at small columns \citep{Padovani2018}, would take part of $\zeta$ and lower the ion flux; the normalisation is in this sense the most favourable one for the melt route. For the spectra of \citet{Padovani2018} with their normalisation constants, Eq.~(\ref{eq:zetaown}) gives \zownH~s$^{-1}$ for model $\mathcal{H}$ and \zownL~s$^{-1}$ for model $\mathcal{L}$ at $N_\mathrm{s}=10^{19}\,\percm$, of which the protons supply \zownHp\ and \zownLp; their fit of the full calculation, which includes helium and the electrons, gives $2.8\times10^{-15}$ and $3.7\times10^{-16}$~s$^{-1}$. For $\zeta=10^{-14}$~s$^{-1}$ at $N_\mathrm{s}=10^{19}\,\percm$ the constant $C$ is therefore \zfacH\ times theirs for model $\mathcal{H}$ and \zfacL\ times for model $\mathcal{L}$. For the radial profiles of particles that enter from outside the cloud the normalisation is instead to a mass-weighted mean $\zeta$ over the cloud (Sect.~\ref{sec:rate}). Figure~\ref{fig:spectra} shows the
eleven spectra as they enter the calculation, for the two limiting columns used in
Sect.~\ref{sec:rate}. The event rate is therefore a function of three inputs: $\zeta$,
the spectral shape at 0.1--100~MeV~u$^{-1}$, and $N_\mathrm{s}$.

Section~\ref{sec:rate} gives the result for the two limiting cases,
particles that enter the cloud from outside and particles that are produced inside it.
Inside the ice-bearing part of the cloud only the second case delivers the long windows at a useful rate, and whether that rate suffices depends on the ionisation rate and on the chemistry.

\paragraph{Window.} The duration and temperature of the liquid phase along one track
are computed with a one-dimensional cylindrical model. The stopping power $S$ of the
ion is deposited radially as a core of uniform dose and a penumbra falling as
$r^{-2}$, each carrying half of $S$ \citep{ChatterjeeSchaefer1976}:
\begin{equation}
D(r)=
\begin{cases}
\dfrac{S}{2\pi r_\mathrm{c}^{2}}, & r\leq r_\mathrm{c},\\[2ex]
\dfrac{S}{4\pi\ln(r_\mathrm{p}/r_\mathrm{c})}\,\dfrac{1}{r^{2}}, & r_\mathrm{c}<r\leq r_\mathrm{p},
\end{cases}
\label{eq:dose}
\end{equation}
where $D$ is the deposited energy per unit volume, $r_\mathrm{c}=\max(11.6\beta\
\mathrm{nm},\,0.5\ \mathrm{nm})$ with $\beta$ the ion velocity in units of $c$, and
$r_\mathrm{p}=\max[40\,(W_\mathrm{max}/\mathrm{keV})^{1.7}\ \mathrm{nm},\,2r_\mathrm{c}]$
approximates the range in water of the most energetic secondary electron,
$W_\mathrm{max}=2m_\mathrm{e}c^{2}\beta^{2}\gamma^{2}$ with $m_\mathrm{e}$ the electron
mass, $c$ the speed of light and $\gamma$ the Lorentz factor of the ion; it lies within 30\% of the ESTAR range at 10~keV
\citep{Berger2017} and extrapolates that law below.

The profile is then evolved by radial heat conduction. We integrate the enthalpy per
unit volume $u$, which carries the latent heat without a separate interface condition,
\begin{equation}
\frac{\partial u}{\partial t}=\frac{1}{r}\frac{\partial}{\partial r}
\left[r\,\kappa(T)\,\frac{\partial T}{\partial r}\right],
\qquad u(r,0)=u_\mathrm{s}(T_{0})+D(r),
\label{eq:heat}
\end{equation}
with $\kappa(T)$ the thermal conductivity and $u_\mathrm{s}(T)$ the enthalpy of the solid defined below. The ice starts at $T_{0}=10$~K; the outer boundary at $r=2\,\mu$m is held at $T_{0}$, and the axis $r=0$ is a symmetry line. The temperature follows from $u$ through
\begin{equation}
T(u)=
\begin{cases}
\sqrt{2u/b}, & u<u_\mathrm{s}(T_\mathrm{m}),\\[1ex]
T_\mathrm{m}, & u_\mathrm{s}(T_\mathrm{m})\leq u\leq u_\mathrm{s}(T_\mathrm{m})+L_\mathrm{f},\\[1ex]
T_\mathrm{m}+\dfrac{u-u_\mathrm{s}(T_\mathrm{m})-L_\mathrm{f}}{(\rho c_p)_\mathrm{liq}}, & u>u_\mathrm{s}(T_\mathrm{m})+L_\mathrm{f},
\end{cases}
\label{eq:TofH}
\end{equation}
where the ice is treated as compact amorphous ice of density 917~kg~m$^{-3}$
(the reasons are given in Sect.~\ref{sec:caveats}). Its
volumetric heat capacity is $\rho c_p(T)=bT$ with $b=6.87\times10^{3}$~J\,m$^{-3}$\,K$^{-2}$,
that is $c_p=7.49\,T$~J~kg$^{-1}$K$^{-1}$, so that $u_\mathrm{s}(T)=\tfrac12 bT^{2}$.
This is the fit of \citet{Klinger1980} to the data of \citet{GiauqueStout1936} without
its constant term. It lies 2--16\% below the ice Ih values of \citet{FeistelWagner2006}
between 50 and 273~K. The melting point is $T_\mathrm{m}=273$~K and the latent heat is
that of ice Ih, $L_\mathrm{f}=3.06\times10^{8}$~J~m$^{-3}$ \citep{FeistelWagner2006}.
The liquid has $(\rho c_p)_\mathrm{liq}=4.2\times10^{6}$~J~m$^{-3}$K$^{-1}$
\citep{WagnerPruss2002} and a conductivity of $0.569$~W\,m$^{-1}$\,K$^{-1}$, within 3\% of
the value at the melting point \citep{Huber2012}; hotter water conducts up to 20\%
better. The conductivity of the solid is $\kappa=0.5+1.5\times10^{-3}T$~W\,m$^{-1}$\,K$^{-1}$.
For vapour-deposited amorphous ice \citet{Kouchi1992} inferred values as low as $10^{-5}$~W\,m$^{-1}$\,K$^{-1}$ from the crystallisation of thin films, and attributed them to microcracks rather than to porosity. \citet{AnderssonSuga1994} measured values several orders of magnitude larger on amorphous ice made under pressure, and \citet{AnderssonSuga2002} list no measurement for compact vapour-deposited ice. We use
the compact values, the conservative choice for the window length
(Sect.~\ref{sec:rate}). For
the amorphous ices made under pressure, the values of \citet{AnderssonSuga1994}, which
\citet{AnderssonSuga2002} assign to the high-density form, rise from 0.56 to
0.64~W\,m$^{-1}$\,K$^{-1}$ between 70 and 125~K. The adopted law has the same slope and lies
7\% above them. The low-density form conducts about twice as well, 1.4 to
1.1~W\,m$^{-1}$\,K$^{-1}$ between 86 and 129~K, and its conductivity falls with temperature
\citep{AnderssonSuga2002}. The effect of these choices on the window is given in
Sect.~\ref{sec:caveats}. Equation~(\ref{eq:TofH}) is the crystallisation limit, in which a refreezing cell
releases $L_\mathrm{f}$ on the plateau at $T_\mathrm{m}$. In the vitrification limit a
cell that has melted keeps $L_\mathrm{f}$ and cools along the liquid branch below
$T_\mathrm{m}$. The two limits bracket the window
lengths. We adopt the crystallisation limit, which gives the shorter windows; the vitrification limit is run as a systematic (Sect.~\ref{sec:rate}). Equations~(\ref{eq:heat}) and (\ref{eq:TofH}) are integrated
with an explicit finite-volume scheme on a radial grid. The cell width
$\Delta r$ from the axis to $r=40$~nm is 0.0625~nm for hydrogen and helium, whose melt radii are 0.6--2.1~nm, and 0.125~nm for the heavier nuclei (708 and 388 cells in all). These widths are numerical, chosen for the convergence of the continuum solution (Appendix~\ref{app:trackcaveats}), and lie below the size of a molecule. Beyond 40~nm each cell edge lies at 1.06
times the radius of the previous edge, so the cells widen from 2.4~nm at 40~nm to
120~nm at the outer boundary. The time step is the same for all
cells and is recomputed at every step as
$\Delta t=0.2\min_j(\Delta r_j^{2}/\alpha_j)$, where $\alpha_j=\kappa/(\rho c_p)$ is
the thermal diffusivity of cell $j$ at its current temperature and phase; the factor
0.2 lies below the stability limit of 0.5 of the explicit scheme. The minimum is set by
the narrow cells of the cold ice around the track, where $\alpha$ is largest
($7.5\times10^{-6}$~m$^{2}$\,s$^{-1}$ at 10~K), so that $\Delta t\approx0.1$ and 0.4~fs for the two widths; the
coarse outer cells never limit the step. The conductivity at a cell face is the
harmonic mean of the two adjacent cells, and the integration stops once the track
holds no liquid.
The liquid lifetime is
\begin{equation}
\tau_\mathrm{liq}=\frac{1}{V_\mathrm{melt}}\int\!\!\int_{\mathrm{liquid},\,T>T_\mathrm{mob}}
\mathrm{d}V\,\mathrm{d}t,
\label{eq:tauliq}
\end{equation}
the volume-weighted time spent above the mobility limit $T_\mathrm{mob}=200$~K,
normalised to the peak melt volume $V_\mathrm{melt}$. The value of $T_\mathrm{mob}$ is
adopted; the mobility-weighted lifetime defined next does not depend on it. The mobility-weighted lifetime
replaces the indicator function of Eq.~(\ref{eq:tauliq}) by a weight $w(T)$,
\begin{equation}
\tau_\mathrm{mob}=\frac{1}{V_\mathrm{melt}}\int\!\!\int_\mathrm{liquid}
w(T)\,\mathrm{d}V\,\mathrm{d}t,
\qquad
w(T)=\left[\frac{T/T_\mathrm{s}-1}{T_\mathrm{m}/T_\mathrm{s}-1}\right]^{\gamma_\mathrm{D}},
\label{eq:taumob}
\end{equation}
where $w(T)$ is the self-diffusion coefficient of liquid water at the temperature $T$
divided by its value at the melting point $T_\mathrm{m}$. It measures how fast the
molecules of the melt move. The distance a molecule diffuses in a given time and the
rate of a diffusion-limited reaction both scale with the diffusion coefficient. One
second of liquid at $T$ therefore allows as many molecular encounters as $w(T)$
seconds of liquid at $T_\mathrm{m}$. The mobility-weighted lifetime is thus the time
that the melt would have to spend at $T_\mathrm{m}$ to give the same transport. The
weight has the form of \citet{SpeedyAngell1976}. In this form $T_\mathrm{s}$ is the
temperature at which the diffusion coefficient of supercooled water extrapolates to
zero, and $\gamma_\mathrm{D}$ is the exponent of the power law. We use
$T_\mathrm{s}=228$~K and $\gamma_\mathrm{D}=1.7$. The weight is zero below
$T_\mathrm{s}$ and is evaluated at $\min(T,550\,\mathrm{K})$. It is 0.3 at 250~K, 1 at
$T_\mathrm{m}$ and 20 at 491~K. A superheated track core therefore counts for more
than its duration, and $\tau_\mathrm{mob}$ exceeds $\tau_\mathrm{liq}$. The measured self-diffusion of water follows the same form
with $T_\mathrm{s}=214$~K and $\gamma_\mathrm{D}=2.08$ \citep{Dehaoui2015}, which gives
weights within 25\% of ours between 250 and 498~K, the upper end of their data, and 31\% above ours when extrapolated to 550~K. The cap of 550~K lies below the
measured limit of superheat of water at atmospheric pressure, 575~K
\citep{Avedisian1985}; caps of 450 and 650~K are run as systematics. The liquid does
not boil, although the window temperature exceeds the atmospheric boiling point. It is
confined at the density of ice by the surrounding cold matrix: on the
saturation line its density at 491~K is 0.84~g\,cm$^{-3}$ and its bulk modulus about 1~GPa \citep{WagnerPruss2002}. Holding it at 0.917~g\,cm$^{-3}$ therefore needs a pressure of order 100~MPa, above the critical pressure of 22~MPa, where no liquid--vapour transition exists. We assume that the cold matrix bears this stress on the nanometre scale of a track. Even at zero external pressure a vapour bubble at 491~K would have to exceed
the critical radius $2\sigma_\mathrm{s}/p_\mathrm{sat}\approx30$~nm (surface tension $\sigma_\mathrm{s}=34$~mN\,m$^{-1}$,
$p_\mathrm{sat}=2.2$~MPa; \citealt{Vargaftik1983,WagnerPruss2002}), larger than any
melt radius; homogeneous nucleation becomes possible only near the limit of superheat,
which is why the cap is placed below it. Where a track reaches the grain surface the
liquid evaporates from its free surface; this loss is the thermal desorption of
Sect.~\ref{sec:ratchet}. The track table
covers 0.1--1000~MeV~u$^{-1}$ with 65 energies per nucleus, 16 per decade, and is interpolated between them; ions outside this range are assigned no melt. The melt
radius and liquid lifetime are 1.4~nm and 8~ps for 100~keV protons, 4.4~nm and
0.08~ns for 1~MeV~u$^{-1}$ oxygen, and 9.5~nm and 0.35~ns for 1~MeV~u$^{-1}$ iron. Protons melt ice up to 2~MeV, helium up to 4~MeV~u$^{-1}$, oxygen up to 20~MeV~u$^{-1}$ and iron up to 90~MeV~u$^{-1}$. A
core radius four times smaller changes these values by less than 1\% (its effect on the
ice age is given in Sect.~\ref{sec:rate}). Over the cosmic-ray spectrum the windows form
a broad distribution, from $4\times10^{-13}$~s for protons to 0.40~ns for nickel at
0.6~MeV~u$^{-1}$, the longest window of the table (Sect.~\ref{sec:rate}). The
chemistry follows this distribution directly: every melting event of the cycle model
draws its window length $\tau$, the liquid lifetime $\tau_\mathrm{liq}$ of Eq.~(\ref{eq:tauliq}), from the reconstructed distribution of the spectrum in question (Sect.~\ref{sec:rate}), by stratified sampling (Appendix~\ref{app:cycle}).
That distribution is weighted by melted volume, so each event of the cycle model stands
for one whole-grain equivalent of melting (Sect.~\ref{sec:rate}).

The temperature inside a track is not uniform and changes during the window, and the
chemistry of each event follows the temperature history of its window. For every
nucleus and energy of the track table we record, along the liquid time of
Eq.~(\ref{eq:tauliq}), the mean temperature of the liquid, weighted by volume and capped
at 550~K, and the mean of the mobility weight $w(T)$ of Eq.~(\ref{eq:taumob}). The liquid
time is divided into 14 equal steps. The histories of all the events whose windows fall in the same length bin of the distribution are averaged, weighted by their whole-grain rate and their window length. Every window of that bin is then solved with the temperature of the bin in each of its 14 steps. The averaging uses the events of the fiducial spectrum, the rising shape for particles produced inside the cloud (Sect.~\ref{sec:rate}). For the windows of 0.1~ns and longer, which make the
C4, the histories of the other spectra are the same; for the shortest windows they
differ by at most 20~K. In the long windows the liquid cools from about 470 to 275~K;
its mean temperature over the liquid time is 350~K, and its mean mobility is a third of
that at 491~K.

The disproportionation fraction $\delta$ of the radical self-reactions depends on
temperature. We evaluate it for each window at the mean temperature of its history,
with the temperature law of \citet{Kanjana2015} and the disproportionation enthalpy
chosen so that $\delta(491\,\mathrm{K})$ takes the value quoted. We quote $\delta$ at the reference temperature of 491~K throughout, the temperature at which we evaluate the literature decomposition (Sect.~\ref{sec:hypothetical}); in the fiducial model $\delta(491\,\mathrm{K})=0.47$ gives
$\delta\approx0.24$ in the long windows. Holding $\delta$ at its 491~K value in every window is run as a systematic (Appendix~\ref{app:restricted}). Solving every window at the single reference temperature $T_\mathrm{w}=491$~K is run as a reference, not a systematic. That temperature is the constant temperature whose Arrhenius factor for an 87~kJ\,mol$^{-1}$ barrier equals the average of that factor over the liquid time of the distribution. It weights the hottest liquid and overstates the rates of the radical chemistry, which carry smaller barriers. The reference temperature is used only for these comparisons and for tabulating rate constants (Table~\ref{tab:reactions}).

\paragraph{Chemistry.} The chemistry of each window is computed in zero dimensions.
The melt is treated as well mixed: each species has a single number density, uniform
over a liquid volume equal to that of the grain, for the duration $\tau$ of the event.
Spatial gradients and transport are not resolved. Each species $i$ with number density
$c_i$ obeys
\begin{equation}
\frac{\mathrm{d}c_i}{\mathrm{d}t} = R_i\big(\{c\},T(t)\big),
\label{eq:rd}
\end{equation}
where $\{c\}$ denotes the set of the number densities $c_j$ of all species, and $R_i$
is the net production rate of species $i$ from the reactions of
Appendix~\ref{app:network}. Rate constants with an activation energy are evaluated at
the temperature $T(t)$ of the step. The rate of the radical self-reactions follows its
measured temperature law (12.7~kJ\,mol$^{-1}$; \citealt{Kanjana2015}) at the mean
temperature of the window, as $\delta$ does. Rate constants without a measured
temperature dependence keep their tabulated values. Multiplying the diffusion-limited
rate constants in addition by $w(T)/w(491\,\mathrm{K})$, which lowers them in the cooling liquid, is run as a systematic (Appendix~\ref{app:restricted}). The radiolytic products of the dose are injected at the
start of the window (Appendix~\ref{app:window}).

\subsection{Network, dose and repeated melting}
\label{sec:network}

The fiducial network comprises 68 species (60 carbon- and nitrogen-bearing species and the 8 species of water radiolysis) and 199 reactions. Its core is water radiolysis
with the rate constants of \citet{Buxton1988}, that of the neutralisation H$_3$O$^+$ + OH$^-$ from \citet{Eigen1964}, and the picosecond yields of
Table~\ref{tab:prov}. It supplies OH, H and the hydrated electron $e^-_\mathrm{aq}$ to the melt. These primary products are injected once, at the start of each window, in proportion to the dose of the window: each species $i$ gains $G_i D_\mathrm{w}/100$ per unit volume (Eq.~(\ref{eq:gsource})). Here $G_i$ is its yield per 100~eV and $D_\mathrm{w}$ the energy per unit volume that the ion deposits inside its melted cylinder, the radial dose of Eq.~(\ref{eq:dose}) integrated to the melt radius. We express it in units of $D_\mathrm{melt}=3.50\times10^{27}$~eV\,m$^{-3}$, the enthalpy that heats the ice from 10~K and melts it. An ion deposits more than $D_\mathrm{melt}$ in the volume that it melts, because the track core is superheated and the heat that spreads from it melts no more: $D_\mathrm{w}$ is \doseinmean$D_\mathrm{melt}$ on average over the fiducial spectrum and \doseinlong$D_\mathrm{melt}$ in the windows of 0.1~ns and longer, each window taking the value of its length bin (Appendix~\ref{app:numerics}). For OH this gives \ohlongM~M in the long windows, \ohlongpct\ of the water, and \ohevent\ radicals per whole-grain event on average. The rest of the energy of the ions is deposited outside the melts and is booked between events (Sect.~\ref{sec:radiolysis}), so that the energy of the model is that of the stopping-power bookkeeping of Sect.~\ref{sec:chain}. We use the picosecond rather than the escape yields because the solutes of the melt scavenge OH within about 45~ps, before the radicals of a spur recombine (Table~\ref{tab:prov}). The fall of the yields with linear energy transfer is treated as a systematic in Sect.~\ref{sec:caveats}. The dose and the yield set are run as systematics: a window dose of $D_\mathrm{melt}$, the melting enthalpy alone, raises the count $\nstar$ of melting events at which ERU/GA reaches its observed value to \Ndmelt, and the escape yields, with half the OH, raise it to \Nescape\ (Sect.~\ref{sec:caveats}). The organic content of the melt is set by the ice inventory: H$_2$CO and CH$_3$OH at 5.0\% and 5.7\% of water, close to the median of their combined abundance in interstellar ice (CH$_3$OH 6--9\%, H$_2$CO 2--7\%), CO at 25\%, and a fixed base at 5.4\% that stands for NH$_3$ \citep{Boogert2015}. The ice of the first event holds water and these four components only; every other species, including the sugars, the acids, CO$_2$, H$_2$ and H$_2$O$_2$, starts at zero (Appendix~\ref{app:cycle}). The H$_2$CO is replenished by CO-hydrogenation resupply between events \citep{WatanabeKouchi2002,Fuchs2009}, capped by the gas-phase carbon share of one grain. In the melt, OH abstracts hydrogen from the hydrated formaldehyde, giving the dihydroxymethyl radical $\bullet$CH(OH)$_2$, and from methanol, giving the hydroxymethyl radical $\bullet$CH$_2$OH, at the room-temperature rates of \citet{Buxton1988}. Two $\bullet$CH(OH)$_2$ react with each other at the rate measured by \citet{StockhausenHenglein1971}, who report the disproportionation to formic acid and formaldehyde. We give that channel the fraction $\delta$ and let the remainder combine to the dihydrate of glyoxal, as for the other hydroxyalkyl pairs. The combination is an assumption; Sect.~\ref{sec:sensitivity} gives the case without it. Sugar chemistry proceeds by radicals. OH abstracts hydrogen from an aldehyde: a
fraction $f_\mathrm{CHO}$ at the carbonyl, giving an acyl radical, and the remainder at
the $\alpha$-carbon, giving a hydroxyalkyl radical \citep{vonSonntag1980}. A ketone has no aldehydic hydrogen,
so OH attack on a ketose gives only the hydroxyalkyl radical. The radicals combine pairwise at the rate measured for $\bullet$CH$_2$OH in
high-temperature water \citep{Kanjana2015}; at the window temperature this rate lies
below the diffusion limit. A fraction $\delta$ of the encounters disproportionates
instead. The self-reaction of $\bullet$CH$_2$OH
itself uses the same rate and the same $\delta$, its disproportionation giving
methanol and formaldehyde (Appendix~\ref{app:window}). Growth is the combination of two hydroxyalkyl radicals (C2$+$C2 $\to$ C4, C2$+$C1 $\to$ C3, \ldots). Every pair of the C1 radicals $\bullet$CH(OH)$_2$ and $\bullet$CH$_2$OH with the C2--C5 sugar radicals is included at the same rate; we call this the complete network. A restricted network, in which the C1 radical $\bullet$CH(OH)$_2$ reacts only with itself, OH and H and methanol is inert, is run as a hypothesis in Sect.~\ref{sec:hypothetical}. In it a single carbon is added to a sugar radical only by the $\bullet$CH$_2$OH that H addition to formaldehyde and the decarbonylation of acyl radicals still supply, and it isolates what the pattern would need from the chemistry. The main destruction of the C2--C6 sugars in the fiducial network is the decarbonylation of the acyl radicals, with barriers from our own saddle-point
calculations (Table~\ref{tab:prov}; method in Appendix~\ref{app:qcmd}), in competition with their repair by H. Which of the
two radicals an OH attack produces depends on whether it meets the free aldehyde or its
gem-diol. The carbonyl availability $h_X$, the fraction of sugar $X$ present as the free
aldehyde or ketone rather than as its hydrate or ring form, is frozen between events
and therefore filters both growth and destruction. The gem-diol radical of GA is
the origin of the ketose C4 (Sect.~\ref{sec:ketose}). Between events, hydrogen atoms, accreted from the gas and produced by the dose outside the melts, reduce the trapped radicals and the dehydro-sugars, the combination products that carry two hydrogen atoms fewer than the parent sugar, back to closed-shell sugars. The dose outside the melts also destroys the closed-shell sugars in the solid; the fiducial model sets this destruction to zero, the most favourable assumption for the route, and Sect.~\ref{sec:radiolysis} applies the measured yields. Closed-shell aldol chemistry is retained in the network with the barrier above. Its absolute rate is anchored to the hydrothermal formose experiment of \citet{Kopetzki2011}, in which the reaction reaches its maximum yield within about a minute at temperatures up to 473~K. With the
87~kJ\,mol$^{-1}$ barrier this corresponds to $k=6.4\times10^{-9}$ M$^{-1}$\,s$^{-1}$ at
273~K for GA$+$GA; the value is therefore an extrapolation of that experiment, not a
measured rate constant. The aldol channels contribute nothing
(Sect.~\ref{sec:meltgated}). The acid--base speciation of formic acid against a fixed base inventory, which stands for the NH$_3$ of the ice, completes the set. The network also carries NH$_3$ and its radiolysis, but the fiducial model sets the explicit NH$_3$ to zero. The provenance of every rate is
given in Table~\ref{tab:prov}. The governing equations of the window chemistry, the
between-event operators, and the complete species and reaction lists are given in
Appendix~\ref{app:network} (Tables~\ref{tab:species} and \ref{tab:reactions}).

\begin{table*}
\caption{Provenance of the principal parameters of the fiducial model. Class:
L $=$ independently published value; D $=$ this work, quantum chemistry; C $=$ computed
from the model chain; R $=$ required by the observations (Sect.~\ref{sec:hypothetical}) and
the subject of the proposed test; T $=$ typical or assumed value; L/R $=$ published and also required. An axis is a parameter varied alone as a systematic (Tables~\ref{tab:nstar} and \ref{tab:nstarres}). These classes are
unrelated to the reaction classes of Table~\ref{tab:reactions}. Rate constants are
in M$^{-1}$\,s$^{-1}$ unless stated.}
\label{tab:prov}
\centering
\scriptsize
\renewcommand{\arraystretch}{0.9}\setlength{\tabcolsep}{5pt}
\begin{tabular}{p{4.4cm}p{3.7cm}cp{8.25cm}}
\hline\hline
Parameter & Value & Class & Source / note \\
\hline
\multicolumn{4}{l}{\emph{Melt window and dose}} \\
Thermal properties of ice and water & Sect.~\ref{sec:chain} & L & compact amorphous ice; sources in Sect.~\ref{sec:chain}, tested in Sect.~\ref{sec:caveats} \\
Window length; temperature history & drawn per event; per-bin $T(t)$ & C & window length (liquid lifetime) from the reconstructed distribution and 14-step temperature history of its bin, both from the cylindrical conduction model (Sect.~\ref{sec:chain}); superheat caps 450 and 650~K as axes; a fixed $T_\mathrm{w}=491$~K as a reference \\
Melting events per Myr & \Rwge\ at $\zeta=10^{-14}$~s$^{-1}$ & C & whole-grain equivalent rate for an isotropic intensity on a sphere, $4\pi\cdot\pi a^{2}J$; cosmic-ray spectrum, attenuation and track model (Sect.~\ref{sec:rate}); local particles, rising shape, all of $\zeta$ carried by the nuclei; $\propto\zeta$ \\
Dose of a window; dose outside the melts per event & \doseinmean$D_\mathrm{melt}$ (\doseinlong\ in the long windows); \doseout$D_\mathrm{melt}$ & C & $D_\mathrm{melt}=3.50\times10^{27}$~eV\,m$^{-3}$, the melting enthalpy at the density of ice; energy inside and outside the melted cylinders from the radial dose of the track model and the stopping-power bookkeeping for a grain of 0.1~$\mu$m (Appendix~\ref{app:numerics}); both evaluated for each spectrum (outside: \doseoutrange); a window dose of $D_\mathrm{melt}$ as axis \\
Grain radius & 0.1~$\mu$m & T & single grain radius of gas--grain chemical models \citep{Hasegawa1992} \\
\multicolumn{4}{l}{\emph{Radiation chemistry}} \\
Primary $G$ values, molecules per 100~eV ($e^-_\mathrm{aq}$, H, OH, H$_2$, H$_2$O$_2$) & 4.8, 1.0, 5.5, 0.15, 0.2 & L$+$T & yields at $\sim$1~ps: 4.8 for $e^-_\mathrm{aq}$ and for OH in \citet{Buxton1988}; published sets span $G$(H)$=0.6$--1.0 and $G$(OH)$=4.8$--5.9, and the H, H$_2$ and H$_2$O$_2$ values are adopted. Escape set (2.7, 0.6, 2.7, 0.45, 0.7; \citealt{Buxton1988} give 2.6 for $e^-_\mathrm{aq}$) as systematic \\
Primary rather than escape yields & scavenging capacity of the melt $2.2\times10^{10}$~s$^{-1}$ & C & escape yields hold for scavenging capacities $\lesssim10^{7}$~s$^{-1}$ \citep{Buxton1988,LaVerne2000}; escape set as a systematic \\
Outside-melt molecular yield $G$(H$_2$O$_2$) & 0.2 per 100~eV & L & within the 0.01--0.6 per 100~eV compiled by \citet{Teolis2017} for penetrating projectiles, where the quoted yields are upper limits; 0.4 and off as axes \\
Cold-phase H: in-situ reductants & $G_\mathrm{H}+G_{e}$ of the dose outside the melts & L & from the $G$ values above \\
Cold-phase H: fraction reaching trapped species, $\varepsilon$ & 0.01 & T & ten times the number ratio ($\sim$$10^{-3}$) of trapped species to H$_2$CO$+$CO$+$GA; $\varepsilon=0$ and $\varepsilon=0.001$ with $\eta=0$ as axes, $\eta$ being the fraction of the in-situ radiolytic H that reduces a trapped species (Appendix~\ref{app:coldh}) \\
\multicolumn{4}{l}{\emph{Radical kinetics in the window}} \\
Radical self-reaction rate (per pair) & $4.9\times10^{9}$ at 491~K & L & \citet{Kanjana2015}, sum of their combination and disproportionation rates (their Arrhenius fit, $2k=8.5\times10^{9}$, gives $k=4.2\times10^{9}$), scaled to the mean temperature of each window with that fit (12.7~kJ\,mol$^{-1}$); cross reactions $\times2$ \\
Disproportionation fraction $\delta$(491~K) & 0.47 (lit.\ 0.73$^{+0.16}_{-0.26}$) & L & decomposition of \citet{Kanjana2015} evaluated at 491~K, at the lower end of its range ($-1\sigma$ of their activation enthalpy); evaluated per window at its mean temperature with their temperature law (0.24 in the long windows); 0.15--0.73 as axes; held fixed as an axis \\
OH $+$ GA, OH $+$ GLD (Arrhenius) & $4.3\times10^{11}e^{-1740/T}$, $3.3\times10^{10}e^{-960/T}$ & L & \citet{Schaefer2023}, aqueous, temperature dependent, extrapolated to 491~K \\
Aldehyde-H fraction of OH attack $f_\mathrm{CHO}$ & 0.5 & L$+$T & experiment and theory give $\sim$0.8 in the gas phase \citep{Butkovskaya2006,Galano2005}; applied to the free aldehyde only (Appendix~\ref{app:window}); 0.7 and 0.8 as axes \\
H $+$ closed-shell carbonyl $k_\mathrm{hadd}$ & $5\times10^{6}$ & L & \citet{Buxton1988}, H $+$ HCHO \\
OH $+$ HCHO, H $+$ HCHO, OH $+$ HCOOH & $1.0\times10^{9}$, $5\times10^{6}$, $1.3\times10^{8}$ & L & \citet{Buxton1988}; tabulated data for total formaldehyde, mostly the gem-diol in water, as in the model \\
OH $+$ CH$_3$OH, H $+$ CH$_3$OH & $9.7\times10^{8}$, $2.6\times10^{6}$ & L & \citet{Buxton1988}; both give $\bullet$CH$_2$OH; methanol inert as the restricted-network hypothesis (Sect.~\ref{sec:hypothetical}) \\
Cross-combination of $\bullet$CH(OH)$_2$ with the sugar radicals & $2k_\mathrm{rad}$, branching $\delta$ & T & no measurement for these pairs; taken as for every other hydroxyalkyl pair (Appendix~\ref{app:window}); switched off as the restricted-network hypothesis \\
Radiolytic destruction of the sugars between events & 0 (fiducial) & T & measured half-life doses of organics in cold H$_2$O ice (Sect.~\ref{sec:radiolysis}) as axes, irreversible and with the measured steady-state fraction \\
OH $+$ glycolic acid & $6\times10^{8}$ (0 as axis) & L$+$T & rate from \citet{Buxton1988}; products lumped as CO$_2$, H$_2$CO and H (a model simplification) \\
OH $+$ ethylene glycol, H $+$ ethylene glycol & $1.8\times10^{9}$, $1.4\times10^{7}$ & L & \citet{Buxton1988}; both give the glycol radical $\bullet$CH(OH)CH$_2$OH \\
OH $+$ CO, H $+$ CO, H $+$ CO$_2$ & $1.3\times10^{9}$, $1.7\times10^{7}$, $1\times10^{6}$ & L & \citet{Buxton1988}; OH$+$CO the middle of 0.79--$2.0\times10^{9}$, H$+$CO$_2$ an upper limit \\
$\bullet$CH(OH)$_2$ self-reaction & $2k=7.2\times10^{8}$ & L & \citet{StockhausenHenglein1971}; combination product carried as glyoxal (Appendix~\ref{app:network}) \\
Acyl decarbonylation $E_\mathrm{a}$ (C2/C3/C4/C5) & 44.6/34.7/41.5/36.2~kJ\,mol$^{-1}$ & D & density-functional saddles (Appendix~\ref{app:qcmd}); pre-exponential factor $10^{13}$~s$^{-1}$; C6 and the C4 and C6 dehydro-sugar acyls reuse the C4 value, the C5 dehydro-sugar acyl the C5 value \\
$\beta$-scission, direct C--C cleavage & 63--101, $>$250~kJ\,mol$^{-1}$ & D & dead channels \\
OH attack on the dehydro-sugars & 0.02 of the parent rate & T & no measurement; 0 and 0.7 as axes (Sect.~\ref{sec:sensitivity}) \\
Aldol barrier; GA$+$GA aldol rate at 273~K & 87~kJ\,mol$^{-1}$; $6.4\times10^{-9}$ M$^{-1}$\,s$^{-1}$ & L$+$C & mid-range of \citet{Delidovich2009} (64--109~kJ\,mol$^{-1}$); rate our extrapolation from the $\lesssim$1~min time scale at $\leq$473~K of \citet{Kopetzki2011}, which reports no rate constants; all aldol steps scaled alike \\
\multicolumn{4}{l}{\emph{Carbonyl availability between events}} \\
Free-aldehyde fraction of GA, $h_\mathrm{GA}$ & 0.10 & L/R & hydration constant $K_\mathrm{hyd}=[\mathrm{gem\textrm{-}diol}]/[\mathrm{aldehyde}]=9$--25, 90--96\% hydrated \citep{Sorensen1972}; assumed to hold in the ice (Sect.~\ref{sec:caveats}); 0.2 and 1 as axes \\
Free-aldehyde fraction of GLD, $h_\mathrm{GLD}$ & 1.0 (aqueous eq.\ 0.05) & T & $K_\mathrm{hyd}=19$, i.e. 95\% hydrated, in D$_2$O at 25~$^{\circ}$C \citep{Nagorski2001}; dehydration rate 0.03~s$^{-1}$ at 25~$^{\circ}$C \citep{Rendina1984}; 0.05--1 as axes (Sect.~\ref{sec:hypothetical}); test in Sect.~\ref{sec:caveats} \\
Cold-phase reduction site of the trapped C4 ketose radical & ketose & R & required by the ketose identity of the C4 (Sect.~\ref{sec:ketose}); half to the aldotetrose as axis \\
Cold-phase reduction site of the trapped C3 radical & ketose & T & assumed, by analogy with the C4 (Sect.~\ref{sec:ketose}); half and all to the aldose as axes \\
Free-carbonyl fraction of DHA, $h_\mathrm{DHA}$ & 1.0 & T & a ketone hydrates weakly; 0.05 as axis (Sect.~\ref{sec:ketose}) \\
Free-aldehyde fraction of the aldotetrose, $h_\mathrm{ald}$ & 0.45 & T & no measurement; a mid-range assumption; 0.02 as axis \\
Open-chain fraction of C5, C6 & 0.02, 0.01 & T & order-of-magnitude values; governs only the OH attack on C5 and C6, their sole sink; hexoses reach \hexcom$\times$GA at $\nstar$ (Sect.~\ref{sec:ratchet}) \\
\multicolumn{4}{l}{\emph{Inventories}} \\
H$_2$CO, CH$_3$OH, CO, base in ice (relative to H$_2$O) & 5.0\%, 5.7\%, 25\%, 5.4\% & L$+$T & \citet{Boogert2015} medians: H$_2$CO 2--7\%, CH$_3$OH 6--9\%, CO 21--25\%, NH$_3$ 6\%; the base stands for NH$_3$ and enters only the acid--base speciation; all 10.7\% as H$_2$CO in the restricted network; CO $\times1/2$, base $\times0$ and $\times2$ as axes \\
Gas density, CO abundance & $10^{4}$~cm$^{-3}$, $10^{-4}$ & T & typical dark-cloud values \\
\hline
\end{tabular}
\end{table*}

Repeated melting is simulated by carrying the surviving inventory of one melt into the
next. Each event starts from the carried inventory, and the radiolytic products of that
event's dose are added to it at once at the start of the window
(Eq.~(\ref{eq:gsource})). The primary radicals OH, H and $e^-_\mathrm{aq}$ are not carried. Those left at the end of a window, which is most of them in the windows of a few picoseconds, are taken to recombine as the melt freezes, and no free ions are carried either; each event injects the primary species afresh. The organic radicals that
remain at the end of a window are reduced by H atoms between events
(Appendix~\ref{app:cycle}). Each event also receives the dose outside the melts. Within a window the water that the radiolysis decomposes and that the reactions consume is debited from the solvent; between events the water is reset to the density of ice (Appendix~\ref{app:window}).

\section{Results}
\label{sec:results}

\subsection{The rate of melting events reconstructed from the cosmic-ray spectrum}
\label{sec:rate}

\begin{figure*}
\centering
\includegraphics[width=0.8\textwidth]{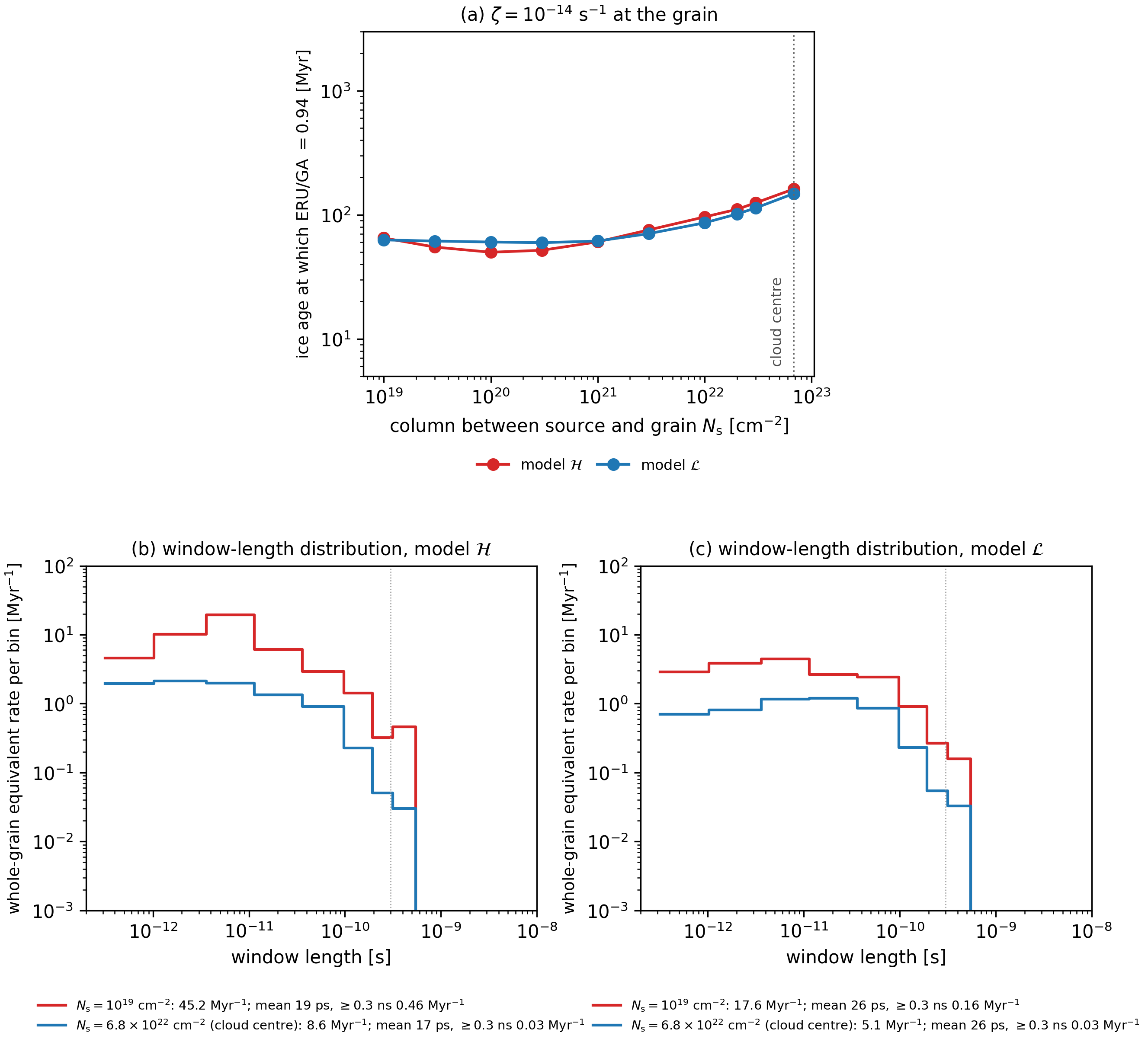}
\caption{Ice age against the column between the source of the particles and the grain. (a) Age at which ERU/GA reaches the observed value at $\zeta=10^{-14}$~s$^{-1}$ at the grain, against $N_\mathrm{s}$, the column along one straight path (red: rising shape, model $\mathcal{H}$; blue: flat shape, model $\mathcal{L}$). Small columns describe particles accelerated near the grain; the dotted line, $N_\mathrm{s}=6.8\times10^{22}\,\percm$, is the column from the surface to the centre (Fig.~\ref{fig:radial} treats all directions). Points: complete network, means of four sequences of windows (eight for the fiducial spectrum). (b) Window-length distribution for model $\mathcal{H}$ as whole-grain equivalent rate per bin at $N_\mathrm{s}=10^{19}\,\percm$ (red) and $6.8\times10^{22}\,\percm$ (blue); the first bin holds every window shorter than 1~ps. (c) The same for model $\mathcal{L}$. The legends give the total rate, the mean window length and the rate of windows of 0.3~ns or longer.}
\label{fig:origin}
\end{figure*}

The melted volume is dominated by protons and helium, which make many short windows.
The chemistry of this paper needs long windows. Those of 0.1~ns and longer make \cfourlong\ of the C4, and those of 0.3~ns and longer \cfourverylong\ of it (see ``Window length and yield'' below). As their diagnostic we quote the rate
of windows of 0.3~ns or longer, which come only from iron-group ions, 87\% of them at 0.3--3~MeV~u$^{-1}$ and the rest slower. The classical picture of cosmic-ray grain heating also singles out the iron group, there at 20--70~MeV~u$^{-1}$, the energies that heat a whole grain \citep{Leger1985,HasegawaHerbst1993}. Whether such ions reach the grains depends on where they come
from. We therefore reconstruct the event rate from the spectrum rather than scale it
from the cosmic-ray ionisation rate $\zeta$. Figure~\ref{fig:origin} summarises the result.
It follows cosmic rays that travel in a straight line from their source to the grain
through gas of uniform density, and takes as the horizontal axis the gas column
$N_\mathrm{s}$ that they cross on the way, which is proportional to the distance
($10^{21}\,\percm$ is 0.03~pc at an H$_2$ number density $n(\mathrm{H_2})=10^{4}$~cm$^{-3}$). For each column
and spectral shape, with $\zeta$ held at $10^{-14}$~s$^{-1}$ at the grain,
Fig.~\ref{fig:origin}a gives the ice age at which ERU/GA reaches its observed value of
0.94. Figures~\ref{fig:origin}b and c give the distribution of window lengths at the
smallest and the largest column, for the rising and the flat shape.

\begin{figure*}
\centering
\includegraphics[width=0.8\textwidth]{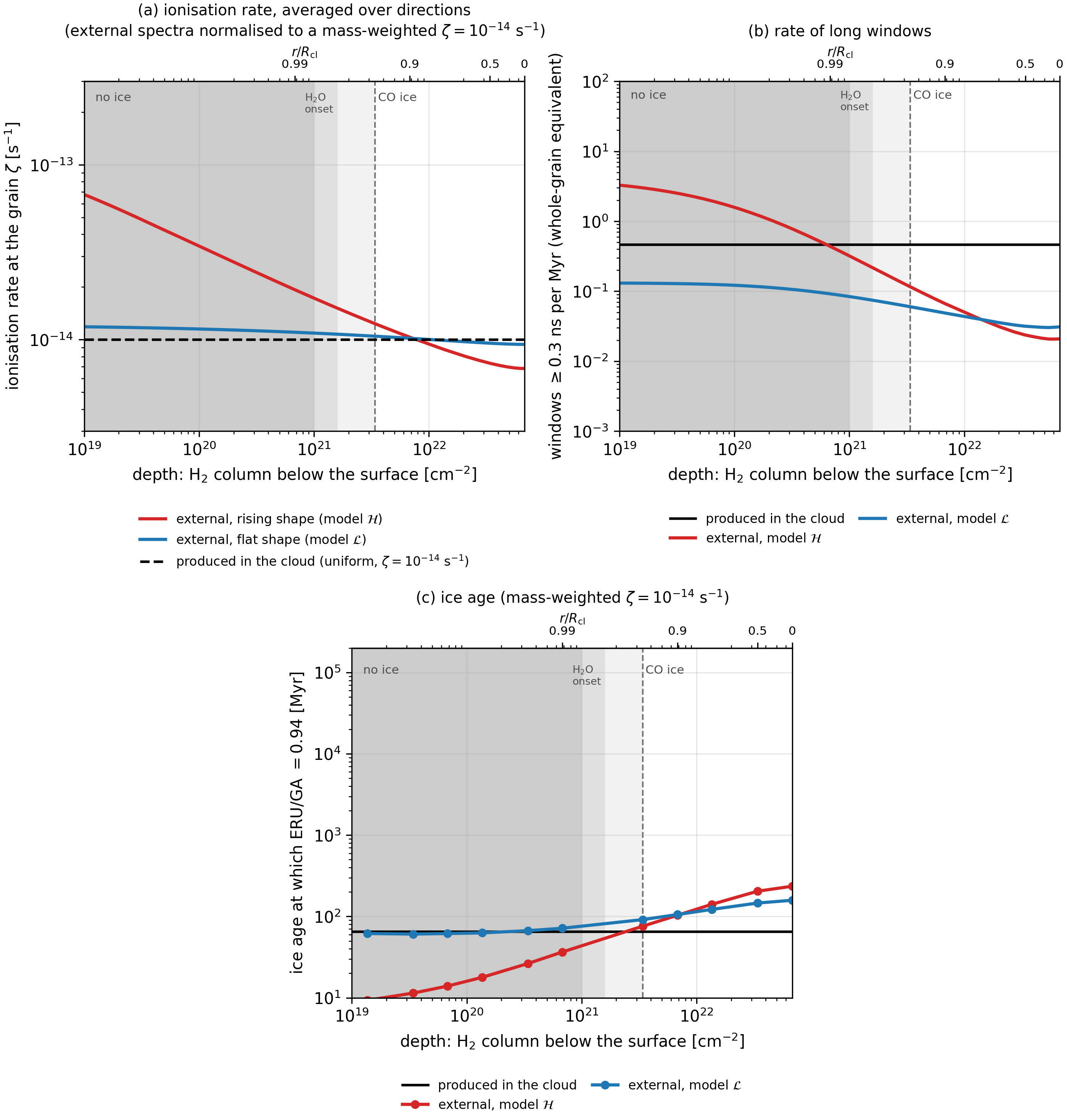}
\caption{Cosmic rays entering a uniform spherical cloud (column $N_\mathrm{c}=6.8\times10^{22}\,\percm$ from centre to surface), against the depth of the grain, $N_\mathrm{c}(1-r/R_\mathrm{cl})$ (upper axis: $r/R_\mathrm{cl}$). External spectra (red: model $\mathcal{H}$; blue: model $\mathcal{L}$) are attenuated along every direction, averaged over directions and normalised to a mass-weighted mean $\zeta=10^{-14}$~s$^{-1}$; black: particles produced inside the cloud ($N_\mathrm{s}=10^{19}\,\percm$; dashed in a). Shading marks the outer layers without the ice that the model assumes: no ice (darkest, below 1.0~mag), onset of water ice (1.0--1.6~mag) and water ice without CO ice (lightest, up to the dashed line at 3.4~mag). These are half of the thresholds measured along lines of sight through a cloud, 2.1--3.2~mag for water ice \citep{Whittet2001,Boogert2013,Boogert2015} and 6.7~mag for CO ice \citep[see also \citealt{Chiar1995}]{Whittet2007}, converted with $10^{21}\,\percm$ per magnitude; in the ultraviolet field of the Central Molecular Zone they are lower limits. (a) Ionisation rate at the grain. (b) Whole-grain equivalent rate of the windows of 0.3~ns or longer. (c) Ice age at which ERU/GA reaches the observed value (complete network; means of four sequences). The central age exceeds that of Fig.~\ref{fig:origin}a for the same column because of the mass-weighted normalisation.}
\label{fig:radial}
\end{figure*}

\begin{figure*}
\centering
\includegraphics[width=\textwidth]{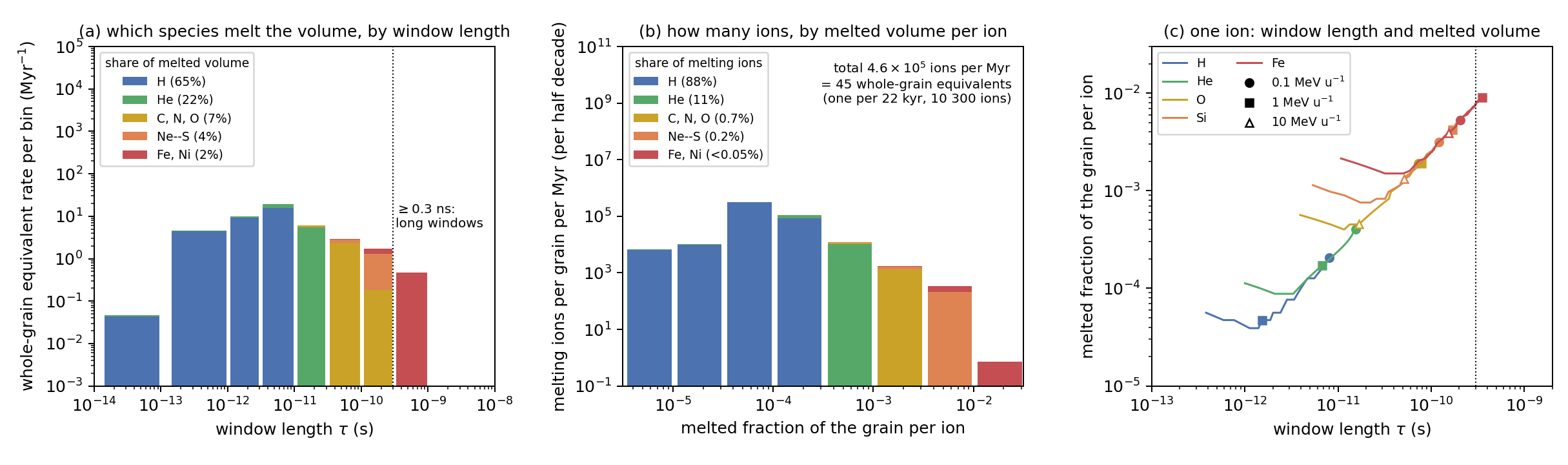}
\caption{Which cosmic-ray species make the windows and the melted volume, for the fiducial spectrum (model $\mathcal{H}$, $N_\mathrm{s}=10^{19}\,\percm$, $\zeta=10^{-14}$~s$^{-1}$). (a) The whole-grain equivalent rate per window-length bin (the red distribution of Fig.~\ref{fig:origin}b), stacked by species group; the legend gives the share of each group in the melted volume. (b) The number of ions per grain per Myr that melt part of the grain, against the fraction of the grain that each melts, in half-decade bins; the legend gives the share of each group in the ions. (c) Single ions of the track table: the melted fraction against the window length, with the energy per nucleon as parameter (filled circle 0.1, square 1, open triangle 10~MeV~u$^{-1}$). The dotted line marks 0.3~ns. Window length and melted volume rise together, and only the iron group reaches the long windows.}
\label{fig:species}
\end{figure*}

Most events melt only a thin cylinder around the track, so we count them by the \emph{whole-grain equivalent rate}. Each event is weighted by the fraction of the grain that it melts. The rate is then the number of times per Myr that a volume equal to the grain is melted, whatever the window length. The grain is the model grain of radius
$a=0.1\,\mu$m (Sect.~\ref{sec:chain}); an event counts as the melted volume, the melted
cross-section of the track times the mean chord $4a/3$ and at most the grain volume,
divided by the grain volume $4\pi a^{3}/3$, so that melting a volume equal to that of a
grain of radius 0.1~$\mu$m counts as one event. For the fiducial spectrum at $\zeta=10^{-14}$~s$^{-1}$ the rate is \Rwge~Myr$^{-1}$, so one whole-grain event is 22~kyr of exposure (Table~\ref{tab:rates} collects the rates and ages of every spectrum of this section). In that time about 10\,300 ions melt part of the grain, 88\% of them protons; the mean melted volume per ion is 410~nm$^{3}$, $1\times10^{-4}$ of the grain. Ions that each melt $10^{-5}$ to $10^{-4}$ of the grain supply 41\% of the melted volume, ions that melt $10^{-4}$ to $10^{-3}$ supply 48\%, and ions that melt more than 0.1\% supply 11\%, through about 2\,100 ions per Myr. The $\nstar$ events of Sect.~\ref{sec:clock} therefore stand for \nionsstar\ melting ions. Figure~\ref{fig:species} resolves these numbers by species. Protons and helium supply 87\% of the melted volume and 99\% of the melting ions, in windows shorter than 30~ps. The windows of 0.3~ns or longer come only from iron-group ions, each of which melts about 0.9\% of the grain. Every event of the chemistry draws its window from the distribution of the spectrum in question (Sect.~\ref{sec:chain}). The count $\nstar$ of whole-grain events at which ERU/GA reaches its observed value is therefore a property of the chemistry and of the window distribution together. It is \Ncom\ for the fiducial spectrum (Sect.~\ref{sec:clock}): the shape that rises towards low energy (model $\mathcal{H}$ of Eq.~(\ref{eq:spec})) for particles produced inside the cloud at a column $N_\mathrm{s}=10^{19}\,\percm$ from the grain, i.e. with almost no attenuation. For each spectrum below we run the complete network with that spectrum's window distribution, four sequences of windows each, and quote the ice age $t^{*}=\nstar$ divided by the whole-grain rate, at $\zeta=10^{-14}$~s$^{-1}$. The C3 excess of Sect.~\ref{sec:c3excess} is present at every one of them. Two shapes are used for the differential intensity $J$: the
shape that rises towards low energy (model $\mathcal{H}$, $J\propto E^{-0.8}$ below the
crossover energy of 650~MeV~u$^{-1}$) and the flat shape (model $\mathcal{L}$, $J\propto
E^{0.1}$), both from Eq.~(\ref{eq:spec}). Each is normalised so that the ionisation rate at the grain is
$\zeta=10^{-14}$~s$^{-1}$, and each is attenuated over the column $N_\mathrm{s}$ between
the source of the particles and the grain. Figure~\ref{fig:spectra} shows these spectra
for the two columns treated next, and Fig.~\ref{fig:origin} what the track model of
Sect.~\ref{sec:chain} and the chemistry make of them.

\begin{table*}
\caption{Melting rates and ice ages for the spectra of Sect.~\ref{sec:rate} (complete network). Rate: whole-grain equivalent melting rate at $\zeta=10^{-14}$~s$^{-1}$; long: the same for windows of 0.3~ns or longer; $\bar\tau$: mean window length weighted by melted volume; $\nstar$ and $t^{*}$: count and ice age at which ERU/GA reaches 0.94.}
\label{tab:rates}
\centering
\small
\begin{tabular}{llrrrrr}
\hline\hline
Particles, shape & $N_\mathrm{s}$ (cm$^{-2}$) or depth & Rate & Long & $\bar\tau$ & $\nstar$ & $t^{*}$ \\
 & & (Myr$^{-1}$) & (Myr$^{-1}$) & (ps) & & (Myr) \\
\hline
Inside the cloud, rising ($\mathcal{H}$) & $10^{19}$ & 45.2 & 0.464 & 19 & 2\,940 & 65 \\
 & $10^{20}$ & 50.1 & 0.583 & 20 & 2\,500 & 50 \\
 & $3\times10^{20}$ & 47.3 & 0.488 & 19 & 2\,450 & 52 \\
 & $10^{21}$ & 37.0 & 0.319 & 19 & 2\,240 & 61 \\
 & $10^{22}$ & 19.2 & 0.0871 & 17 & 1\,840 & 96 \\
\quad porous mantle, $\kappa\times0.1$ & $10^{19}$ & 79.7 & 8.77 & 123 & 3\,370 & 42 \\
\quad refractory nuclei $\times$3 & $10^{19}$ & 46.9 & 1.31 & 34 & 1\,330 & 28 \\
Inside the cloud, flat ($\mathcal{L}$) & $10^{19}$ & 17.6 & 0.157 & 26 & 1\,100 & 63 \\
 & $10^{21}$ & 13.0 & 0.138 & 30 & 800 & 61 \\
 & $3\times10^{21}$ & 11.0 & 0.0934 & 30 & 770 & 70 \\
From outside, rising ($\mathcal{H}$)\tablefootmark{a} & centre & 6.0 & 0.0208 & 16 & 1\,400 & 240 \\
 & $r/R_\mathrm{cl}=0.9$ & 15.8 & 0.0665 & 17 & 1\,630 & 100 \\
 & $r/R_\mathrm{cl}=0.99$ & 59.4 & 0.438 & 18 & 2\,170 & 37 \\
 & surface ($r/R_\mathrm{cl}=0.9998$) & 282.8 & 3.1 & 20 & 2\,630 & 9 \\
From outside, flat ($\mathcal{L}$)\tablefootmark{a} & centre & 4.8 & 0.0309 & 26 & 760 & 160 \\
 & surface & 14.7 & 0.13 & 27 & 910 & 62 \\
From outside, normalised at the grain\tablefootmark{b} & centre, $\mathcal{H}$ & 8.6 & 0.0303 & 17 & 1\,390 & 160 \\
 & centre, $\mathcal{L}$ & 5.1 & 0.0327 & 26 & 750 & 150 \\
\hline

\end{tabular}
\tablefoot{\tablefoottext{a}{Attenuated along every direction and averaged, normalised to a mass-weighted mean $\zeta=10^{-14}$~s$^{-1}$ over the cloud (Fig.~\ref{fig:radial}).}
\tablefoottext{b}{The centre-to-surface column along one path, normalised to $\zeta=10^{-14}$~s$^{-1}$ at the grain (right end of Fig.~\ref{fig:origin}a).}}
\end{table*}

\paragraph{Particles from outside the cloud.} If the cosmic rays are accelerated elsewhere, for example by supernova remnants in the Central Molecular Zone, they cross the gas of G+0.693 before they reach a grain, and $N_\mathrm{s}$ depends on the depth of the grain. The observed $N(\mathrm{H_2})=1.35\times10^{23}\,\percm$ \citep[from the C$^{18}$O column of][]{Martin2008} runs through the whole cloud; for a uniform sphere the column from the centre to the surface is $6.8\times10^{22}\,\percm$ in every direction. Iron ions that enter at 1--10~MeV~u$^{-1}$ have ranges of $1.9\times10^{20}$--$2.3\times10^{21}\,\percm$ and do not reach the centre. Those that arrive there at 1--10~MeV~u$^{-1}$ entered within a narrow band near 84~MeV~u$^{-1}$, which the slowing spreads over a decade, so the arriving intensity is a tenth to a fifth of the source intensity. The particles that do arrive give a low rate of melting events and few long windows (Table~\ref{tab:rates}; Fig.~\ref{fig:origin}b, c), and the observed ratio is reached only after \tecom~Myr for the rising shape and \tecl~Myr for the flat shape. These rates require a fine grid in source energy, because the arriving particles come from a narrow band of source energies. Figure~\ref{fig:radial} follows the depth dependence for a uniform sphere of radius $R_\mathrm{cl}$ irradiated isotropically. For a grain at radius $r$ the column to the surface in the direction at angle $\theta$ to the outward radius is $N_\mathrm{s}=N_\mathrm{c}\,[-x\mu+(x^{2}\mu^{2}+1-x^{2})^{1/2}]$, with $x=r/R_\mathrm{cl}$, $\mu=\cos\theta$ and $N_\mathrm{c}=6.8\times10^{22}\,\percm$. The spectrum is attenuated along each direction with Eq.~(\ref{eq:csda}), averaged over 81 directions and normalised to a mass-weighted mean $\zeta=10^{-14}$~s$^{-1}$. For the rising shape $\zeta$ falls from $8.6\times10^{-14}$~s$^{-1}$ at the very surface ($6.2\times10^{-14}$~s$^{-1}$ at $r/R_\mathrm{cl}=0.9998$, the outermost row of Table~\ref{tab:rates}) to $7\times10^{-15}$~s$^{-1}$ at the centre, and the rate of windows of 0.3~ns or longer falls much faster (Fig.~\ref{fig:radial}b), because the iron-group ions stop in the outer layer whereas the protons that carry the ionisation go through. The ice age therefore falls towards the surface: \trcentre~Myr at the centre, \trninety~Myr at $r/R_\mathrm{cl}=0.9$, \trniner~Myr at 0.99, where the extinction from the surface is 0.7~mag and no ice mantle exists \citep{Whittet2001,Boogert2015}, and \trsurf~Myr at the surface (Table~\ref{tab:rates}). The observed ERU/GA is reached within 7~Myr at no depth, and every age below 40~Myr lies in the layers that hold no ice. The flat shape is slower in the outer layers and reaches \trlcentre~Myr at the centre.

\paragraph{Particles produced inside the cloud.} If the ionising particles are accelerated inside G+0.693, in the shocks and turbulence that also sputter its ice \citep{RequenaTorres2006,Zeng2020}, $N_\mathrm{s}$ is the column between the local acceleration sites and the grain. For $N_\mathrm{s}=10^{19}\,\percm$ and the rising shape the whole-grain equivalent rate is \Rwge~Myr$^{-1}$, the mean window 19~ps, and the windows of 0.3~ns or longer carry 0.46~Myr$^{-1}$ (Table~\ref{tab:rates}): about 53 iron-group ions per grain per Myr, each melting about 0.9\% of the grain. Of these, 46 are among the 55 iron-group ions at 0.3--3~MeV~u$^{-1}$ that strike the grain per Myr, and the rest are slower. These are the longest of the windows in which the sugar chemistry runs. The age has a minimum of \tNab~Myr between $N_\mathrm{s}=10^{20}$ and $3\times10^{20}\,\percm$, the range of iron at 1--2~MeV~u$^{-1}$ (Fig.~\ref{fig:origin}a). Two effects combine there. The low-energy protons that carry most of the ionisation of the rising shape stop, so holding $\zeta$ at the grain fixed raises the whole spectrum, by a factor of 2.0 at $10^{20}\,\percm$; and the iron-group ions that melt most are not yet lost, because ions that enter at 3--10~MeV~u$^{-1}$ slow into the melting band. At larger columns the melting band is fed only by the few ions accelerated to higher energies, and the age rises steadily to the value of the external particles at the cloud centre (Table~\ref{tab:rates}). The flat shape holds fewer slow nuclei for each fast one, but the normalisation assigns all of $\zeta$ to the nuclei and so raises the flat spectrum by the larger factor (Sect.~\ref{sec:chain}). Near its source it gives 18~Myr$^{-1}$, with 0.16~Myr$^{-1}$ in the windows of 0.3~ns or longer, and the count is \Nflat. The age is \tlcom~Myr, equal to that of the rising shape within the scatter of the sequences. At $N_\mathrm{s}\geq10^{21}\,\percm$ the age is \tflatfar~Myr, \flatfarfac\ times that of the rising shape. For a given $\zeta$ carried by nuclei the spectral shape therefore matters little. It matters through the share of $\zeta$ that the nuclei carry. For the flat model the full calculation of \citet{Padovani2018}, in which primary electrons dominate, gives a $\zeta$ nine times that of our integral over its nuclei (Sect.~\ref{sec:chain}). If electrons supplied that difference, the age at a given total $\zeta$ would be about nine times longer. For the rising model our integral exceeds their value by a factor of 1.7, so the two calculations agree only to that level. Depleting the refractory nuclei (Mg, Si, Fe, Ni) by a factor of ten lengthens the age by a factor of \depfac. Enriching them by three, as the metallicity of the Galactic centre or an origin of the low-energy ions in sputtered dust might, shortens it by a factor of \depthreefac\ (Table~\ref{tab:rates}). A track core four times narrower changes the age by \corefac, and the vitrification limit of Sect.~\ref{sec:chain} by \vitfac.

\paragraph{Window length and yield.} The yield of a window grows steeply with its
length. In runs of 60 identical windows at 491~K with a dose of $D_\mathrm{melt}$ each, GLD scales as $\tau^{1.7}$ between 0.3 and 0.70~ns, as $\tau^{3}$ between 0.1 and 0.3~ns and as $\tau^{5}$ below 30~ps, and flattens above 2~ns ($\tau^{0.4}$). Glycolic acid and ERU are steeper ($\tau^{2}$, $\tau^{3.4}$ and $\tau^{8}$ in the same three ranges), and the hexoses steeper still. GA is linear between 0.1 and 0.3~ns, flattens above and scales as $\tau^{2}$--$\tau^{2.6}$ below 0.1~ns. In the fiducial run, up to $\nstar$, the windows of 0.1~ns or longer, \evlong\ of the events, make \cfourlong\ of the C4, whereas the windows shorter than 30~ps, \evshort\ of the events, make \cfourshort\ of it and deliver \ohgashort\ of the OH attacks on GA. The
short windows therefore process the inventory without building it: they carry the
radiolytic dose that eventually consumes GA, and they favour GA over the
larger sugars. The temperature history matters as much as the length. With every window held at the fixed $T_\mathrm{w}=491$~K the count is \Ntw\ events instead of \Ncom\ along the histories, because the liquid spends most of its time well below 491~K (Sect.~\ref{sec:chain}). Crediting the superheated core at that fixed temperature with its full diffusivity, the mobility-weighted lifetime of Eq.~(\ref{eq:taumob}), would cut that count to \Ntwmob. The same weight applied along the histories also slows the cooling liquid; run as a systematic in the restricted network it raises the count from \Nres\ to \Nresmob\ (Appendix~\ref{app:restricted}). What these rates imply for G+0.693 is examined
in Sect.~\ref{sec:origin}.

\paragraph{Sensitivity to the thermal conductivity of the ice.} The window length is
set by conduction of the deposited energy into the surrounding ice, so it scales
inversely with the thermal conductivity $\kappa$ of that ice (as $\kappa^{-0.8}$ in
the track model). The values above use the conductivity measured for compact
amorphous ice (Sect.~\ref{sec:chain}). Lowering $\kappa$ to 0.3 of that law lengthens
the window of 1~MeV~u$^{-1}$ iron from 0.35 to 0.86~ns and shortens the ice age from \tcom\ to \tkthree~Myr; lowering it to 0.1 gives 2.1~ns and \tkcom~Myr. A porous, vapour-deposited mantle with a conductivity ten times below that of compact amorphous ice would therefore shorten the age by a factor of \kfac\ at the same $\zeta$. Whether such a mantle can exist under cosmic-ray irradiation is discussed in
Sect.~\ref{sec:origin}.

\subsection{Within the window the radical route dominates, and the thermal route contributes nothing}
\label{sec:meltgated}

Two routes can make sugars in ice. The thermal route is the classical formose reaction,
in which closed-shell sugars condense with one another over an activation barrier, so
that temperature alone sets its rate. The radical route, used everywhere else in this
paper, instead lets radiolysis make radicals and lets the radicals combine. Irradiated cold ices also yield sugars without any melt (Sects.~\ref{sec:radiolysis} and \ref{sec:c3disc}); the claim of this section concerns the chemistry inside a window. The rest of
this paper assumes that the radical route runs, and that it runs only while the ice is
liquid. This subsection tests both assumptions. We ask what the same grain would make
if it were never melted, and how much the thermal route contributes inside a window at
491~K. A large answer to the first would make the melt window unnecessary and would
free the chemistry from the event rate of Sect.~\ref{sec:rate}. A large answer to the
second would mean that the rest of this paper follows the wrong chemistry. The first
question needs a description of the solid as well as of the liquid.

Melting acts on the chemistry through two gates. The first is mobility. For this
comparison the state of the ice is described by a solid fraction $\phi$, 1 in the solid
and 0 in the liquid, uniform over the grain. It enters the rates through one factor,
\begin{equation}
\mathrm{mob}(\phi)=(1-\phi)+\phi f_\mathrm{s},
\label{eq:mob}
\end{equation}
with $f_\mathrm{s}$ the ratio of the solid to the liquid diffusivity: every
diffusion-limited rate coefficient in $R_i$ of Eq.~(\ref{eq:rd}) is multiplied by
$\mathrm{mob}(\phi)$. The chemistry is otherwise the zero-dimensional model of
Eq.~(\ref{eq:rd}). In the liquid $\mathrm{mob}=1$, which is the model used everywhere
else in this paper. In the solid $\mathrm{mob}=f_\mathrm{s}$. Even co-located radicals then cannot react, and spatially
separated reactants can meet only during a window. The second gate is the Arrhenius
factor of the aldol steps, $E_a=87$~kJ\,mol$^{-1}$, which is negligible at 10~K.

The parameter $f_\mathrm{s}$ is the least well known in the model. Molecular dynamics
with a machine-learned H--O potential (Appendix~\ref{app:qcmd}) gives upper bounds
$f_\mathrm{s}<6\times10^{-5}$ for H and $<1.5\times10^{-3}$ for OH, measured at 0.57
of the melting point of the potential; the true 10\,K values are probably orders of magnitude lower. The
results of this paper do not depend on $f_\mathrm{s}$, because the window chemistry is
computed in the liquid. It enters only the comparison made here.

\begin{figure*}
\centering
\includegraphics[width=\textwidth]{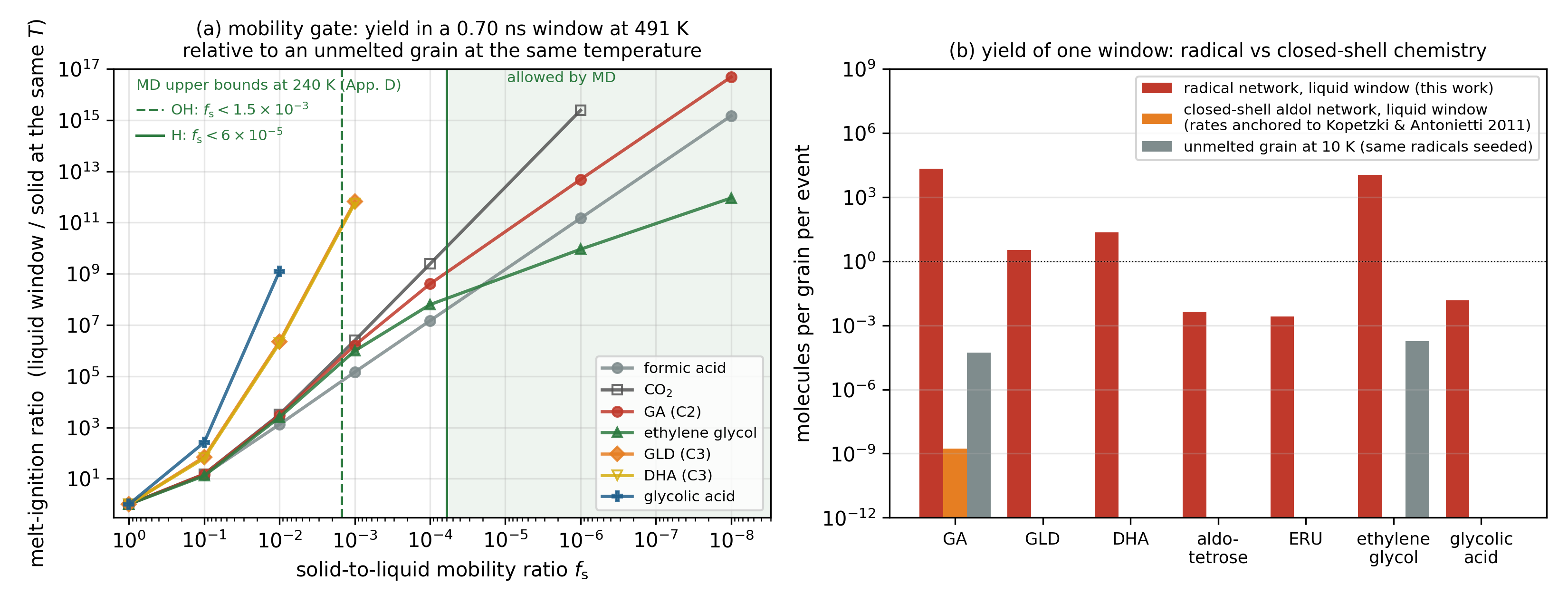}
\caption{One long window (0.70~ns at 491~K, with the dose of the longest windows, \doseinlong$D_\mathrm{melt}$).
(a) Melt-ignition ratio: the number of molecules of each product made per grain in
the liquid window, divided by the number made in an unmelted grain at the same
temperature with the same radicals, against the solid mobility $f_\mathrm{s}$. The ratio is 1 at $f_\mathrm{s}=1$, where the two runs are
identical by construction. Each curve stops at the smallest $f_\mathrm{s}$ for which
the unmelted grain still makes something; below it the yield falls under the numerical precision of the calculation, is written as zero, and the
ratio is not finite. GLD and DHA coincide down to $f_\mathrm{s}=10^{-3}$, and GA and CO$_2$ down to $10^{-2}$. Vertical lines: the upper
bounds on $f_\mathrm{s}$ from the molecular dynamics of Appendix~\ref{app:qcmd} at
240~K (0.57 of the melting point of the potential, 420~K), for OH (dashed) and H (solid); the shaded side is the region they allow. The
dynamics give no lower bound. (b) Molecules per grain per event for the radical network
(this work), for the closed-shell aldol network with rates anchored to the hydrothermal
experiment of \citet{Kopetzki2011}, and for an unmelted grain at 10~K seeded with the
same radicals. In (a) $f_\mathrm{s}$ decreases to the right, from the liquid ($f_\mathrm{s}=1$) towards the cold solid. Bars below $10^{-12}$ are zero.}
\label{fig:ignition}
\end{figure*}

Figure~\ref{fig:ignition} isolates the two gates with one long window held at a fixed temperature, 0.70~ns at 491~K, with the dose of the longest windows of the cycle model, \doseinlong$D_\mathrm{melt}$ (Sect.~\ref{sec:network}), and the initial ice inventory. This is a generous case in length and temperature. The window is 1.7 times the longest liquid window of the track table, and it is hotter than the liquid of the long windows, which cools from about 470 to 275~K. Panel (a) considers seven products: formic acid, CO$_2$, GA, ethylene glycol, GLD, DHA and glycolic acid. For each it compares the number of molecules made per grain in the liquid window with the number made in an unmelted grain. The unmelted grain receives the same radicals at the same temperature but keeps the solid mobility $f_\mathrm{s}$. We call the ratio of the two the melt-ignition ratio. At small $f_\mathrm{s}$ the ratio scales as $f_\mathrm{s}^{-n_\mathrm{d}}$, with $n_\mathrm{d}$ the
number of diffusion-limited steps between the radiolytic radicals and the product:
$n_\mathrm{d}=2$ for formic acid and GA, 1 for ethylene glycol and 3 for CO$_2$ at $f_\mathrm{s}\leq10^{-4}$. Even at $f_\mathrm{s}=10^{-2}$ the unmelted grain makes at most $1\times10^{-5}$ molecules of the C3 sugars per event and $1\times10^{-11}$ of glycolic acid, and nothing above the numerical precision for $f_\mathrm{s}\leq10^{-4}$. At the 10~K of the quiescent ice the largest yield of any species is $1\times10^{-2}$ molecules per grain per event (formic acid), and no sugar beyond C2 is made at all. In this network, and with the radicals of one event, sugar formation beyond C2 therefore requires the liquid; the value of $f_\mathrm{s}$ only sets how completely the solid is frozen. Cold radiolysis over much larger doses does make C3 sugars in the laboratory (Sect.~\ref{sec:c3disc}); that route lies outside this model.

Panel (b) compares, in the same window, the radical network with the closed-shell aldol network at the laboratory-anchored rates. In that run the OH attack on the sugars is represented by chain scission (class S of Table~\ref{tab:reactions}) instead of the radical channels. The combination of two $\bullet$CH(OH)$_2$ gives the dihydrate of glyoxal, which becomes GA only between events (Appendix~\ref{app:network}). Inside the window GA comes from the combination of $\bullet$CH(OH)$_2$ with $\bullet$CH$_2$OH, at $2\times10^{4}$ molecules per grain in this first event of the radical network. Beyond C2 the aldol network yields less than $10^{-12}$ molecules of every sugar per event, whereas the radical network yields 3.4, 23 and 0.015 for GLD, DHA and glycolic acid in the first event, before any C2 has accumulated, and $7\times10^{-3}$ of the C4. The gap is the ratio of the window to the time
the formose reaction needs to run at this temperature in the laboratory (about a
minute at 473~K, \citealt{Kopetzki2011}), $10^{-11}$ for this window. The number of
events does not close it either. Summed over the \Ncom\ events of the fiducial exposure the aldol channels make \aldolgld\ GLD and \aldolery\ aldotetrose molecules per grain, against \cthreeborn\ C3 and \cfourborn\ C4 molecules made by the radical channels, and they carry less than $10^{-9}$ of the production of every sugar. At the
10~K of the quiescent ice the Arrhenius factor of the aldol barrier removes the
thermal route entirely. The closed-shell channels are retained in the network for
completeness; the results of this paper are those of the radical network.

The finite window also limits the radical yield. Holding the liquid of
Fig.~\ref{fig:ignition} for ten times as long, 7.0~ns instead of 0.70~ns, with the same dose injected once at the start, raises the C2 yield by a factor of 2.4, the C3 yields by factors of 2--4 and the C4 yields by factors of 3--4, because growth beyond C2 needs the C2 radicals that the window itself has to make. A single window
is therefore not enough, and Sect.~\ref{sec:ratchet} follows the inventory across many.

\subsection{One window is not enough: the accumulation ratchet}
\label{sec:ratchet}

A single window converts only a small fraction of the GA into larger sugars: even
the long window of Fig.~\ref{fig:ignition} leaves ERU/GA at about $10^{-7}$ after one event. Growth becomes possible only across
repeated events, through three properties that together act as a ratchet.

First, retention. The sugars stay in the ice between events because they are bound to
it. The desorption energy $E_\mathrm{des}$ is the energy a molecule must acquire to
leave the ice surface. Written as a temperature, $E_\mathrm{des}/\kb$ with $\kb$ the
Boltzmann constant, computed values for GA on amorphous water ice span
3500--11000\,K \citep{Kakkar2025}. We adopt 5500\,K for GA and scale with the square
root of the number of hydroxyl groups, which gives 9500\,K for ERU; these values enter only the retention
estimate made here. The rate at which a molecule in the surface monolayer desorbs is
given by the Polanyi--Wigner expression $k_\mathrm{des}=\nu_0\exp(-E_\mathrm{des}/\kb T)$,
with $\nu_0$ the attempt frequency. With $\nu_0=10^{12}$~s$^{-1}$, typical of molecules
bound to a surface, $k_\mathrm{des}$ at 491~K removes $1.4\times10^{-2}$ of
the GA and $4\times10^{-6}$ of the ERU in the surface monolayer per nanosecond
of liquid; averaged over a grain of radius 0.1~$\mu$m, whose outer monolayer holds
1.2\% of the volume, this is $2\times10^{-4}$ and $5\times10^{-8}$ per nanosecond. The \Ncom\ events of the fiducial exposure add up to \liqtot~ns of liquid (the mean window, weighted by melted volume, is \taubarps~ps), so the losses are \gaevap\ and \eruevap\ over the whole exposure.
These are upper limits, because the liquid is on average at 350~K rather than 491~K. The ice is a
cumulative reactor.

Second, open-chain persistence. GLD, DHA and ERU cannot form five- or six-membered rings; the argument concerns the aldotetrose and the C5 and C6 sugars. Ring closure and its reverse proceed by proton
transfer and are slow. As an analogue we use the measured dehydration rate of the GLD hydrate in
water, 0.03~s$^{-1}$ at 25~$^{\circ}$C \citep{Rendina1984}. We extrapolate it to the window temperature with barriers of 40--80~kJ\,mol$^{-1}$, which bracket the 60~kJ\,mol$^{-1}$ measured for that reaction \citep{Rendina1984}. A fraction of order $10^{-8}$--$10^{-5}$ of the sugars then cyclises per nanosecond of liquid, and at most \cyclmax\ over the \liqtot~ns of the whole exposure. At the 10~K between events the same step is frozen for any barrier in that range. The sugars therefore refreeze in their reactive open-chain forms
and can react again in the next event. The network does not rely on this for the C5 and C6 sugars: OH attacks them only through small open-chain fractions (Table~\ref{tab:prov}), a choice that affects those sugars alone.

Third, superlinearity. The radicals that build C4 are made from the sugars already
present, so the C2$\to$C4 conversion per window grows with the sugar inventory. Each
successive window converts C2 to C4 faster than the last, and the same pairs continue
the ladder: the gem-diol radical of GA combines with the radical of
ERU to give a C6, so hexoses follow ERU.

The mean sugar size grows from 2.0 carbons, when GA is the only sugar, to \meancom\ at the count $\nstar$ at which the observed ERU/GA is reached (Table~\ref{tab:main}). The C2:C4 ratio inverts: ERU/GA passes 1 soon after $\nstar$ and keeps rising. This monotonic rise makes the ratio a count of melting events, which Sect.~\ref{sec:clock} calibrates: along the temperature histories of the windows, the observed ERU/GA is reached after $\nstar=\Ncom$ whole-grain melting events. Across the systematics of Table~\ref{tab:nstar} that keep the fate of the trapped radicals and do not destroy the sugars between events (Sect.~\ref{sec:clock}), $\nstar$ ranges from \nstarlo\ to \nstarhi. The inventory at $\nstar$ is given in Table~\ref{tab:main}.

\begin{table}
\caption{The model at the count $\nstar$ at which ERU/GA reaches the observed value, for the complete network (fiducial; means of 8 sequences of windows), compared with observations. With $\delta(491\,\mathrm{K})=0.47$, $h_\mathrm{GLD}=1$ and $h_\mathrm{GA}=0.1$.}
\label{tab:main}
\centering
\scriptsize
\setlength{\tabcolsep}{3pt}
\begin{tabular}{lcc}
\hline\hline
Quantity & Model, $\nstar=\Ncom$ & Observed \\
\hline
ERU/GA                                   & 0.94 & $0.94\pm0.09$ \\
GLD/GA                                   & 0.11 & $\leq$0.054 \\
DHA/GA                                   & 1.6 & $\leq$0.108\tablefootmark{a} \\
Glycolic acid/GA                         & 0.029 & $\leq$0.15--0.7\tablefootmark{b} \\
Aldotetrose/GA                           & 0.29 & \ldots \\
Hexoses/GA                               & 0.74 & \ldots \\
Pentoses/GA                              & 0.81 & \ldots \\
Ethylene glycol/GA                       & 0.45 & 2.5\tablefootmark{c} \\
Methanol/GA                              & 9.9 & \ldots \\
HCOOH/$\HHO$                             & 15\% & 0.5--6\%\tablefootmark{d} \\
$\COO/\HHO$                              & 11\% & 19--28\%\tablefootmark{d} \\
ERU/C3                                   & 0.54 & $\geq$5.8 \\
Ketose fraction of the C4                & 76\% & \ldots \\
GA per grain                             & $4.8\times10^{5}$ & \ldots \\
ERU per grain                            & $4.5\times10^{5}$ & \ldots \\
Mean sugar size [C]\tablefootmark{e}     & 3.7 & \ldots \\
Sugars with $\geq$C6\tablefootmark{e}    & 13\% & (undetected) \\
\hline

\end{tabular}
\tablefoot{Observed sugar values from \citet{JimenezSerra2026}; ice HCOOH and $\COO$
bands from \citet{Boogert2015}. ERU/C3 is ERU/(GLD$+$DHA).
\tablefoottext{a}{From $N(\mathrm{DHA})\leq1.0\times10^{13}\,\percm$, obtained with the
excitation temperature fixed at 10~K and the line width at 21~km\,s$^{-1}$
\citep{JimenezSerra2026}; the confidence level of the limit is not stated there
(the same survey uses line-integrated $3\sigma$ limits for other non-detections;
\citealt{SanzNovo2026}). With an excitation temperature of 20--22~K, as measured for
GA, the limit would be about twice as high. Sect.~\ref{sec:c3excess} shows that the limit is firm.}
\tablefoottext{b}{From $N\leq2.3\times10^{13}\,\percm$ \citep{JimenezSerra2020}: 0.7 relative to the GA column of that work ($3.3\times10^{13}\,\percm$), both at an excitation temperature of 8~K; 0.15 relative to the GA column used here if the acid is excited at 22~K like GA (our evaluation; Sect.~\ref{sec:predictions}).}
\tablefoottext{c}{The aGg$'$ conformer of ethylene glycol at an excitation temperature
of 10~K \citep{JimenezSerra2026}.}
\tablefoottext{d}{The HCOOH and $\COO$ entries are ceilings for the sugar network, not
targets (Sect.~\ref{sec:limits}).}
\tablefoottext{e}{Number-weighted over the closed-shell sugars C2--C12, ERU
included.}}
\end{table}

\subsection{Why the surviving C4 is a ketose, and why the C3 ketose outlives the aldose}
\label{sec:ketose}

\begin{figure*}
\centering
\includegraphics[width=\textwidth]{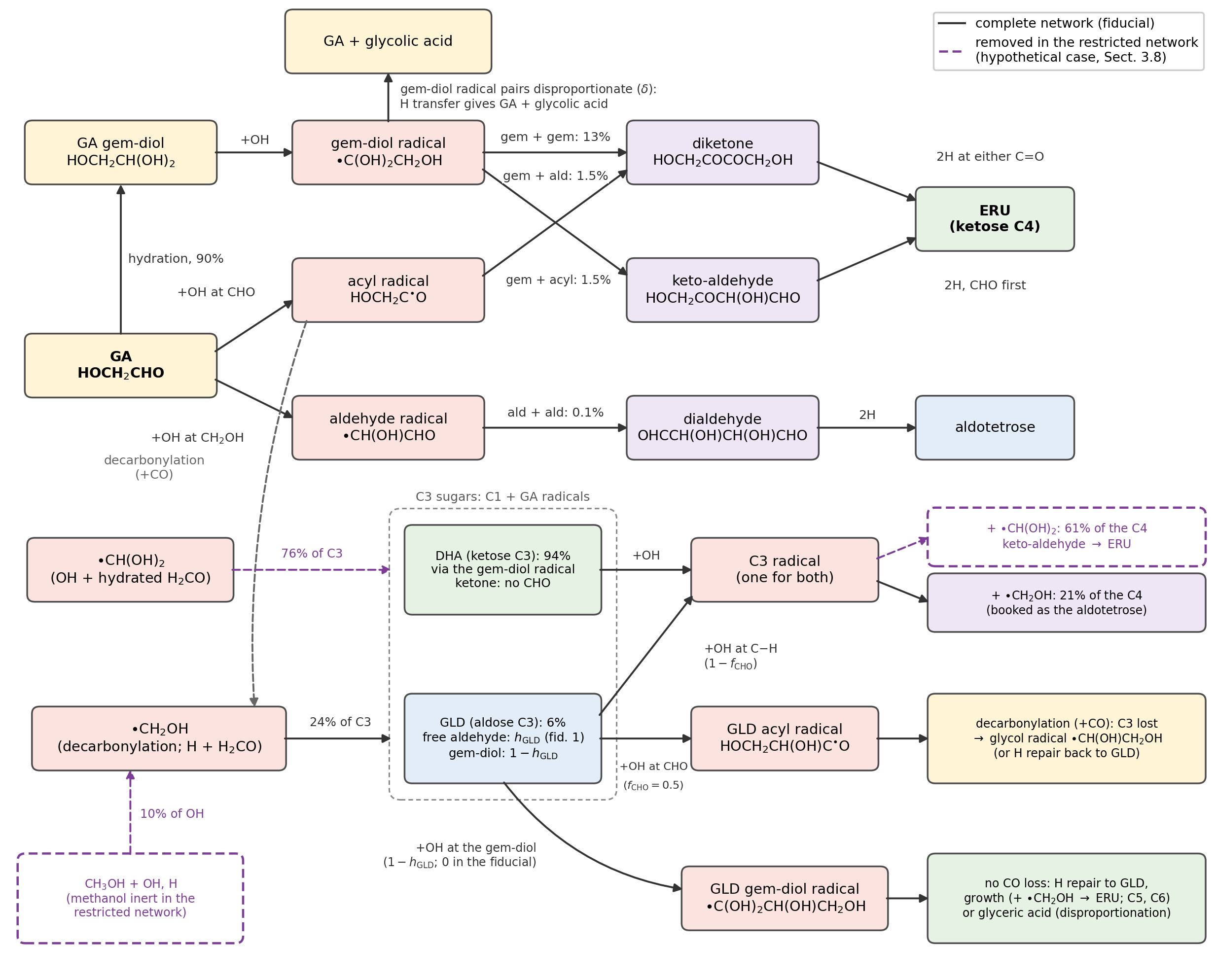}
\caption{The radical chemistry of the C3 and C4 sugars in a window. OH attacks the hydrated GA (90\%) at its gem-diol carbon and the free aldehyde at its CHO or CH$_2$OH group; radical pairs combine (fraction $1-\delta$) or disproportionate ($\delta$), the gem-diol radical acting as the H donor and giving glycolic acid. Percentages: shares of the C3 and C4 births by each channel up to $\nstar$ in the fiducial run. The C1 radicals, $\bullet$CH(OH)$_2$ from hydrated formaldehyde and $\bullet$CH$_2$OH from methanol and from the decarbonylation of the acyl radical, make the C3 with the GA radicals and most of the C4 with the C3 radical (Sect.~\ref{sec:ketose}). Carbonyl compounds and trapped radicals are reduced by H between events. Lower part: the two C3 isomers under OH attack. DHA gives only the C3 radical; GLD reacts according to its free-aldehyde fraction $h_\mathrm{GLD}$ (1 in the fiducial model), and the decarbonylation of its acyl radical is the only channel that destroys a C3 sugar. Dashed violet: the reactions removed in the restricted network of the hypothetical case (Sect.~\ref{sec:hypothetical}); its third assumption, no radiolysis between events, concerns the cold phase and is not shown.}
\label{fig:scheme}
\end{figure*}

In liquid water GA is 90--96\% hydrated, i.e. its hydration equilibrium
constant $K_\mathrm{hyd}=[\mathrm{gem\textrm{-}diol}]/[\mathrm{aldehyde}]$ is 9--25
\citep{Sorensen1972}; the lower value is the kinetic measurement, the upper an empirical
estimate. Between
melting events the ice preserves the hydration state that each aldehyde acquired
(Sect.~\ref{sec:caveats}). This is an untested assumption about the birth state of GA (Sect.~\ref{sec:caveats}). Nine of ten OH attacks on GA therefore strike the
gem-diol (Fig.~\ref{fig:scheme}) and produce the radical $\bullet$C(OH)$_2$CH$_2$OH rather than the aldehyde
radical $\bullet$CH(OH)CHO (abstraction at the CH$_2$OH carbon of the gem-diol is not
included in the fiducial model). Table~\ref{tab:nstar} gives the effect of putting 20\% and 50\% of the attack there. When two gem-diol radicals combine, or one combines with
$\bullet$CH(OH)CHO, the product carries a C(OH)$_2$ group. (The gem-diol radical could also lose water to give the acyl radical before it combines. Water elimination from dihydroxyalkyl radicals is catalysed by acid and base \citep{Steenken1986,vonSonntag1980}, but no rate has been measured for this 1,1-dihydroxyalkyl radical. With a first-order dehydration at $10^{8}$ and $10^{9}$~s$^{-1}$ the count becomes \Ngemdehyda\ and \Ngemdehydb\ and the ketose fraction \ketgemdehyda\ and \ketgemdehydb; Table~\ref{tab:nstar}. The acyl radical combines to the same ketone, so the dehydration does not move the carbonyl.) Its dehydration places the
carbonyl on the former aldehyde carbon. The C4 is therefore born with a carbonyl at
C2, which the cold-phase hydrogen addition (Sect.~\ref{sec:network}) reduces to
ERU. Combination of two aldehyde radicals instead gives the dialdehyde, which
reduces to the aldotetrose. Two gem-diol radicals, or a gem-diol radical and the acyl
radical of GA, HOCH$_2$C$^{\bullet}$O, give the symmetric diketone HOCH$_2$COCOCH$_2$OH.
Hydrogen added to either of its carbonyls gives ERU, so this part of the ketose
does not depend on where the hydrogen adds. A gem-diol and an aldehyde radical give
the keto-aldehyde HOCH$_2$COCH(OH)CHO, which gives ERU if the aldehyde is reduced
first, as the network assumes. In the
complete network a third path dominates: a C1 radical adds to the trapped radical of
DHA, the leading C3 sugar, and the product carries the carbonyl of DHA. Up to $\nstar$
in the fiducial run, \cfourdhm\ of the C4 is born from a C1 radical and the C3 radical (\cfourdhmd\ from $\bullet$CH(OH)$_2$, which gives the keto-aldehyde, and \cfourmeoh\ from $\bullet$CH$_2$OH). The rest is born from two gem-diol radicals (\cfourgem), from a gem-diol radical with the acyl or the aldehyde radical (\cfourgemx) and as the dialdehyde (\cfourald; Fig.~\ref{fig:scheme}); the remaining \cfourglyc\ comes from the glycol radical. \ketcom\ of the closed-shell C4 is the ketose; the remainder is the aldotetrose, which
the network books for the product of $\bullet$CH$_2$OH with the C3 radical
(Appendix~\ref{app:window}). If that product is booked as the ketose, consistent with
the return of the C3 radical as DHA between events, the share is \ketcomwin. The ketose, having no aldehydic hydrogen, is not destroyed by
OH: the attack gives its hydroxyalkyl radical, which is repaired, combines into a C5 or
C6, or returns to ERU between events.
The C3 follows the same logic. The gem-diol radical of GA with either C1 radical gives a C3 whose carbonyl sits on the former aldehyde carbon, i.e. the C3 ketose, DHA, which is the main C3 product of a single window (Sect.~\ref{sec:meltgated}). With $\bullet$CH$_2$OH the product is DHA itself; with $\bullet$CH(OH)$_2$ it is the dehydro-C3 OHC$-$CO$-$CH$_2$OH, which becomes DHA when the cold-phase hydrogen reduces its aldehyde, an assumption of the network. The aldose, GLD, comes from the aldehyde radical and is the minor product. A ketone hydrates weakly, so DHA sits
in the ice as the free carbonyl and OH attacks it at the full rate. Its radical is a
semidione, which the network carries as one species for the ketose and the aldose,
and the site at which the cold-phase hydrogen adds decides which tautomer returns.
The network returns it as the ketose, and returns the radical of ERU as
ERU (Appendix~\ref{app:coldh}). This is an assumption. Addition at the carbon restores the ketose for the radicals born from the ketoses. Addition at the oxygen gives the enediol, the common intermediate of both sugars. Its tautomerisation is a proton transfer, which is frozen at 10--20~K, as the hydration steps are. In the basic melt, moreover, the radical is largely the semidione anion, and its protonation site on cooling is not known. For ERU only abstraction at C1 gives a radical that could
return as the aldotetrose, so the alternative run below, which returns half of all its
radicals as the aldotetrose, overstates that path. Inside the window the network still returns the C3 radical as the aldose (H repair
and the disproportionations); the radicals that survive the window return as the
ketose. The combination of the C3 radical with $\bullet$CH(OH)$_2$, by contrast, is booked as the keto-aldehyde that reduces to the ketose: the new carbon is an aldehyde for $\bullet$CH(OH)$_2$, and for the formyl radical HCO in the free-H$_2$CO rows of Table~\ref{tab:nstar}, and the radical of DHA keeps its ketone. The row of Table~\ref{tab:nstar} that writes all in-window products of the C3 radical as the ketose removes this asymmetry. Structurally that bookkeeping is the natural one for the radical of DHA: $\bullet$CH$_2$OH added to $\bullet$CH(OH)$-$CO$-$CH$_2$OH gives HOCH$_2-$CH(OH)$-$CO$-$CH$_2$OH, erythrulose itself. Its addition to the C2 radical of GLD gives a branched C4, and a linear aldotetrose arises only from the radical at the CH$_2$OH carbon of GLD. Since \dhashare\ of the C3 is DHA, the fiducial \ketcom\ is a lower bound with respect to this bookkeeping and that row the natural value. DHA is therefore the leading C3 sugar: it stands at \dhacom\ times its limit and GLD at \gldcom\ times (Sect.~\ref{sec:c3excess}). The reduction site sets the partition of the C3 between its two isomers, not its total: returning the C3 radical wholly as the aldose moves the excess from DHA to GLD (\cthreeald\ times the summed limit). Returning the ERU radical half as the aldotetrose makes the C4 mostly the aldotetrose, and ERU/GA stays far below the observed value (Table~\ref{tab:nstar}). The ketose identity of the C4 therefore requires that the
trapped C4 radical returns as the ketose; it holds whether or not DHA is hydrated.

The hydration of GA selects the ketose. If GA in the ice were the free aldehyde
($h_\mathrm{GA}\to1$, with the aldotetrose also unhydrated), the gem-diol radical would
not form. The C4 born from two GA radicals would then be the dialdehyde, i.e. the aldotetrose, and ERU/GA would rise only because radiolysis destroys GA; without the C1 radicals the observed value is reached only when GA is gone (Appendix~\ref{app:restricted}). The C1 radicals keep a second ketose path open, the addition of $\bullet$CH(OH)$_2$ to the C3 radical, so the aldotetrose share is \hnonecom\ and \hnonecomq. The ketose then rests on the same C1 addition that makes the C3 excess. A ketose C4 with the C3 sugars below their limits therefore requires a hydrated GA. The same
gem-diol radical pairs are the source of glycolic acid by disproportionation: the
ketose and the acid are products of one branch. The selection also assumes that OH attacks the hydrated GA at its gem-diol carbon. Attack at the CH$_2$OH carbon gives the $\alpha$-radical, whose products are aldoses. With 20\% and 50\% of the attack on the hydrated form at that carbon the ketose fraction is \ketgemtwo\ and \ketgemfive, and the count \Ngemtwo\ and \Ngemfive\ (Table~\ref{tab:nstar}). Rate constants three times higher for the free aldehyde forms of GA and GLD, which react faster than their hydrates, give \ketkfree\ and \Nkfree.

Two channels make the ketose, and they differ in robustness. The gem-diol channel, the combination of two gem-diol radicals or of a gem-diol radical with the acyl or $\alpha$-radical of GA, depends only on the hydration of GA. It gives all of the C4 in the restricted network, where the ketose fraction is \ketres, and \cfourgemall\ of the C4 births in the complete one. The C1 channel, a C1 radical added to the C3 radical, gives another \cfourdhm\ of the births in the complete network, and the identity of its product follows the bookkeeping of the C3 radical discussed above. The fiducial fraction of \ketcom\ therefore rests on the gem-diol channel for \cfourgemall\ of the C4 and on that bookkeeping for most of the rest; across the axes of Table~\ref{tab:nstar} that reach the observed ratio the fraction spans \ketrange. Only the gem-diol channel is robust to every assumption about the ice, and it is the one that the restricted network isolates.

\subsection{The complete network overproduces the C3 sugars}
\label{sec:c3excess}

\begin{figure*}
\centering
\includegraphics[width=\textwidth]{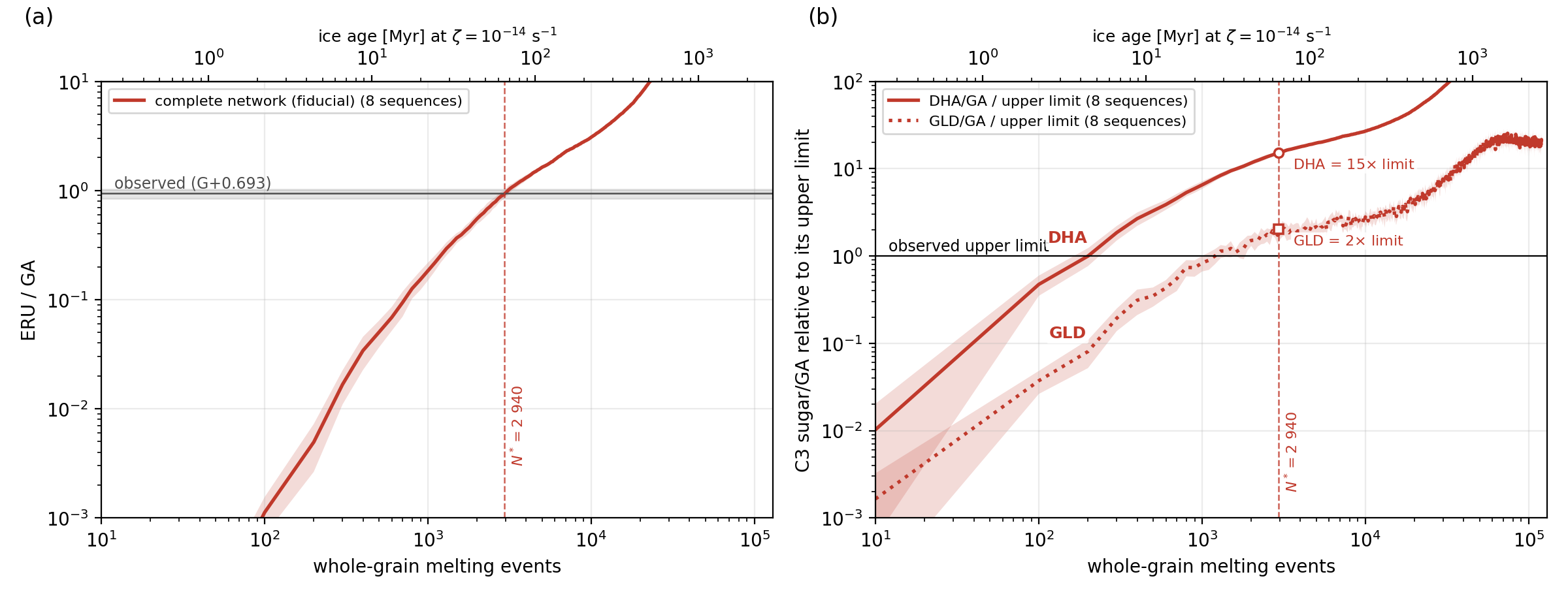}
\caption{(a) ERU/GA versus the number of whole-grain melting events, every event
drawing its window from the reconstructed distribution (local particles, rising
shape). Means of 8 sequences of windows (fiducial run), bands their standard deviation. The upper axis converts the count into time at $\zeta=10^{-14}$~s$^{-1}$. The observed ratio (band) is crossed after \Ncom\ events. (b) DHA/GA (solid) and GLD/GA (dotted) relative to their observed upper limits (horizontal line at 1; logarithmic axis). The dashed vertical line in both panels marks $\nstar$; the open symbols give the two C3 ratios there, DHA at \dhacom\ and GLD at \gldcom\ times its limit. The DHA limit is crossed after \ndhalim\ events, long before ERU/GA reaches its observed value.}
\label{fig:clockdelta}
\end{figure*}

The window makes a C3 sugar by adding one carbon to a C2 radical. In the complete
network the carbon comes from two C1 radicals. The dihydroxymethyl radical
$\bullet$CH(OH)$_2$ is made by OH attack on the hydrated formaldehyde, the substrate
from which the melt makes GA in the first place. The hydroxymethyl radical
$\bullet$CH$_2$OH is made by OH and H attack on methanol and by H addition to
formaldehyde. Per event in the fiducial run, up to $\nstar$, \ohfrachcho\ of the OH attacks fall on the hydrated formaldehyde, \ohfracmeoh\ on methanol and \ohfracga\ on GA, while formic acid and formate take \ohfracfa, so the C1 radicals outnumber the C2 radicals
\cratio\ to one. Every encounter of a C1 radical with a C2 radical that combines gives
a C3, at the same rate as the C2$+$C2 encounters that give the C4. The C3 therefore
grows in step with the C4, with the larger partner concentration, and the ladder is climbed one rung at a time: C2 $\to$ C3 $\to$ C4. Figure~\ref{fig:clockdelta}b shows the result. When
ERU/GA reaches its observed value, after $\nstar=\Ncom$ events, DHA stands at
\dhacom\ times its observed upper limit and GLD at \gldcom\ times its limit
(Table~\ref{tab:main}). The excess grows with exposure, because the C3 keeps being
made from the C2 while the C4 accumulates from both. The limit itself is firm. At the abundance of the model DHA would show lines at about a quarter of the strength of the GA lines, because its dipole moment is smaller and its partition function larger ($\mu_b=1.77$~D and $Q(9.375\,\mathrm{K})=822$ against $\mu_b=2.33$~D and 200; \citealt{Lovas2003,Marstokk1970,Pickett1998,Endres2016}). That is \dhacom\ times the threshold of the survey, beyond the factor of 2--3 by which the excitation temperature and the higher conformers can raise the limit (Table~\ref{tab:main}, note a). Even with the limit doubled, for an excitation temperature of 20--22~K, DHA stands at \dhacomhalf\ times it.

\begin{table*}
\caption{The count $\nstar$ of whole-grain melting events at which ERU/GA
$=0.94$, the two C3 ratios GLD/GA and DHA/GA relative to their upper limits, and
glycolic acid/GA at that count, across the systematics of the complete network. Every
event draws its window from the reconstructed distribution (local particles, rising
shape) and follows the temperature history of its bin. Fiducial:
$\delta(491\,\mathrm{K})=0.47$, $h_\mathrm{GLD}=1$, $h_\mathrm{GA}=0.1$, OH destruction of
glycolic acid included, no radiolysis of the sugars between events, all other parameters as in Table~\ref{tab:prov}. Each entry is the mean over 4 sequences of windows (8 for the
fiducial). The standard deviation of $\nstar$ between sequences is \sdrange. Rows with
four sequences are to be compared with the same four sequences of the fiducial:
$\nstar=\Ncomfour$, GLD/GA and DHA/GA at \gldcomfour\ and \dhacomfour\ of their
limits, glycolic acid/GA \glycomfour.}
\label{tab:nstar}
\centering
\scriptsize
\setlength{\tabcolsep}{4pt}\begin{tabular}{p{6.1cm}ccccc}
\hline\hline
Model variant & $\nstar$ & GLD/GA / limit & DHA/GA / limit & glycolic/GA & ketose (\%) \\
\hline
Fiducial, complete network (8 sequences of windows) & 2\,940 & 2.0 & 15 & 0.029 & 76 \\
Methanol inert (substrate all H$_2$CO) & 2\,450 & 1.9 & 14 & 0.018 & 99 \\
No cross-combination of $\bullet$CH(OH)$_2$ with the sugar radicals (methanol active) & 10\,800 & 1.4 & 15 & 0.016 & 37 \\
$\delta=0.15$ / 0.73 & 2\,600 / 3\,250 & 1.7 / 2.3 & 15 / 15 & 0.018 / 0.039 & 77 / 75 \\
$h_\mathrm{GLD}=0.05$ (aqueous $K_\mathrm{hyd}$) & 2\,740 & 10 & 10 & 0.027 & 82 \\
Literature pair ($\delta=0.73$, $h_\mathrm{GLD}=0.05$) & 3\,040 & 12 & 9.6 & 0.036 & 83 \\
No hydration of GA and the aldotetrose ($h=1$)\tablefootmark{b} & 3\,770 & 2.1 & 16 & 0 & 61 \\
Temperature cap of the superheated liquid 450 / 650~K & 4\,690 / 2\,270 & 1.6 / 2.6 & 14 / 16 & 0.015 / 0.042 & 74 / 77 \\
Escape rather than primary yields & 4\,930 & 1.6 & 15 & 0.023 & 77 \\
Trapped small radicals survive to the next event & 850 & 2.1 & 17 & 0.089 & 90 \\
Trapped radicals largely unreduced: a tenth of the accretion term, no in-situ reductants ($\varepsilon=0.001$, $\eta=0$; Appendix~\ref{app:coldh}) & 510 & 17 & 14 & 0.50 & 82 \\
No H$_2$CO resupply between events (substrate closed) & 4\,980 & 2.1 & 17 & 0.017 & 58 \\
Base inventory 0 / $\times$2 & 2\,200 / 3\,290 & 1.9 / 2.1 & 15 / 15 & 0.034 / 0.027 & 78 / 75 \\
C3 radical reduced to the aldose (all) & 3\,770 & 51 & 0.48 & 0.025 & 76 \\
Products of the C3 radical in the window as the ketose (Sect.~\ref{sec:ketose}) & 2\,420 & 0.20 & 14 & 0.031 & 99.8 \\
C4 ketose radical reduced half to the aldotetrose\tablefootmark{c} & $>$120\,000 & 2.6\tablefootmark{a} & 15\tablefootmark{a} & 0.036\tablefootmark{a} & 8\tablefootmark{a} \\
DHA hydrated ($h_\mathrm{DHA}=0.05$) & 8\,890 & 1.1 & 177 & 0.012 & 75 \\
Dose of the window $\times$2 / $\times$0.5 & 1\,290 / 7\,070 & 2.5 / 1.4 & 15 / 16 & 0.036 / 0.022 & 78 / 74 \\
Dose of the window equal to the melting enthalpy $D_\mathrm{melt}$ (Sect.~\ref{sec:caveats}) & 12\,000 & 1.1 & 16 & 0.015 & 72 \\
No OH destruction of glycolic acid & 2\,890 & 2.1 & 15 & 0.29 & 76 \\
Sugars destroyed between events, $D_{1/2}=2300$ / 690 / 69~eV per molecule (Sect.~\ref{sec:radiolysis}) & 3\,470 / $>$120\,000 / $>$120\,000 & 1.7 / 1.5\tablefootmark{a} / 0.77\tablefootmark{a} & 15 / 12\tablefootmark{a} / 3.4\tablefootmark{a} & 0.022 / 0.016\tablefootmark{a} / 0.0081\tablefootmark{a} & 75 / 72\tablefootmark{a} / 80\tablefootmark{a} \\
\quad $D_{1/2}=2300$~eV, carbon of the destroyed sugars returned as GA units / H$_2$CO / CO & $>$120\,000 / 3\,590 / 3\,480 & 1.5\tablefootmark{a} / 1.7 / 1.7 & 11\tablefootmark{a} / 14 / 15 & 0.039\tablefootmark{a} / 0.022 / 0.022 & 76\tablefootmark{a} / 76 / 75 \\
Sugars destroyed between events, $D_{1/2}=2.8$~eV, reversible with $c=0.25$ & 10\,400 & 1.9 & 18 & 0.019 & 74 \\
Radicals that survive a window recombine at its end (Sect.~\ref{sec:caveats}; 4 sequences) & 180 & 9.0 & 10 & 0.23 & 85 \\
Free H$_2$CO: OH abstracts the aldehydic H and HCO adds as an acyl, fraction 0.5 / 1 & 1\,410 / 1\,810 & 2.5 / 2.6 & 17 / 19 & 0.036 / 0.019 & 77 / 66 \\
\quad the same with the in-window products of the C3 radical as the ketose & 1\,250 / 1\,340 & 0.23 / 0.31 & 16 / 16 & 0.032 / 0.017 & 99.9 / 99.9 \\
OH attack of hydrated GA and GLD at the CH$_2$OH carbon, fraction 0.2 / 0.5 & 2\,980 / 3\,120 & 2.0 / 2.3 & 15 / 16 & 0.024 / 0.017 & 75 / 70 \\
OH rate constants of the free aldehyde forms $\times$3 & 2\,500 & 0.83 & 17 & 0.030 & 77 \\
Gem-diol radical of GA dehydrates to the acyl radical at $10^{8}$ / $10^{9}$~s$^{-1}$ (Sect.~\ref{sec:ketose}) & 2\,890 / 2\,900 & 2.1 / 2.1 & 15 / 15 & 0.030 / 0.028 & 76 / 76 \\
$\bullet$CH(OH)$_2$ + GA $\alpha$-radical booked as hydroxymalonaldehyde ($\to$ GLD) & 2\,890 & 2.2 & 15 & 0.030 & 76 \\
Cold-route GA and EG, (GA/H$_2$CO, EG/CH$_3$OH) = (0.04, 0.03) / (0.04, 0.12) / (0.01, 0.05) (Sect.~\ref{sec:limits}) & 2\,170 / 1\,610 / 2\,310 & 2.3 / 3.0 / 2.8 & 16 / 17 / 16 & 0.031 / 0.037 / 0.036 & 79 / 81 / 78 \\
$\bullet$CH(OH)$_2$ self-reaction by disproportionation only \citep{StockhausenHenglein1971} & 2\,910 & 2.2 & 16 & 0.025 & 76 \\
\hline

\end{tabular}
\tablefoot{\tablefoottext{a}{Value at the peak of ERU/GA within the run; the
ratio is not reached.}
\tablefoottext{b}{The C1 radicals still make a ketose from the C3 radical: \kethnone\ of the C4 is the ketose (Sect.~\ref{sec:ketose}).}
\tablefoottext{c}{The ketose fraction of the C4 is \keterysym, and ERU/GA peaks at \pkerysym.}

Glycolic acid is destroyed by OH ($k=6\times10^{8}$ M$^{-1}$\,s$^{-1}$) except in the row
that removes that channel.}
\end{table*}

Table~\ref{tab:nstar} shows that no parameter of the model brings the C3 sugars below
their limits. The disproportionation fraction $\delta$, the hydration of GLD, the
site of the cold-phase reduction, the hydration of DHA, the radiolytic yields, the
dose per event and the temperature of the window leave the two C3 sugars together at \cthreesumrange\ times their summed limit (\cthreesumhdha\ with DHA hydrated). On every axis at least one of the two C3 sugars stays above its limit. Holding methanol
inert leaves DHA at \dhameohoff\ times its limit, and removing the cross-combination
of $\bullet$CH(OH)$_2$ while methanol stays active leaves it at \dhadhmoff\ times:
either C1 radical alone supplies the excess. Only the removal of both, the hypothesis of Sect.~\ref{sec:hypothetical}, brings both C3 sugars below their limits.

Two statements hold for any choice of the decarbonylation and repair rates within the network and show why
no destruction channel can rescue the C3.

\emph{Destruction cannot remove more C3 than OH attack supplies.} Every channel that
destroys a closed-shell sugar in the melt begins with hydrogen abstraction by OH, the
only oxidant present at useful concentration. The acyl radical so formed either
decarbonylates or is repaired by H. The C3 removal rate is therefore bounded by the OH
attack rate on the free-aldehyde fraction, $k_\mathrm{OH}[\mathrm{OH}]\,h_\mathrm{GLD}$
with $k_\mathrm{OH}$ the rate constant of OH attack on GLD, however fast the downstream
decarbonylation is. Returning the whole trapped C3 radical as the free aldehyde, the
case most exposed to this sink, leaves the C3 sugars at \cthreeald\ times the summed
limit (Table~\ref{tab:nstar}).

\emph{Isomerisation cannot remove C3 either.} The Lobry de Bruyn--Alberda van
Ekenstein aldose--ketose interconversion redistributes carbon between GLD and DHA but
conserves their sum. No choice of its rate or equilibrium constant can bring both
below their limits if the sum exceeds the sum of the limits, so the argument does not depend on the rate. In the fiducial model the sum (GLD$+$DHA)/GA at
$\nstar$ is \cthreesum, against the summed limit of 0.162.

The C3 excess is the same failure as that of the grain-surface model of
\citet{JimenezSerra2026}, which overproduces both C3 sugars by 25--70$\times$, and we
attribute it to the same cause: a radical chemistry that adds carbons one at a time does not skip
C3. The melt route adds two filters that the surface route lacks, the hydration of GA
(which selects the ketose) and the decarbonylation of the free aldehyde (which removes
part of GLD), but neither acts on the supply of C3 from the C1 radicals. What would
suppress the C3 is a C1 radical that does not combine with the C2 radicals. No
measurement supports such a selectivity for $\bullet$CH(OH)$_2$ or $\bullet$CH$_2$OH
(Sect.~\ref{sec:caveats}), so we treat it as a hypothesis in
Sect.~\ref{sec:hypothetical}.

\subsection{The exposure that the C2$\approx$C4 equality needs}
\label{sec:clock}

\begin{figure}
\centering
\includegraphics[width=\columnwidth]{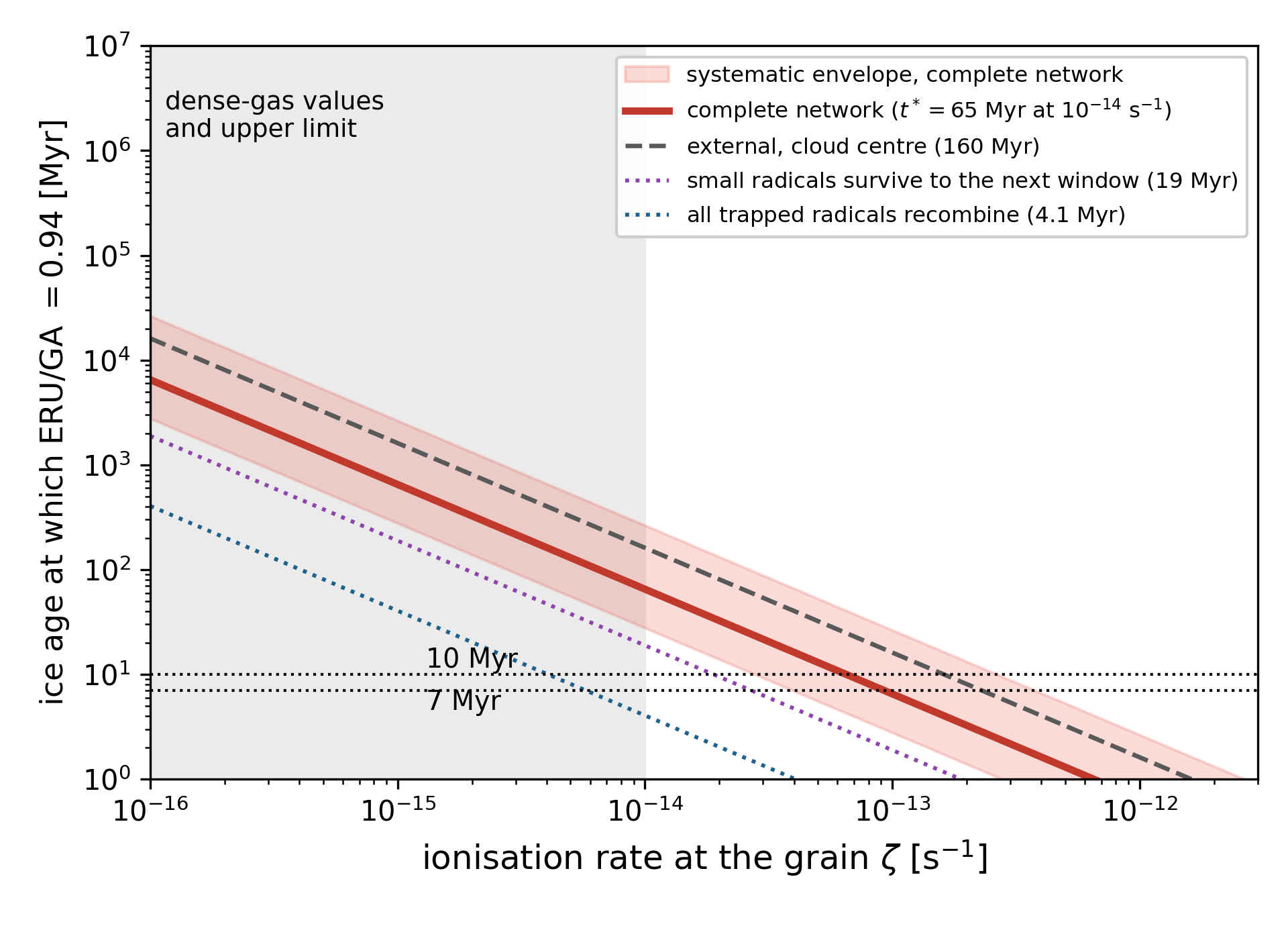}
\caption{Ice age at which ERU/GA reaches the observed value, as a function of the
ionisation rate $\zeta$ at the grain, for particles produced inside the cloud with the
rising spectral shape. Red: the fiducial model (\tcom~Myr at $\zeta=10^{-14}$~s$^{-1}$); the shaded band is the systematic envelope of Table~\ref{tab:nstar}
(\tlo--\thi~Myr at $\zeta=10^{-14}$~s$^{-1}$). The grey region marks the values inferred for dense gas (Sgr~B2 envelope and hot cores; \citealt{vanderTak2006,Bonfand2019}) and
the upper limit on the dense gas of the Central Molecular Zone \citep{Ginsburg2016}.
Horizontal lines mark cloud lifetimes of 7 and 10~Myr. Dashed: cosmic rays entering
from outside, attenuated to the centre of the cloud and normalised to $\zeta$ there. The age scales as $1/\zeta$. The dotted lines give two alternative fates of the trapped radicals (Sect.~\ref{sec:caveats}): small radicals that survive to the next window, and complete recombination at the end of every window.}
\label{fig:zeta}
\end{figure}

Because carry-over converts C2 into C4, ERU/GA rises monotonically with the number of
melting events (Fig.~\ref{fig:clockdelta}a). Every event draws its window from the
reconstructed distribution of Sect.~\ref{sec:rate} (local particles, rising shape,
$N_\mathrm{s}=10^{19}\,\percm$). The count $\nstar$ is expressed in whole-grain
equivalent melting events (Sect.~\ref{sec:rate}). For this distribution they occur
\Rwge\ times per Myr at $\zeta=10^{-14}$~s$^{-1}$, the sum of the bins of the red curve
in Fig.~\ref{fig:origin}b. The ratio reaches the observed value as ERU approaches its
plateau and GA, which the radiolysis of later events consumes, has fallen to \gafrac\ of its maximum; it continues to rise afterwards. The observed ratio $0.94\pm0.09$ is reached after $\nstar=\Ncom$ events (standard deviation \Ncomsd\ over 8 random sequences of windows), while DHA is already far above its limit (Sect.~\ref{sec:c3excess}). The count depends little on the disproportionation fraction $\delta$: over $\delta(491\,\mathrm{K})=0.15$--0.73 it spans \Ncomdx, because $\delta$ is evaluated at the mean temperature of each window between two anchors, 0.15 at 298~K and the quoted value at 491~K. In the long
windows, which make the C4, the three settings give 0.15, 0.24 and 0.33. Across the systematics of Table~\ref{tab:nstar}, excluding the runs that do not reach the ratio, the radiolysis rows and the three rows that change the fate of the trapped radicals (no reduction, survival of the small radicals, recombination), the count spans \nstarlo--\nstarhi.

The conversion of a count into a time is
$t=\nstar/(\Rwge\,\mathrm{Myr^{-1}}\,\zeta/10^{-14}\,\mathrm{s^{-1}})$ for particles
produced inside the cloud: \tcom~Myr at $\zeta=10^{-14}$~s$^{-1}$, with the envelope \tlo--\thi~Myr. The age scales as
$1/\zeta$ (Fig.~\ref{fig:zeta}). Molecular clouds in nearby disc galaxies live 10--30~Myr \citep{Chevance2020}. In the Central Molecular Zone the models of \citet{Jeffreson2018} give 1.4--9~Myr, so 7~Myr lies in the upper part of that range. The fiducial age is longer than either lifetime, and the whole envelope is longer than the lifetime in the Central Molecular Zone. An ice younger than 10~Myr requires $\zeta\gtrsim\zetacomten$~s$^{-1}$, and one younger than 7~Myr requires $\zeta\gtrsim\zetacom$~s$^{-1}$.
At these ionisation rates the interval between events is shorter, but the atomic
hydrogen density of the gas rises in proportion to $\zeta$, so the cold-phase hydrogen
budget stays complete (Appendix~\ref{app:coldh}). This scaling holds only while the
gas stays molecular, i.e. below $\zeta\approx3\times10^{-13}$~s$^{-1}$
(Sect.~\ref{sec:origin}). Both requirements lie below that limit. Whether the dense gas of G+0.693 reaches these values is
examined in Sect.~\ref{sec:origin}. At the local ionisation rate,
$\zeta\approx10^{-17}$~s$^{-1}$, particles produced inside the cloud melt a grain 0.05
times per Myr, and a grain receives 0.5 melting events in 10\,Myr; the ratchet then
does not turn at all, and sugar growth stalls at C2.

Reading the gas-phase ratio as the ice-phase ratio assumes a desorption that does not
depend on binding energy. With $\Delta E_\mathrm{des}\simeq4000$\,K between ERU and GA,
thermal desorption would suppress the gas-phase ratio relative to the ice by a factor
$\sim$$10^{-17}$ even at 100\,K, so the assumption requires sputtering, which is the
desorption mechanism established for the Galactic-centre clouds
\citep{RequenaTorres2006} and for G+0.693 itself \citep{Zeng2020}. The absolute
columns then require \desfrac\ of the ice GA and of the ice ERU at $\nstar$ to be in
the gas phase (Appendix~\ref{app:column}). The two fractions are equal because the
model reproduces the observed ratio, and they are small enough to be supplied by
sputtering.

\subsection{Radiolysis between events}
\label{sec:radiolysis}

\begin{figure*}
\centering
\includegraphics[width=\textwidth]{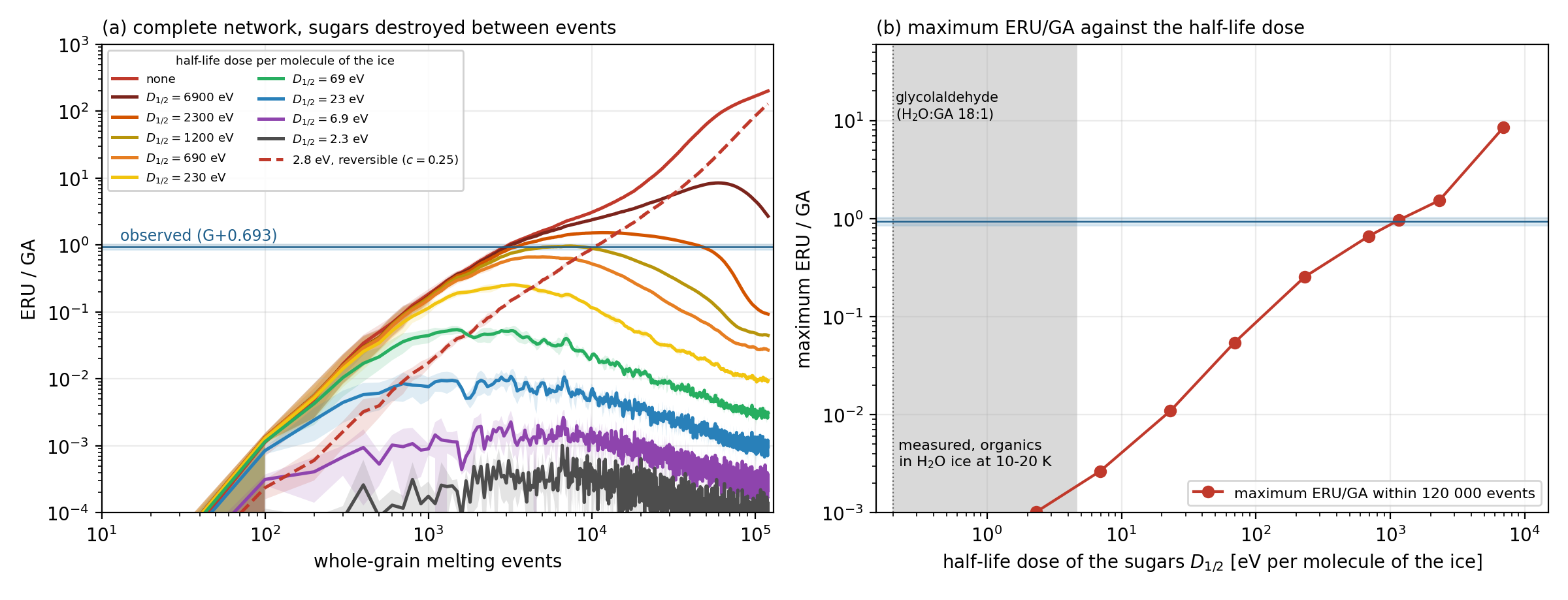}
\caption{Radiolytic destruction of the sugars between events by the dose deposited outside the melts,
in the complete network. (a) ERU/GA against the number of whole-grain melting events for half-life doses $D_{1/2}$ of the sugars from none to 2.3~eV per molecule of the ice (means of four sequences of windows; 8 for the case without destruction), and for the reversible case with
$D_{1/2}=2.8$~eV and a steady-state surviving fraction $c=0.25$ (dashed). (b) The
maximum ERU/GA reached within 120\,000 events against $D_{1/2}$. The band marks the
observed ratio; the grey region the half-life doses measured for organics diluted in
H$_2$O ice at 10--20~K, and the dotted line the value for glycolaldehyde in
H$_2$O:GA $=18$:1 at 10~K \citep{Hudson2005}.}
\label{fig:radiolysis}
\end{figure*}

The particles that melt the grain deposit most of their energy outside the volumes that they melt. Per whole-grain melting event the eleven nuclei deposit \dosetotfac\ times the melting enthalpy of the grain, \evtot~eV per water molecule (Appendix~\ref{app:numerics}). Of this, \doseinmean\ times the melting enthalpy lies inside the melted cylinders and drives the window chemistry (Sect.~\ref{sec:network}); the remaining \doseout\ times, \evout~eV per molecule, is deposited in ice that stays solid. That part has two sources. The secondary electrons of the melting ions carry \doseoutmeltions\ times the melting enthalpy beyond the melt radius. The ions that melt nothing, mostly protons above 2~MeV and helium nuclei above 4~MeV~u$^{-1}$, add \dosenonmelt\ times: such a proton loses 0.03--2~keV in the grain, a helium nucleus four times more at the same speed. Over the \Ncom\ events of the fiducial exposure the dose outside the melts adds up to about \doseouttot~eV per molecule. The fiducial model books this dose only as H$_2$ and
H$_2$O$_2$ and as a source of the reductants between events (Appendix~\ref{app:coldh});
it assumes that the closed-shell sugars survive it. That assumption can be tested
against laboratory measurements.

The radiolytic stability of an organic solute in cold water ice is measured as a
half-life dose $D_{1/2}$, the dose per molecule of the ice at which half of the solute
is destroyed, with MeV protons at 10--20~K and in-situ infrared spectroscopy. For
small organics diluted to 5\% or less in H$_2$O ice, the concentrations of the sugars here, the measured values are 0.2--4.6~eV
per molecule: 0.2~eV for glycolaldehyde in H$_2$O:GA $=18$:1 at 10~K
\citep[4.4~eV in the pure solid;][]{Hudson2005}, 1.2~eV for ethylene glycol at 20:1
(same work), 1.7--2.8~eV for glycine at 50:1--200:1 at 15~K \citep{Gerakines2013},
about 0.9~eV for thymine at 50:1 \citep{Materese2020}, 0.5--4.4~eV for uracil at 30:1--270:1 \citep{Gerakines2022}
and 2.6--4.6~eV for benzene and the azines at 50:1 \citep{Tribbett2026}. More concentrated mixtures are more stable: uracil at 10:1 survives beyond 11~eV per molecule \citep{Gerakines2022}. Dilution in
water makes every molecule measured more fragile, not less, because the radicals of the
water attack the solute. For most of the glycine and thymine samples the destruction rate falls as the temperature rises.
The destruction is not complete: at 13--20~K the decay curves approach an equilibrium
fraction $c\approx0.15$--0.4 of the parent for doses up to 10--17~eV per molecule
\citep{Materese2020,Gerakines2022,Tribbett2026}. We treat this equilibrium below as a return of the destroyed molecules to the parent.

We apply these measurements with the operator of Appendix~\ref{app:cycle}: each event
destroys the fraction $1-2^{-\evout\,\mathrm{eV}/D_{1/2}}$ of every sugar, irreversibly or with the
measured steady-state return. Figure~\ref{fig:radiolysis} shows the result. \radreached. \radnot. The observed ratio is reached
only for $D_{1/2}\gtrsim\dhalfthr$~eV per molecule, \dhalffac\ times above the most stable organic measured in dilute water ice and \dhalfgafac\ times above glycolaldehyde. The
reason is simple: the sugars must survive the \Ncom\ events of the count, each of which delivers
\evoutgly\ of the half-life dose of glycine at 200:1 (2.8~eV) and \evoutga\ times that of GA itself. The destruction does not remove the C3 excess: at the peak of the ratio DHA or GLD is still above its limit (Table~\ref{tab:nstar}). Nor does it help to keep the carbon of the destroyed sugars in the ice. We returned it to the inventory as glycolaldehyde units, as H$_2$CO or as CO, for each of the half-life doses above. The threshold stays between about 1000 and 2300~eV for H$_2$CO and CO, and it rises above 2300~eV for glycolaldehyde units, which replenish the denominator of the ratio (Table~\ref{tab:nstar}); at the measured half-life doses the ratio stays below one per cent in every case. A C4 sugar that is destroyed has to be rebuilt through many windows, and it is destroyed again on the way. What limits the ratio is the rate at which the windows build the C4 against the rate at which the dose removes it, not the loss of carbon.

The measured steady state is the one laboratory result that could keep the sugars.
If the fragments return to the parent at the measured rate, the sugar inventory settles
at the fraction $c$ of what it would otherwise be, and the ratio keeps rising. With
$D_{1/2}=2.8$~eV, the glycine value at 200:1 and 15~K, and $c=0.25$, the observed ratio is reached after \Nrevd\ events, \revfacd\ times the count without destruction, so the ionisation rate that an ice younger than 7~Myr requires rises to $\zetarevd$~s$^{-1}$, at the limit above which the gas does not stay molecular (Sect.~\ref{sec:origin}). This case rests on an extrapolation: the steady state has been
followed to 17~eV per molecule in the laboratory, and the exposure here is \dosefaclo--\dosefachi\ times larger (\doseouttot~eV per molecule at $\nstar$, \doserevd~eV at the count of the reversible case). Such doses would decompose the water itself many times over, a regime the measurements do not reach. The measurements also use 0.8--1~MeV protons, whose
ionisation is 6--40 times denser along the track than that of the 10--100~MeV protons and of the secondary electrons that carry the dose outside the melts here. For heavy ions the destruction cross
section of methanol ice rises as the 1--1.5 power of the stopping power
\citep{deBarros2011}, so destruction per unit dose falls at lower ionisation density,
by a factor of a few over that range; that does not close a gap of a factor of \dhalffac\ or more. Neither the concentration nor the temperature calls for a further correction. At $\nstar$ GA, ERU and DHA are each at \sugfracrange\ of the water (all sugars together \sugfractot), within the 0.3--5\% of the measurements, and the dust of G+0.693 is at their temperatures. A realistic half-life dose for the sugars is therefore at most about 10--30~eV per molecule, the largest measured value times the factor for the ionisation density. That is \dhalfgap\ below the threshold of \dhalfthr~eV.

\paragraph{Synthesis by the same dose.} The dose outside the melts may also make sugars in cold ice. Irradiated ices yield sugars after warm-up, with ultraviolet light \citep{deMarcellus2015,Meinert2016} and with electrons at 40~K \citep{Zhang2024}; the radiolysis runs book only the destruction. The omission does not favour the melt route. In the electron-irradiated ices of \citet{Zhang2024} the C3 sugars are at least as abundant as the C4 ones, and the only quantitative model of the cold route, that of \citet{JimenezSerra2026}, overproduces both C3 sugars by 25--70 times when it reaches ERU/GA $\approx1$ (Sect.~\ref{sec:intro}). A cold contribution would therefore shorten the count but raise the C3 excess, and the C3 limits would exclude it by the same argument. The radiolysis obstacle is stated for the sugars made in the melt: the dose that destroys them between events is the same whether or not it also makes others.

\subsection{The hypothetical case: what the pattern would require}
\label{sec:hypothetical}

Sections~\ref{sec:c3excess} and \ref{sec:radiolysis} each exclude the route on its own, and Sect.~\ref{sec:clock} constrains it. This section asks
the converse question: if the obstacles were removed, would the pattern follow, and
what would the chemistry then predict? The answer matters for two reasons. It shows
that the ketose selection of Sect.~\ref{sec:ketose} is not an artefact of the failure,
and it states, in the form of numbers, what a future measurement would have to find
for the route to be revived.

The hypothetical case is defined by three assumptions, each contrary to a
measurement or to a rate that the complete network takes at face value. (a) The C1
radicals do not add to the sugar radicals: the restricted network, in which
$\bullet$CH(OH)$_2$ reacts only with itself, OH and H, methanol is
inert and the whole 10.7\% substrate is H$_2$CO. No measured selectivity supports this
(Sect.~\ref{sec:caveats}); it is the network in which one carbon is added only by
$\bullet$CH$_2$OH, the species that the cold-phase hydrogenation of formaldehyde
supplies. (b) The sugars survive the dose between events: either the destruction is
set to zero, or the measured steady state of Sect.~\ref{sec:radiolysis} holds to the
full exposure. (c) The grains receive the melting events of a locally produced,
rising spectrum at $\zeta\gtrsim\zetares$~s$^{-1}$ (Sect.~\ref{sec:origin}).

\begin{figure*}
\centering
\includegraphics[width=\textwidth]{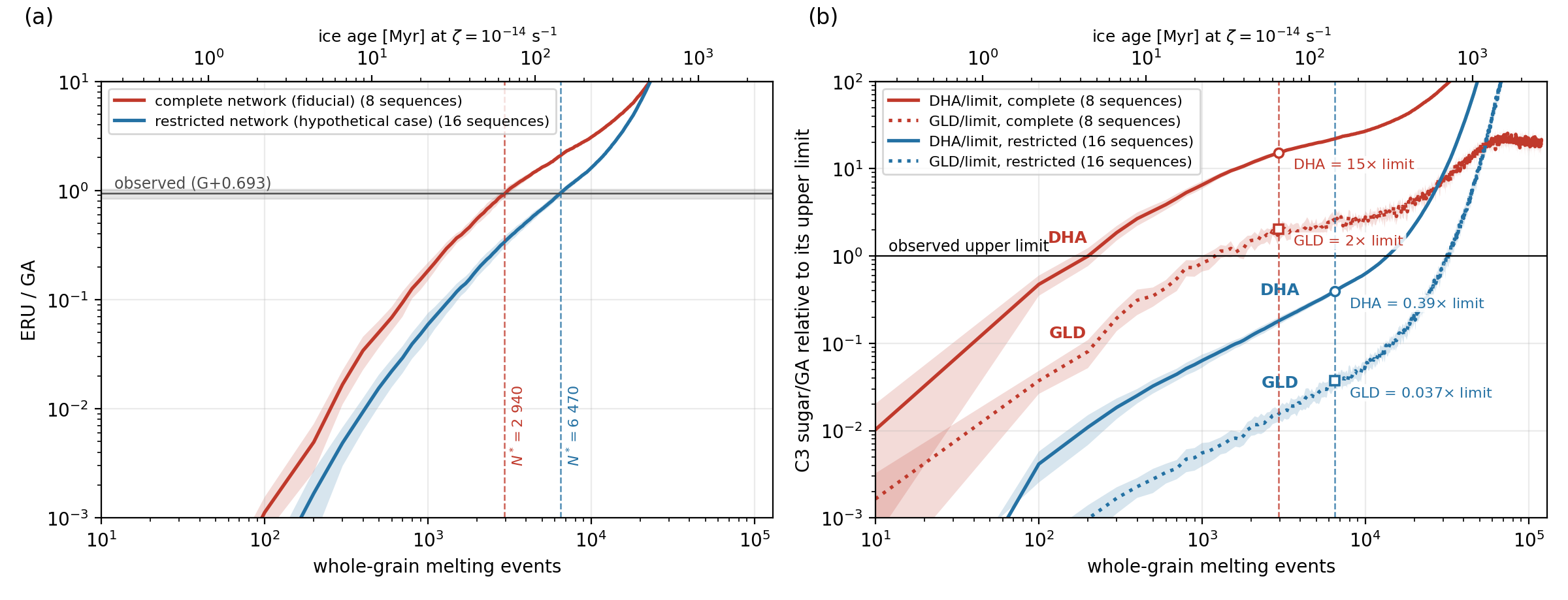}
\caption{The hypothetical case against the fiducial model. (a) ERU/GA against the
number of whole-grain melting events: red, the complete network (fiducial; means of 8
sequences of windows, bands their standard deviation); blue, the restricted network of
the hypothetical case (\nresseq\ sequences). The upper axis converts the count into time at
$\zeta=10^{-14}$~s$^{-1}$. The observed ratio (band) is crossed after \Ncom\ and
\Nres\ events. (b) DHA/GA (solid) and GLD/GA (dotted) relative to their observed upper limits for the two networks. The dashed vertical lines mark $\nstar$ of each network, and the open symbols the C3 ratios there.}
\label{fig:clockboth}
\end{figure*}

\begin{figure*}
\centering
\includegraphics[width=\textwidth]{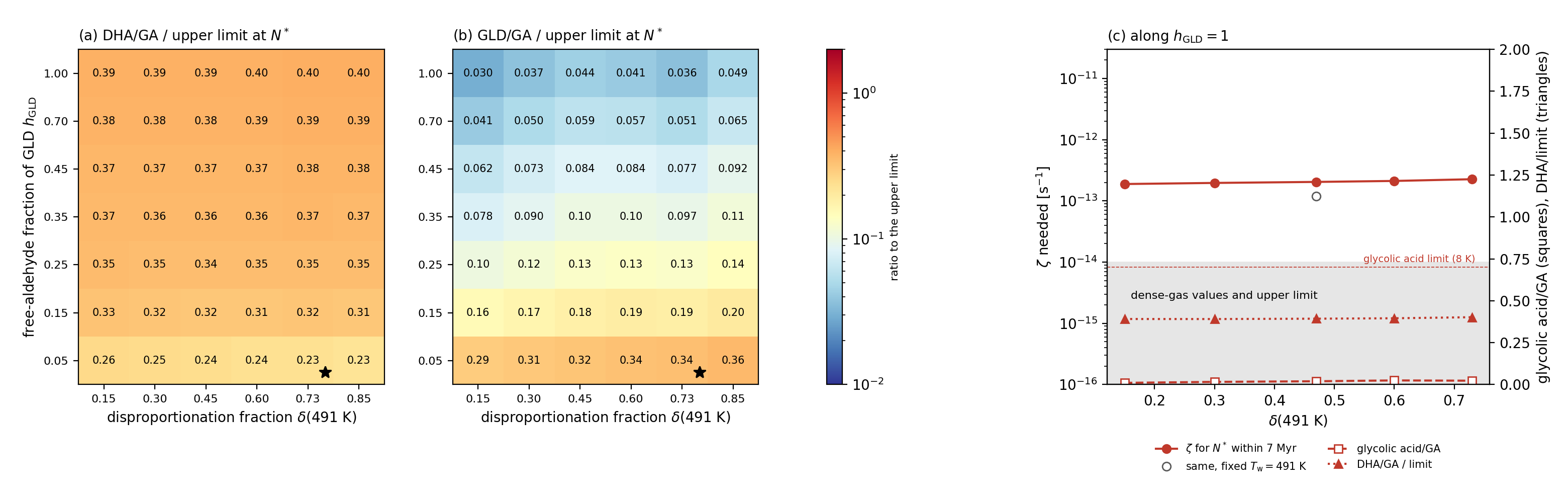}
\caption{The hypothetical case: the restricted network without radiolysis of the sugars between events. (a) DHA, the leading C3 sugar, relative to its observed upper limit (DHA/GA divided by 0.108), and (b) GLD relative to its limit (GLD/GA divided by 0.054), on the same logarithmic colour scale, with the value in each cell. The grids span the disproportionation fraction $\delta(491\,\mathrm{K})=0.15$--0.85 (the fiducial value 0.47 lies next to the column at 0.45) and the free-aldehyde fraction of GLD between events $h_\mathrm{GLD}=0.05$--1, with all other parameters at the literature values. Each cell is read at the count $\nstar$ at which ERU/GA first reaches 0.94 (means of four sequences of windows per cell). The star marks the literature
pair ($\delta=0.73$, $h_\mathrm{GLD}=0.05$). Lowering $h_\mathrm{GLD}$ moves the C3 from DHA to GLD; their sum stays at \sumgrid\ of the summed limit. (c) Along $h_\mathrm{GLD}=1$. Left axis: the ionisation rate needed to reach $\nstar$ within a cloud lifetime of 7~Myr (filled circles). The open circle is the same chemistry with every window at a fixed $T_\mathrm{w}=491$~K. The grey band marks the values measured in dense gas and the upper limit on the dense gas of the Central Molecular Zone. Right axis: glycolic acid/GA (squares, with its OH destruction) and DHA/GA relative to its limit (triangles), at $\nstar$. The
dashed line is the published limit on glycolic acid/GA for an excitation temperature
of 8~K.}
\label{fig:require}
\end{figure*}

Under (a) and (b) the pattern is reproduced (Fig.~\ref{fig:clockboth}). Table~\ref{tab:mainres} lists the model state at the count $\nstar=\Nres$ at which ERU/GA reaches 0.94 (\nresseq\ sequences of windows; standard deviation \Nressd): GLD/GA at \gldres\ and DHA/GA at \dhares\ of their limits, \ketres\ of the C4 as the ketose, hexoses at \hexres$\times$GA, glycolic acid at \glyres$\times$GA and CO$_2$/H$_2$O at \cotwores; HCOOH/H$_2$O is \hcoohres, above its observed ceiling (Sect.~\ref{sec:limits}). The count is \comfacinv\ times that of the complete network, because the C1 radicals no longer add to the C3 radical to make C4, and the ketose selection of Sect.~\ref{sec:ketose} holds without the C1 path: 90\% of the C4 is born from two gem-diol radicals or a gem-diol and an acyl radical, 9\% as the keto-aldehyde and 0.3\% as the dialdehyde. Figure~\ref{fig:require} maps the C3
sugars and the required ionisation rate over a grid in the two least certain chemical
quantities. The first is the disproportionation fraction $\delta$ of the radical
self-reactions: the transition-state decomposition of \citet{Kanjana2015}, evaluated at 491~K, gives
$\delta(491\,\mathrm{K})=0.73^{+0.16}_{-0.26}$, against a measured disproportionation-to-combination ratio of 0.17, i.e. $\delta=0.15$, at 298~K
\citep{Wang1996}. The second is the fraction $h_\mathrm{GLD}$ of GLD that survives the
interval between events as the free aldehyde rather than the gem-diol; the aqueous
hydration equilibrium $K_\mathrm{hyd}=19$ \citep{Nagorski2001} gives
$h_\mathrm{GLD}=0.05$, and only the free aldehyde carries the acyl-decarbonylation
channel that destroys C3. Each ratio is read at the count at which ERU/GA first crosses
0.94, because GLD/GA itself changes with exposure. Three results follow.

(i) \emph{The count depends little on either quantity.} $\nstar$ rises from \Nreslo\
events at $\delta(491\,\mathrm{K})=0.15$ to \Nres\ at 0.47, \Nresdxmid\ at 0.60 and \Nreshi\ at the literature central value, and it stays between \Nresgrid\ over the whole grid; $h_\mathrm{GLD}$ moves it by less than 1\%. An ice age inside a cloud lifetime of
7~Myr therefore requires $\zeta\gtrsim\zetareslo$~s$^{-1}$ at the
room-temperature branching, $\zetares$~s$^{-1}$ at $\delta(491\,\mathrm{K})=0.47$ and
$\zetareshi$~s$^{-1}$ at the literature central value (Fig.~\ref{fig:require}c).
With the measured steady state of the radiolysis between events
(Sect.~\ref{sec:radiolysis}) the requirement rises to $\zetarevresd$~s$^{-1}$. All of these
exceed the values inferred for dense gas by two orders of magnitude or more and the upper limit on the dense gas of the zone by a factor of 19 or more; the last also exceeds
the $3\times10^{-13}$~s$^{-1}$ at which the gas ceases to be molecular
(Sect.~\ref{sec:origin}). Condition (c) is what they state.

(ii) \emph{The C3 limits are met everywhere on the grid.} DHA lies at \dhagrid\ of its limit (Fig.~\ref{fig:require}a) and GLD at \gldgrid\ (Fig.~\ref{fig:require}b), so neither the hydration of
GLD nor $\delta$ is constrained by the C3 sugars. The hydration of GLD moves the C3 between the two isomers rather than changing its amount: OH attack on the free aldehyde passes through the C3 radical, which returns as DHA, whereas the hydrated GLD returns as GLD. The sum stays at \sumgrid\ of the summed limit over the whole grid. Within the restricted network, what the pattern then requires of the
chemistry is limited to two points: the trapped C4 radical returns as the ketose, and
GA is hydrated (Sect.~\ref{sec:ketose}). Both C3 limits were derived with the
excitation temperature fixed at 10~K; at the 20--22~K measured for GA they would be of
order twice as high (Table~\ref{tab:main}). The systematics of the restricted network
(Appendix~\ref{app:restricted}) keep DHA at \dhasys\ and GLD at \gldsys\ of their limits on every axis that follows the temperature histories and reaches the observed ratio, and the count within \Nsys. The lowest DHA and the highest GLD belong to the row that returns the C3 radical as the aldose.

(iii) \emph{Glycolic acid would test whether the acid survives the window.} With its
OH destruction ($k=6\times10^{8}$ M$^{-1}$\,s$^{-1}$; \citealt{Buxton1988}; products
taken as CO$_2$, H$_2$CO and H, so that no carbon-centred radical is re-injected),
glycolic acid/GA at $\nstar$ is \glyresdx\ over $\delta(491\,\mathrm{K})=0.15$--0.73 (\glyres\ at 0.47): over thousands of events the acid made in the long windows is
destroyed in all the others. Without the destruction channel the acid reaches \glyresnooh\ at $\delta(491\,\mathrm{K})=0.47$ and \glyresnoohhi\ at 0.73, and the sugar ratios are unchanged
within 5\%. \citet{JimenezSerra2020} searched for the acid towards G+0.693, and their
limit corresponds to glycolic acid/GA $\leq0.7$ with both columns from that work at 8~K, or $\leq0.15$ if the acid is excited like GA (Sect.~\ref{sec:predictions}). A destroyed acid meets both; a surviving acid
meets the looser limit only for $\delta(491\,\mathrm{K})\lesssim0.5$, and not the
stricter one.

The hypothetical case is therefore internally consistent and testable, and its
predictions are collected in Sect.~\ref{sec:predictions}. It is not, however, a model
of G+0.693: assumption (a) contradicts the diffusion-limited kinetics of every
hydroxyalkyl radical pair that has been measured, assumption (b) contradicts every
half-life dose measured in cold water ice, and assumption (c) contradicts every
ionisation rate measured in dense gas. Unless one of these measurements is found not
to apply to the ice of G+0.693, the ERU must be supplied by chemistry outside the melt
window, such as the cold surface route proposed in the
detection paper \citep{JimenezSerra2026}, and the C2:C4 ratio cannot be read as a
count of melting events.

\subsection{The model abundances, and the acids as ceilings}
\label{sec:limits}

Table~\ref{tab:main} gives the model state at $\nstar$. HCOOH and $\COO$ are not products exclusive to the sugar
network. Interstellar ice $\COO$ forms mainly by CO$+$OH on grains \citep{Ioppolo2011},
and both species are abundant ice constituents in their own right
\citep{Boogert2015}. For the sugar network they are therefore ceilings, not targets.
The $\COO$ ceiling is met: the network supplies $\COO/\HHO$ of \cotwocom\ at $\nstar$, below the medians of 19\% and 28\% observed towards massive and low-mass young stellar objects \citep{Boogert2015}. The HCOOH ceiling is
not: the network supplies HCOOH/$\HHO$ of \hcoohcom, \hcoohover\ times the upper end of the 0.5--6\% range
reported towards massive young stellar objects for a species whose identification is
not secure. The formic acid comes from the radiolytic oxidation of CO and H$_2$CO over the many events, not from the sugars. Halving the CO inventory (a systematic run in the restricted network, Appendix~\ref{app:restricted}) lowers it to \hcoohcohalf\ and changes $\nstar$ by \Ncohalfpct, GLD by \cohalfgld, glycolic acid by \cohalfgly\ and DHA by \cohalfdha. The base inventory, which controls the dissociation of the acid, moves it as well: in the restricted network HCOOH/$\HHO$ is \hcoohresbasezero\ without the base and \hcoohresbasetwo\ with twice the base (Appendix~\ref{app:network}). The excess therefore bounds the CO that the melts process and the base of the ice rather than the sugar chemistry. A CO inventory below the adopted 25\% of water is physically
plausible in G+0.693. CO is the most volatile of the major ice components, and its
abundance relative to water varies widely between lines of sight \citep{Boogert2015};
the dust of the Galactic-centre clouds is at 10--30~K \citep{RequenaTorres2006}, and
G+0.693 has been shocked \citep{Zeng2020}, so the outer layers of the mantle may have
lost part of their CO in past warm phases, or never frozen it out fully. Ethylene glycol
is the third entry of Table~\ref{tab:main} with an established route outside the sugar
network: the surface hydrogenation of CO makes it together with GA
\citep{Fedoseev2015}. In the window it is made by the combination of two
$\bullet$CH$_2$OH radicals and destroyed by OH. At $\nstar$ it stands at
\egcom$\times$GA, below the 2.5 observed for one conformer. The melts therefore supply a minority of the observed glycol. The shortfall is a measure of the cold route. In the simulations of \citet{Fedoseev2015} the surface hydrogenation of CO makes GA at 3--5\% of the H$_2$CO and ethylene glycol at 2--4\% of the methanol, and their experiments give yields of 1--9\%. We supply GA at 4\% of the initial H$_2$CO and EG at 3\% of the initial CH$_3$OH, and the same fractions of every H$_2$CO resupply. EG/GA is then \egcoldc\ at $\nstar$, still below the observed 2.5. The count is \Ncoldc\ (\tcoldc~Myr), DHA \dhacoldc\ times its limit and the ketose fraction \ketcoldc\ (Table~\ref{tab:nstar}). An EG yield of 12\%, three times the simulated range, brings EG/GA to \egcolda, at or above the observed value (a third pair is given in the table). The cold route at laboratory yields thus supplies part of the glycol and of the GA, shortens the clock to \coldcfac\ of its fiducial length, and leaves the C3 excess unchanged, because the melt makes the C3 from whatever GA it finds. The radiolysis of the melts, not the supply, limits the GA inventory. Every ratio that is compared with an observation is read at
$\nstar$, because a constraint derived at one exposure does not transfer to another.

\section{Discussion}
\label{sec:discussion}

\subsection{Conditions on the ionising particles, and how to test them}
\label{sec:origin}

The exposure of Sect.~\ref{sec:clock} is a condition on the ionising particles that depends on the chemistry only through the count: even the C2$\approx$C4 equality alone is reached only if the grains receive iron-group ions at 0.3--3~MeV~u$^{-1}$ at the rate that a locally produced spectrum with $\zeta\gtrsim\zetacom$~s$^{-1}$ delivers. We state three conditions on the particles (Conditions~1--3 below; they are distinct from the three conditions on the route as a whole in Sects.~\ref{sec:hypothetical} and \ref{sec:conclusions}), what the literature says about each, how each can be tested, and what could narrow the gap.

\paragraph{Condition 1: the ionisation rate in the dense gas.} The count of \Ncom\ melting events is reached within 7~Myr only for $\zeta\gtrsim\zetacom$~s$^{-1}$, and within 10~Myr only for $\zeta\gtrsim\zetacomten$~s$^{-1}$ (the hypothetical case of Sect.~\ref{sec:hypothetical} needs \Nres\ events and $\zeta\gtrsim\zetares$~s$^{-1}$ for 7~Myr). The count assumes that the radicals trapped at the end of a window are reduced between events; if all of them recombined, it would fall to \Nec\ events and $\zeta\gtrsim\zetaec$~s$^{-1}$ would suffice (Sect.~\ref{sec:caveats}).
Values of $\zeta=1$--$11\times10^{-14}$~s$^{-1}$ are inferred for the Central Molecular Zone, but for its diffuse gas, from H$_3^+$ absorption and models of photon-dominated regions \citep{LePetit2016}. In dense gas
the evidence points lower. The H$_3$O$^+$/H$_2$O ratio in the envelope of Sgr~B2 gives
$\zeta\approx4\times10^{-16}$~s$^{-1}$, with two independent uncertainties of a factor of
two each \citep{vanderTak2006}, and chemical models of the Sgr~B2(N) hot cores prefer
$\zeta\approx7\times10^{-16}$~s$^{-1}$ \citep{Bonfand2019}. For G+0.693 itself the
constraints come from chemical models of individual detections. Shock models reproduce PO$^+$ for
$\zeta=10^{-15}$--$10^{-14}$~s$^{-1}$ \citep{Rivilla2022PO}, models of HOCS$^+$ need $10^{-15}$--$10^{-14}$~s$^{-1}$ \citep{SanzNovo2024}, and the grain-surface models of the sugars match best for $\zeta\geq1.3\times10^{-14}$~s$^{-1}$ \citep{JimenezSerra2026}. All three are model-dependent. The high gas temperatures of the dense gas in the Central Molecular Zone do not require a high $\zeta$; turbulent heating explains them. The same analysis sets an upper limit $\zeta<10^{-14}$~s$^{-1}$ on the dense gas of the zone \citep{Ginsburg2016}. The limit holds if the cosmic-ray heating is uniform across the zone, and it is quoted without stating whether it is per H$_2$ or per hydrogen nucleus; a local enhancement in one cloud is constrained by it only through the gas temperature of that cloud. For gas at $n(\mathrm{H_2})=5\times10^{3}$~cm$^{-3}$ \citet{YusefZadeh2013} find equilibrium temperatures of 130 and 280~K for $\zeta=10^{-14}$ and $10^{-13}$~s$^{-1}$ per hydrogen nucleus. The highest value inferred in the region that still stands is the $\zeta\approx10^{-13}$~s$^{-1}$ per hydrogen atom, about $2\times10^{-13}$~s$^{-1}$ per H$_2$, inferred for the Arches cluster region from its
6.4~keV iron line, if low-energy ions produce that line \citep{Tatischeff2012}.
Condition~1 therefore requires the dense gas of G+0.693 to be ionised at the level inferred for the diffuse gas of the zone, \facgins\ times above the upper limit of \citet{Ginsburg2016} on its dense gas and \facdense\ times above the values inferred for the denser gas of Sgr~B2. The required value lies below that inferred for the Arches region and below the $\zeta\approx3\times10^{-13}$~s$^{-1}$ above which gas at $n(\mathrm{H_2})=10^{4}$~cm$^{-3}$ cannot stay molecular (see below). No measurement in the dense gas of the Galactic centre meets the condition, so the standard case fails it; against the upper limit the margin is a factor of \facgins, smaller than that of DHA and far smaller than that of the radiolysis between events. Of the three obstacles of Sect.~\ref{sec:results} this one is also the least certain, because it follows the count. With the escape yields it rises to $\zetaesc$~s$^{-1}$ and with a window dose equal to the melting enthalpy to $\zetadmelt$~s$^{-1}$, above the limit at which the gas stays molecular (Sect.~\ref{sec:caveats}); in the recombination limit of the trapped radicals it falls to $\zeta\gtrsim\zetaec$~s$^{-1}$, which the upper limit on the dense gas and the chemical models of this cloud allow and which is \facecdense\ times the values inferred for the dense gas of Sgr~B2. Condition~1 therefore stands or falls with the fate of the trapped radicals (Sect.~\ref{sec:caveats}).

\paragraph{Condition 2: the particles are cosmic rays, produced locally.} An external
spectrum attenuated to the centre of the cloud cannot give $\zeta=10^{-14}$~s$^{-1}$
unless the whole spectrum is enhanced over the local interstellar value. The factor
is 49 for the rising shape and 370 for the flat shape, relative to the spectra of \citet{Padovani2018}. The GeV $\gamma$-ray emission of the Central Molecular Zone shows no excess
corresponding to such an enhancement, so any excess of cosmic rays must be confined to
low energies \citep{Ravikularaman2025}. If $\zeta$ is as high as Condition~1 requires, the ionising
particles must therefore be a low-energy population that does not come from outside. \citet{Ravikularaman2025} propagated such populations through the Central
Molecular Zone and found that the power needed to sustain
$\zeta=2\times10^{-14}$~s$^{-1}$ over the whole zone, $10^{40}$--$10^{41}$~erg~s$^{-1}$,
is too large; they conclude that cosmic rays alone cannot explain the zone-wide
ionisation. Ionisation by X-ray photons would not help, because photons do not melt
grains. For a single cloud the budget is less severe than for the whole zone, but it is still
demanding. Sustaining $\zeta=10^{-14}$~s$^{-1}$ in a cloud of mass $M$ costs
$2.4\times10^{37}\,(M/10^{5}\,M_\odot)$~erg~s$^{-1}$ in ionisation, at 35~eV per ion
pair (as in Appendix~\ref{app:numerics}) and a mean mass of 2.8 hydrogen-atom masses per H$_2$
molecule (helium included), hence
$2.4\times10^{38}\,(M/10^{5}\,M_\odot)$~erg~s$^{-1}$ of mechanical power at an
acceleration efficiency of 10\%, and \powerfac\ times more at the $\zetacom$~s$^{-1}$ of Condition~1. The turbulent dissipation power of the cloud,
$M\sigma_v^{3}/\ell$, is $7\times10^{37}\,(M/10^{5}\,M_\odot)$~erg~s$^{-1}$ for a line width
of 20~km~s$^{-1}$ (full width at half maximum, i.e. a three-dimensional velocity
dispersion $\sigma_v=15$~km~s$^{-1}$) and a size $\ell=3$~pc. That is less than a third of
the power at $10^{-14}$~s$^{-1}$, and \powerfrac\ of that at $\zetacom$~s$^{-1}$. The cloud--cloud collision that is thought to shock this cloud involves
two components separated by $\sim$20~km~s$^{-1}$ \citep{Zeng2020}; for two equal
masses the kinetic energy in the centre-of-mass frame is
$10^{50}\,(M/10^{5}\,M_\odot)$~erg, which would sustain the mechanical power at
$10^{-14}$~s$^{-1}$ for $1.3\times10^{4}$~yr, about 500 times shorter than the 7~Myr
cloud lifetime. The cloud mass is not
measured; we quote every number per $10^{5}\,M_\odot$. Condition~2 therefore requires
that far more than the mechanical power of the cloud is converted into MeV particles,
and continuously replenished. The known budget falls short of this by more than an order of magnitude, and the condition has no direct observational support in this
cloud.

\paragraph{Condition 3: nuclei carry the ionisation and fill the cloud.}
The ages above assign all of $\zeta$ to nuclei. With that assignment the two spectral shapes give the same age to within 10\%, near the source and after attenuation through
$N_\mathrm{s}\geq10^{21}\,\percm$ (Fig.~\ref{fig:origin}a). If primary electrons supplied most of the ionisation, as in the flat model of \citet{Padovani2018}, the age would lengthen in proportion, by up to a factor of about nine (Sect.~\ref{sec:rate}). The ions that melt the grains, at
0.3--3~MeV~u$^{-1}$, have ranges of $8\times10^{19}$--$5\times10^{20}\,\percm$, i.e.
0.003--0.02~pc at
$n(\mathrm{H_2})=10^{4}$~cm$^{-3}$. The acceleration must therefore be distributed
through the cloud on that scale, as turbulent re-acceleration would be, rather than
confined to a few shock fronts. The ions this requires, at 0.3--3~MeV~u$^{-1}$, lie
below the threshold for $\pi^{0}$ production and are not constrained by the
$\gamma$-rays. Depletion of the refractory nuclei in the accelerated gas lengthens the age by a factor of \depfac\ (Sect.~\ref{sec:rate}).

\paragraph{What decides.} Condition~1 is measurable with existing data. The
ionisation rate of dense gas can be inferred from the column densities of HCO$^+$,
N$_2$H$^+$ and CO isotopologues with the analytic method of \citet{Luo2024}. The method needs the abundances of these species, which the existing single-dish surveys of G+0.693 provide, the gas density, to which the result is proportional, and assumed abundances of N$_2$ and O. It was developed for warm dense gas and is quoted as accurate to 50\% for hydrogen-nucleus densities of $10^{4}$~cm$^{-3}$ or more, gas temperatures of 50--100~K, visual extinctions of 4~mag or more and ages above $2\times10^{4}$~yr at solar and twice solar metallicity. It was tested up to $\zeta=10^{-14}$~s$^{-1}$. At a density of $10^{4}$~cm$^{-3}$ it underestimates $\zeta$ above $5\times10^{-15}$~s$^{-1}$, by up to a factor of three at $10^{-14}$~s$^{-1}$, because CO is destroyed; at $10^{5}$~cm$^{-3}$ and above it shows no such bias \citep{Luo2024}. At the density of G+0.693 a high value is therefore a lower limit, and the method has not been tested at the values that Condition~1 requires. It can exclude the route but not confirm it: a value well below $10^{-14}$~s$^{-1}$ would exclude the route for this cloud with standard assumptions, unless the bias grows steeply beyond the tested range. A higher value would call for an independent tracer of the dense-gas ionisation. If the route is excluded, the ERU would have to be made outside the window, for example
by the cold surface route of \citet{JimenezSerra2026}. In that case the chemistry
of Sects.~\ref{sec:ketose}--\ref{sec:hypothetical} still applies wherever windows of
$\gtrsim$0.3~ns do occur, near any local source of MeV ions. Glycolic acid
(Sect.~\ref{sec:predictions}) tests the survival of the acid in the window
independently of the spectrum.

Four ways of avoiding Conditions~1--3 can be considered; two fail and two lie outside
this model. A cloud whose sugars come only from its outer, cosmic-ray-exposed layer
fails because that layer holds no ice. At $r/R_\mathrm{cl}=0.99$, where the visual
extinction from the surface is 0.7~mag, the count takes \trniner~Myr; the outermost layer needs \trsurf~Myr but holds no ice either (Sect.~\ref{sec:rate}). Small grains melted as a whole by
one ion need fewer events but fail as well. A grain of radius $\lesssim$15~nm is melted through by a single iron-group ion at 0.2--7~MeV~u$^{-1}$; the drop then cools by evaporation in about a microsecond if the grain is bare ice, $10^{3}$ times longer than a track window. (A cold refractory core would conduct the heat away within nanoseconds.) In one run of the restricted network with 1-$\mu$s windows at 273~K and the room-temperature disproportionation fraction, the chemistry in such a drop reaches the observed ERU/GA after about \Nus\ events, with hexoses at \hexus$\times$GA. But such ions strike a 15~nm grain 2.2 times per Myr at $\zeta=10^{-14}$~s$^{-1}$ (5.0 times per Myr for 10~nm, which the ions from neon to sulphur also melt through), so the count needs \tusrange~Myr, or $\zeta\gtrsim\zetausrange$~s$^{-1}$ within 7~Myr. That is \usfacrange\ times above the upper limit on the dense gas. Each melt, moreover, evaporates about a fifth of the mantle, so the mantle does not survive the \Nus\ melts that the ratio needs. A power-law size distribution of the \citet{Mathis1977} type, extended to 5~nm, puts about half of the grain surface on such grains, but their mantles would be lost
long before the count is reached. Liquid windows of microseconds from grain--grain
collisions in the shocks are possible energetically and would supply exposure
independently of the cosmic rays; by the same small-grain run, the chemistry in them would reach the observed ratio after a few hundred events. An order-of-magnitude budget shows how narrow that route is. Melting both grains of a collision takes 0.11~eV per molecule from 10~K. Two equal grains colliding at relative speed $v$ have $mv^{2}/8$ per molecule in the centre-of-mass frame, with $m$ the mass of a water molecule. With half of that energy thermalised, $v$ must exceed about 3~km\,s$^{-1}$; above about 6.5~km\,s$^{-1}$ the energy exceeds the sublimation enthalpy and the collision vaporises the ice rather than melting it. Collisions at such speeds also shatter grains \citep{Tielens1994}, which this budget ignores. In the band between, a grain of radius 0.1~$\mu$m in gas of $n(\mathrm{H_2})=10^{4}$~cm$^{-3}$ collides $1.4\times10^{-11}$ times per second at 3~km\,s$^{-1}$. That rate assumes a grain density of $4\times10^{-8}$~cm$^{-3}$, from a dust-to-gas mass ratio of 0.01 in grains of density 3~g\,cm$^{-3}$, and a cross-section of $1.3\times10^{-9}$~cm$^{2}$. The grain then collides about 0.4 times in $10^{3}$~yr, our order-of-magnitude estimate of the time for which grains drift at such speeds through a C-type shock. A few hundred microsecond windows would then need several hundred shock passages, or a grain--grain velocity dispersion of 3~km\,s$^{-1}$ sustained for about \tuscoll~Myr, which would also strip the mantle at the same rate. A model of the collision rate and of the mantle lost per collision is the natural next step for this chemistry, because it does not depend on $\zeta$. A second possibility combines the two
routes of Sect.~\ref{sec:c3disc}: cold radical chemistry builds the inventory, and a
smaller number of melt windows then filters the C3 sugars through the decarbonylation
of the free aldehyde. Both are left to future work.

\paragraph{What could narrow the gap.} Two departures from the standard case could
narrow the gap, and both can be tested; neither closes it. The first is the ice itself. The
window length scales as $\kappa^{-0.8}$ with the thermal conductivity of the mantle
(Sect.~\ref{sec:rate}). With a conductivity ten times below that of compact
amorphous ice, the observed ratio is reached within 7~Myr for $\zeta\gtrsim\zetakcom$~s$^{-1}$, a factor of \kfac\ below the compact case. That is within the range measured in the diffuse gas of the zone, but still \kgins\ times the upper limit on its dense gas. Vapour-deposited amorphous ice can be porous or cracked. Conductivities as low as $10^{-5}$~W\,m$^{-1}$\,K$^{-1}$, five orders of magnitude below the compact value, have been inferred for it \citep{Kouchi1992}, who attribute them to microcracks; the bulk amorphous ices measured directly conduct like the compact law (Sect.~\ref{sec:chain}). Relative to that report a factor of ten is a conservative assumption. The mantle would also have to stay porous under the
irradiation that melts it, because swift heavy ions compact porous ice
\citep{Dartois2013}. A porous mantle is therefore possible only if accretion rebuilds
the porosity between events faster than the ions remove it. The scaling of the window
length with $\kappa$ alone does not model a porous mantle, whose density, melt
enthalpy per volume and window temperature would all differ; it bounds the direction
and the order of the effect. The second is the ionisation history. The chemistry depends
on the number of melting events, not on when they occurred, so an episode of higher
ionisation in the past would count. The Galactic centre has had such episodes. The Fermi bubbles \citep{Su2010} record an energy injection at the Galactic centre, dated to an outburst of Sgr~A* 1--3~Myr ago by \citet{BlandHawthorn2013}. X-ray reflection in the molecular clouds records a period of high luminosity of Sgr~A* that illuminated Sgr~B2 about 300~yr ago \citep{Koyama1996} and faded about 100~yr ago \citep{Ponti2010}. Two constraints bound this possibility. The cloud must remain molecular, which at $n(\mathrm{H_2})=10^{4}$~cm$^{-3}$ requires $\zeta\lesssim3\times10^{-13}$~s$^{-1}$, in proportion to the density. The destruction of H$_2$ by cosmic rays, on a timescale $\sim$$1/\zeta$, would otherwise outrun its formation on grains, whose rate coefficient is $3\times10^{-17}$~cm$^{3}$~s$^{-1}$ \citep{Jura1975} times the hydrogen density. This order-of-magnitude balance agrees with the transition from atomic to molecular gas at $\zeta/n\simeq1.5\times10^{-17}$~cm$^{3}$~s$^{-1}$ at 100~K, with $n$ the density of hydrogen nuclei, in the models of \citet{BialySternberg2015}. About a fifth of the hydrogen is already atomic at the limit, and CO is destroyed at lower ionisation rates \citep{Luo2024}. The iron-group ions that make the windows also
sputter the mantle, at $10^{3}$--$10^{4}$ molecules per ion near the stopping maximum \citep[see also \citealt{BringaJohnson2004}]{Dartois2015}. Such ions, at 0.3--3~MeV~u$^{-1}$, strike a grain of radius 0.1~$\mu$m about 55 times per
Myr at $\zeta=10^{-14}$~s$^{-1}$ (Sect.~\ref{sec:rate}). With the yield proportional to the square of the stopping power and normalised to the measurement for nickel \citep{Dartois2015}, the whole fiducial spectrum removes \sputrate\ molecules per Myr at each surface crossing, \sputfeband\ of them by those iron-group ions and \sputlight\ by protons and helium nuclei. Over the \tcom~Myr of the count this is \sputlo--\sputhi\ of a mantle of
$1.3\times10^{8}$ molecules, for one and two surface crossings per ion. A shorter interval between events does not starve the cold-phase
reduction, because the atomic hydrogen density of the gas rises in proportion to
$\zeta$ (Appendix~\ref{app:coldh}). Within
these bounds, an episode would have to last \tepmol~Myr at the molecular limit of $3\times10^{-13}$~s$^{-1}$ and \tepthirteen~Myr at $10^{-13}$~s$^{-1}$, a third to nearly the whole of the cloud lifetime. The recorded activity of the Galactic centre gives no evidence of an episode of that length in the dense gas. The ice mantle would then
record the integrated ionisation history of the cloud rather than its present state.

\subsection{The two routes fail the C3 sugars for the same reason}
\label{sec:c3disc}

Two routes can build the sugars of G+0.693: recombination on cold grain surfaces, and
the melt window of this paper. Both reproduce the C2$\approx$C4 equality, and both
overproduce the C3 sugars. The published grain-surface model does so by the recombination of C1 and C2 radicals on cold amorphous solid water, in the network that also builds the C4 from GA- and ethylene-glycol-derived radicals \citep{JimenezSerra2026}, and exceeds the C3 limits by factors of 25--70 (their Sect.~2.3). The melt-window model does so by the same radical
chemistry in a liquid window, and exceeds the limit of DHA by \dhacom\ times
(Sect.~\ref{sec:c3excess}). The cause is common to both: the C1 radicals of
formaldehyde and methanol add one carbon to any sugar radical, so no rung of the ladder is skipped. The melt adds two filters, the hydration of GA and the decarbonylation of
the free aldehyde, that the surface route lacks. They select the C4 ketose and remove
part of the aldose GLD, but they do not touch the supply of C3. Cold radiolysis in the
laboratory gives the same picture: electron irradiation of methanol ice at 40~K leaves GLD 20--120 times more abundant than DHA in the residues analysed after warm-up, over doses of 1--50~eV per atomic mass unit \citep{Zhang2024}, i.e. C3 sugars in quantity.

The observed scarcity of the C3 sugars therefore excludes both routes as they stand,
and it asks for a chemistry that skips C3. Within the melt route the only way is the
hypothetical restricted network, in which $\bullet$CH(OH)$_2$ does not combine with
the sugar radicals and methanol is inert. The C3 then comes only from the
$\bullet$CH$_2$OH of the H addition to formaldehyde and stays below the limits, with DHA as the leading C3 sugar and GLD \gldoverdha\ times lower
(Sect.~\ref{sec:hypothetical}). A deep search for the two C3 sugars, whose spectra
are known, remains the most direct test of any radical route. A detection of either
at or above its present limit would confirm the monotonic ladder, and a limit ten times deeper would exclude the hypothetical restricted network, as the present limits exclude the complete network and the surface model.

\subsection{Predictions of the complete network and of the hypothetical case}
\label{sec:predictions}

The first two predictions hold for the complete network and are what a melt route, as
the measured kinetics describe it, would show. The remaining ones are those of the
hypothetical case of Sect.~\ref{sec:hypothetical}: they state what would be observed
if the three obstacles were removed, and they are the tests that would revive the
route.

\begin{enumerate}
\item \textbf{A melt route gives a monotonic sugar ladder.} In the complete network the
C3 sugars reach \dhacom\ (DHA) and \gldcom\ (GLD) times their limits when ERU/GA
reaches 0.94, with DHA the leading C3 sugar because its radical returns as the ketose
(Sect.~\ref{sec:c3excess}). A detection of DHA near its present limit, or of GLD, would
be consistent with a radical ladder of either route; the present limits exclude both.
\item \textbf{The C4 is a ketose in any melt in which GA is hydrated.} In the complete
network \ketcom\ of the C4 is the ketose, most of it born by the addition of a C1 radical to the trapped radical of DHA; in the restricted network \ketres\ is
(Sect.~\ref{sec:ketose}). Without the hydration of GA the C4 is the aldotetrose in
the restricted network and \hnonecom\ the aldotetrose in the complete one
(Table~\ref{tab:nstar}). The grain-surface model of \citet{JimenezSerra2026} also makes the ketose ERU, so the ketose alone does not separate the two routes.
\item[] \emph{Hypothetical case (restricted network, no radiolysis between events):}
\item \textbf{Glycolic acid would test whether the acid survives the window.} The same gem-diol radical pairs that make the C4 ketose make glycolic acid by disproportionation. If OH destroys it at the aqueous rate, it lies at \glyres$\times$GA at $\nstar$ (\glyresdx\ across $\delta$, and \glyressys\ across the other systematics of Appendix~\ref{app:restricted} along the temperature histories, except \glyresch\ without the cold-phase reduction); the complete network gives \glycom. If it survives the
window, it is \glyresnooh$\times$GA at the fiducial $\delta$ and \glyresnoohhi$\times$GA at the
literature value (\glynoohcom\ in the complete network). \citet{JimenezSerra2020} found $N\leq2.3\times10^{13}\,\percm$ towards G+0.693, for
an excitation temperature $T_\mathrm{ex}=8$~K. Their GA column at the same temperature gives glycolic
acid/GA $\leq0.7$. If the acid is excited like GA in the more recent analyses
($T_\mathrm{ex}=22$~K; \citealt{Rivilla2022,JimenezSerra2026}), the same line gives
$\leq0.15\times$GA (our evaluation in local thermodynamic equilibrium, LTE). The destroyed acid lies below both limits.
The surviving acid lies just below the limit of 0.7 at the fiducial $\delta$, exceeds it at the literature value, and exceeds the limit of 0.15 at both. A detection near 0.02$\times$GA would support the destruction of the acid in the window. A detection near the published limit would require a surviving acid and a $\delta(491\,\mathrm{K})$ at or above the fiducial value.
\item \textbf{The C3 ketose lies a factor of \dhabelow\ below its limit; the C3 aldose far below it.} DHA lies at \dhares\ of its upper limit (\dhafix\ across the systematics of Appendix~\ref{app:restricted} along the temperature histories, at
fixed reduction site and hydration), because its trapped radical returns as the
ketose (Sect.~\ref{sec:ketose}), whereas GLD lies at \gldres\ of its limit (\gldreshyd\ with its aqueous hydration). The first C3 sugar to be detected should therefore be DHA, at \dhagares$\times$GA (\dhagafix), with GLD at about \gldgares$\times$GA (\gldgareshyd$\times$GA with its aqueous hydration); this prediction
rests on the return of the trapped C3 radical as the ketose
(Sect.~\ref{sec:ketose}). A detection of GLD at or above the level of
DHA would falsify the reduction site adopted here, and a DHA
limit well below 0.01$\times$GA would require a partial return of the C3 radical as the
aldose (Sect.~\ref{sec:ketose}).
\item \textbf{The melt resets the deuteration of the sugars to that of the water.} Over the exposure each sugar molecule is attacked by OH and repaired by H many times, so its C--H hydrogen is drawn from the radiolytic pool of the melt, whose D/H is that of the water ice. ERU and GA made in melts should share that value, with no enhancement in ERU. The cold route builds ERU from HCO and $\bullet$CH$_2$OH formed on grains at 10~K, which inherit the elevated D/H of cold grain chemistry. In G+0.693 the D/H ratio differs between species. In the broad component that carries the complex organics it is at most $6\times10^{-5}$ for HCO$^+$ and N$_2$H$^+$ and $(4$--$13)\times10^{-4}$ for HCN and HNC \citep{Colzi2022}. At a ratio of $10^{-3}$ the lines of a singly deuterated GA are some $10^{3}$ times weaker than those of GA, below the present surveys. This test therefore needs deeper data or a source with a higher D/H.
\end{enumerate}

Seven further predictions of the hypothetical case are collected in Table~\ref{tab:pred}.

\begin{table*}
\caption{Further predictions of the hypothetical case (restricted network, at $\nstar$ unless stated), and what would test each.}
\label{tab:pred}
\centering
\small
\begin{tabular}{p{0.30\textwidth}p{0.30\textwidth}p{0.34\textwidth}}
\hline\hline
Prediction & Model value & Test \\
\hline
Hexoses accompany ERU & $\approx$\hexres$\times$GA (\hexresdx\ across $\delta$); \bigres\ of the sugar molecules are C6 or larger; \hexrestwo$\times$GA at $2\nstar$ & No interstellar hexose is detected; a limit well below 0.1$\times$GA towards G+0.693 would falsify the ratchet \\
The aldotetrose accompanies ERU & Erythrose plus threose at $\approx$\aldres$\times$GA, growing with exposure & The aldehyde-radical channel cannot be switched off \\
The sugar network makes methanol & \meohressys$\times$GA across the axes that reach the ratio, apart from the two that change the cold-phase budget and the fate of the small radicals (\meohres\ fiducial; \meohresdx\ across $\delta$), from the cold-phase reduction of $\bullet$CH$_2$OH \citep{WatanabeKouchi2002} and its disproportionation & Methanol column of G+0.693 \\
GA in the ice is the gem-diol & Without its hydration \hnoneresald\ of the C4 is the aldotetrose and the ratio is reached only after \Nreshnone\ events, when GA is gone (Sect.~\ref{sec:ketose}); free H$_2$CO gives \kethfone--\kethfhalf\ ketose in the complete network (\kethfonek\ if the C3 radical returns as the ketose); hydrated DHA rises from \dhares\ to \dhareshyd\ of its limit & Transmission Fourier-transform infrared (FTIR) spectroscopy of H$_2$O:carbonyl ices at 10--20~K (Sect.~\ref{sec:caveats}) \\
HCOOH and CO$_2$ from the melts & HCOOH/H$_2$O $\approx$\hcoohrange, above the observed upper end, from the oxidation of CO and H$_2$CO; CO$_2$/H$_2$O \cotwocom\ (complete) and \cotwores\ (restricted), against observed medians of 19--28\% (Sect.~\ref{sec:limits}) & An ice HCOOH abundance securely below 6\% would require less CO processing by the melts or more base in the ice \\
The dense-gas ionisation rate decides & Within 7~Myr the count needs $\zeta\gtrsim\zetacom$--$\zetares$~s$^{-1}$ (Sect.~\ref{sec:origin}); $\zetaec$~s$^{-1}$ in the recombination limit of the trapped radicals; a porous mantle still needs $\zetakcom$~s$^{-1}$ & $\zeta$ from HCO$^+$, N$_2$H$^+$ and CO \citep{Luo2024}: a value well below $10^{-14}$~s$^{-1}$ excludes the route with standard assumptions; a higher value is only a lower limit \\
Desorption must be non-thermal & Equal desorption fractions for GA and ERU (Sect.~\ref{sec:clock}); thermal selection would suppress the C4 by orders of magnitude & A G+0.693-like pair in a quiescent, shock-free cloud would contradict the framework \\
\hline
\end{tabular}
\end{table*}

\emph{Observability.} Glycolic acid and DHA can both be sought in the existing
single-dish surveys of G+0.693, made with the Yebes 40\,m telescope at 31--50~GHz and
the 30\,m telescope of the Institut de Radioastronomie Millim\'etrique (IRAM) in three bands between 83 and 174~GHz, with an rms of 0.25--2.5~mK per channel
\citep{JimenezSerra2026}. The emission of the complex organics there is extended over
the beams, sub-thermally excited ($T_\mathrm{ex}\lesssim15$~K for most species, 22~K for GA) and about
20~km\,s$^{-1}$ wide \citep{JimenezSerra2020,JimenezSerra2026}. We evaluated the lines
in LTE at these temperatures with the catalogues of the Cologne Database for Molecular
Spectroscopy (CDMS) and of the Jet Propulsion Laboratory (JPL)
\citep{Endres2016,Pickett1998}; the same evaluation reproduces the four strongest GA lines of \citet{JimenezSerra2020} to within 1\% (Appendix~\ref{app:lines}). Glycolic acid at $10^{13}\,\percm$
(0.1$\times$GA) gives line peaks of \obsglyrefseven~mK at 31--50~GHz and \obsglyrefthree~mK at
83--115~GHz, with 16--19 lines within a factor of two of the strongest at 7~mm and 7--27 at 3~mm, depending on $T_\mathrm{ex}$. Components closer than the line width are counted as one line (Appendix~\ref{app:lines}). A
multi-line fit of the existing spectra can therefore reach this level, which separates
the surviving acid (\glyresnooh--\glyresnoohhi$\times$GA) from the destroyed one. The destroyed acid, at \glyres$\times$GA, gives peaks \obsglyweak\ times weaker, \obsglypk~mK; the same significance in a single line then needs about \obsglytime\ times the integration time. DHA at \dhagares$\times$GA gives peaks of about \obsdhapk~mK. A single \obssinglepk~mK line at $5\sigma$ needs an rms of \obssinglerms~mK, about \obssingle\ times the present integration at the best 7~mm rms of 0.25~mK. A stack recovers most of this. At 7~mm, \obsglyn\ lines of glycolic acid lie within a factor of two of the strongest (\obsdhan\ for DHA); Table~\ref{tab:lines} lists the four strongest of each band. A stack of these lines, with their signal-to-noise ratios added in quadrature, reaches $5\sigma$ with \obsglystack\ times the present integration for the destroyed acid and \obsdhastack\ times for DHA, at $T_\mathrm{ex}=8$--22~K. Where the rms is 0.9~mK the factors are \obsdhastackhi\ for DHA and \obsglystackhi\ for the acid. The stack must leave out lines blended with other species, which in a survey this rich will remove part of them. The
strongest lines of both molecules lie at 7 and 3~mm. Interferometers are less suited,
because maps of this cloud lose the extended flux without single-dish data
\citep{Zeng2020}.

\subsection{The clock in other reservoirs}
\label{sec:reservoirs}

The ERU/GA ratio of the complete network is a monotonic function of the count of whole-grain melting events. If the sugars survived between events, it could therefore be read as a clock wherever the ionisation rate, the spectral shape and the exposure time are known. Table~\ref{tab:reservoirs} applies it to other reservoirs of ice, with the melting rates of Table~\ref{tab:rates} and the mean curves of Fig.~\ref{fig:clockdelta}, extrapolated below 100 events as power laws. The substrate is that of G+0.693 (H$_2$CO and CH$_3$OH at 10.7\% of the water), so the values are indicative. Two conclusions are robust. In the dense gas of the solar neighbourhood and in prestellar ices the count stays below one, and the melt route makes no C4 at all; a detection of ERU in such a source would point to another route. The outer 10~$\mu$m of a comet nucleus, the range of the iron-group ions that melt, is exposed to the unshielded flat spectrum for 4.5~Gyr. The nuclei of model $\mathcal{L}$ of \citet{Padovani2018}, unattenuated, give $\zeta=4\times10^{-17}$~s$^{-1}$ by Eq.~(\ref{eq:zetaown}). At that rate the layer receives about \Ncomet\ events, which gives ERU/GA of \erucomet\ and DHA/GA of \dhacomet\ on the fiducial curves. These curves belong to the window distribution of the rising shape; with that of the flat shape, whose count is \Nflat, ERU/GA would be about 0.1. The melt route therefore predicts erythrulose well below glycolaldehyde in comets. At comet 67P/Churyumov--Gerasimenko ethylene glycol and glycolaldehyde appear in the fit of the mass spectrum of the COSAC instrument on the Philae lander \citep{Goesmann2015}, and the ROSINA mass spectrometer on Rosetta finds a signal that may belong to ethylene glycol \citep{Schuhmann2019}; erythrulose has not been reported, although mass spectrometry does not separate it from its isomers. At the diffuse-gas value of $10^{-13}$~s$^{-1}$ sustained for 7~Myr the ratio reaches \erudiffuse, with DHA \dhadiffuse\ times above its limit.

\begin{table*}
\caption{The ERU/GA clock applied to other reservoirs (complete network, fiducial chemistry). $N$ is the count of whole-grain melting events for the ionisation rate $\zeta$, spectral shape and exposure given; ERU/GA and DHA/GA are read from the fiducial curves, which belong to the window distribution of the rising shape (the flat rows are therefore lower limits; Table~\ref{tab:rates}) and do not include the radiolysis of the sugars between events (Sect.~\ref{sec:radiolysis}); the entries are therefore what the melt chemistry alone would give.}
\label{tab:reservoirs}
\centering
\small
\begin{tabular}{llllrll}
\hline\hline
Reservoir & Spectrum & $\zeta$ (s$^{-1}$) & Exposure (Myr) & $N$ & ERU/GA & DHA/GA \\
\hline
G+0.693 (this paper) & rising, inside & $1\times10^{-14}$ & 7 & 317 & 0.019 & 0.21 \\
G+0.693, diffuse-gas $\zeta$ & rising, inside & $1\times10^{-13}$ & 7 & 3\,170 & 1 & 1.7 \\
Sgr B2 envelope & rising, inside & $4\times10^{-16}$ & 7 & 13 & $2\times10^{-6}$ & $2\times10^{-3}$ \\
Dense cloud, solar neighbourhood & flat, from outside (centre) & $3\times10^{-17}$ & 10 & 0.15 & --- & --- \\
Comet nucleus, outer 10~$\mu$m, Oort cloud & flat, inside & $4\times10^{-17}$ & 4500 & 324 & 0.02 & 0.22 \\
Prestellar ice before collapse & flat, inside & $1\times10^{-16}$ & 1 & 0.18 & --- & --- \\
\hline

\end{tabular}
\end{table*}

\subsection{Sensitivity to the model parameters}
\label{sec:sensitivity}

Table~\ref{tab:nstar} collects the response of the five observables of the complete
network to every axis we varied. Ten axes move the count by a factor of two or more. A window dose equal to the melting enthalpy raises it to \Ndmelt, half the window dose to \Ndosehalf, the reversible radiolysis between events to \Nrevd, the removal of the $\bullet$CH(OH)$_2$ cross-combination with methanol left active to \Ndhmoff\ and the hydration of DHA to \Nhdha. Doubling the window dose lowers it to \Ndosetwo, the survival of the trapped small radicals to \Nsmalloff, and the absence of their reduction to \Nchzero. Free H$_2$CO, for half of the H$_2$CO, lowers it to \Nhfhalf\ (\Nhfone\ for all of it; \Nhfhalfk\ and \Nhfonek\ with the C3 radical booked as the ketose). The tenth is the recombination limit of Sect.~\ref{sec:caveats}, \Nec\ events. The escape yields raise the count by a factor of \escfac, to \Nescape, the closed substrate by a similar factor, and the base inventory moves it by \pctbasezero\ (no base) and \pctbasetwo\ (twice the base). The percentages of this section are relative to the four sequences that the axis runs share with the fiducial model, whose count is \Ncomfour. The envelope of Fig.~\ref{fig:zeta} runs from
$\nstar=\nstarlo$ to \nstarhi; the three rows that change the fate of the trapped radicals lie below it and are drawn or quoted separately. The temperature cap moves the count by \capfac\ over
450--650~K, the disproportionation fraction $\delta$ by \dxpctlo\ to \dxpcthi\ over 0.15--0.73, and the remaining axes by less than \restpct\ (most for the cold-route C2, the unhydrated GA and the return of the C3 radical as the aldose). The irreversible destruction of the sugars reaches the ratio only for $D_{1/2}\gtrsim\dhalfthr$~eV per molecule (Sect.~\ref{sec:radiolysis}), and the return of the C4 radical half as the aldotetrose does not reach it (ERU/GA peaks at \pkerysym).

The C3 excess is the robust result. DHA stays at \dharange\ times its limit on every axis that reaches the ratio, except two that move the partition of the C3 rather than its amount. Its own hydration raises it to \dhahdha, and the return of the C3 radical as the aldose moves the excess to GLD (\gldcthreeald\ times its limit). GLD is at \gldmostrange\ times its limit on most axes. It is at \gldec\ in the recombination limit, at \gldhgld--\gldlit\ with its aqueous hydration and at \gldch\ without the cold-phase reduction. The sum of the two C3 sugars lies at \cthreesumrange\ times the sum of the limits on every axis, and at \cthreesumhdha\ with DHA hydrated. Glycolic acid, destroyed in the window, stays at or below \glymax$\times$GA except without the cold-phase reduction (\glych) and in the recombination limit (\glyec); without its OH destruction it is \glynoohcom$\times$GA, against \glyresnooh\ in the restricted network.

The ketose share of the C4 is \ketmostrange\ on most axes, \ketsmalloff\ with surviving small radicals, \kethfone--\kethfhalf\ with free H$_2$CO and \ketclosed\ with the substrate closed. It is \ketmeohoff\ with methanol inert and \ketcomwin\ when the window products of the C3
radical are booked as the ketose, because the C4 born from $\bullet$CH$_2$OH and the
C3 radical, which the network books as the aldotetrose, is then absent or rebooked
(Sect.~\ref{sec:ketose}). It falls to \ketdhmoff\ without the $\bullet$CH(OH)$_2$ cross-combination, when $\bullet$CH$_2$OH makes most of the C4, to \kethnone\ without the hydration of GA and to \keterysym\ when the C4 radical returns half as the aldotetrose. The
order in which the windows arrive matters little: the standard deviation of $\nstar$
between sequences is \sdrange.

\subsection{Caveats and parameter provenance}
\label{sec:caveats}

Table~\ref{tab:experiments} lists the laboratory measurements that would fix the assumptions on which the conclusions depend most, with the effect that each has in the model. The paragraphs below discuss them in turn.

\begin{table*}
\caption{Laboratory measurements that would fix the principal assumptions, and their leverage in the model (Table~\ref{tab:nstar}).}
\label{tab:experiments}
\centering
\small
\begin{tabular}{p{0.25\textwidth}p{0.19\textwidth}p{0.25\textwidth}p{0.21\textwidth}}
\hline\hline
Quantity & Assumed value & Measurement & Leverage \\
\hline
Hydration state of GA and H$_2$CO in H$_2$O ice at 10--20~K & GA 90\% gem-diol, H$_2$CO hydrated (birth states) & Transmission FTIR of H$_2$O:GA and H$_2$O:H$_2$CO ices; C$=$O band on warming; D$_2$O labelling & Ketose fraction \ketcom\ $\to$ \kethnone\ (GA free) or \kethfone--\kethfhalf\ (H$_2$CO free); count $\nstar$ $\to$ \Nhnone\ or \Nhfhalf--\Nhfone\ \\
Disproportionation fraction $\delta$ of the gem-diol radical pairs at 350--490~K & 0.47 (transition-state extrapolation of the 298~K value) & Pulse radiolysis of GA and H$_2$CO solutions at elevated temperature; product analysis & Count \dxpctlo\ to \dxpcthi\ over 0.15--0.73; glycolic acid $\times$2 \\
Dehydration of the gem-diol radicals to acyl radicals & Neglected (rate not measured); run at $10^{8}$ and $10^{9}$~s$^{-1}$ & Pulse radiolysis or electron spin resonance of $\bullet$C(OH)$_2$CH$_2$OH in water & Count \Ngemdehyda\ and \Ngemdehydb; ketose fraction \ketgemdehyda\ and \ketgemdehydb \\
Half-life dose of GA and ERU in H$_2$O ice at 10--20~K beyond 20~eV per molecule & 0.2--4.6~eV per molecule (other organics at $\leq$5\%) & Proton irradiation with infrared and mass-spectrometric monitoring & Ratio peaks at \pkthirty\ for $D_{1/2}=2.3$~eV and \pkten\ for 6.9~eV, and is reached only for $D_{1/2}\gtrsim\dhalfthr$~eV; reversible survival of 25\% raises $\nstar$ to \Nrevcom \\
Radical yields at the linear energy transfer of iron at 1~MeV~u$^{-1}$ & Primary yields of radiolysis at low linear energy transfer & Heavy-ion radiolysis of frozen dilute solutions with scavengers & Count $\times$\escfac\ with escape yields; $\times$\dosehalffac\ with half the dose \\
Mobility of H, OH and small radicals in compact amorphous ice below 200~K & $f_\mathrm{s}<6\times10^{-5}$ for H and $<1.5\times10^{-3}$ for OH (molecular-dynamics upper bounds at 240~K) & Isotope-labelled diffusion in ice films; molecular dynamics with the potential of Appendix~\ref{app:qcmd} & Whether any chemistry runs without melting (Sect.~\ref{sec:meltgated}) \\
Melt radius and liquid lifetime of an iron track & Continuum track model, crystallisation limit & Molecular dynamics of a 1~MeV~u$^{-1}$ iron track with the machine-learned potential & Window length; the vitrification limit changes the age by \vitfac \\
\hline
\end{tabular}
\end{table*}

\paragraph{The hydration state of the small aldehydes.} This assumption decides whether
the C4 is the ketose, and where the C3 aldose sits relative to its limit. It is
testable. The
model assigns each aldehyde a fixed carbonyl availability: GA at its
aqueous equilibrium (90\% gem-diol) and GLD as the free aldehyde. Under
the frozen kinetics of Sect.~\ref{sec:ratchet} neither value can be reached by
equilibration in the ice; both are birth-state assumptions, the first taken from
water. The aqueous hydration constants of the two are similar ($K_\mathrm{hyd}=9$--25 for GA and 19 for GLD; \citealt{Sorensen1972,Nagorski2001}), so if the ice realises this difference it must
be kinetic. The state that each molecule acquires when it is formed, or during the
brief liquid, persists. The interval between melting events, about 22~kyr at \Rwge\ whole-grain events per Myr, is spent at 10--20~K,
where both hydration and dehydration are frozen. The dehydration of the GLD hydrate,
measured at 0.03~s$^{-1}$ at 25~$^{\circ}$C \citep{Rendina1984}, is slower than the
age of the Universe at 10~K for any barrier above 20~kJ\,mol$^{-1}$. A transmission-FTIR experiment on H$_2$O:aldehyde
ices at 10--20~K can decide this. (i) Depositing GA as its hydrate and
showing that the C$=$O band at $\sim$1740~cm$^{-1}$ does not appear on warming to 20~K
tests the freezing of the reverse reaction. (ii) Co-depositing GLD with
H$_2$O and following the C$=$O band and the 1050--1100~cm$^{-1}$ C--O region tests
whether forward hydration proceeds. (iii) D$_2$O labelling confirms the mechanism. The
outcomes bear on the model as follows. If neither species hydrates, \hnoneresald\ of the C4 is the aldotetrose in the restricted network and the observed ratio is reached only when GA is gone; in the
complete network \hnonecom\ is the aldotetrose and the ratio is reached later, with the
C3 excess unchanged (Sect.~\ref{sec:ketose}, Table~\ref{tab:nstar}). If both hydrate, the C3 aldose of the restricted network stays below its limit (\gldreshyd); in the complete network the hydration of GLD moves the C3 excess between the isomers (Table~\ref{tab:nstar}).
The same experiment should include DHA, which the model holds as the free ketone;
hydrated, it would rise from \dhares\ to \dhareshyd\ of its limit in the restricted network (Appendix~\ref{app:restricted}) and by \hdhafac\ in the complete one. The network writes the closed-shell products of the radical pairs in their dehydrated form. This is a bookkeeping of the birth state, not a dehydration in the window. It is the same assumption as the fixed carbonyl availabilities, applied at birth. One asymmetry follows. GA made inside
the window joins the pool of the inherited molecule and is attacked with the same
$h_\mathrm{GA}=0.1$ at the next event, so its birth state is taken to be the same
90\% gem-diol, whereas GLD made in the window is taken as the free
aldehyde. The row $h_\mathrm{GA}=0.20$ of Table~\ref{tab:nstarres} bounds the effect of
a partly unhydrated GA pool: $\nstar$ changes by 1\%, the aldotetrose rises by a factor of 3.7 and the ketose fraction falls to \hgareket, with the other observables within 3\%. The same question applies to H$_2$CO, which the model takes as hydrated: its OH attack gives $\bullet$CH(OH)$_2$, the C1 radical that supplies \cthreedhm\ of the C3. The 5.8~$\mu$m band of interstellar ice is attributed to the free aldehyde \citep{Boogert2015}, and in the window the free form keeps its carbonyl: its uncatalysed hydration takes longer than a millisecond even at 491~K, extrapolated from the kinetics of \citet{Winkelman2002} in near-neutral water, and the acid and base catalysis of the melt would have to shorten that by more than six orders of magnitude to hydrate it within a window. We therefore ran the network with part of the H$_2$CO as the free aldehyde, whose OH attack gives the formyl radical HCO; HCO adds to the sugar radicals as an acyl, so that the new carbon carries an aldehyde (Appendix~\ref{app:network}). With half of the H$_2$CO free the count is \Nhfhalf\ and DHA stands at \dhahfhalf\ times its limit, GLD at \gldhfhalf; with all of it free the three are \Nhfone, \dhahfone\ and \gldhfone, and the ketose fraction is \kethfhalf\ and \kethfone\ (Table~\ref{tab:nstar}). Free H$_2$CO therefore lowers the count, to \thfhalf--\thfone~Myr at $\zeta=10^{-14}$~s$^{-1}$ or $\zeta\gtrsim\zetahf$--$\zetahfone$~s$^{-1}$ within 7~Myr. The formyl radical combines with $\bullet$CH$_2$OH and the sugar radicals without the competing dimerisation of $\bullet$CH(OH)$_2$ to GA. The C3 excess stays in place. The ketose fraction falls because the formyl radical adds to the $\alpha$-radical of GA as an aldose precursor; its addition to the C3 radical is booked as the keto-aldehyde, as for $\bullet$CH(OH)$_2$ (Sect.~\ref{sec:ketose}). With all in-window products of the C3 radical booked as the ketose the fraction is \kethfonek, the count \Nhfhalfk\ and \Nhfonek, and DHA \dhahfonek\ times its limit (Table~\ref{tab:nstar}). The ketose selection by hydrated GA thus holds for free H$_2$CO as well, with a fraction of \kethfone--\kethfonek\ set by the bookkeeping of the C3 radical.

\paragraph{Dose and radical yield per window.} The window chemistry is driven by the energy that the ion deposits inside its melted cylinder, \doseinmean\ times the melting enthalpy on average and \doseinlong\ times in the long windows (Sect.~\ref{sec:network}; Appendix~\ref{app:numerics}), with the picosecond yields spread uniformly over the melt. This conserves the energy of the track, and it is an upper bound on the radical supply for two reasons. The dose is concentrated in the track core, where the radicals are born at a density above the mean of the melt and recombine among themselves. And the yields of the radicals that escape a track fall steeply with the linear energy transfer: the Fricke yield, which follows the yields of the hydrated electron and the H atom, drops almost tenfold from fast electrons to uranium ions \citep{LaVerneSchuler1996,LaVerne2000track}. For nickel ions at up to 5000~keV\,$\mu$m$^{-1}$ the H-atom yield is about a tenth of that of fast electrons \citep{LaVerneSchuler1996}; a complete set of radical yields at the linear energy transfer of iron at 1~MeV~u$^{-1}$ (about 4000~keV\,$\mu$m$^{-1}$) has not been measured. Table~\ref{tab:nstar} gives two cases with a smaller supply. With the escape yields of low-linear-energy-transfer radiation the count is \Nescape, \facesc\ times the fiducial one. With a window dose equal to the melting enthalpy alone, as if the excess energy of the core left the melt before its radicals reacted, it is \Ndmelt, \facdmelt\ times the fiducial one, which is \tdmelt~Myr at $\zeta=10^{-14}$~s$^{-1}$ and $\zeta\gtrsim\zetadmelt$~s$^{-1}$ within 7~Myr. Yields below the escape set would raise the count further. The count depends steeply on the dose of the window: doubling it lowers the count to \Ndosetwo, and halving it raises it to \Ndosehalf\ (in the restricted network the factors for half the dose and for the escape yields are \resdosefac\ and \resescfac; Appendix~\ref{app:restricted}). The fiducial count is therefore the smallest that an energy-conserving bookkeeping of the dose allows, the choice most favourable to the route. The $G$ values of Sect.~\ref{sec:network} are those of low-linear-energy-transfer radiation; their fall with linear energy transfer enters only through this systematic, as in the radiolysis-driven grain chemistry of \citet{Shingledecker2018}, which also uses a single set of yields. The escape yields lower GLD/GA by \gldescdrop\ and leave DHA/GA unchanged (Table~\ref{tab:nstar}), and the C3 excess remains in every one of these cases. The dose outside the melts, \evout~eV per molecule per event, adds H$_2$ and H$_2$O$_2$ to the inventory, and its H atoms and electrons enter the cold-phase hydrogen budget (Appendix~\ref{app:coldh}); its destruction of the sugars is the subject of Sect.~\ref{sec:radiolysis}.

\paragraph{The cross-combination of the C1 radicals.} The C3 excess of the complete
network rests on the rate at which $\bullet$CH(OH)$_2$ and $\bullet$CH$_2$OH combine
with the C2 radicals. No measurement exists for these pairs. The model takes the
cross rate $2k_\mathrm{rad}$, twice the measured self-reaction rate of
$\bullet$CH$_2$OH \citep{Kanjana2015}, as it does for every other hydroxyalkyl pair,
and the same branching $\delta$. The self-reaction of $\bullet$CH(OH)$_2$ is measured
at $2k=7.2\times10^{8}$~M$^{-1}$\,s$^{-1}$ at room temperature \citep[who report only its disproportionation]{StockhausenHenglein1971}, within an order of magnitude of
the diffusion limit, and the radical is small. A cross rate an order of magnitude below
the assumed one would still leave DHA above its limit, because the C1
radicals outnumber the C2 radicals \cratio\ to one. The hypothesis of
Sect.~\ref{sec:hypothetical}, that the cross rate is negligible, therefore requires a
selectivity of two orders of magnitude for which we know no precedent in aqueous
radical chemistry. It also requires that methanol, whose OH rate constant is measured,
does not feed $\bullet$CH$_2$OH into the melt: with that channel alone DHA stays at
\dhadhmoff\ times its limit (Table~\ref{tab:nstar}). Pulse radiolysis of formaldehyde--glycolaldehyde solutions at
300--500~K, following the C3 products, would measure it directly.

\paragraph{The disproportionation fraction $\delta$ at the window temperature.} The literature value is
an extrapolation. \citet{Kanjana2015} measured the total self-reaction rate of
$\bullet$CH$_2$OH and decomposed it assuming a barrierless combination and the room-temperature branching of \citet{Wang1996}; their fit gives an activation enthalpy of $17.4\pm7$~kJ\,mol$^{-1}$ for the disproportionation. Evaluated at 491~K, which is our step, this gives $\delta=0.73$ (0.47--0.89). Our own relaxed scans with unrestricted Kohn--Sham M06-2X for the hydrogen-bonded pairs (Appendix~\ref{app:qcmd}) place the maximum of the disproportionation profile $\approx$12~kJ\,mol$^{-1}$ above the pre-reaction complex for both $\bullet$CH$_2$OH and
$\bullet$CH(OH)$_2$. They also give a comparable barrier (15--20~kJ\,mol$^{-1}$) for the
C--C combination of the bulky gem-diol radical pair. A branching of order one-half at
491~K is therefore supported qualitatively. The absolute value requires a dynamical
(variational or trajectory) treatment, because both channels are barrierless with
respect to the separated radicals. The count depends little on it: over
$\delta(491\,\mathrm{K})=0.15$--0.73 it spans \Ncomdx\ in the complete network and \Nreslo--\Nreshi\ in the restricted one, because in the long
windows, which make the C4, the cooler liquid compresses the range of $\delta$ to
0.15--0.33. We
therefore treat $\delta$ as a quantity to be measured through glycolic acid, once its
survival in the window is known, rather than computed.

Table~\ref{tab:prov} sorts the principal inputs of the fiducial model into the five classes of its caption. An input is computed here by quantum chemistry, taken from an independent publication, computed from the model chain, required by the observations, or a typical value varied as a systematic. Three further caveats are material. The first,
and the largest, is the ionisation rate and the origin of the ionising particles in
the dense gas of G+0.693 (Sect.~\ref{sec:origin}). Every time in this paper is the
count divided by an event rate proportional to $\zeta$, and the chemistry itself exists
only if the particles are locally produced heavy ions. The second is the temperature of the window. Each window follows the temperature
history of its length bin, under a 550~K cap on the superheated liquid in the track
core. Caps of 450 and 650~K move the count by \capfac\ (Table~\ref{tab:nstar}; in the restricted network to \Nrescaplo\ and \Nrescaphi\ from \Nres, Appendix~\ref{app:restricted}). The temperature enters the Arrhenius constants, the rate of the radical self-reactions and $\delta$. Holding $\delta$ at its 491~K value in every window raises the restricted-network count to \Nresdfix. Solving every window at a fixed 491~K instead gives \Nrestw, because the liquid spends most of its time well below that temperature (Sect.~\ref{sec:chain}). The third is the treatment of transport in the cooling liquid. The rate constants of the model follow their measured temperature laws; scaling the diffusion-limited ones in addition by the mobility of the water, $w(T)/w(491\,\mathrm{K})$, raises the restricted-network count to \Nresmob. The weight is normalised at the reference temperature, so it lowers every diffusion-limited constant in the cooler liquid, including those measured at room temperature. The axis is therefore a deliberately pessimistic bound on the count, not an estimate; the temperature law used for the radical self-reactions is the measured one, which already contains the slowing of their diffusion.

What an exposure beyond $\nstar$ does is known from the model itself. At twice the count it converts the inventory into hexoses and larger sugars: \bigcomtwo\ of the sugar molecules are at C6 or above, hexoses/GA $=\hexcomtwo$, ERU/GA $=\erucomtwo$ and DHA is at \dhacomtwo\ times its limit (in the restricted network \bigrestwo, \hexrestwo\ and \erurestwo). A grain exposed for several times $\nstar$ would therefore not show the observed C2$\approx$C4 pattern even in the hypothetical case.

The methods of the quantum-chemical barriers (class D in Table~\ref{tab:prov}) and of
the molecular dynamics used for $f_\mathrm{s}$ and for the conductivity check are given
in Appendix~\ref{app:qcmd}.

\paragraph{Thermal properties and the window.} We reran the whole track table with
literature properties throughout: the heat capacity and latent heat of ice Ih
\citep{FeistelWagner2006}, the heat capacity and conductivity of liquid water
\citep{WagnerPruss2002,Huber2012}, and one of the two measured amorphous-ice
conductivities (Sect.~\ref{sec:chain}). With the high-density law, which the adopted one
follows, the mean liquid lifetime changes by 2\% and the ice age by \hdafac. With the
low-density law, which conducts about twice as well, the mean liquid lifetime falls by
26\%, the windows of 0.3~ns or longer almost vanish, and the ice age rises by a factor of \ldafac; the required ionisation rate rises in proportion.
The structure of the ice matters far more. Crystalline ice Ih conducts heat 5--15 times better at 70--130~K \citep{Slack1980,AnderssonInaba2005}; with its conductivity the
liquid lifetime is about four times shorter and the melt radius 40\% smaller, which
would remove the long windows on which this paper rests.

\paragraph{A closed inventory.} Apart from the H$_2$CO resupply and the loss of H$_2$,
the model keeps the inventory of the grain closed. At $n(\mathrm{H_2})=10^{4}$~cm$^{-3}$
gas-phase molecules freeze out within a few $10^{5}$~yr, far less than the ice ages
found here, and sputtering removes \sputlo--\sputhi\ of a mantle over the ice age
(Sect.~\ref{sec:origin}). Both exchange mainly the volatile ice. The sugars, which do
not sublimate at 10~K, leave the grain only by sputtering. A mantle that is renewed
faster than the count would never reach the observed ratio, so these exchanges can
only raise the exposure that the pattern requires.

\paragraph{Finite grain and amorphous ice.} The finite size of the grain and the amorphous state of the ice, including the glassy refreezing of the melt, are discussed in Appendix~\ref{app:trackcaveats}. The finite grain does not change the windows. Glassy refreezing, the vitrification limit, would shorten the ice age by \vitfac\ (Sect.~\ref{sec:rate}); the fiducial model uses the crystallisation limit.

\paragraph{Freezing-accelerated chemistry.} Reactions in a freezing solution can run up to $10^{5}$ times faster than in the liquid \citep{Takenaka1992,KimChoi2011,OConcubhairSodeau2013}. The growing ice rejects the solutes into the shrinking liquid, where their concentrations rise by orders of magnitude, and the pH shifts with them. This freeze concentration needs a crystallisation front that advances more slowly than the solutes diffuse away from it, below the diffusive speed $D/\lambda$ \citep{Aziz1982}, about 1~m~s$^{-1}$ for a liquid diffusivity of $10^{-9}$~m$^{2}$~s$^{-1}$ and an interface width of 1~nm. Laboratory freezing takes seconds to hours, and even at its fastest, near 255--260~K, crystalline ice grows at only about 0.1~m~s$^{-1}$ \citep{Xu2016}. The melts of this paper, 1 to 10~nm in radius, close in 10~ps to 0.4~ns (Sect.~\ref{sec:chain}), at interface speeds of 25 to 190~m~s$^{-1}$ (up to 1400~m~s$^{-1}$ in shorter windows). They cannot crystallise; they vitrify (Appendix~\ref{app:trackcaveats}), with every solute frozen in place. Freeze concentration therefore does not operate in a cosmic-ray track. It can operate in a track held near the melting point, as in the setup of our earlier work \citep{Saito2024}, which with the corrected code refreezes within about a microsecond with a front of order 1~cm~s$^{-1}$ (Appendix~\ref{app:trackcaveats}); that is one reason not to transfer yields between the two settings. What remains of the non-equilibrium chemistry of the quench is the fate of the radicals that survive the window. The fiducial model traps them in the glass and reduces them between events (Appendix~\ref{app:coldh}). As an upper bound on their recombination we let every surviving radical react with the others at the end of each window, through the radical--radical reactions of the network run to completion with the rate constants and the branching of the window just closed, before the cold-phase reduction. About $\ecrad$ radicals per window, mostly $\bullet$CH(OH)$_2$ and $\bullet$CH$_2$OH, then combine instead of returning to H$_2$CO and CH$_3$OH. The observed ratio is reached at \Nec\ events (\tec~Myr), with DHA at \dhaec\ times its limit and GLD at \gldec\ (Table~\ref{tab:nstar}). The count falls by a factor of \ecfacinv, because the C1 radicals that the fiducial model returns to H$_2$CO and CH$_3$OH now combine into GA and add to the sugar radicals. An ionisation rate of $\zetaec$~s$^{-1}$ would then suffice within 7~Myr, and the exposure condition of Sect.~\ref{sec:origin} would no longer fail on its own. The C3 excess remains, and GLD, at \gldcom\ times its limit in the fiducial model, rises to \gldec. The limit is extreme, because the radicals of a glass at 10~K cannot move (Sect.~\ref{sec:meltgated}), but it shows which conclusion depends on the fate of the trapped radicals: the exposure, not the C3 excess.

\paragraph{The fate of the trapped radicals.} Three treatments of the radicals left at the end of a window bracket the exposure. In the fiducial model they are trapped and reduced by the hydrogen of the cold phase (Appendix~\ref{app:coldh}), and the count is \Ncom. If the small radicals instead survive to the next window and react there, the count is \Nsmalloff\ and $\zeta\gtrsim\zetasmalloff$~s$^{-1}$ suffices within 7~Myr (Table~\ref{tab:nstar}). If every radical recombines as the window closes, the count is \Nec\ and $\zeta\gtrsim\zetaec$~s$^{-1}$ suffices. The fiducial choice rests on the immobility of the glass. The binding energy of $\bullet$CH$_2$OH to amorphous water ice has not been measured. That of CH$_3$OH is about 5000~K for a standard prefactor of $10^{12}$~s$^{-1}$, and diffusion barriers are usually taken as 0.3--0.8 of the binding energy \citep{Minissale2022}, so a hop of a radical of this size costs of order 1500~K or more. At 10~K no hop occurs. In our track model an iron-group ion that crosses the whole grain deposits about 500~keV and heats the $0.1\,\mu$m grain to about 75~K (0.4~MeV and 70~K in the classical estimate), from which it cools by evaporation within about 10~$\mu$s \citep{HasegawaHerbst1993}. At 75~K a barrier of 1500~K gives about $2\times10^{3}$ hops per second, so in the 10~$\mu$s the radicals make at most a few hundredths of a hop before the grain is cold again. The hydrogen atoms that reduce them tunnel and need no such hop. \citet{JenniskensBlake1994} attribute the recombination of radicals in photolysed ices at 38--68~K to the restructuring of high-density into low-density amorphous ice; a mantle that an iron-group ion has once warmed to about 75~K has made that transition (Appendix~\ref{app:trackcaveats}), so it does not recur between later events. Nor does the liquid itself offer the time: after the window the melt crosses the supercooled range from 200 to 136~K within about a nanosecond, too short for a viscous liquid to bring two radicals together. The conduction time is about 0.1~ns for $r\approx10$~nm and a thermal diffusivity of $10^{-6}$~m$^{2}$~s$^{-1}$, and the phase-field model of Appendix~\ref{app:trackcaveats} reaches 200~K after 0.6~ns. The recombination limit is therefore an extreme, kept to show which conclusion it would move.

\paragraph{Few molecules in a small melt.} The window chemistry is solved as mean-field rate equations in the whole-grain volume, but a single ion melts only part of it (Fig.~\ref{fig:species}). At $\nstar$ the grain holds \gagrain\ GA and \cthreegrain\ C3 molecules, and each window receives \ohevent\ OH per whole-grain equivalent on average. A proton melts $7\times10^{-5}$ of the grain on average, so its melt holds about \gaproton\ GA, about 140 OH and a few sugar radicals; the second-order reactions between sugar radicals are then overestimated by the rate equations. An iron-group ion at 0.3--3~MeV~u$^{-1}$ melts 0.5--1\% of the grain, so its melt holds several thousand GA and OH and hundreds of sugar radicals, enough for the mean-field treatment. The windows of 0.1~ns and longer, which belong to the ions that melt $10^{-3}$ of the grain or more, make \cfourlong\ of the C4 (Sect.~\ref{sec:rate}). A stochastic treatment would lower the small contribution of the proton windows and lengthen the count by at most that share. For the longest window of the fiducial distribution at $\nstar$, in the melt of one iron ion (0.9\% of the grain, \gillnev\ reaction events), a direct Gillespie simulation with the same reaction table, initial state and temperature history reproduces the mean-field production within its scatter over \gillnrep\ replicates: \gillcthree\ dehydro-C3 molecules against \gillcthreemf, \gillcfour\ dehydro-C4 against \gillcfourmf, \gillcsix\ dehydro-C6 against \gillcsixmf, and the C1 radicals left at the end within \gillcone. The melt is also not uniform: with $D=10^{-9}$--$10^{-8}$~m$^{2}$~s$^{-1}$ and $\tau=0.35$~ns the diffusion length $\sqrt{6D\tau}$ is 1.4--4.6~nm, comparable to the 7--10~nm melt radius of the iron-group tracks, and the radicals are born in the core. Recombination among them is therefore favoured over attack on the sugars, which is the effect that the escape-yield axis of Table~\ref{tab:nstar} represents.

\section{Conclusions}
\label{sec:conclusions}

We have asked whether transient melt windows, opened in interstellar ice by cosmic-ray
heavy ions, can make the sugar pattern of G+0.693$-$0.027, with every rate that
controls the result held to an independently published value where one exists. The answer is negative on two firm counts and conditionally on a third, and positive on one mechanism. The counts rest on different inputs: the C1 radical chemistry, the particle spectrum together with the radical supply of a window and the fate of the trapped radicals, and the radiolytic stability of the sugars. The C3 excess and the radiolysis are the firm ones.

\begin{enumerate}
\item \textbf{The thermal route is excluded, and within the window the radical route is the only chemistry
a window can drive.} The classical formose reaction, anchored to hydrothermal
laboratory rates, makes fewer than $10^{-3}$ molecules of any C3 or C4 sugar
per grain over the whole exposure, ten to thirteen orders of magnitude below the
radical route (Sect.~\ref{sec:meltgated}).
\item \textbf{The C4 is a ketose if GA in the ice is hydrated.} Nine of ten OH attacks on GA in ice then strike its gem-diol, and the C4 is born with its carbonyl on the former aldehyde carbon: \ketcom\ of the C4 is the ketose in the complete network and \ketres\ in the restricted one, provided the trapped C4 radical is reduced back to the ketose (Sect.~\ref{sec:ketose}). Only the gem-diol channel, which gives \cfourgemall\ of the C4 births in the complete network and all of them in the restricted one, is independent of how the C3 radical is booked. Without the hydration the C4 is the aldotetrose in the restricted network; in the complete network the C1 addition keeps \kethnone\ of it the ketose. With free H$_2$CO the formyl radical replaces $\bullet$CH(OH)$_2$ and the fraction is \kethfone--\kethfonek, set by the bookkeeping of the C3 radical (Sect.~\ref{sec:caveats}). This selection does not depend on the exposure. The grain-surface route also makes the ketose ERU, so the ketose alone does not identify a melt.
\item \textbf{The melt overproduces the C3 sugars.} The dihydroxymethyl radical of hydrated formaldehyde and the hydroxymethyl radical of methanol add one carbon to any sugar radical, so no rung of the ladder is skipped. When ERU/GA reaches its observed value, DHA stands at \dhacom\ times its limit and GLD at \gldcom\ times. No parameter of the model brings the C3 below the limits, two bounds show that no destruction or isomerisation channel can, and the grain-surface model fails for the same reason.
\item \textbf{The exposure is demanding unless the trapped radicals recombine.} The C2$\approx$C4 equality needs \Ncom\ whole-grain melting events, which particles produced inside the cloud at $\zeta=10^{-14}$~s$^{-1}$ supply in \tcom~Myr and external cosmic rays in \tecom--\trcentre~Myr at the cloud centre, depending on where $\zeta$ is normalised. Within 7~Myr the count is reached only for $\zeta\gtrsim\zetacom$~s$^{-1}$. That is the level inferred for the diffuse gas of the Central Molecular Zone, \facgins\ times above the upper limit on its dense gas and \facdense\ times above the values inferred for the dense gas of Sgr~B2. This is the weakest of the three obstacles and the least certain, because it follows the count. The count scales steeply with the radical supply of a window: with a window dose equal to the melting enthalpy, instead of the energy that the track deposits in the melt, the requirement is $\zetadmelt$~s$^{-1}$, and with the escape yields $\zetaesc$~s$^{-1}$. It also rests on the reduction of the trapped radicals between events, as a glass at 10~K requires. If all of them recombined, \Nec\ events and $\zeta\gtrsim\zetaec$~s$^{-1}$ would suffice, a value that the dense gas of the zone may have; DHA and GLD would then stand at \dhaec\ and \gldec\ times their limits. A porous mantle lowers the threshold by a factor of \kfac, and cold-route GA and glycol at laboratory yields shorten the count to \Ncoldc\ (Sect.~\ref{sec:limits}).
\item \textbf{The sugars do not survive the dose between events.} The same particles deposit \evout~eV per molecule per event in the ice that they do not melt, about \doseouttot~eV per molecule over the exposure, against half-life doses of 0.2--4.6~eV per molecule measured for organics diluted to about 5\% or less in H$_2$O ice at 10--20~K (0.2~eV for glycolaldehyde itself). At these values the ratio stays below \pkten. It is reached only if the sugars are \dhalffac\ times more stable than any organic measured in dilute cold ice, or if the measured steady-state survival of 25\% holds to doses \dosefaclo--\dosefachi\ times beyond the laboratory, which raises the required ionisation rate to $\zetarevd$~s$^{-1}$, the limit at which the gas ceases to be molecular. This obstacle does not depend on the ionisation rate: the dose per melting event is fixed by the spectrum.
\item \textbf{The hypothetical case is consistent and testable.} If the C1 radicals did not add to the sugar radicals, the sugars survived the solid-phase dose and the grains saw $\zeta\gtrsim\zetares$~s$^{-1}$ from a local source of MeV ions, the model would reproduce the pattern. ERU/GA reaches 0.94 after \Nres\ events, both C3 sugars stay below their limits, \ketres\ of the C4 is the ketose, the hexoses stand at \hexres$\times$GA and glycolic acid at \glyres$\times$GA (\glyresnooh--\glyresnoohhi$\times$GA without its OH destruction). Each condition can be tested: by pulse radiolysis of formaldehyde and glycolaldehyde solutions, by proton irradiation of glycolaldehyde and erythrulose in H$_2$O ice to doses well beyond 20~eV per molecule, and by the dense-gas ionisation rate of G+0.693.
\end{enumerate}

With standard assumptions the sugars of G+0.693 are therefore not made in cosmic-ray
melt windows, and the ERU must be supplied by chemistry outside the window, such as
the cold surface route of the detection paper, which faces the same C3 excess. Two
results of this paper stand on their own: the ketose selection by hydrated GA in any
hot radical chemistry of ice, and the reconstruction of the whole-grain melting rate
from the cosmic-ray spectrum, which shows that long liquid windows exist only where
iron-group ions at 0.3--3~MeV~u$^{-1}$ reach the grains.

\section*{Data availability}

The simulation codes (the zero-dimensional window chemistry \texttt{chem0d}, and the
grid code \texttt{pfice}, used only for the checks of Appendices~\ref{sec:verification} and \ref{app:trackcaveats}), the chemical network definition, the machine-learned potentials, cells and inputs of the molecular dynamics of Appendix~\ref{app:qcmd}, the analysis
scripts and the output files needed to reproduce every figure and table in this paper
are archived on Zenodo at \url{https://doi.org/10.5281/zenodo.23282574} (the snapshot at submission) and developed at \url{https://github.com/seiki-saito/ice-sugar-clock}. The archive and the repository will be made public when the paper is published; until then they are available to the referees on request.

\begin{acknowledgements}
This work was supported by JSPS KAKENHI
Grant Number JP26K22392. Part of the computations was carried out on the Plasma
Simulator of the National Institute for Fusion Science under project NIFS25KIST066.
Generative-AI tools (Google Gemini and Anthropic Claude Code) were used for language editing, for generating and executing analysis code under the authors' direction, and as an aid in the literature audit. All scientific content, model choices and conclusions are the authors' own and were verified by them.
\end{acknowledgements}

%\bibliographystyle{aa}
%\bibliography{refs}

\begin{appendix}

\section{The reaction network}
\label{app:network}

\subsection{The cycle as a discrete map}
\label{app:cycle}

The repeated-melting calculation advances a per-grain inventory vector
$\mathbf{N}=\{N_i\}$ (molecules per grain) through $N_\mathrm{run}$ melting events. The model
grain is a sphere of radius $a=0.1\,\mu$m (Sect.~\ref{sec:chain}), of volume
$V=4.19\times10^{-21}$~m$^{3}$, and holds $N_\mathrm{w}=1.29\times10^{8}$ water molecules. The initial inventory is $6.45\times10^{6}$ molecules of H$_2$CO, $7.37\times10^{6}$ of CH$_3$OH and $3.27\times10^{7}$ of CO (5.0, 5.7 and 25\% of the water), with a fixed base of $6.9\times10^{6}$ (5.4\%); in the restricted network the H$_2$CO and CH$_3$OH are replaced by $1.38\times10^{7}$ molecules of H$_2$CO (10.7\%). Every other species starts at zero. Interstellar ice also holds CO$_2$, at about a quarter of the water \citep{Boogert2015}; in the model CO$_2$ arises only in the melts (\cotwocom\ of the water at $\nstar$). Starting instead with CO$_2$ at 28\% of the water raises the count by 7\% and leaves the C3 excess and the ketose fraction unchanged (four sequences); its reduction in the melts then raises HCOOH/H$_2$O at $\nstar$ to 40\%. One event $k\to k+1$ applies five operators in order.
\begin{enumerate}
\item \emph{Window chemistry} (Appendix~\ref{app:window}). The grain is a single well-mixed liquid volume $V$, and the radiolytic radicals are injected. The reaction network is integrated for the window length $\tau$ of this event in 14 steps of
$\tau/14$, each at the temperature of the history of its length bin. The integration is zero-dimensional: there is no grid,
and the number densities $c_i=N_i/V$ are the only state. The window lengths are the quantiles $(j-\tfrac12)/N_\mathrm{b}$, $j=1\ldots N_\mathrm{b}$, of the
reconstructed distribution of Sect.~\ref{sec:rate}, weighted by melted volume, in blocks of $N_\mathrm{b}=120\,000$ events, each block shuffled into a random order. Each seed of the shuffle is one sequence of windows; different
sequences differ only in that order, and runs of different length with the same seed share their first events. The temperature histories are taken from the
cylindrical conduction model (Sect.~\ref{sec:chain}).
\item \emph{Cold-phase reduction} (Appendix~\ref{app:coldh}): the trapped radicals and
dehydro-sugars are reduced to closed-shell species by the H atoms available between
events.
\item \emph{H$_2$ loss}: $N(\mathrm{H_2})\to0.5\,N(\mathrm{H_2})$ (desorption of the volatile product). The fraction is not measured. It is our estimate from the grid code, in which 0.47 of the H$_2$ (binding energy 440~K) desorbs thermally from a melt that reaches the surface. Losing all of the H$_2$ leaves the count unchanged, and retaining all of it raises the count by 27\% with the C3 excess unchanged (four sequences).
\item \emph{Outside-melt radiolysis}: the dose deposited in the grain outside the melted
volumes, $E_\mathrm{nm}=\doseout\,D_\mathrm{melt}V$ per event for the fiducial spectrum, with $D_\mathrm{melt}$ the melting enthalpy per unit volume
(Eq.~(\ref{eq:gsource}); Appendix~\ref{app:numerics}), produces
the molecular pair $2\mathrm{H_2O}\to\mathrm{H_2}+\mathrm{H_2O_2}$ with
$G_\mathrm{om}=0.2$ per 100~eV: $\Delta N_{\mathrm{H_2O_2}}=G_\mathrm{om}E_\mathrm{nm}/100$
and $\Delta N_{\mathrm{H}_2}=0.5\,\Delta N_{\mathrm{H_2O_2}}$, the factor 0.5 being the retained fraction of operator 3. The same dose destroys closed-shell sugars in the solid. With a half-life dose $D_{1/2}$ per water molecule, i.e. a destruction constant $G_\mathrm{d}=100\ln2/D_{1/2}$ per 100~eV absorbed per molecule of the grain, each sugar species loses the fraction $1-\exp(-G_\mathrm{d}E_\mathrm{nm}/100N_\mathrm{w})$ per event, where $E_\mathrm{nm}/N_\mathrm{w}=\evout$~eV is that dose per water molecule. The destruction is applied before operator 3; the two commute. In the reversible variant the destroyed molecules enter a damaged pool from which they return to the parent at the rate $k_{-1}=k_1c/(1-c)$, with $k_1$ the destruction rate above and $c$ the steady-state surviving fraction measured in the laboratory (Sect.~\ref{sec:radiolysis}). In a further variant the carbon of the destroyed sugars is returned to the inventory as glycolaldehyde units (with one H$_2$CO for an odd carbon number, and two H$_2$CO for glycolaldehyde itself), as H$_2$CO or as CO; the sugars are (CH$_2$O)$_n$, so the atoms balance. The fiducial model sets $G_\mathrm{d}=0$.
\item \emph{Substrate resupply}: $\Delta N_{\mathrm{H_2CO}}=\min(5.72\times10^{5},
\Sigma_\mathrm{max}-\Sigma_k)$, where $\Sigma_k$ is the cumulative supply and
$\Sigma_\mathrm{max}=1.93\times10^{7}$ is the H$_2$CO that the gas-phase CO share of one
grain, $6.43\times10^{7}$ molecules, yields at the retained fraction 0.3 (CO
hydrogenation between events; the cap is reached after 34 events). The per-event
supply is the CO accreted at $n(\mathrm{H_2})=10^{4}$~cm$^{-3}$, $x_\mathrm{CO}=10^{-4}$
and 10~K, of which we take a fraction 0.3 to be retained as H$_2$CO, the steady-state H$_2$CO yield per CO in the hydrogenation experiments of \citet{Fuchs2009} at 12~K; the share
$\Sigma_\mathrm{max}$ divides the gas-phase CO among grains at an ice water abundance of
$10^{-4}$ per H nucleus. With the $4\times10^{-5}$ of Appendix~\ref{app:column} the
share would be 2.5 times larger; the substrate axes of Table~\ref{tab:nstarres} cover this range.
\end{enumerate}
The species carried from one event to the next are all closed-shell organics, the acids, CO, CO$_2$, H$_2$, H$_2$O$_2$ and NH$_3$. The radicals HOCO, HCO and NH$_2$, which have no entry in Table~\ref{tab:coldh}, are carried as well. The HCO that H~$+$~CO makes in the fiducial network reacts only with H, OH, itself and $\bullet$CH$_2$OH; its addition to the sugar radicals is part of the free-H$_2$CO variant (Sect.~\ref{sec:caveats}). The primary radicals and the free ions are not. The $e^-_\mathrm{aq}$, H, OH, H$_3$O$^+$ and OH$^-$ left at the end of a window are taken to recombine as the melt freezes, so that no net charge passes from one event to the next. The carbon-centred radicals and the dehydro-sugars are carried only through operator 2; when the hydrogen budget falls short, the unreduced remainder enters the next window. Methanol is part of the initial inventory and is formed by operator 2 from $\bullet$CH$_2$OH and in the window by the disproportionation of two $\bullet$CH$_2$OH (row 79). In the window OH and H abstract its hydrogen and give $\bullet$CH$_2$OH (rows 84 and 85; \citealt{Buxton1988}); the restricted network of Sect.~\ref{sec:hypothetical} holds it inert. The observables are
read from $\mathbf{N}$ at the end of each event. Hydration states enter only
through the fixed fractions $h_X$ of Appendix~\ref{app:window}; no hydration or
dehydration reaction is integrated, which is the frozen-state assumption of
Sect.~\ref{sec:caveats}.

\subsection{Window chemistry}
\label{app:window}

Within the window the concentrations $c_i=N_i/V$ (number densities; the rate
constants of Table~\ref{tab:reactions} are in M$^{-1}$\,s$^{-1}$ and are converted to
m$^{3}$\,s$^{-1}$ with Avogadro's number and the litre) obey
\begin{equation}
\frac{\mathrm{d}c_i}{\mathrm{d}t}=\sum_r \nu_{ir}R_r,\qquad
R_r = k_r\,c_{s_1(r)}\,c_{s_2(r)}.
\label{eq:ode}
\end{equation}
Here $r$ runs over the 199 reactions of Table~\ref{tab:reactions}, $R_r$ is the
rate of reaction $r$, $s_1(r)$ and $s_2(r)$ are its two reactant species, with $c_{s_2(r)}\equiv1$ for a unimolecular reaction. The stoichiometric coefficient $\nu_{ir}$ of species $i$ in reaction $r$ is negative for a reactant and positive for a product. It is read from the reaction equation of Table~\ref{tab:reactions} (for reaction 0, $\nu=-2$ for e$^-_\mathrm{aq}$, $+1$ for H$_2$ and $+2$ for OH$^-$). The column $k$(491~K) of that table gives $k_r$ at the reference temperature of 491~K. For a self-reaction, $s_1(r)=s_2(r)$, the rate is $k_r c_{s_1}^{2}$ and two molecules of $s_1$ are consumed, so $k_r$ is the per-pair rate constant. The rate constant is
\begin{equation}
k_r = k_r^{0}\,\exp\!\left(-\frac{E_{\mathrm{a},r}}{k_\mathrm{B}T}\right)
      \,\mathrm{mob}(\phi)^{[\dagger]},
\label{eq:rate}
\end{equation}
where $k_r^{0}$ is the pre-exponential factor, $E_{\mathrm{a},r}$ the activation
energy per molecule (listed per mole in Table~\ref{tab:reactions}), $k_\mathrm{B}$ the Boltzmann constant,
$\mathrm{mob}(\phi)=(1-\phi)+\phi f_\mathrm{s}$, with $f_\mathrm{s}$ the solid-to-liquid mobility ratio of Sect.~\ref{sec:meltgated}, multiplies the reactions marked
$\dagger$ in Table~\ref{tab:reactions} (the exponent $[\dagger]$ is 1 for them and 0
otherwise).
In the cycling calculation $\phi=0$, so $\mathrm{mob}=1$ and
Eq.~(\ref{eq:rate}) is a plain Arrhenius law at the temperature of each step. The pre-exponential
$k_r^{0}$ of the closed-shell aldol class (A) is set so that $k_r(273\,\mathrm{K})$
matches the hydrothermal anchor of \citet{Kopetzki2011} ($6.4\times10^{-9}$
M$^{-1}$\,s$^{-1}$ for GA$+$GA, all other aldol steps scaled by the same factor); Table~\ref{tab:reactions} lists the resulting
$k_r(491\,\mathrm{K})$.

\emph{Initial condition.} At $t=0$ the carried inventory is placed in the melt,
$c_i(0)=N_i/V$, and the radiolytic products of the dose of the window are added at once:
\begin{equation}
c_i(0)\;\mathrel{+}=\;\frac{G_i}{100}\,D_\mathrm{w},
\qquad D_\mathrm{w}=x_\mathrm{in}(\tau)\,D_\mathrm{melt}.
\label{eq:gsource}
\end{equation}
Here $G_i$ is the radiolytic yield of species $i$ in molecules per 100~eV of
deposited energy and $D_\mathrm{w}$ is the dose density of the window, the
energy per unit volume that the ion deposits inside its melted cylinder. $D_\mathrm{melt}=3.50\times10^{27}$~eV\,m$^{-3}$ is the enthalpy per unit volume that heats
ice from 10~K to the melting point and melts it, with the properties of Sect.~\ref{sec:chain} (latent heat $3.06\times10^{8}$~J\,m$^{-3}$ at the density of ice): $5.6\times10^{8}$~J\,m$^{-3}$. The factor $x_\mathrm{in}$ is tabulated for each window-length bin (Appendix~\ref{app:numerics}); it is \doseinmean\ on average and \doseinlong\ in the long windows. For a dose of $D_\mathrm{melt}$ the injected density of OH is $1.9\times10^{26}$~m$^{-3}$ (\ohmeltM~M), \oheventmelt\ radicals in the grain volume; for $e^-_\mathrm{aq}$ and H it is $7.0\times10^{5}$ and $1.5\times10^{5}$. These numbers are multiplied by $x_\mathrm{in}$. $G_i$ is non-zero
only for the six primary products of water radiolysis, for which
$G$($e^-_\mathrm{aq}$, H, OH, H$_2$, H$_2$O$_2$, H$_3$O$^+$)
$=(4.8, 1.0, 5.5, 0.15, 0.2, 4.8)$ per 100~eV (escape yields 2.7, 0.6, 2.7, 0.45,
0.7, 2.7 as a systematic); $G_i=0$ for every other species. The water consumed or released by each reaction
($\Delta$H$_2$O column of Table~\ref{tab:reactions}) is debited from or credited to the
solvent within the window. The water itself is not carried between events: every window starts from the water density of ice, $3.07\times10^{28}$~m$^{-3}$. The grain is thus treated as open in water. In the fiducial run the oxygen that has passed from water into CO$_2$, formic acid, the sugars and H$_2$O$_2$ by $\nstar$ amounts to about 0.3 of the water molecules of the grain. The model does not debit this oxygen between events. It therefore overstates the water concentration, which enters the rates at most linearly, by up to that fraction at $\nstar$.

\emph{Acid speciation.} Formic acid is carried as one species; the fraction $\chi$
present as formate is obtained at every evaluation from the local proton balance,
\begin{align}
\chi&=\frac{K_\mathrm{a}}{K_\mathrm{a}+[\mathrm{H}^+]},\\
[\mathrm{H}^+]&=\tfrac12\Big[(P-K_\mathrm{a})+\sqrt{(K_\mathrm{a}-P)^2+4K_\mathrm{a}(P+C_\mathrm{FA})}\Big],
\end{align}
where $C_\mathrm{FA}=[\mathrm{HCOOH}]_\mathrm{tot}$, $P=[\mathrm{H_3O^+}]-[\mathrm{OH^-}]-B$, $B$ is
a fixed base inventory of $1.65\times10^{27}$~m$^{-3}$ (5.4\% of the water, standing for NH$_3$) and $\mathrm{p}K_\mathrm{a}=3.75$. Where the
base exceeds the acid and the net H$_3$O$^+$ ($P+C_\mathrm{FA}<0$), the root is negative and
$[\mathrm{H}^+]$ is set to zero, i.e. the acid is fully dissociated ($\chi=1$). In the fiducial run this holds for about the first \nchi\ events. The formic acid then exceeds the base, and $\chi\approx B/C_\mathrm{FA}$ is about \chistar\ at $\nstar$. The
OH~$+$~formic acid rate is $(1-\chi)\,1.3\times10^{8}+\chi\,3.2\times10^{9}$
M$^{-1}$\,s$^{-1}$ \citep{Buxton1988}; Table~\ref{tab:reactions} lists the value for the undissociated acid.

\emph{Carbonyl availability.} A closed-shell aldehyde $X$ is attacked by OH at the
Arrhenius rate $k_X(T)$ of Table~\ref{tab:prov}; the C4--C6 sugars and DHA
reuse the GLD value. A fraction $h_X$ of the molecules is
the free aldehyde; of the attacks on these, a fraction $f_\mathrm{CHO}=0.5$ strikes the
aldehydic hydrogen and gives the acyl radical, the remainder the $\alpha$-carbon and
gives the hydroxyalkyl radical. The remaining fraction $1-h_X$ is the gem-diol; for
GA and GLD its attack gives the gem-diol radical, and for the
C4--C6 sugars, whose gem-diol radicals are not followed, it is routed to the
hydroxyalkyl radical. Table~\ref{tab:reactions} lists these as separate
reactions whose rate constants already include $h_X$ and $f_\mathrm{CHO}$
($h_\mathrm{GA}=0.10$, $h_\mathrm{GLD}=1$, $h_\mathrm{DHA}=1$, $h_\mathrm{ald}=0.45$,
$h_\mathrm{C5}=0.02$, $h_\mathrm{C6}=0.01$). The network does not follow hydrates as
separate species. Closed-shell products of radical combination are written in their
carbonyl form, and at the next event every molecule of a species is attacked with the
same $h_X$, whatever its birth. GA made in a window is therefore treated as 90\%
gem-diol, like the inherited GA, and GLD as the free aldehyde. Following the hydrates
and the free carbonyls as separate species, with the birth state set by the channel
that makes them, would remove this assumption (Sect.~\ref{sec:caveats}). A ketone has no aldehydic hydrogen, so the
attack on ERU (row 97) proceeds at the full rate and gives only the ketose radical. That radical is reduced back to ERU in the window (rows 99 and 101) or between events (Table~\ref{tab:coldh}). It also combines with $\bullet$CH$_2$OH or the aldehyde radical of GA to give a C5 or C6 (rows 98 and 100), and with the gem-diol radical of GA to give a C6 (row 149). The dehydro-tetrose
born from the gem-diol radical pairs (ERYketm2) carries the keto group. The network carries the symmetric diketone and the keto-aldehyde as this one species. OH attack on it, at 0.02 of the parent rate, gives its keto-acyl radical (row 133; glyoxal and the dehydro-C3 are not attacked in the window), which is repaired to it (row 185), loses CO to
the C3 radical (row 195), or is reduced to ERU between events. For DHA the attack (row 72) gives
the C3 radical that the network shares with GLD, the two parents being
related by the enediol tautomerism; between events it is reduced back to
DHA (Table~\ref{tab:coldh}; Sect.~\ref{sec:ketose}).

\emph{Radical--radical reactions.} The pairs of carbon-centred radicals listed in
Table~\ref{tab:reactions} react at the per-pair rate
$k_\mathrm{rad}$ (self reactions) or $2k_\mathrm{rad}$ (cross reactions), the sum of
the combination and disproportionation rates of \citet{Kanjana2015}. It is
$4.86\times10^{9}$ M$^{-1}$\,s$^{-1}$ at 491~K, the sum of the activation-controlled rates of their decomposition, 15\% above their Arrhenius fit of the measured rate ($k=4.2\times10^{9}$), and is scaled with the temperature law of that measurement (12.7~kJ\,mol$^{-1}$) to the mean temperature of each window. For the pairs
that build the C3 to C6 sugars a fraction $\delta$, evaluated at the same mean
temperature ($\delta(491\,\mathrm{K})=0.47$ in the fiducial model; Sect.~\ref{sec:chain}),
disproportionates (O--H hydrogen transfer, giving a closed-shell sugar
plus an aldehyde or acid) and $1-\delta$ combines; both branches are listed
separately. The pair list covers every pair of the hydroxyalkyl radicals, of the two gem-diol radicals and of the acyl radical of GA whose product has at most six carbons (rows 70--71, 78--83, 89--96, 98--100, 102--110, 134--157, 162--167 and 168--169). The hydroxyalkyl radicals are $\bullet$CH$_2$OH, the aldehyde radicals of GA and GLD, the glycol radical, the radicals of ERU and of the aldotetrose, and the C5 radical. There are six exceptions: the acyl radical of GA with itself and with the aldehyde
radical, and the acyl radical of GA and the glycol radical with the two C4
radicals.
The acyl radicals of GLD and of the C4 sugars are given no combination
partner in the fiducial network; their fates are decarbonylation and repair. Giving
them, and the six pairs above, the same partners as the acyl radical of GA
(30 further rows, run as a systematic in the restricted network) changes $\nstar$ by \acylfullN\ and GLD/GA by \acylfullgld\ relative to the same sequences (Table~\ref{tab:nstarres}). The cold-phase reduction returns the trapped C3 radical,
which the network carries as one species for both C3 isomers, as DHA,
and the trapped radical of ERU, carried separately from that of the
aldotetrose, as ERU (Table~\ref{tab:coldh}); the aldose alternatives are run
as systematics (Sect.~\ref{sec:ketose}). The dihydroxymethyl radical $\bullet$CH(OH)$_2$, the most abundant carbon radical of the window, reacts with itself, OH and H (rows 43--45 and 49). In the complete network it also reacts with the C1--C5 hydroxyalkyl radicals, the two gem-diol radicals and the acyl radical of GA at the cross rate $2k_\mathrm{rad}$ with the branching $\delta$ (rows 111--130; row 131 is the H addition to the dehydro-C3). Its combination products carry the new carbon as a gem-diol, which dehydrates to a carbonyl on that carbon. With $\bullet$CH$_2$OH it gives GA, with the radicals of GA the dehydro-C3 OHC$-$CO$-$CH$_2$OH (species C3m2, reduced to DHA between events), and with the glycol radical GLD. With the C3, C4 and C5 radicals it gives the dehydro-C4 ERYketm2, the dehydro-C5 RIBm2 and the dehydro-C6 HEXm2. For the $\alpha$-radical of GA the bookkeeping rule would instead give hydroxymalonaldehyde and GLD. We keep the first choice as the fiducial one. The alternative changes the count by less than 2\% and the C3 ratios at $\nstar$ by up to 6\%: DHA is \dhadhmgld\ and GLD \glddhmgld\ times their limits at \Ndhmgld\ events (Table~\ref{tab:nstar}). In the disproportionation branch $\bullet$CH(OH)$_2$ is the hydrogen donor and gives formic acid with the closed-shell partner. The restricted network of Sect.~\ref{sec:hypothetical} omits these cross reactions. As a sensitivity (Sect.~\ref{sec:caveats}), a fraction of the H$_2$CO can be taken as the free aldehyde. Its OH attack then gives the formyl radical HCO instead of $\bullet$CH(OH)$_2$, and HCO adds to the same radicals as an acyl, so that the new carbon carries an aldehyde. With the gem-diol radical of GA the product is the dehydro-C3, and with its $\alpha$-radical it is hydroxymalonaldehyde, reduced to GLD between events. With the C3 radical it is the keto-aldehyde ERYketm2, the same skeleton as for $\bullet$CH(OH)$_2$. In the disproportionation branch HCO gives CO. A second sensitivity puts a fraction of the OH attack on the hydrated GA and GLD at the CH$_2$OH carbon, which gives the $\alpha$-radical rather than the gem-diol radical. The gem-diol radical of GLD has the same growth channels as
that of GA (rows 134--146): with $\bullet$CH$_2$OH it gives ERU,
with the C2 radicals a C5, with itself or the C3 radical a C6, and its
disproportionation gives glyceric acid. The C5 products of these pairs carry two
hydrogen atoms fewer than a pentose (species RIBm2, rows 158--161), except the product
of the glycol radical (row 141), and are reduced to the pentose between events
(Table~\ref{tab:coldh}), as the corresponding C4 and C6 species are. Pairs whose disproportionation product is not followed by the network are listed with their combination branch only, at the full rate. They are the self-reactions of the aldehyde radicals of GA and of the C3 radical, the C3 radical with the aldehyde or the acyl radical of GA, and the pairs of a C4 radical with the aldehyde radical of GA. In every pair of a gem-diol radical with a
C2 or larger hydroxyalkyl or acyl radical the gem-diol radical is taken as the hydrogen
donor, so that the disproportionation gives the acid; with $\bullet$CH$_2$OH it is the
acceptor (rows 106 and 135). The reverse transfers are not listed. The self-reaction of $\bullet$CH$_2$OH uses the same $k_\mathrm{rad}$ and
$\delta$ as the sugar radicals, its disproportionation giving methanol and
formaldehyde (rows 78 and 79); that of $\bullet$CH(OH)$_2$ uses the room-temperature
value measured by \citet{StockhausenHenglein1971}, $2k=7.2\times10^{8}$~M$^{-1}$\,s$^{-1}$ ($3.6\times10^{8}$ per pair). That measurement reports only the disproportionation; the split with $\delta$ is our assumption (Sect.~\ref{sec:network}).
The combination of two $\bullet$CH(OH)$_2$ (row 43) gives the tetrol (HO)$_2$CH$-$CH(OH)$_2$, the dihydrate of glyoxal, which is carried as GLYOX and reduced to GA by two H atoms between events, charged to the cold-phase budget (Table~\ref{tab:coldh}). Ethylene glycol is attacked by OH
and H at the room-temperature \citet{Buxton1988} rates (rows 87 and 88), returning the glycol radical. The OH$+$H$_2$CO and $\bullet$CH(OH)$_2$ rates are likewise
room-temperature values, whereas OH$+$GA and OH$+$GLD are extrapolated with their
measured Arrhenius dependence (Table~\ref{tab:prov}). Pairs of two
C4 or larger radicals, and pairs of a gem-diol radical with a C4 radical other than
those listed, would give products beyond C6 and are not included. Pairs involving
HCO, the cross reactions of the C1 radicals, and the H-atom repairs (class R of
Table~\ref{tab:reactions}) use the
\citet{Buxton1988} values where they exist and, otherwise, a generic
diffusion-limited $10^{9}$~M$^{-1}$\,s$^{-1}$. The CO reactions take the values of \citet{Buxton1988}: OH$+$CO $1.3\times10^{9}$ (the middle of the measured 0.79--$2.0\times10^{9}$), H$+$CO $1.7\times10^{7}$ and H$+$CO$_2$ $1\times10^{6}$~M$^{-1}$\,s$^{-1}$, the last an upper limit. The NH$_3$ side reactions of Table~\ref{tab:reactions}, which carry no citation in Table~\ref{tab:prov}, use assumed values of $9\times10^{7}$ to $10^{10}$~M$^{-1}$\,s$^{-1}$; they are inactive in the fiducial model, which has no explicit NH$_3$. None of these rates controls a sugar observable
(Sect.~\ref{sec:sensitivity}).

\emph{Integration.} The window of length $\tau$ is covered by 14 outer steps of $\tau/14$,
each with the rate constants at the temperature of the corresponding step of the
history (Sect.~\ref{sec:chain}).
Within each outer step Eq.~(\ref{eq:ode}) is integrated with the explicit
two-stage Heun scheme (an Euler predictor followed by a trapezoidal corrector) and an
adaptive substep
\begin{equation}
\Delta t_\mathrm{s}=\min\!\Big[\Delta t_\mathrm{rem},\;
f_\mathrm{st}\min_{s}\frac{c_s}{|\mathrm{d}c_s/\mathrm{d}t|}\Big],\qquad f_\mathrm{st}=0.05.
\end{equation}
Here $\Delta t_\mathrm{rem}$ is the time left in the outer step and the inner minimum
runs over the species that are decreasing and lie above an absolute floor of
$10^{5}$~m$^{-3}$. No species therefore loses more than 5\% in one substep. The shortest
depletion times at the start of a window are of order picoseconds (OH and
$e^-_\mathrm{aq}$ against the ice inventory), so the substeps are shorter still. Concentrations
are clamped at zero after each stage, a species below the floor is set to zero and does
not control the step, and at most $2\times10^{5}$ substeps are allowed per outer step, a
limit that is not reached. Tightening $f_\mathrm{st}$ to 0.002 changes the abundances by
$10^{-4}$, and removing the floor by $5\times10^{-6}$.

\subsection{Cold-phase reduction between events}
\label{app:coldh}

Between events, H atoms reduce the trapped radicals and dehydro-sugars to their
closed-shell parents (Table~\ref{tab:coldh}). The number of H atoms available per
event is
\begin{equation}
H_\mathrm{b}=\varepsilon\,n_\mathrm{H}\bar v\,\pi a^{2}\,\Delta t_\mathrm{ev}
 \;+\;\eta\,\frac{E_\mathrm{nm}}{100}\,(G_\mathrm{H}+G_{e}),\qquad
 E_\mathrm{nm}=\doseout\,D_\mathrm{melt}V,
\label{eq:hbudget}
\end{equation}
where the first term is the accretion of gas-phase H ($n_\mathrm{H}=3$~cm$^{-3}$,
$\bar v$ the thermal speed at 10~K, $a$ the grain radius and $\Delta t_\mathrm{ev}$ the interval
between events, $2.2\times10^{4}$~yr at \Rwge\ events per Myr, Sect.~\ref{sec:rate}),
$3.0\times10^{7}$ atoms per event,
of which a fraction $\varepsilon$ reaches the trapped species. The second term is the
number of reductants made in the ice by the dose outside the melts $E_\mathrm{nm}$:
$G_\mathrm{H}=1.0$ and $G_{e}=4.8$ per 100~eV are the primary yields of H and
$e^-_\mathrm{aq}$ of Eq.~(\ref{eq:gsource}) (0.6 and 2.7 in the escape set), and $\eta$
is the fraction of them that reduces a trapped species. The hydrated electron counts
because it becomes H on meeting H$_3$O$^+$ or reduces a radical directly. Neither
$\varepsilon$ nor $\eta$ is measured. We take $\varepsilon=0.01$. In the runs the number ratio of
the trapped species to all the species in the ice that react with H (H$_2$CO, CO and
GA) is about $10^{-3}$, so $\varepsilon=0.01$ assumes that an accreted atom prefers a radical tenfold; H adds to a radical without a barrier, but to CO and H$_2$CO only through one. We
take $\eta=1$, an upper limit, because these reductants are made inside the ice next to
the trapped species; with the dose outside the melts of the local spectrum this term is \hinsitu\ per event, ten times the accretion term. If the demand $\sum_j m_{\mathrm{H},j}N_j$ over
Table~\ref{tab:coldh} exceeds $H_\mathrm{b}$, all entries are reduced by the common
factor $H_\mathrm{b}/\sum_j m_{\mathrm{H},j}N_j$. In the fiducial model the in-situ
term and the accretion term ($3.0\times10^{5}$ per event) together
exceed the mean demand per event by a factor of about \hmargin, and the budget is never short. The reduction of the trapped species is therefore complete, and the result does not depend on $n_\mathrm{H}$,
$\varepsilon$ or $\Delta t_\mathrm{ev}$: runs of the restricted network with $\varepsilon=0.001$--0.007, and with $\varepsilon=0$ (in-situ term only), give the count of the fiducial sequences unchanged. The density $n_\mathrm{H}=3$~cm$^{-3}$ is the atomic hydrogen density of a dark cloud at $\zeta\approx10^{-16}$~s$^{-1}$. The runs keep this low value at every $\zeta$, so the accretion term is a lower limit. In a
molecular cloud it is set by the balance of production by cosmic rays and formation
on grains, $n_\mathrm{H}\approx\zeta/(3\times10^{-17}\,\mathrm{cm^{3}\,s^{-1}})$ with the formation rate coefficient of \citet{Jura1975}, and
rises in proportion to $\zeta$. The
same interval enters the substrate resupply, which is capped by the gas-phase carbon
share of one grain (Appendix~\ref{app:cycle}). What matters is $\eta$, the fraction of the in-situ reductants that reach a trapped species. Half of the radiolytic products of that dose are oxidants, which the model does not follow in the solid, so $\eta=1$ is the limit of complete reduction. The opposite limit is the row with $\varepsilon=0.001$ and $\eta=0$ of Tables~\ref{tab:nstar} and \ref{tab:nstarres} (accretion only, $3.0\times10^{4}$ per event). The budget then covers a small part of the demand, most of the trapped radicals enter the next window unreduced, and the count falls to \Nchzero\ (\Nresch\ in the restricted network), with GLD at \gldch\ times its limit. The axis ``trapped small radicals survive'' excludes
$\bullet$CH(OH)$_2$ and $\bullet$CH$_2$OH from the operator and gives \Nsmalloff. A short budget
therefore lowers the count, and the fiducial value is the conservative one.

\begin{table}
\caption{Cold-phase reduction table. Each trapped species is converted to its parent
consuming $m_\mathrm{H}$ hydrogen atoms.}
\label{tab:coldh}
\centering
\small
\begin{tabular}{llc}
\hline\hline
Trapped species & Parent & $m_\mathrm{H}$ \\
\hline
GAyl, GAac, GEMR2 & GA & 1 \\
C3yl & DHA & 1 \\
GLDac, GEMR3 & GLD & 1 \\
ERYm1, ERYac & ERY & 1 \\
ERYm2 & ERY & 2 \\
ERYm2ac & ERY & 3 \\
ERYkyl & ERYket & 1 \\
ERYketm2 & ERYket & 2 \\
ERYketm2ac & ERYket & 3 \\
C5yl, RIBac & RIB & 1 \\
RIBm2 & RIB & 2 \\
RIBm2ac & RIB & 3 \\
C6yl, HEXac & HEX & 1 \\
HEXm2 & HEX & 2 \\
HEXm2ac & HEX & 3 \\
POL2 / POL3 / POL4 / POL5 & EG / GLYC / ERT / PTL & 1 \\
DHM / CH2OH & HCHO / CH3OH & 1 \\
C3m2 & DHA & 2 \\
GLYOX & GA & 2 \\
GLDm2 & GLD & 2 \\
\hline
\end{tabular}
\tablefoot{Species names as in Table~\ref{tab:species}. GLDm2 occurs only in the variants with free H$_2$CO and with the hydroxymalonaldehyde booking. The row DHM / CH2OH is switched off
in the ``trapped small radicals survive'' systematic.}
\end{table}

\subsection{Species and reactions}
\label{app:species}

Table~\ref{tab:species} lists the 60 species of the fiducial network other than the water radiolysis set, with their formulae. Table~\ref{tab:reactions} lists all 199 reactions with the rate constant
evaluated at the reference temperature of 491~K, the activation energy where one is used, the
reaction class and the water stoichiometry. Both tables are generated from the
reaction list that the code writes at run time, so they are the network as
integrated. Reactions listed with $k=0$ are present in the code but switched off in
the fiducial model. They are of two kinds. The OH chain-scission class S is not used for
the C2--C6 sugars, whose OH attack is the abstraction class O, and remains active only
for the C7, C8 and C12 condensates. The OH attack on the GLD gem-diol is switched off
because its factor $1-h_\mathrm{GLD}$ is zero. The OH destruction of glycolic acid
(reaction 86) gives CO$_2$, H$_2$CO and H, so that no carbon-centred radical is
re-injected. The aldose--ketose isomerisation rows carry rate constants scaled from the aldol
anchor (0.1 and 0.03 of the GA$+$GA rate, with the aldol barrier); they transfer
nothing within a window, and the C3 argument of Sect.~\ref{sec:c3excess} does not depend on their rate. Species of the nitrogen network (HCN, glycine, adenine)
present in the code are not populated in the sugar calculations and are omitted.

\onecolumn
\begin{table}[!h]
\caption{Species of the sugar network. The code label ERY denotes the aldotetrose
(erythrose and threose) and ERYket ERU.}
\label{tab:species}
\centering
\footnotesize
\begin{tabular}{llp{10cm}}
\hline\hline
Code & Formula & Description \\
\hline
\texttt{N2H4} & H$_{4}$N$_{2}$ & hydrazine \\
\texttt{NH2} & H$_{2}$N & amino radical \\
\texttt{NH3} & H$_{3}$N & ammonia \\
\texttt{CH2OH} & CH$_{3}$O & hydroxymethyl radical \\
\texttt{CH3OH} & CH$_{4}$O & methanol \\
\texttt{CO} & CO & carbon monoxide \\
\texttt{CO2} & CO$_{2}$ & carbon dioxide \\
\texttt{DHM} & CH$_{3}$O$_{2}$ & dihydroxymethyl radical $\bullet$CH(OH)$_2$ \\
\texttt{FA} & CH$_{2}$O$_{2}$ & formic acid / formate \\
\texttt{HCHO} & CH$_{2}$O & formaldehyde \\
\texttt{HCO} & CHO & formyl radical \\
\texttt{HOCO} & CHO$_{2}$ & HOCO radical \\
\texttt{CANNP} & C$_{2}$H$_{6}$O$_{3}$ & Cannizzaro product pair \\
\texttt{EG} & C$_{2}$H$_{6}$O$_{2}$ & ethylene glycol \\
\texttt{GA} & C$_{2}$H$_{4}$O$_{2}$ & glycolaldehyde \\
\texttt{GAac} & C$_{2}$H$_{3}$O$_{2}$ & $\bullet$C(O)CH$_2$OH (acyl radical of GA) \\
\texttt{GAyl} & C$_{2}$H$_{3}$O$_{2}$ & $\bullet$CH(OH)CHO ($\alpha$-radical of GA) \\
\texttt{GEMR2} & C$_{2}$H$_{5}$O$_{3}$ & $\bullet$C(OH)$_2$CH$_2$OH (gem-diol radical of GA) \\
\texttt{GLYAC} & C$_{2}$H$_{4}$O$_{3}$ & glycolic acid \\
\texttt{GLYOX} & C$_{2}$H$_{2}$O$_{2}$ & glyoxal (as its dihydrate; combination product of two $\bullet$CH(OH)$_2$, reduced to GA between events) \\
\texttt{POL2} & C$_{2}$H$_{5}$O$_{2}$ & $\bullet$CH(OH)CH$_2$OH (glycol radical) \\
\texttt{SACC2} & C$_{2}$H$_{3}$O$_{2}$$^-$ & saccharinate C2 \\
\texttt{C3m2} & C$_{3}$H$_{4}$O$_{3}$ & dehydro-C3 OHC$-$CO$-$CH$_2$OH (combination product of $\bullet$CH(OH)$_2$ with a C2 radical; reduced to DHA between events) \\
\texttt{C3yl} & C$_{3}$H$_{5}$O$_{3}$ & C3 semidione radical shared by GLD and DHA \\
\texttt{DHA} & C$_{3}$H$_{6}$O$_{3}$ & dihydroxyacetone \\
\texttt{GEMR3} & C$_{3}$H$_{7}$O$_{4}$ & gem-diol radical of GLD \\
\texttt{GLCAC} & C$_{3}$H$_{6}$O$_{4}$ & glyceric acid \\
\texttt{GLD} & C$_{3}$H$_{6}$O$_{3}$ & glyceraldehyde \\
\texttt{GLDac} & C$_{3}$H$_{5}$O$_{3}$ & acyl radical of GLD \\
\texttt{GLYC} & C$_{3}$H$_{8}$O$_{3}$ & glycerol \\
\texttt{POL3} & C$_{3}$H$_{7}$O$_{3}$ & glycerol radical \\
\texttt{SACC3} & C$_{3}$H$_{5}$O$_{3}$$^-$ & saccharinate C3 \\
\texttt{ERT} & C$_{4}$H$_{10}$O$_{4}$ & erythritol \\
\texttt{ERY} & C$_{4}$H$_{8}$O$_{4}$ & aldotetrose (erythrose/threose) \\
\texttt{ERYac} & C$_{4}$H$_{7}$O$_{4}$ & acyl radical of C4 \\
\texttt{ERYket} & C$_{4}$H$_{8}$O$_{4}$ & erythrulose \\
\texttt{ERYketm2} & C$_{4}$H$_{6}$O$_{4}$ & 2-keto-tetrose-al (GEMR2 combination product) \\
\texttt{ERYketm2ac} & C$_{4}$H$_{5}$O$_{4}$ & keto-acyl radical of ERYketm2 (returns to erythrulose between events) \\
\texttt{ERYkyl} & C$_{4}$H$_{7}$O$_{4}$ & $\alpha$-radical of erythrulose \\
\texttt{ERYm1} & C$_{4}$H$_{7}$O$_{4}$ & C4 hydroxyalkyl radical \\
\texttt{ERYm2} & C$_{4}$H$_{6}$O$_{4}$ & dehydro-tetrose (dialdehyde) \\
\texttt{ERYm2ac} & C$_{4}$H$_{5}$O$_{4}$ & acyl radical of ERYm2 \\
\texttt{POL4} & C$_{4}$H$_{9}$O$_{4}$ & erythritol radical \\
\texttt{SACC4} & C$_{4}$H$_{7}$O$_{4}$$^-$ & saccharinate C4 \\
\texttt{C5yl} & C$_{5}$H$_{9}$O$_{5}$ & $\alpha$-radical of C5 \\
\texttt{POL5} & C$_{5}$H$_{11}$O$_{5}$ & pentitol radical \\
\texttt{PTL} & C$_{5}$H$_{12}$O$_{5}$ & pentitol \\
\texttt{RIB} & C$_{5}$H$_{10}$O$_{5}$ & pentose (ribose) \\
\texttt{RIBac} & C$_{5}$H$_{9}$O$_{5}$ & acyl radical of C5 \\
\texttt{RIBm2} & C$_{5}$H$_{8}$O$_{5}$ & dehydro-pentose (GEMR3 and C3 radical combination products) \\
\texttt{RIBm2ac} & C$_{5}$H$_{7}$O$_{5}$ & acyl radical of RIBm2 \\
\texttt{SACC5} & C$_{5}$H$_{9}$O$_{5}$$^-$ & saccharinate C5 \\
\texttt{C6yl} & C$_{6}$H$_{11}$O$_{6}$ & $\alpha$-radical of C6 \\
\texttt{HEX} & C$_{6}$H$_{12}$O$_{6}$ & hexose \\
\texttt{HEXac} & C$_{6}$H$_{11}$O$_{6}$ & acyl radical of C6 \\
\texttt{HEXm2} & C$_{6}$H$_{10}$O$_{6}$ & dehydro-hexose \\
\texttt{HEXm2ac} & C$_{6}$H$_{9}$O$_{6}$ & acyl radical of HEXm2 \\
\texttt{HEP} & C$_{7}$H$_{14}$O$_{7}$ & C7 condensate \\
\texttt{OCT} & C$_{8}$H$_{16}$O$_{8}$ & C8 condensate \\
\texttt{DODEC} & C$_{12}$H$_{22}$O$_{11}$ & C12 disaccharide \\
\hline

\end{tabular}
\tablefoot{Radicals are written with their formula; the site is given in the
description. The water species e$^-_\mathrm{aq}$, H, OH, H$_2$, H$_2$O$_2$,
H$_3$O$^+$, OH$^-$ and H$_2$O are not listed.}
\end{table}

\footnotesize
\begin{longtable}{r p{8.4cm} c c c c}
\caption{The reaction network as integrated. $k$ is evaluated at 491~K; second-order
rate constants in M$^{-1}$\,s$^{-1}$, first-order in s$^{-1}$. $E_\mathrm{a}$ in
kJ\,mol$^{-1}$ where an Arrhenius factor is used. Class: W water radiolysis; A closed-shell aldol, retro-aldol, Cannizzaro and saccharinic channels (anchored to
$k(273\,\mathrm{K})=6.4\times10^{-9}$ M$^{-1}$\,s$^{-1}$ for GA$+$GA); S OH chain scission (active only for C7, C8 and C12); O OH abstraction
(rate includes $h_X$ and $f_\mathrm{CHO}$); C radical--radical combination or
disproportionation; R reduction by H or $e^-_\mathrm{aq}$; D unimolecular
decarbonylation or $\beta$-scission. $\dagger$: multiplied by $\mathrm{mob}(\phi)$
when $\phi>0$ (Eq.~(\ref{eq:rate})). $\Delta$H$_2$O: water molecules produced ($+$) or
consumed ($-$) per reaction step.}
\label{tab:reactions}\\
\hline\hline
\# & Reaction & $k$(491 K) & $E_\mathrm{a}$ & Class & $\Delta$H$_2$O \\
\hline
\endfirsthead
\caption{continued.}\\
\hline\hline
\# & Reaction & $k$(491 K) & $E_\mathrm{a}$ & Class & $\Delta$H$_2$O \\
\hline
\endhead
\hline
\endfoot
0 & e$^-_\mathrm{aq}$ $+$ e$^-_\mathrm{aq}$ $\to$ H$_2$ $+$ 2\,OH$^-$ & $5.50\times10^{9}$ &  & W$^\dagger$ & -2 \\
1 & e$^-_\mathrm{aq}$ $+$ H $\to$ H$_2$ $+$ OH$^-$ & $2.50\times10^{10}$ &  & W$^\dagger$ & -1 \\
2 & e$^-_\mathrm{aq}$ $+$ OH $\to$ OH$^-$ & $3.00\times10^{10}$ &  & W$^\dagger$ & 0 \\
3 & e$^-_\mathrm{aq}$ $+$ H$_3$O$^+$ $\to$ H & $2.30\times10^{10}$ &  & W$^\dagger$ & +1 \\
4 & e$^-_\mathrm{aq}$ $+$ H$_2$O$_2$ $\to$ OH $+$ OH$^-$ & $1.10\times10^{10}$ &  & W$^\dagger$ & 0 \\
5 & H $+$ H $\to$ H$_2$ & $7.80\times10^{9}$ &  & W$^\dagger$ & 0 \\
6 & H $+$ OH $\to$ H$_2$O & $7.00\times10^{9}$ &  & W$^\dagger$ & +1 \\
7 & H $+$ H$_2$O$_2$ $\to$ OH & $9.00\times10^{7}$ &  & W & +1 \\
8 & OH $+$ OH $\to$ H$_2$O$_2$ & $5.50\times10^{9}$ &  & W$^\dagger$ & 0 \\
9 & OH $+$ H$_2$ $\to$ H & $4.20\times10^{7}$ &  & W & +1 \\
10 & H$_3$O$^+$ $+$ OH$^-$ $\to$ H$_2$O & $1.40\times10^{11}$ &  & W$^\dagger$ & +2 \\
11 & \texttt{HCHO} $+$ \texttt{HCHO} $\to$ \texttt{GA} & $1.57\times10^{-7}$ & 87.0 & A$^\dagger$ & 0 \\
12 & \texttt{GA} $+$ \texttt{HCHO} $\to$ \texttt{GLD} & $7.87\times10^{-6}$ & 87.0 & A$^\dagger$ & 0 \\
13 & \texttt{GLD} $+$ \texttt{HCHO} $\to$ \texttt{ERY} & $7.87\times10^{-6}$ & 87.0 & A$^\dagger$ & 0 \\
14 & \texttt{ERY} $+$ \texttt{HCHO} $\to$ \texttt{RIB} & $7.87\times10^{-8}$ & 87.0 & A$^\dagger$ & 0 \\
15 & \texttt{ERY} $\to$ 2\,\texttt{GA} & $1.57\times10^{-7}$ & 87.0 & A$^\dagger$ & 0 \\
16 & \texttt{GA} $+$ \texttt{GA} $\to$ \texttt{ERY} & $1.57\times10^{-1}$ & 87.0 & A$^\dagger$ & 0 \\
17 & \texttt{GA} $+$ \texttt{GLD} $\to$ \texttt{RIB} & $7.87\times10^{-8}$ & 87.0 & A$^\dagger$ & 0 \\
18 & \texttt{GLD} $+$ OH$^-$ $\to$ \texttt{DHA} $+$ OH$^-$ & $1.57\times10^{-2}$ & 87.0 & A$^\dagger$ & 0 \\
19 & \texttt{DHA} $+$ OH$^-$ $\to$ \texttt{GLD} $+$ OH$^-$ & $4.72\times10^{-3}$ & 87.0 & A$^\dagger$ & 0 \\
20 & \texttt{DHA} $+$ \texttt{HCHO} $\to$ \texttt{ERY} & $7.87\times10^{-6}$ & 87.0 & A$^\dagger$ & 0 \\
21 & \texttt{GLD} $+$ \texttt{DHA} $\to$ \texttt{HEX} & $7.87\times10^{-6}$ & 87.0 & A$^\dagger$ & 0 \\
22 & \texttt{ERY} $+$ \texttt{GA} $\to$ \texttt{HEX} & $7.87\times10^{-6}$ & 87.0 & A$^\dagger$ & 0 \\
23 & \texttt{GLD} $+$ \texttt{ERY} $\to$ \texttt{HEP} & $7.87\times10^{-6}$ & 87.0 & A$^\dagger$ & 0 \\
24 & \texttt{GA} $+$ \texttt{RIB} $\to$ \texttt{HEP} & $7.87\times10^{-6}$ & 87.0 & A$^\dagger$ & 0 \\
25 & \texttt{ERY} $+$ \texttt{ERY} $\to$ \texttt{OCT} & $7.87\times10^{-6}$ & 87.0 & A$^\dagger$ & 0 \\
26 & \texttt{GA} $+$ \texttt{HEX} $\to$ \texttt{OCT} & $7.87\times10^{-6}$ & 87.0 & A$^\dagger$ & 0 \\
27 & \texttt{HEX} $+$ \texttt{HEX} $\to$ \texttt{DODEC} & $7.87\times10^{-6}$ & 87.0 & A$^\dagger$ & +1 \\
28 & \texttt{HCHO} $+$ \texttt{HCHO} $\to$ \texttt{CANNP} & $1.57\times10^{-7}$ & 87.0 & A$^\dagger$ & -1 \\
29 & \texttt{GA} $+$ OH$^-$ $\to$ \texttt{SACC2} & $1.57\times10^{-7}$ & 87.0 & A$^\dagger$ & +1 \\
30 & \texttt{GLD} $+$ OH$^-$ $\to$ \texttt{SACC3} & $1.57\times10^{-7}$ & 87.0 & A$^\dagger$ & +1 \\
31 & \texttt{ERY} $+$ OH$^-$ $\to$ \texttt{SACC4} & $1.57\times10^{-7}$ & 87.0 & A$^\dagger$ & +1 \\
32 & \texttt{RIB} $+$ OH$^-$ $\to$ \texttt{SACC5} & $1.57\times10^{-7}$ & 87.0 & A$^\dagger$ & +1 \\
33 & \texttt{GA} $+$ OH $\to$ 2\,\texttt{HCHO} $+$ 0.5\,H$_2$O$_2$ & 0 &  & S$^\dagger$ & 0 \\
34 & \texttt{GLD} $+$ OH $\to$ \texttt{GA} $+$ \texttt{HCHO} $+$ 0.5\,H$_2$O$_2$ & 0 &  & S$^\dagger$ & 0 \\
35 & \texttt{DHA} $+$ OH $\to$ \texttt{GA} $+$ \texttt{HCHO} $+$ 0.5\,H$_2$O$_2$ & 0 &  & S$^\dagger$ & 0 \\
36 & \texttt{ERY} $+$ OH $\to$ 2\,\texttt{GA} $+$ 0.5\,H$_2$O$_2$ & 0 &  & S$^\dagger$ & 0 \\
37 & \texttt{RIB} $+$ OH $\to$ \texttt{GA} $+$ \texttt{GLD} $+$ 0.5\,H$_2$O$_2$ & 0 &  & S$^\dagger$ & 0 \\
38 & \texttt{HEX} $+$ OH $\to$ 2\,\texttt{GLD} $+$ 0.5\,H$_2$O$_2$ & 0 &  & S$^\dagger$ & 0 \\
39 & \texttt{HEP} $+$ OH $\to$ \texttt{GLD} $+$ \texttt{ERY} $+$ 0.5\,H$_2$O$_2$ & $1.50\times10^{9}$ &  & S$^\dagger$ & 0 \\
40 & \texttt{OCT} $+$ OH $\to$ 2\,\texttt{ERY} $+$ 0.5\,H$_2$O$_2$ & $1.50\times10^{9}$ &  & S$^\dagger$ & 0 \\
41 & \texttt{DODEC} $+$ OH $\to$ 2\,\texttt{HEX} $+$ 0.5\,H$_2$O$_2$ & $1.50\times10^{9}$ &  & S$^\dagger$ & -1 \\
42 & OH $+$ \texttt{HCHO} $\to$ \texttt{DHM} & $1.00\times10^{9}$ &  & O$^\dagger$ & 0 \\
43 & \texttt{DHM} $+$ \texttt{DHM} $\to$ \texttt{GLYOX} & $1.91\times10^{8}$ &  & C$^\dagger$ & +2 \\
44 & \texttt{DHM} $+$ \texttt{DHM} $\to$ \texttt{HCHO} $+$ \texttt{FA} & $1.69\times10^{8}$ &  & C$^\dagger$ & +1 \\
45 & \texttt{DHM} $+$ OH $\to$ \texttt{FA} & $1.00\times10^{9}$ &  & O$^\dagger$ & +1 \\
46 & H $+$ \texttt{HCHO} $\to$ \texttt{CH2OH} & $5.00\times10^{6}$ &  & R$^\dagger$ & 0 \\
47 & e$^-_\mathrm{aq}$ $+$ \texttt{HCHO} $\to$ \texttt{CH2OH} $+$ OH$^-$ & $1.00\times10^{7}$ &  & R$^\dagger$ & -1 \\
48 & \texttt{CH2OH} $+$ OH $\to$ \texttt{HCHO} & $1.00\times10^{9}$ &  & O$^\dagger$ & +1 \\
49 & \texttt{DHM} $+$ H $\to$ \texttt{HCHO} & $1.00\times10^{9}$ &  & R$^\dagger$ & +1 \\
50 & OH $+$ \texttt{FA} $\to$ \texttt{CO2} $+$ H & $1.30\times10^{8}$ &  & O$^\dagger$ & +1 \\
51 & e$^-_\mathrm{aq}$ $+$ \texttt{CO2} $\to$ \texttt{HOCO} $+$ OH$^-$ & $7.70\times10^{9}$ &  & R$^\dagger$ & -1 \\
52 & H $+$ \texttt{CO2} $\to$ \texttt{HOCO} & $1.00\times10^{6}$ &  & R$^\dagger$ & 0 \\
53 & \texttt{HOCO} $+$ \texttt{HOCO} $\to$ \texttt{FA} $+$ \texttt{CO2} & $1.00\times10^{9}$ &  & C$^\dagger$ & 0 \\
54 & OH $+$ \texttt{NH3} $\to$ \texttt{NH2} & $9.00\times10^{7}$ &  & O & +1 \\
55 & H $+$ \texttt{NH2} $\to$ \texttt{NH3} & $1.00\times10^{10}$ &  & R$^\dagger$ & 0 \\
56 & e$^-_\mathrm{aq}$ $+$ \texttt{NH2} $\to$ \texttt{NH3} $+$ OH$^-$ & $1.00\times10^{10}$ &  & R$^\dagger$ & -1 \\
57 & \texttt{NH2} $+$ \texttt{NH2} $\to$ \texttt{N2H4} & $1.00\times10^{9}$ &  & C$^\dagger$ & 0 \\
58 & OH $+$ \texttt{CO} $\to$ \texttt{HOCO} & $1.30\times10^{9}$ &  & O$^\dagger$ & 0 \\
59 & H $+$ \texttt{CO} $\to$ \texttt{HCO} & $1.70\times10^{7}$ &  & R$^\dagger$ & 0 \\
60 & \texttt{HCO} $+$ H $\to$ \texttt{HCHO} & $5.00\times10^{9}$ &  & R$^\dagger$ & 0 \\
61 & \texttt{HCO} $+$ \texttt{HCO} $\to$ \texttt{HCHO} $+$ \texttt{CO} & $1.00\times10^{9}$ &  & C$^\dagger$ & 0 \\
62 & \texttt{HCO} $+$ OH $\to$ \texttt{CO} & $1.00\times10^{9}$ &  & O$^\dagger$ & +1 \\
63 & \texttt{HCO} $+$ \texttt{CH2OH} $\to$ \texttt{GA} & $1.00\times10^{9}$ &  & C$^\dagger$ & 0 \\
64 & OH $+$ \texttt{GA} $\to$ 0.5\,\texttt{GAac} $+$ 0.5\,\texttt{GAyl} & $1.24\times10^{9}$ & 14.5 & O$^\dagger$ & +1 \\
65 & OH $+$ \texttt{GA} $\to$ \texttt{GEMR2} & $1.12\times10^{10}$ & 14.5 & O$^\dagger$ & 0 \\
66 & OH $+$ \texttt{GLD} $\to$ 0.5\,\texttt{GLDac} $+$ 0.5\,\texttt{C3yl} & $4.67\times10^{9}$ & 8.0 & O$^\dagger$ & +1 \\
67 & OH $+$ \texttt{GLD} $\to$ \texttt{GEMR3} & 0 & 8.0 & O$^\dagger$ & 0 \\
68 & \texttt{GEMR2} $+$ H $\to$ \texttt{GA} & $1.00\times10^{9}$ &  & R$^\dagger$ & +1 \\
69 & \texttt{GEMR3} $+$ H $\to$ \texttt{GLD} & $1.00\times10^{9}$ &  & R$^\dagger$ & +1 \\
70 & \texttt{GEMR2} $+$ \texttt{GEMR2} $\to$ \texttt{GA} $+$ \texttt{GLYAC} & $2.28\times10^{9}$ &  & C$^\dagger$ & +1 \\
71 & \texttt{GEMR3} $+$ \texttt{GEMR3} $\to$ \texttt{GLD} $+$ \texttt{GLCAC} & $2.28\times10^{9}$ &  & C$^\dagger$ & +1 \\
72 & OH $+$ \texttt{DHA} $\to$ \texttt{C3yl} & $4.67\times10^{9}$ & 8.0 & O$^\dagger$ & +1 \\
73 & OH $+$ \texttt{ERYm2} $\to$ \texttt{ERYm2ac} & $9.34\times10^{7}$ & 8.0 & O$^\dagger$ & +1 \\
74 & OH $+$ \texttt{HEXm2} $\to$ \texttt{HEXm2ac} & $9.34\times10^{7}$ & 8.0 & O$^\dagger$ & +1 \\
75 & OH $+$ \texttt{ERY} $\to$ 0.225\,\texttt{ERYac} $+$ 0.775\,\texttt{ERYm1} & $4.67\times10^{9}$ & 8.0 & O$^\dagger$ & +1 \\
76 & OH $+$ \texttt{RIB} $\to$ 0.01\,\texttt{RIBac} $+$ 0.99\,\texttt{C5yl} & $4.67\times10^{9}$ & 8.0 & O$^\dagger$ & +1 \\
77 & OH $+$ \texttt{HEX} $\to$ 0.005\,\texttt{HEXac} $+$ 0.995\,\texttt{C6yl} & $4.67\times10^{9}$ & 8.0 & O$^\dagger$ & +1 \\
78 & \texttt{CH2OH} $+$ \texttt{CH2OH} $\to$ \texttt{EG} & $2.58\times10^{9}$ &  & C$^\dagger$ & 0 \\
79 & \texttt{CH2OH} $+$ \texttt{CH2OH} $\to$ \texttt{CH3OH} $+$ \texttt{HCHO} & $2.28\times10^{9}$ &  & C$^\dagger$ & 0 \\
80 & \texttt{GAyl} $+$ \texttt{CH2OH} $\to$ \texttt{GLD} & $5.15\times10^{9}$ &  & C$^\dagger$ & 0 \\
81 & \texttt{GAyl} $+$ \texttt{CH2OH} $\to$ \texttt{GA} $+$ \texttt{HCHO} & $4.57\times10^{9}$ &  & C$^\dagger$ & 0 \\
82 & \texttt{GAac} $+$ \texttt{CH2OH} $\to$ \texttt{DHA} & $5.15\times10^{9}$ &  & C$^\dagger$ & 0 \\
83 & \texttt{GAac} $+$ \texttt{CH2OH} $\to$ \texttt{GA} $+$ \texttt{HCHO} & $4.57\times10^{9}$ &  & C$^\dagger$ & 0 \\
84 & OH $+$ \texttt{CH3OH} $\to$ \texttt{CH2OH} & $9.70\times10^{8}$ &  & O$^\dagger$ & +1 \\
85 & H $+$ \texttt{CH3OH} $\to$ \texttt{CH2OH} $+$ H$_2$ & $2.60\times10^{6}$ &  & R & 0 \\
86 & OH $+$ \texttt{GLYAC} $\to$ \texttt{CO2} $+$ \texttt{HCHO} $+$ H & $6.00\times10^{8}$ &  & O$^\dagger$ & +1 \\
87 & OH $+$ \texttt{EG} $\to$ \texttt{POL2} & $1.80\times10^{9}$ &  & O$^\dagger$ & +1 \\
88 & H $+$ \texttt{EG} $\to$ \texttt{POL2} $+$ H$_2$ & $1.40\times10^{7}$ &  & R$^\dagger$ & 0 \\
89 & \texttt{GAac} $+$ \texttt{POL2} $\to$ \texttt{ERYket} & $5.15\times10^{9}$ &  & C$^\dagger$ & 0 \\
90 & \texttt{GAac} $+$ \texttt{POL2} $\to$ 2\,\texttt{GA} & $4.57\times10^{9}$ &  & C$^\dagger$ & 0 \\
91 & \texttt{GAyl} $+$ \texttt{POL2} $\to$ \texttt{ERY} & $5.15\times10^{9}$ &  & C$^\dagger$ & 0 \\
92 & \texttt{GAyl} $+$ \texttt{POL2} $\to$ 2\,\texttt{GA} & $4.57\times10^{9}$ &  & C$^\dagger$ & 0 \\
93 & \texttt{POL2} $+$ \texttt{POL2} $\to$ \texttt{ERT} & $2.58\times10^{9}$ &  & C$^\dagger$ & 0 \\
94 & \texttt{POL2} $+$ \texttt{POL2} $\to$ \texttt{EG} $+$ \texttt{GA} & $2.28\times10^{9}$ &  & C$^\dagger$ & 0 \\
95 & \texttt{POL2} $+$ \texttt{CH2OH} $\to$ \texttt{GLYC} & $5.15\times10^{9}$ &  & C$^\dagger$ & 0 \\
96 & \texttt{POL2} $+$ \texttt{CH2OH} $\to$ \texttt{EG} $+$ \texttt{HCHO} & $4.57\times10^{9}$ &  & C$^\dagger$ & 0 \\
97 & OH $+$ \texttt{ERYket} $\to$ \texttt{ERYkyl} & $4.67\times10^{9}$ & 8.0 & O$^\dagger$ & +1 \\
98 & \texttt{ERYkyl} $+$ \texttt{CH2OH} $\to$ \texttt{RIB} & $5.15\times10^{9}$ &  & C$^\dagger$ & 0 \\
99 & \texttt{ERYkyl} $+$ \texttt{CH2OH} $\to$ \texttt{ERYket} $+$ \texttt{HCHO} & $4.57\times10^{9}$ &  & C$^\dagger$ & 0 \\
100 & \texttt{ERYkyl} $+$ \texttt{GAyl} $\to$ \texttt{HEXm2} & $9.72\times10^{9}$ &  & C$^\dagger$ & 0 \\
101 & \texttt{ERYkyl} $+$ H $\to$ \texttt{ERYket} & $1.00\times10^{9}$ &  & R$^\dagger$ & 0 \\
102 & \texttt{GEMR2} $+$ \texttt{GAyl} $\to$ \texttt{ERYketm2} & $5.15\times10^{9}$ &  & C$^\dagger$ & +1 \\
103 & \texttt{GEMR2} $+$ \texttt{GAyl} $\to$ \texttt{GA} $+$ \texttt{GLYAC} & $4.57\times10^{9}$ &  & C$^\dagger$ & 0 \\
104 & \texttt{GEMR2} $+$ \texttt{GEMR2} $\to$ \texttt{ERYketm2} & $2.58\times10^{9}$ &  & C$^\dagger$ & +2 \\
105 & \texttt{GEMR2} $+$ \texttt{CH2OH} $\to$ \texttt{DHA} & $5.15\times10^{9}$ &  & C$^\dagger$ & +1 \\
106 & \texttt{GEMR2} $+$ \texttt{CH2OH} $\to$ \texttt{GA} $+$ \texttt{HCHO} & $4.57\times10^{9}$ &  & C$^\dagger$ & +1 \\
107 & \texttt{GEMR2} $+$ \texttt{POL2} $\to$ \texttt{ERYket} & $5.15\times10^{9}$ &  & C$^\dagger$ & +1 \\
108 & \texttt{GEMR2} $+$ \texttt{POL2} $\to$ \texttt{GLYAC} $+$ \texttt{EG} & $4.57\times10^{9}$ &  & C$^\dagger$ & 0 \\
109 & \texttt{GEMR2} $+$ \texttt{GAac} $\to$ \texttt{ERYketm2} & $5.15\times10^{9}$ &  & C$^\dagger$ & +1 \\
110 & \texttt{GEMR2} $+$ \texttt{GAac} $\to$ \texttt{GLYAC} $+$ \texttt{GA} & $4.57\times10^{9}$ &  & C$^\dagger$ & 0 \\
111 & \texttt{DHM} $+$ \texttt{GEMR2} $\to$ \texttt{C3m2} & $5.15\times10^{9}$ &  & C$^\dagger$ & +2 \\
112 & \texttt{DHM} $+$ \texttt{GEMR2} $\to$ \texttt{FA} $+$ \texttt{GA} & $4.57\times10^{9}$ &  & C$^\dagger$ & +1 \\
113 & \texttt{DHM} $+$ \texttt{GAyl} $\to$ \texttt{C3m2} & $5.15\times10^{9}$ &  & C$^\dagger$ & +1 \\
114 & \texttt{DHM} $+$ \texttt{GAyl} $\to$ \texttt{FA} $+$ \texttt{GA} & $4.57\times10^{9}$ &  & C$^\dagger$ & 0 \\
115 & \texttt{DHM} $+$ \texttt{GAac} $\to$ \texttt{C3m2} & $5.15\times10^{9}$ &  & C$^\dagger$ & +1 \\
116 & \texttt{DHM} $+$ \texttt{GAac} $\to$ \texttt{FA} $+$ \texttt{GA} & $4.57\times10^{9}$ &  & C$^\dagger$ & 0 \\
117 & \texttt{DHM} $+$ \texttt{CH2OH} $\to$ \texttt{GA} & $5.15\times10^{9}$ &  & C$^\dagger$ & +1 \\
118 & \texttt{DHM} $+$ \texttt{CH2OH} $\to$ \texttt{FA} $+$ \texttt{CH3OH} & $4.57\times10^{9}$ &  & C$^\dagger$ & 0 \\
119 & \texttt{DHM} $+$ \texttt{POL2} $\to$ \texttt{GLD} & $5.15\times10^{9}$ &  & C$^\dagger$ & +1 \\
120 & \texttt{DHM} $+$ \texttt{POL2} $\to$ \texttt{FA} $+$ \texttt{EG} & $4.57\times10^{9}$ &  & C$^\dagger$ & 0 \\
121 & \texttt{DHM} $+$ \texttt{C3yl} $\to$ \texttt{ERYketm2} & $5.15\times10^{9}$ &  & C$^\dagger$ & +1 \\
122 & \texttt{DHM} $+$ \texttt{C3yl} $\to$ \texttt{FA} $+$ \texttt{GLD} & $4.57\times10^{9}$ &  & C$^\dagger$ & 0 \\
123 & \texttt{DHM} $+$ \texttt{GEMR3} $\to$ \texttt{ERYketm2} & $5.15\times10^{9}$ &  & C$^\dagger$ & +2 \\
124 & \texttt{DHM} $+$ \texttt{GEMR3} $\to$ \texttt{FA} $+$ \texttt{GLD} & $4.57\times10^{9}$ &  & C$^\dagger$ & +1 \\
125 & \texttt{DHM} $+$ \texttt{ERYkyl} $\to$ \texttt{RIBm2} & $5.15\times10^{9}$ &  & C$^\dagger$ & +1 \\
126 & \texttt{DHM} $+$ \texttt{ERYkyl} $\to$ \texttt{FA} $+$ \texttt{ERYket} & $4.57\times10^{9}$ &  & C$^\dagger$ & 0 \\
127 & \texttt{DHM} $+$ \texttt{ERYm1} $\to$ \texttt{RIBm2} & $5.15\times10^{9}$ &  & C$^\dagger$ & +1 \\
128 & \texttt{DHM} $+$ \texttt{ERYm1} $\to$ \texttt{FA} $+$ \texttt{ERY} & $4.57\times10^{9}$ &  & C$^\dagger$ & 0 \\
129 & \texttt{DHM} $+$ \texttt{C5yl} $\to$ \texttt{HEXm2} & $5.15\times10^{9}$ &  & C$^\dagger$ & +1 \\
130 & \texttt{DHM} $+$ \texttt{C5yl} $\to$ \texttt{FA} $+$ \texttt{RIB} & $4.57\times10^{9}$ &  & C$^\dagger$ & 0 \\
131 & \texttt{C3m2} $+$ H $\to$ \texttt{C3yl} & $5.00\times10^{6}$ &  & R$^\dagger$ & 0 \\
132 & \texttt{ERYketm2} $+$ H $\to$ \texttt{ERYkyl} & $5.00\times10^{6}$ &  & R$^\dagger$ & 0 \\
133 & OH $+$ \texttt{ERYketm2} $\to$ \texttt{ERYketm2ac} & $9.34\times10^{7}$ & 8.0 & O$^\dagger$ & +1 \\
134 & \texttt{GEMR3} $+$ \texttt{CH2OH} $\to$ \texttt{ERYket} & $5.15\times10^{9}$ &  & C$^\dagger$ & +1 \\
135 & \texttt{GEMR3} $+$ \texttt{CH2OH} $\to$ \texttt{GLD} $+$ \texttt{HCHO} & $4.57\times10^{9}$ &  & C$^\dagger$ & +1 \\
136 & \texttt{GEMR3} $+$ \texttt{GAyl} $\to$ \texttt{RIBm2} & $5.15\times10^{9}$ &  & C$^\dagger$ & +1 \\
137 & \texttt{GEMR3} $+$ \texttt{GAyl} $\to$ \texttt{GLCAC} $+$ \texttt{GA} & $4.57\times10^{9}$ &  & C$^\dagger$ & 0 \\
138 & \texttt{GEMR3} $+$ \texttt{GEMR2} $\to$ \texttt{RIBm2} & $5.15\times10^{9}$ &  & C$^\dagger$ & +2 \\
139 & \texttt{GEMR3} $+$ \texttt{GEMR2} $\to$ 0.5\,\texttt{GLCAC} $+$ 0.5\,\texttt{GA} $+$ 0.5\,\texttt{GLYAC} $+$ 0.5\,\texttt{GLD} & $4.57\times10^{9}$ &  & C$^\dagger$ & +1 \\
140 & \texttt{GEMR3} $+$ \texttt{GEMR3} $\to$ \texttt{HEXm2} & $2.58\times10^{9}$ &  & C$^\dagger$ & +2 \\
141 & \texttt{GEMR3} $+$ \texttt{POL2} $\to$ \texttt{RIB} & $5.15\times10^{9}$ &  & C$^\dagger$ & +1 \\
142 & \texttt{GEMR3} $+$ \texttt{POL2} $\to$ \texttt{GLCAC} $+$ \texttt{EG} & $4.57\times10^{9}$ &  & C$^\dagger$ & 0 \\
143 & \texttt{GEMR3} $+$ \texttt{GAac} $\to$ \texttt{RIBm2} & $5.15\times10^{9}$ &  & C$^\dagger$ & +1 \\
144 & \texttt{GEMR3} $+$ \texttt{GAac} $\to$ \texttt{GLCAC} $+$ \texttt{GA} & $4.57\times10^{9}$ &  & C$^\dagger$ & 0 \\
145 & \texttt{GEMR3} $+$ \texttt{C3yl} $\to$ \texttt{HEXm2} & $5.15\times10^{9}$ &  & C$^\dagger$ & +1 \\
146 & \texttt{GEMR3} $+$ \texttt{C3yl} $\to$ \texttt{GLCAC} $+$ \texttt{GLD} & $4.57\times10^{9}$ &  & C$^\dagger$ & 0 \\
147 & \texttt{GEMR2} $+$ \texttt{C3yl} $\to$ \texttt{RIBm2} & $5.15\times10^{9}$ &  & C$^\dagger$ & +1 \\
148 & \texttt{GEMR2} $+$ \texttt{C3yl} $\to$ \texttt{GLYAC} $+$ \texttt{GLD} & $4.57\times10^{9}$ &  & C$^\dagger$ & 0 \\
149 & \texttt{GEMR2} $+$ \texttt{ERYkyl} $\to$ \texttt{HEXm2} & $5.15\times10^{9}$ &  & C$^\dagger$ & +1 \\
150 & \texttt{GEMR2} $+$ \texttt{ERYkyl} $\to$ \texttt{GLYAC} $+$ \texttt{ERYket} & $4.57\times10^{9}$ &  & C$^\dagger$ & 0 \\
151 & \texttt{GEMR2} $+$ \texttt{ERYm1} $\to$ \texttt{HEXm2} & $5.15\times10^{9}$ &  & C$^\dagger$ & +1 \\
152 & \texttt{GEMR2} $+$ \texttt{ERYm1} $\to$ \texttt{GLYAC} $+$ \texttt{ERY} & $4.57\times10^{9}$ &  & C$^\dagger$ & 0 \\
153 & \texttt{C3yl} $+$ \texttt{C3yl} $\to$ \texttt{HEXm2} & $4.86\times10^{9}$ &  & C$^\dagger$ & 0 \\
154 & \texttt{GAyl} $+$ \texttt{C3yl} $\to$ \texttt{RIBm2} & $9.72\times10^{9}$ &  & C$^\dagger$ & 0 \\
155 & \texttt{POL2} $+$ \texttt{C3yl} $\to$ \texttt{RIB} & $5.15\times10^{9}$ &  & C$^\dagger$ & 0 \\
156 & \texttt{POL2} $+$ \texttt{C3yl} $\to$ \texttt{GLD} $+$ \texttt{GA} & $4.57\times10^{9}$ &  & C$^\dagger$ & 0 \\
157 & \texttt{GAac} $+$ \texttt{C3yl} $\to$ \texttt{RIBm2} & $9.72\times10^{9}$ &  & C$^\dagger$ & 0 \\
158 & \texttt{RIBm2} $+$ H $\to$ \texttt{C5yl} & $5.00\times10^{6}$ &  & R$^\dagger$ & 0 \\
159 & OH $+$ \texttt{RIBm2} $\to$ \texttt{RIBm2ac} & $9.34\times10^{7}$ & 8.0 & O$^\dagger$ & +1 \\
160 & \texttt{RIBm2ac} $+$ H $\to$ \texttt{RIBm2} & $1.00\times10^{9}$ &  & R$^\dagger$ & 0 \\
161 & \texttt{RIBm2ac} $\to$ \texttt{ERYkyl} $+$ \texttt{CO} & $1.41\times10^{9}$ & 36.2 & D & 0 \\
162 & \texttt{C3yl} $+$ \texttt{CH2OH} $\to$ \texttt{ERY} & $5.15\times10^{9}$ &  & C$^\dagger$ & 0 \\
163 & \texttt{C3yl} $+$ \texttt{CH2OH} $\to$ \texttt{GLD} $+$ \texttt{HCHO} & $4.57\times10^{9}$ &  & C$^\dagger$ & 0 \\
164 & \texttt{GAyl} $+$ \texttt{GAyl} $\to$ \texttt{ERYm2} & $4.86\times10^{9}$ &  & C$^\dagger$ & 0 \\
165 & \texttt{ERYm1} $+$ \texttt{CH2OH} $\to$ \texttt{RIB} & $5.15\times10^{9}$ &  & C$^\dagger$ & 0 \\
166 & \texttt{ERYm1} $+$ \texttt{CH2OH} $\to$ \texttt{ERY} $+$ \texttt{HCHO} & $4.57\times10^{9}$ &  & C$^\dagger$ & 0 \\
167 & \texttt{ERYm1} $+$ \texttt{GAyl} $\to$ \texttt{HEXm2} & $9.72\times10^{9}$ &  & C$^\dagger$ & 0 \\
168 & \texttt{C5yl} $+$ \texttt{CH2OH} $\to$ \texttt{HEX} & $5.15\times10^{9}$ &  & C$^\dagger$ & 0 \\
169 & \texttt{C5yl} $+$ \texttt{CH2OH} $\to$ \texttt{RIB} $+$ \texttt{HCHO} & $4.57\times10^{9}$ &  & C$^\dagger$ & 0 \\
170 & \texttt{ERYm2} $+$ H $\to$ \texttt{ERYm1} & $5.00\times10^{6}$ &  & R$^\dagger$ & 0 \\
171 & \texttt{HEXm2} $+$ H $\to$ \texttt{C6yl} & $5.00\times10^{6}$ &  & R$^\dagger$ & 0 \\
172 & \texttt{ERYm1} $+$ H $\to$ \texttt{ERY} & $1.00\times10^{9}$ &  & R$^\dagger$ & 0 \\
173 & \texttt{C3yl} $+$ H $\to$ \texttt{GLD} & $1.00\times10^{9}$ &  & R$^\dagger$ & 0 \\
174 & \texttt{GAyl} $+$ H $\to$ \texttt{GA} & $1.00\times10^{9}$ &  & R$^\dagger$ & 0 \\
175 & \texttt{GAac} $+$ H $\to$ \texttt{GA} & $1.00\times10^{9}$ &  & R$^\dagger$ & 0 \\
176 & \texttt{GLDac} $+$ H $\to$ \texttt{GLD} & $1.00\times10^{9}$ &  & R$^\dagger$ & 0 \\
177 & \texttt{ERYac} $+$ H $\to$ \texttt{ERY} & $1.00\times10^{9}$ &  & R$^\dagger$ & 0 \\
178 & \texttt{POL2} $+$ H $\to$ \texttt{EG} & $1.00\times10^{9}$ &  & R$^\dagger$ & 0 \\
179 & \texttt{POL3} $+$ H $\to$ \texttt{GLYC} & $1.00\times10^{9}$ &  & R$^\dagger$ & 0 \\
180 & \texttt{C5yl} $+$ H $\to$ \texttt{RIB} & $1.00\times10^{9}$ &  & R$^\dagger$ & 0 \\
181 & \texttt{C6yl} $+$ H $\to$ \texttt{HEX} & $1.00\times10^{9}$ &  & R$^\dagger$ & 0 \\
182 & \texttt{RIBac} $+$ H $\to$ \texttt{RIB} & $1.00\times10^{9}$ &  & R$^\dagger$ & 0 \\
183 & \texttt{HEXac} $+$ H $\to$ \texttt{HEX} & $1.00\times10^{9}$ &  & R$^\dagger$ & 0 \\
184 & \texttt{ERYm2ac} $+$ H $\to$ \texttt{ERYm2} & $1.00\times10^{9}$ &  & R$^\dagger$ & 0 \\
185 & \texttt{ERYketm2ac} $+$ H $\to$ \texttt{ERYketm2} & $1.00\times10^{9}$ &  & R$^\dagger$ & 0 \\
186 & \texttt{HEXm2ac} $+$ H $\to$ \texttt{HEXm2} & $1.00\times10^{9}$ &  & R$^\dagger$ & 0 \\
187 & \texttt{POL4} $+$ H $\to$ \texttt{ERT} & $1.00\times10^{9}$ &  & R$^\dagger$ & 0 \\
188 & \texttt{POL5} $+$ H $\to$ \texttt{PTL} & $1.00\times10^{9}$ &  & R$^\dagger$ & 0 \\
189 & \texttt{GAac} $\to$ \texttt{CH2OH} $+$ \texttt{CO} & $1.80\times10^{8}$ & 44.6 & D & 0 \\
190 & \texttt{GLDac} $\to$ \texttt{POL2} $+$ \texttt{CO} & $2.03\times10^{9}$ & 34.7 & D & 0 \\
191 & \texttt{ERYac} $\to$ \texttt{POL3} $+$ \texttt{CO} & $3.85\times10^{8}$ & 41.5 & D & 0 \\
192 & \texttt{RIBac} $\to$ \texttt{POL4} $+$ \texttt{CO} & $1.41\times10^{9}$ & 36.2 & D & 0 \\
193 & \texttt{HEXac} $\to$ \texttt{POL5} $+$ \texttt{CO} & $3.85\times10^{8}$ & 41.5 & D & 0 \\
194 & \texttt{ERYm2ac} $\to$ \texttt{C3yl} $+$ \texttt{CO} & $3.85\times10^{8}$ & 41.5 & D & 0 \\
195 & \texttt{ERYketm2ac} $\to$ \texttt{C3yl} $+$ \texttt{CO} & $3.85\times10^{8}$ & 41.5 & D & 0 \\
196 & \texttt{HEXm2ac} $\to$ \texttt{C5yl} $+$ \texttt{CO} & $3.85\times10^{8}$ & 41.5 & D & 0 \\
197 & \texttt{ERYm1} $\to$ \texttt{GA} $+$ \texttt{GAyl} & $1.94\times10^{6}$ & 63.1 & D & 0 \\
198 & \texttt{C3yl} $\to$ \texttt{HCHO} $+$ \texttt{GAyl} & $1.80\times10^{2}$ & 101.0 & D & 0 \\

\end{longtable}
\normalsize
\twocolumn

\subsection{Anchoring of the topology}
Three features of the topology are essential to the results, and each is anchored to
an independent quantity. The decomposition sinks are the acyl decarbonylations with
computed barriers (Table~\ref{tab:prov}); without them the C4/C2 ordering is not
reproduced by growth alone. The carbonyl-availability filter $h_X$ takes the measured
hydration equilibrium of GA; the values for GLD,
DHA, the aldotetrose and the C5 and C6 sugars are assumptions varied as
axes (Table~\ref{tab:prov}); without the filter \hnoneresald\ of the C4 of the restricted network is the aldotetrose.
The cold-phase reduction between events has as its budget the
accreted and in-situ radiolytic H, which exceeds the mean demand by a factor of about \hmargin\ (Appendix~\ref{app:coldh}); without it most of the modelled C4 remains as the
dehydro-tetroses and their radicals rather than as ERU. The gem-diol radical channel follows from the
hydration filter.

\subsection{Acid--base speciation}
Formic acid is produced by the disproportionation of the C1 radicals and by the
oxidation of the dihydroxymethyl radical. It is destroyed by OH about 25 times faster as
formate than as the acid \citep{Buxton1988}. Its speciation is set by the base
inventory, taken as 5.4\% NH$_3$ relative to water \citep{Boogert2015}. The base matters over the whole exposure, because the OH that the formate takes is lost to the sugars. In the restricted network the count is \Nresbasezero\ without the base and \Nresbasetwo\ with twice the base (\pctresbasezero\ and \pctresbasetwo), and the ice HCOOH/H$_2$O at $\nstar$ is \hcoohresbasezero\ and \hcoohresbasetwo\ against \hcoohres\ in the fiducial model (Table~\ref{tab:nstarres}). Table~\ref{tab:nstar} gives the same axis for the complete network.

\subsection{Carbon-flux audit}
When the network is built, every reaction is checked for the conservation of carbon, hydrogen and oxygen, with water counted through its $\Delta$H$_2$O entry. The primary radiolysis yields of Appendix~\ref{app:window} are not exactly balanced in redox: they leave about $7\times10^{4}$ reducing equivalents per event, $3\times10^{-4}$ of the hydrogen of the grain. Summed over the count this excess is of the order of the hydrogen of the grain. Like the oxygen taken from the water (Appendix~\ref{app:window}), it is not closed against the water of the grain; in the model it ends in the H$_2$ that is lost between events and in the reduced organics. The fluxes through each channel are integrated over the run.
At $\nstar$ in the fiducial model, we consider the carbon held in species other than
H$_2$CO, CH$_3$OH and CO. Of this carbon, \audox\ is in CO$_2$, HOCO and formic acid (predominantly the radiolytic oxidation of CO and H$_2$CO, not of sugars) and \audsug\ in closed-shell sugars; the polyols hold \audpol\ and glycolic and glyceric acid \audacid. In the restricted network the first two shares are \audoxres\ and \audsugres, and the last two are below 0.1\%. The trapped radicals hold a negligible share between events because the
cold-phase hydrogen closes them.

\section{The restricted network}
\label{app:restricted}

The restricted network is the hypothesis of Sect.~\ref{sec:hypothetical}: the
dihydroxymethyl radical reacts only with itself, OH and H, methanol is inert, and the
whole 10.7\% substrate is H$_2$CO. Everything else is as in the complete network.
Table~\ref{tab:mainres} gives its state at $\nstar=\Nres$ against the observations, and Table~\ref{tab:nstarres} its systematics, computed in the same way as Table~\ref{tab:nstar}; the text below summarises them. Without the C1 path \ketres\ of the closed-shell C4 is the ketose (\ketresrange\ across the variants below that reach the observed ratio) and the aldotetrose is at $\approx$\aldres$\times$GA. Without the hydration of GA \hnoneresald\ of the C4 is the aldotetrose, ERU/GA rises only as radiolysis destroys GA, and it reaches the observed value only after \Nreshnone\ events, by which time DHA stands at \hnoneresdha\ times its limit and the sugars are mostly hexoses. With every window at the fixed $T_\mathrm{w}=491$~K that case still gives a ketose (\kettwhnone), through the combination of the acyl radical of GA with the glycol radical, a channel that no longer dominates in the cooler liquid of the temperature histories. Returning the C3 radical wholly as the aldose lowers DHA/GA to \dharescthreeald\ of its limit and raises GLD/GA to \gldrescthreeald\ of its limit; returning the C4 radical half as the aldotetrose leaves \keterysymres\ of the C4 as the ketose. The radiolysis between events (Sect.~\ref{sec:radiolysis}) acts on this network as on the complete one. \radreachedres. \radnotres. The reversible case with $c=0.25$ reaches the ratio after \Nrev\ events.

\begin{table}
\caption{As Table~\ref{tab:main}, for the restricted network of the hypothetical case (Sect.~\ref{sec:hypothetical}; means of \nresseq\ sequences of windows). The footnote marks refer to the footnotes of Table~\ref{tab:main}.}
\label{tab:mainres}
\centering
\scriptsize
\setlength{\tabcolsep}{3pt}
\begin{tabular}{lcc}
\hline\hline
Quantity & Restricted, $\nstar=\Nres$ & Observed \\
\hline
ERU/GA                                   & 0.94 & $0.94\pm0.09$ \\
GLD/GA                                   & 0.0020 & $\leq$0.054 \\
DHA/GA                                   & 0.042 & $\leq$0.108\tablefootmark{a} \\
Glycolic acid/GA                         & 0.020 & $\leq$0.15--0.7\tablefootmark{b} \\
Aldotetrose/GA                           & 0.0040 & \ldots \\
Hexoses/GA                               & 0.57 & \ldots \\
Pentoses/GA                              & 0.045 & \ldots \\
Ethylene glycol/GA                       & 0.0016 & 2.5\tablefootmark{c} \\
Methanol/GA                              & 13 & \ldots \\
HCOOH/$\HHO$                             & 17\% & 0.5--6\%\tablefootmark{d} \\
$\COO/\HHO$                              & 12\% & 19--28\%\tablefootmark{d} \\
ERU/C3                                   & 21 & $\geq$5.8 \\
Ketose fraction of the C4                & 100\% & \ldots \\
GA per grain                             & $5.2\times10^{5}$ & \ldots \\
ERU per grain                            & $4.9\times10^{5}$ & \ldots \\
Mean sugar size [C]\tablefootmark{e}     & 3.7 & \ldots \\
Sugars with $\geq$C6\tablefootmark{e}    & 22\% & (undetected) \\
\hline

\end{tabular}
\end{table} The spectrum variants of Sect.~\ref{sec:rate} (Figs.~\ref{fig:origin} and \ref{fig:radial}) were also run with this network. Its ages are longer by factors of \resagefac, against the factor \comfacinv\ of the fiducial count: \tresecom~Myr against \tecom\ for the external rising spectrum at the cloud centre, normalised at the grain, \treslcom~Myr against \tlcom\ for the flat local spectrum, and \treskcom~Myr against \tkcom\ for the porous mantle with a tenth of the compact conductivity, at $\zeta=10^{-14}$~s$^{-1}$.

\begin{table*}
\caption{As Table~\ref{tab:nstar}, for the restricted network (the hypothetical case
of Sect.~\ref{sec:hypothetical}). Each entry is the mean over 4 sequences of windows
(\nresseq\ for the fiducial). Rows with
four sequences are to be compared with the same four sequences of the fiducial:
$\nstar=\Nresfour$, GLD/GA and DHA/GA at \gldresfour\ and \dharesfour\ of their limits, glycolic acid/GA \glyresfour. The row at a fixed $T_\mathrm{w}$, with its two entries, is a reference for the treatment of the window temperature, not a systematic.}
\label{tab:nstarres}
\centering
\scriptsize
\setlength{\tabcolsep}{4pt}\begin{tabular}{p{6.1cm}ccccc}
\hline\hline
Model variant & $\nstar$ & GLD/GA / limit & DHA/GA / limit & glycolic/GA & ketose (\%) \\
\hline
Fiducial (16 sequences of windows) & 6\,470 & 0.037 & 0.39 & 0.020 & 99.6 \\
$\delta=0.15$ / 0.30 / 0.60 & 5\,980 / 6\,230 / 6\,710 & 0.030 / 0.037 / 0.041 & 0.39 / 0.39 / 0.40 & 0.010 / 0.016 / 0.025 & 99.6 / 99.6 / 99.6 \\
$\delta=0.73$ (literature central) & 7\,160 & 0.036 & 0.40 & 0.023 & 99.5 \\
$h_\mathrm{GLD}=0.05$ (aqueous $K_\mathrm{hyd}$) / 0.70 & 6\,450 / 6\,450 & 0.32 / 0.060 & 0.24 / 0.38 & 0.023 / 0.023 & 99.6 / 99.6 \\
Literature pair ($\delta=0.73$, $h_\mathrm{GLD}=0.05$) & 7\,170 & 0.34 & 0.23 & 0.023 & 99.5 \\
$h_\mathrm{GA}=0.20$ & 6\,510 & 0.042 & 0.39 & 0.022 & 98 \\
No hydration of GA and the aldotetrose ($h=1$)\tablefootmark{b} & 252\,800 & 1.5 & 16 & 0 & 4 \\
Temperature cap of the superheated liquid 450 / 650~K & 9\,440 / 5\,210 & 0.027 / 0.039 & 0.34 / 0.43 & 0.0080 / 0.026 & 99.6 / 99.6 \\
$\delta$ held at its 491~K value in every window & 8\,570 & 0.041 & 0.41 & 0.026 & 99 \\
Diffusion-limited rate constants scaled by $w(T)/w(491\,\mathrm{K})$ & 23\,300 & 0.0096 & 0.25 & 0.0057 & 99.6 \\
Window chemistry at a fixed $T_\mathrm{w}=491$~K: liquid lifetime / mobility-weighted lifetime & 3\,770 / 620 & 0.066 / 0.22 & 0.78 / 1.3 & 0.11 / 0.39 & 99 / 99 \\
Escape rather than primary yields & 11\,200 & 0.021 & 0.30 & 0.015 & 99.6 \\
Trapped small radicals survive to the next event & 1\,150 & 0.042 & 0.39 & 0.084 & 99.6 \\
Cold-phase H: in-situ term only ($\varepsilon=0$) / a tenth of the accretion term and no in-situ term ($\varepsilon=0.001$, $\eta=0$; Appendix~\ref{app:coldh}) & 6\,450 / 820 & 0.044 / 0.29 & 0.39 / 0.069 & 0.023 / 0.89 & 99.6 / 99.5 \\
Molecular yield of the dose outside the melts 0.4 per 100~eV / off (in-situ reductants kept) & 5\,900 / 7\,460 & 0.027 / 0.040 & 0.35 / 0.46 & 0.017 / 0.016 & 99.6 / 99.6 \\
No H$_2$CO resupply between events (substrate closed) & 10\,300 & 0.027 & 0.30 & 0.0062 & 99.6 \\
H$_2$CO inventory $\times$1/3 / $\times$3 & 7\,340 / 5\,720 & 0.034 / 0.039 & 0.36 / 0.50 & 0.013 / 0.027 & 99.6 / 99.5 \\
Base inventory 0 / $\times$2 & 4\,790 / 7\,340 & 0.022 / 0.045 & 0.30 / 0.44 & 0.026 / 0.017 & 99.6 / 99.6 \\
CO inventory $\times$1/2 & 6\,080 & 0.036 & 0.37 & 0.020 & 99.6 \\
$f_\mathrm{CHO}=0.7$ / 0.8 & 6\,410 / 6\,390 & 0.066 / 0.088 & 0.38 / 0.36 & 0.023 / 0.023 & 99.8 / 99.9 \\
H repair of the sugar radicals off (rows 68 and 174 of Table~\ref{tab:reactions} and their analogues) & 6\,350 & 0.0053 & 0.42 & 0.023 & 99.6 \\
Acyl$+$alkyl combination off (rows 82--83) & 6\,450 & 0.042 & 0.37 & 0.023 & 99.6 \\
OH attack on dehydro-sugars 0 / 0.7 & 6\,450 / 6\,450 & 0.044 / 0.044 & 0.39 / 0.39 & 0.023 / 0.023 & 99.6 / 99.6 \\
$h_\mathrm{ald}=0.02$ / $k_\mathrm{hadd}=10^{9}$ & 6\,450 / 6\,450 & 0.044 / 0.044 & 0.39 / 0.39 & 0.023 / 0.023 & 99.6 / 99.6 \\
Acyl radicals of C3 and C4 given combination partners (Appendix~\ref{app:window}) & 6\,490 & 0.043 & 0.39 & 0.023 & 99.6 \\
C3 radical reduced to the aldose, half / all & 6\,440 / 6\,440 & 0.54 / 0.79 & 0.13 / 0.0037 & 0.023 / 0.023 & 99.6 / 99.6 \\
C4 ketose radical reduced half to the aldotetrose\tablefootmark{c} & $>$204\,000 & 0.022\tablefootmark{a} & 0.18\tablefootmark{a} & 0.025\tablefootmark{a} & 10\tablefootmark{a} \\
DHA hydrated ($h_\mathrm{DHA}=0.05$) & 6\,450 & 0.0049 & 0.71 & 0.023 & 99.6 \\
Pairs of the GLD gem-diol radical, of the C3 radical and of the dehydro-pentose off (28 rows), $h_\mathrm{GLD}=0.05$ & 5\,140 & 0.32 & 0.30 & 0.012 & 99.6 \\
Dose of the window $\times$2 / $\times$0.5 & 2\,880 / 15\,400 & 0.049 / 0.022 & 0.37 / 0.38 & 0.028 / 0.011 & 99.6 / 99.6 \\
Dose of the window equal to the melting enthalpy $D_\mathrm{melt}$ & 25\,200 & 0.017 & 0.37 & 0.0080 & 99.6 \\
No OH destruction of glycolic acid, $\delta=0.47$ / 0.73 & 6\,420 / 7\,070 & 0.044 / 0.037 & 0.39 / 0.40 & 0.68 / 1.0 & 99.6 / 99.5 \\
\hline

\end{tabular}
\tablefoot{\tablefoottext{a}{Value at the peak of ERU/GA within the run; the
ratio is not reached.}
\tablefoottext{b}{\hnoneresald\ of the C4 is the aldotetrose; ERU/GA rises only as GA is destroyed.}
\tablefoottext{c}{The ketose fraction of the C4 is \keterysymres, and ERU/GA stays near \pkerysymres.}}
\end{table*}

Table~\ref{tab:nstarres} collects the response of the five observables to every axis we
varied. Six axes move the count by about a factor of two or more. They are the survival of the trapped small radicals, the absence of their reduction, the dose of the window, the hydration of GA, the mobility scaling of the diffusion-limited constants and the reversible radiolysis between events (the last is given in the text above, not in the table). Apart from the row without the hydration of GA and the fixed-temperature references, the counts run from \Nresch\ and 1\,150, for the two rows that leave trapped radicals unreduced, to \Ndmeltres, for a window dose equal to the melting enthalpy. Relative to the same four sequences, the disproportionation fraction $\delta$ moves the count by only \resdxpctlo\ to \resdxpcthi\ over 0.15--0.73, and the temperature cap of the superheated liquid by \rescappctlo\ to \rescappcthi\ over 650--450~K. The C3 aldose spans \gldfix$\times$ its limit at fixed reduction site and hydration (\gldres$\times$ in the fiducial model), and \gldreshyd$\times$ with the aqueous hydration of GLD. Relative to the same four sequences, eleven axes move it by factors of two or more, up to \gldfixfac\ for the site of the cold-phase reduction. They are $h_\mathrm{GLD}$, that site, the hydration of DHA, the H back-reduction of the GA radical, the absence of the cold-phase reduction, $f_\mathrm{CHO}$, the mobility scaling, the escape yields, the base inventory, half the window dose and a window dose equal to the melting enthalpy. Six more move it by 15--40\%: the temperature cap, the closed substrate, the molecular yield of the dose outside the melts, $\delta$, the H$_2$CO inventory and the CO inventory. The C3 ketose spans \dhafix$\times$ its limit at fixed reduction site and hydration (\dhares$\times$ in the fiducial model); the absence of the cold-phase reduction gives \dhachres, the escape yields \dhaescres, and the H$_2$CO inventory and the survival of the small radicals \dhahchores\ and \dhasmallres. The reduction site and the hydration of DHA move it outside
that span (Sect.~\ref{sec:ketose}). Glycolic acid/GA, destroyed in the window, stays
at \glyressys\ on every axis except the absence of the cold-phase reduction, the rows without its OH destruction and the fixed-$T_\mathrm{w}$ reference rows (and it is absent without hydration); whether it survives is the larger question (Sect.~\ref{sec:predictions}).
Some axes test the network itself: the combination partners of the C3 and C4 acyl radicals, the site of the cold-phase reduction of the C3 radical and the hydration of DHA. They leave $\nstar$ within about \acylfullN\ of the same four sequences of the fiducial. Returning the C4 radical half as the aldotetrose, in contrast, prevents the observed ratio from being reached. The reduction sites and the
hydration of DHA move the partition of the C3 between its isomers and the ketose
fraction of the C4, as Sect.~\ref{sec:ketose} describes. Two axes move no observable by more than 1\% relative to the same sequences: the OH attack on the dehydro-sugars and the closed-shell carbonyl hydrogenation rate $k_\mathrm{hadd}$. They are listed for completeness. So is the hydration of the aldotetrose $h_\mathrm{ald}$, which moves only the aldotetrose itself (by \aldfacres). The base inventory moves the count by \pctresbasezero\ (none) and \pctresbasetwo\ (twice the base; Appendix~\ref{app:network}).
The order in which the windows arrive matters little: the standard deviation of
$\nstar$ over \nresseq\ sequences drawn from the same distribution is \sdres.

\section{Numerical scheme and verification}
\label{app:numerics}

\subsection{Solver}

The window chemistry is solved by a zero-dimensional program. Its state is one vector of
number densities for the well-mixed melt of volume $V$; there is no grid and no
transport. It integrates Eq.~(\ref{eq:ode}) with the adaptive Heun substepping of
Appendix~\ref{app:window}, positivity clamping at each stage, and an absolute floor
below which a species is excluded from step control. The substeps reach picoseconds at
the start of a window, so explicit single-step integration diverges. The same
network and integrator are one source file, which a three-dimensional grid code also
uses; that code serves only as the check of Appendix~\ref{sec:verification}.

\subsection{Energy bookkeeping}
\label{app:dose}

The energy deposited per ion is $\min[AE,S(E)\,4a/3]$ as in Sect.~\ref{sec:chain}, at a hit rate of $4\pi\cdot\pi a^{2}J$ for the isotropic intensity $J$. For the fiducial spectrum at $\zeta=10^{-14}$~s$^{-1}$ the eleven nuclei deposit \evmyr~eV per $\HHO$ per Myr, which is \dosetotfac\ times the melting enthalpy of the grain, or \evtot~eV per molecule, per whole-grain melting event. The dose and the ionisation rate are tied by construction, because the normalisation of Eq.~(\ref{eq:zetaown}) uses the same particles and the stopping tables of the same compilation. The energy that the nuclei lose in the gas is $\zeta W_\mathrm{ion}$, 11~eV per H$_2$ per Myr at that $\zeta$. The energy deposited per $\HHO$ molecule of the grain is \zwratioH\ times larger. This factor is the ratio of the stopping power of water per molecule to that of H$_2$, averaged over the spectrum. For protons it is 2.1 at 0.1~MeV, 3.4 at 1~MeV and 4.3 at 100~MeV: water has five times the electrons, at a higher mean excitation energy. For the flat shape it is \zwratioL, and for the attenuated spectra \zwratioHfar.

The energy of a melting ion is divided with the radial dose of Eq.~(\ref{eq:dose}). Per unit track length the part inside the largest melt radius $r_\mathrm{melt}$ is $S/2+(S/2)\ln(r_\mathrm{melt}/r_\mathrm{c})/\ln(r_\mathrm{p}/r_\mathrm{c})$, or $(S/2)(r_\mathrm{melt}/r_\mathrm{c})^{2}$ if the melt lies inside the core. Divided by the melting enthalpy of the melted cylinder, $D_\mathrm{melt}\pi r_\mathrm{melt}^{2}$, it is the factor $x_\mathrm{in}$ of Eq.~(\ref{eq:gsource}). Averaged over the ions of a window-length bin with their whole-grain rates it is \doseinbins\ for the bins of the fiducial distribution, from the shortest to the longest, and \doseinmean\ overall. A value below one marks windows that the interpolation of the track table creates between an ion that melts and one that does not (Appendix~\ref{app:trackcaveats}). The remainder of the deposited energy, \doseout\ times the melting enthalpy of the grain per event, is the dose outside the melts of Sect.~\ref{sec:radiolysis} and of operator~4 (Appendix~\ref{app:cycle}): \doseoutmeltions\ from the melting ions and \dosenonmelt\ from the ions that melt nothing. Inside and outside the melts the model thus books the whole deposited energy. For the other spectra of Sect.~\ref{sec:rate} the same evaluation gives a dose outside the melts of \doseoutrange\ times the melting enthalpy of the grain per event. It is largest for the flat shape and for the attenuated spectra, whose fast protons deposit energy without melting. The secondary electrons that leave the grain are not subtracted (Appendix~\ref{app:trackcaveats}).

\subsection{Repeated melting}

Uniform carry-over between events is justified because each sub-volume of the grain is
melted about $\nstar=\Ncom$ times over the exposure. Poisson fluctuations in the number of melts
per sub-volume therefore bias sugar--sugar products by about $1/\nstar$, less than 0.1\%. Each event deposits the window dose of Eq.~(\ref{eq:gsource}). The dose outside the melts is added as
$2\HHO\rightarrow\mathrm{H_2}+\mathrm{H_2O_2}$ with $G=0.2$ per 100\,eV (0.4 and
off as systematic axes; the in-situ reductants of Appendix~\ref{app:coldh} are kept in both). Switching this molecular channel off, or doubling its yield, changes
$\nstar$ by \omN, GLD/GA by up to \omgld\ and the other observables of Table~\ref{tab:main} by less than \omother, relative to the same sequences (a check made with the restricted network; Table~\ref{tab:nstarres}). The resupply and event-interval
constants of Appendix~\ref{app:network} are bounded in Appendices~\ref{app:cycle}
and \ref{app:coldh}. Decomposed water is debited from the solvent reservoir at 10.70 $\HHO$ per 100\,eV.
Each event
takes its window length from the reconstructed distribution (Appendix~\ref{app:cycle})
and its temperature from the cylindrical conduction model of Sect.~\ref{sec:chain}.
The resulting sugar inventories in the fiducial model are summarised in Table~\ref{tab:main}.

\subsection{Verification}
\label{sec:verification}

The grid code, which shares the source of the network and of the integrator with the zero-dimensional program, passes a suite of 63 tests with the present network. They cover analytic reaction solutions, atom and charge conservation, the melt cycle and the heat solver. As a check of the zero-dimensional solver inside the complete cycle, we repeated cycle runs of the present model with the window chemistry solved by the grid code instead, on a uniform box in which every cell follows the same equations. Everything else is the same driver: the sequence of windows, the dose of each window, its 14-step temperature history, the dose outside the melts and the operators between events. The comparison therefore tests how the two programs embed the shared network and integrator in a window, with the inputs of the present model. It does not test the network itself or the operators between events.

Six runs were compared on a box of one cell, each beyond its $\nstar$: two sequences of the fiducial model (4000 events each), the restricted network (8000 events), $\delta=0.73$ (4500), every window at the fixed temperature of 491~K (1500) and the irreversible radiolysis between events with $D_{1/2}=2300$~eV (4500). A seventh run followed the fiducial model for 400 events on $24^{3}$ cells. In all seven, every inventory at every written event is identical in the two solvers to the seven digits written, and so is $\nstar$. The operators between events are checked separately by the carbon balance: at the end of the fiducial run of 120\,000 events the carbon of the inventory equals the initial carbon plus the resupply to eight digits.

\subsection{Absolute column check}
\label{app:column}

The ice water column towards G+0.693 is
$N(\HHO)_\mathrm{ice}=x_\mathrm{ice}\,2N(\mathrm{H_2})=1.1\times10^{19}\,\percm$,
adopting the median ice water abundance towards background stars,
$x_\mathrm{ice}=4\times10^{-5}$ per H nucleus \citep[the full range is
$0.9$--$6\times10^{-5}$;][]{Boogert2015}, and $N(\mathrm{H_2})=1.35\times10^{23}\,\percm$
\citep{Martin2008}. The modelled grain holds $1.29\times10^{8}$ water molecules, so the per-grain inventories at $\nstar$ (Table~\ref{tab:main}) correspond to ice-phase columns of \gacolcom\ (GA) and \erucolcom\ (ERU) in the complete network, and \gacolres\ (GA), \erucolres\ (ERU), \gldcolres\ (GLD) and \dhacolres$\,\percm$ (DHA) in the restricted network. Matching the observed gas-phase columns then requires desorbed fractions of \desfrac\ (complete) or \desfracres\ (restricted) for both GA and ERU. These are equal because $\nstar$ is defined by the observed ratio; the check is that both are small enough to be supplied by sputtering. In the restricted network the same fraction would place GLD at \gldres\ and DHA at \dhares\ of their observed upper limits; in the complete network it places DHA at \dhacom\ times its limit.

\subsection{Thermal conductivity check}
\label{app:kappa}

Table~\ref{tab:kappa} lists the conductivities obtained by reverse non-equilibrium
molecular dynamics with the machine-learned potential of Appendix~\ref{app:qcmd}, for
the four temperatures and three cell lengths of that calculation. The mean over all
twelve runs, $1.03\pm0.22\,\wmk$, is the value quoted in Appendix~\ref{app:qcmd}.

\begin{table}
\caption{Thermal conductivity of amorphous ice from reverse non-equilibrium molecular dynamics (machine-learned potential,
$\rho=0.917$\,g\,cm$^{-3}$), in $\wmk$.}
\label{tab:kappa}
\centering
\begin{tabular}{lcccc}
\hline\hline
$T$ [K] & $L_z=47.9$\,\AA & 76.7\,\AA & 115.0\,\AA & mean \\
\hline
100 & 1.006 & 1.379 & 0.824 & $1.07\pm0.28$ \\
150 & 0.971 & 1.178 & 0.899 & $1.02\pm0.15$ \\
200 & 0.950 & 1.074 & 1.484 & $1.17\pm0.28$ \\
250 & 0.957 & 0.895 & 0.734 & $0.86\pm0.12$ \\
\hline
\end{tabular}
\tablefoot{Production 600\,ps for the two shorter cells and 350\,ps for the longest.
The $T=250$\,K, $L_z=115$\,\AA\ run has $\Delta T=28$\,K, slightly above the
linear-response band, and should be discounted accordingly.}
\end{table}

\section{Quantum chemistry and molecular dynamics}
\label{app:qcmd}

\subsection{Saddle points}

The barriers of class D in Table~\ref{tab:prov} are saddle points of the isolated
radicals, computed for this work. Energies are unrestricted Kohn--Sham M06-2X
\citep{Zhao2008} with the def2-TZVP basis, on geometries optimised with def2-SVP
\citep{Weigend2005}, in implicit water described by the conductor-like polarisable continuum model \citep{Barone1998} with
relative permittivity $\epsilon_\mathrm{r}=78.4$, and with zero-point corrections from the def2-SVP frequencies. For each
radical we generate six conformers with the experimental-torsion distance-geometry method of \citet{Riniker2015} in its third version \citep{Wang2020etkdg} and take the lowest. We locate the barrier by a relaxed scan along the breaking C--C bond, optimise the transition state, and keep it only if it has exactly one imaginary frequency. The calculations use PySCF \citep{Sun2020} with the geomeTRIC optimiser
\citep{Wang2016geom}. These barriers enter the network as the rate coefficients of the radical destruction channels, and are separate from the potentials described next.

The radical-pair profiles quoted in Sect.~\ref{sec:caveats} use the same functional and solvent model in broken-symmetry form. The singlet pair starts from the two separated doublet radicals, and the scan constrains the transferring O--H$\cdots$C distance (disproportionation) or the C$\cdots$C distance (combination), relaxing all other coordinates with def2-SVP; energies are recomputed with def2-TZVP. For $\bullet$CH$_2$OH and $\bullet$CH(OH)$_2$ the disproportionation profile rises by about 12~kJ\,mol$^{-1}$ from the hydrogen-bonded pre-reaction complex to its maximum. For the gem-diol radical pair of GA the combination profile rises by 15~kJ\,mol$^{-1}$ (def2-SVP) to 20~kJ\,mol$^{-1}$ (def2-TZVP). All these maxima lie below the energy of the separated radicals. The scans are not converged transition-state searches, and we use them only for the qualitative statement of Sect.~\ref{sec:caveats}.

\subsection{Molecular dynamics}

Two properties of the ice come from molecular dynamics. The first is the solid-to-liquid
mobility ratio $f_\mathrm{s}$, which Sect.~\ref{sec:meltgated} needs for the comparison
with an unmelted grain. The second is the thermal conductivity of amorphous ice, which
we compute to check the value used in the window model of Sect.~\ref{sec:chain}.
Neither enters the fiducial chemistry. Both use machine-learned interatomic potentials
for the H--O system. Their reference data are plane-wave density functional theory single points computed with
VASP \citep{Kresse1996} with the Perdew--Burke--Ernzerhof (PBE) functional \citep{Perdew1996}, with a 600~eV cutoff, a
$k$-point spacing of 0.3~\AA$^{-1}$, and spin polarisation for every structure that
holds an open-shell species. The potential itself is a sparse Gaussian process built
with FLARE \citep{Vandermause2020,Vandermause2022}, with B2 descriptors of 5~\AA{}
cutoff and a normalised dot-product kernel, grown by on-the-fly active learning. Successive rounds added structures that hold H, OH, H$_2$ and H$_2$O$_2$, and structures at the density of the ice. Three versions are used here: two for $f_\mathrm{s}$ (see below) and one, retrained at the density of the ice, for the conductivity. The version for H, OH and H$_2$ is trained on 200 structures: 94 frames of ab initio molecular dynamics of liquid and amorphous water, with and without the solutes, and 106 structures selected by active learning. Its root-mean-square force residual on these two groups is 0.28 and 0.29~eV~\AA$^{-1}$, and its energy offset on the first is 32~meV per atom. The version for the conductivity is trained on 218 structures (94 and 124), with force residuals of 0.26~eV~\AA$^{-1}$ on both groups. The version for H$_2$O$_2$ is trained on 212 structures (94 and 118), with force residuals of 0.27 and 0.29~eV~\AA$^{-1}$; the second value includes six active-learning structures added after its training. Apart from those six, the residuals are measured on the training structures; no held-out set was evaluated. The molecular dynamics are run in LAMMPS \citep{Thompson2022} with the
mapped potential and a short-range repulsive term.
Radical-diffusion molecular dynamics gives the $f_\mathrm{s}$ upper bounds quoted in
Sect.~\ref{sec:meltgated}. The H, OH and H$_2$ runs use the potential of the radical
data set; the H$_2$O$_2$ runs use a version retrained in a 10.7~\AA{} cell, larger than
twice the descriptor cutoff, because a smaller training cell let a single H$_2$O$_2$
see its own periodic image. Both are combined with the screened Coulomb repulsion of
\citet{ZBL1985} between 0.5 and 0.9~\AA.

\begin{figure}
\centering
\includegraphics[width=\columnwidth]{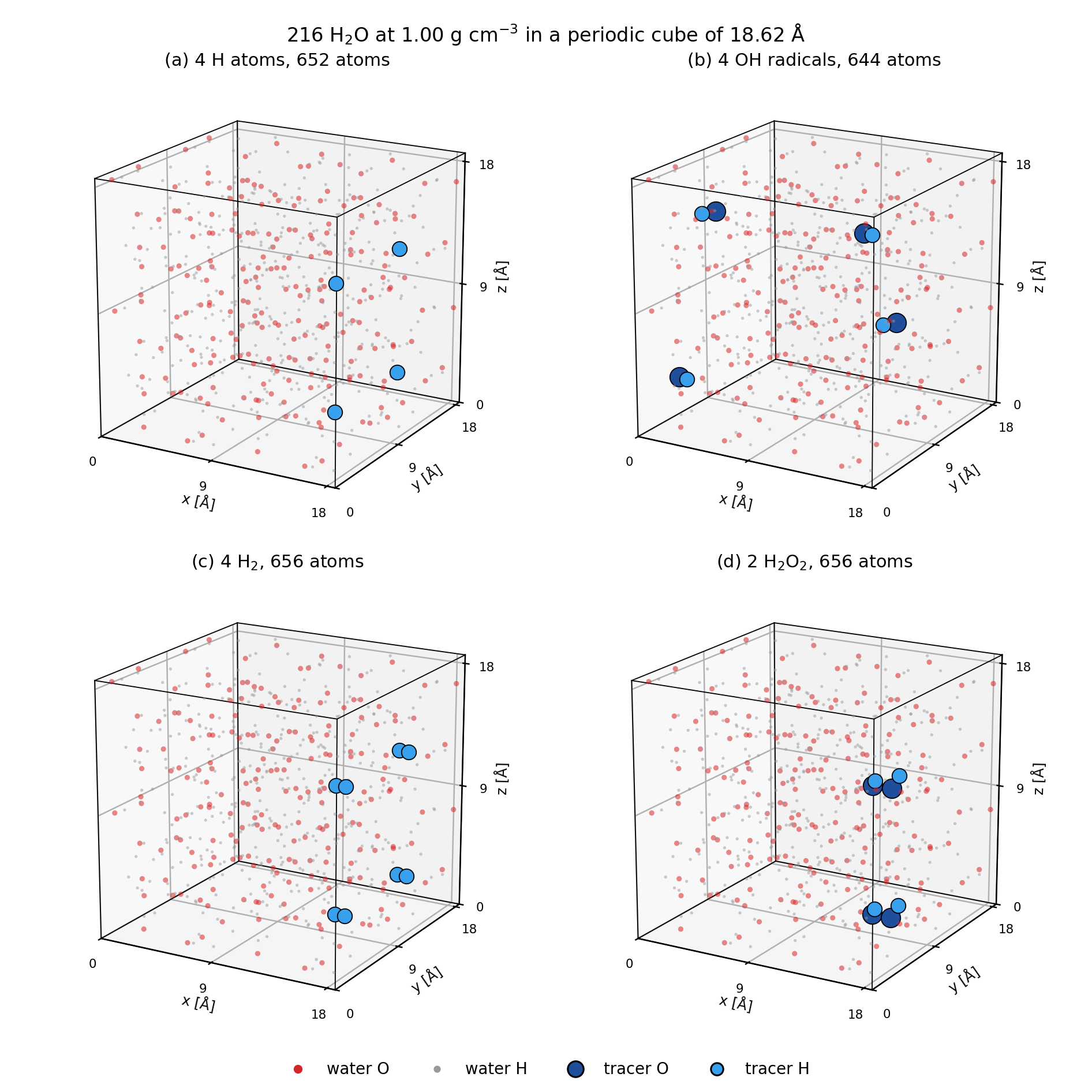}
\caption{The cells of the radical-diffusion molecular dynamics: 216 water molecules at
1.00~g\,cm$^{-3}$ in a periodic cube of 18.62~\AA, with (a) four H atoms, (b) four
OH radicals, (c) four H$_2$ and (d) two H$_2$O$_2$ as tracers (large symbols). Water
O atoms are shown in red, water H atoms in grey.}
\label{fig:mdcell}
\end{figure}

\emph{Cell.} The solvent is amorphous water of 216 molecules (648 atoms) at
1.00~g\,cm$^{-3}$ in a periodic cube of 18.62~\AA. It is a 27-molecule cell from the
ab initio melt-and-quench part of the training set, replicated $2\times2\times2$. One
tracer species is added per run (Fig.~\ref{fig:mdcell}): four H atoms placed in the
four largest voids, four OH made by removing one H atom from each of four water
molecules far apart, four H$_2$ molecules in voids, or two H$_2$O$_2$ molecules in
voids. The tracers start at least 5~\AA{} apart.

\emph{Run.} The energy is first minimised. Velocities are then drawn from a Gaussian
at the target temperature with zero total linear and angular momentum, and the cell is
equilibrated for 10~ps at constant volume with a Nos\'e--Hoover thermostat
\citep{Nose1984,Hoover1985} of 50~fs damping time. The production run continues with
the same thermostat; it is not microcanonical. The total linear momentum is removed
every 200 steps, because a drift of the cell would otherwise add to the displacement of
the few tracer atoms. The time step is 0.5~fs. We used LAMMPS of 22~July~2025 (update~4) with the FLARE pair style built from its development source. The mean-square displacement of the
tracer atoms, all atoms of a tracer molecule (the O atom alone for OH), is written every 0.1~ps, and the
positions of the tracers are written every 1~ps to check that no pair recombines. The two mapped potentials of the tracer runs, the four cells, the cell builder, the LAMMPS inputs and the mean-square displacements are part of the data archive.

\emph{Result.} Runs of 100~ps at 240, 320 and 360~K, with three velocity seeds each for H, OH and H$_2$ and one for H$_2$O$_2$, show no diffusive regime (runs at 400~K were discarded because the potential left its training range): the mean-square displacement levels off at
0.1--0.6\,\AA$^2$, the amplitude of the motion in the cage of neighbours. A run of
500~ps at 240~K, 0.57 of the melting point of PBE water ($417\pm3$~K; \citealt{Yoo2009}), ends at
0.13--1.06\,\AA$^2$ (H, H$_2$O$_2$, H$_2$, OH; OH reaches 2.7\,\AA$^2$ in between).
The bound $D<\langle r^2(t)\rangle/6t$ at its end gives $4.4\times10^{-9}$ (H),
$3.5\times10^{-8}$ (OH), $2.5\times10^{-8}$ (H$_2$) and
$1.5\times10^{-8}\,\mathrm{cm^2\,s^{-1}}$ (H$_2$O$_2$). The liquid branch cannot be
simulated, because PBE water is glassy below its melting point. We therefore divide by
the diffusion coefficients in liquid water at 298~K. For OH and H no direct
measurement in liquid water is available. We take the self-diffusion coefficient of
water, $2.3\times10^{-5}\,\mathrm{cm^2\,s^{-1}}$ \citep{Mills1973}, for OH, and
$7.0\times10^{-5}\,\mathrm{cm^2\,s^{-1}}$, the value adopted in diffusion-kinetic models
of water radiolysis, for H. The range of $(2.2$--$2.8)\times10^{-5}$ used for OH in
such models changes its bound by less than 25\%. For H$_2$ we take $5\times10^{-5}$,
near the 25~$^{\circ}$C value of the measurements, which span $(4$--$6.5)\times10^{-5}$ over 5--35~$^{\circ}$C in the data of \citet{Jahne1987} (see also \citealt{FerrellHimmelblau1967}), and for H$_2$O$_2$
$1.4\times10^{-5}\,\mathrm{cm^2\,s^{-1}}$, measured in 0.1~M phosphate buffer at 25~$^{\circ}$C \citep{vanStroeBiezen1993}.
This gives $f_\mathrm{s}<6\times10^{-5}$ (H), $1.5\times10^{-3}$ (OH),
$5\times10^{-4}$ (H$_2$) and $1\times10^{-3}$ (H$_2$O$_2$). These bounds reject
$f_\mathrm{s}=10^{-2}$, which places a floor under the melt-ignition ratio. The cell is
at 0.57 of the melting point rather than the 0.04 of the 10~K ice, so the true values
are probably far smaller. Two limitations of the set-up remain. The OH run follows the
original O atom, so a transfer of the radical between molecules,
H$_2$O$\,+\,\bullet$OH$\,\to\,\bullet$OH$\,+\,$H$_2$O, would move the radical without
being seen; the site of the unpaired electron was not monitored. The thermostat acts
on every atom during the production run; a microcanonical production run would be the
cleaner choice. Since no diffusion is resolved, the bound is set by the amplitude of
the motion in the cage, which is fixed by the temperature rather than by the
thermostat.

Reverse non-equilibrium molecular dynamics gives the thermal conductivity. The cell is
amorphous water at 0.917\,g\,cm$^{-3}$, $9.59\times9.59$\,\AA{} in cross-section and
elongated along $z$ to $L_z=47.9$, 76.7 and 115.0\,\AA. It is annealed at 350~K,
quenched to the target temperature, and equilibrated at constant volume and
temperature. A heat flux is then imposed in the microcanonical ensemble by the velocity
swap of \citet{MullerPlathe1997} between 20 slabs, applied to oxygen atoms only so that
the swap conserves momentum and energy exactly. Four temperatures give
$\kappa=1.03\pm0.22\,\wmk$ (Table~\ref{tab:kappa}). No dependence on temperature or
system size is resolved. The fitted temperature coefficient $\mathrm{d}\kappa/\mathrm{d}T$ is
$-9\times10^{-4}\pm1.2\times10^{-3}\,\wmk\,\mathrm{K^{-1}}$. The slopes of $1/\kappa$
against $1/L_z$ differ from zero by 0.1 to 2.6 standard errors and are inconsistent in sign between temperatures,
so no finite-size extrapolation is applied. This conductivity is within a factor of 1.7 of
the values adopted in the window model of Sect.~\ref{sec:chain}. It is flat in
temperature, whereas the measured conductivity rises slightly for the high-density
amorphous form and falls for the low-density one \citep{AnderssonSuga2002}. At the
simulated density of 0.917\,g\,cm$^{-3}$ it lies closest to the low-density value.
These are classical values and therefore upper limits below the Debye temperature, and
the PBE functional over-structures water, so we use the simulation for the order of
magnitude only.

\section{Caveats on the track model}
\label{app:trackcaveats}

\paragraph{The shortest windows.} The windows of the protons and of helium, which supply most of the melted volume, are the least certain part of the track table. Their melt radii, 0.6--2.1~nm, are a few molecular diameters, and a continuum heat equation is at its limit there. Two numerical choices matter at this scale. The first is the energy grid. A proton melts a cylinder of 0.75~nm radius, the track core, at 2.05~MeV and nothing at 2.37~MeV, the next energy of the table, and the melted volume is interpolated between the two. With the 16 energies per decade used here this band is 15\% wide in energy. The second is the cell width. Halving the 0.0625~nm cells changes the melted volume of a proton or a helium nucleus by at most 10\% at single energies, with either sign, and by 2\% on average; the liquid lifetime changes by at most 14\%. With cells of 0.25~nm the changes reach 35\%. For iron at 1~MeV~u$^{-1}$ cells of 0.25, 0.125 and 0.0625~nm give the same melted volume. The whole-grain rate is therefore uncertain by about 10\% from these choices, and the windows of 0.1~ns and longer, which make \cfourlong\ of the C4, by 1\%. A lower rate of short windows lowers the count and the rate together, so the age changes by less than the rate does.

\paragraph{The track in a finite grain.} The track model deposits the energy of the ion
at once and follows its conduction in an unbounded medium. Three consequences of the
finite grain are small. First, heat reaches the surface of a 0.1~$\mu$m grain in about
10~ns, after the longest windows have closed, so the grain boundary does not change the
windows. Second, the secondary electrons of an iron ion at 1--3~MeV~u$^{-1}$ travel
150--1000~nm, beyond the grain. They carry 4--21\% of the deposited energy out of it,
depending on the ion energy and on how close the track passes to the surface. This
lowers the dose outside the melts, whose effect on the count is bounded by the
dose outside the melts axis of Table~\ref{tab:nstarres} (\omN). The melt itself lies within
10~nm of the axis and is unchanged. Third, the energy that stays in the grain, about
0.5~MeV for such an ion, warms the whole grain to about 75~K before it cools, close to the 0.4~MeV and 70~K of the whole-grain heating of \citet{HasegawaHerbst1993}. During that time some CO may
sublimate and the trapped radicals may become mobile. The loss of CO is bounded by the
CO axis of Table~\ref{tab:nstarres} (a \Ncohalfpct\ change of the count for half the CO). Radical
combinations at that temperature would add a route outside the window, of the kind
computed by \citet{JimenezSerra2026}; the model does not include it. The continuum
track model itself (instant deposition, Fourier conduction, equilibrium melting) has not been checked against molecular dynamics of an iron track, beyond the comparison with the phase-field model below. The machine-learned potential
of Appendix~\ref{app:qcmd} makes such a check possible, and we leave it to future work.

\paragraph{Amorphous rather than crystalline ice.} We model the ice as amorphous for
four reasons. First, the ice of quiescent clouds is observed to be amorphous, and
crystalline ice is seen only where young stars have heated it \citep{Boogert2015}.
G+0.693 has no detected heating source other than its low-velocity shocks
\citep{Zeng2020}, and the
dust of Galactic-centre clouds is at 10--30~K \citep{RequenaTorres2006}. Ice deposited at
15~K is, moreover, the high-density amorphous form, which converts to the low-density
form only on warming to 38--68~K \citep{JenniskensBlake1994}; this is why the adopted conductivity follows the high-density measurement. The whole-grain warming to about 75~K after each iron-group strike (see above) may, however, convert the mantle to the low-density form. The low-density law raises the ice age by a factor of \ldafac\ (Sect.~\ref{sec:caveats}), which strengthens the conclusion. The track model nevertheless uses the
density of 0.917~g\,cm$^{-3}$, whereas the high-density form is up to 1.1~g\,cm$^{-3}$ \citep{JenniskensBlake1996}. The dose per molecule and the melting enthalpy per molecule do not depend on the density. The energy lost per unit length rises in proportion to it. To first order the melted cross-section is therefore unchanged. We have not recomputed the track table at the higher density. The ambient temperature $T_0=10$~K is at the
lower end of the dust temperatures; at 20~K the enthalpy needed to melt the ice falls
by 0.2\%, which leaves the windows unchanged, but accreted H atoms stay on the ice for
a shorter time, which lowers the fraction $\varepsilon$ of accreted H atoms that reach
the trapped species; this does not change the count, because the in-situ reductants dominate the budget (Appendix~\ref{app:coldh}). Second,
amorphous ice does not crystallise at these temperatures. The onset time measured by
\citet{JenniskensBlake1996}, $3.3\times10^{-15}\,\mathrm{s}\times\exp(44\,\mathrm{kJ\,mol^{-1}}/R_\mathrm{g}T)$ with
$R_\mathrm{g}$ the gas constant,
is about 10~yr at 100~K and exceeds the age of the Universe below 70~K. Even at 200~K,
near a track, it is about 1~ms, a million times longer than a window. Third, the
particles that make the windows also amorphise crystalline ice. Below 100~K, swift heavy
ions turn crystalline ice into compact amorphous ice after about 0.8~eV per molecule
\citep{Dartois2015}, and lighter ions do the same \citep{Fama2010}. That value is the dose at which the crystalline fraction has fallen to $1/e$. The grain of this paper receives \evtot~eV per molecule per whole-grain event, inside and outside the melts (Appendix~\ref{app:dose}), so any crystalline ice is amorphised within a few events.
Fourth, the melt itself refreezes as a glass. (The track model books the latent heat in the crystallisation limit, which gives the shorter windows; the vitrification limit, closer to this glassy refreezing, changes the ice age by \vitfac; Sect.~\ref{sec:rate}.) Water vitrifies when it is cooled faster
than $6.4\times10^{6}$~K~s$^{-1}$ \citep{Mowry2025}, and a melted track cools below the
200~K mobility limit within about a nanosecond (Sect.~\ref{sec:chain}), at more than
$10^{10}$~K~s$^{-1}$. Each window therefore leaves its volume amorphous. A grain whose
ice had been crystallised, for example in the warm envelope of a protostar, would not
run the chemistry of this paper.

\paragraph{Check against the phase-field model.} The published calculation of \citet{Saito2024} contained three coding errors: a periodic-boundary fault in the cascade target, a wrong gradient coefficient of the phase field and an unset thermal boundary. An erratum has been submitted to the journal. The corrected phase-field code, with an optional temperature-dependent interface mobility added and rerun with the properties of this paper (compact amorphous ice, the dose profile of Eq.~(\ref{eq:dose}) for iron at 1~MeV~u$^{-1}$, a 512~nm box of 1~nm cells, the interface mobility of that work), gives the following for a 10~K background. The core falls below the melting point after 0.4~ns and below 200~K after 0.6~ns. The phase field, which that code relaxes with a constant mobility, recovers half of the melt after 3~ns and 95\% after 7~ns. With a mobility that follows the 44~kJ\,mol$^{-1}$ barrier of crystallisation it does not recover at all, leaving a glass. With the 272~K background of the earlier work the same track stays at the melting point; half of the melt is recovered after 0.6~$\mu$s and 95\% after 1.2~$\mu$s (0.8 and 2.5~$\mu$s with the properties and the track of that work). The microsecond refreezing of that work therefore belongs to its warm boundary condition, not to the track; backgrounds of 20~K, a freezing-point depression of 18~K, more than the 11~K of an ideal solution with 10~mol\% of solute, and 0.5~nm cells (in a 256~nm box) change the 10~K result by about 10\%. The calculation is a two-dimensional cross-section of the track.

\section{Lines for the observational tests}
\label{app:lines}

Table~\ref{tab:lines} lists, for each survey band, the four strongest lines of glycolic acid and of DHA at the columns of the hypothetical case, evaluated in LTE with the CDMS and JPL catalogues \citep[the glycolic acid entry rests on the laboratory work of \citealt{Kisiel2016}]{Endres2016,Pickett1998} for a line width of 20~km\,s$^{-1}$, optically thin emission and a filling factor of one, as in Sect.~\ref{sec:predictions}. The catalogues list the components of a line, such as the two members of an asymmetry doublet, as separate entries. Components closer than one line width form one observed line. We merge them: the peak is the maximum of the summed Gaussian profiles, and the entry carries the frequency of that maximum and the transitions $J_{K_a,K_c}$ of its strongest components (at most two are listed). The line counts and the stacks of Sect.~\ref{sec:predictions} use the merged lines. The same evaluation reproduces the GA intensities of \citet{JimenezSerra2020} to within 1\% for the four strongest lines.

\begin{table*}
\caption{Strongest predicted lines of glycolic acid and DHA in the bands of the existing surveys of G+0.693, at two excitation temperatures. Each entry gives the frequency in MHz, the transition $J'_{K_a',K_c'}$--$J''_{K_a'',K_c''}$ of the strongest component or components, and in parentheses the upper-level energy and the peak brightness temperature in mK. The columns are those of the restricted network (the hypothetical case of Sect.~\ref{sec:hypothetical}); the bands are those of the Yebes 40~m and IRAM 30~m surveys.}
\label{tab:lines}
\centering
\scriptsize
\setlength{\tabcolsep}{4pt}\begin{tabular}{p{0.15\textwidth}p{0.135\textwidth}rp{0.605\textwidth}}
\hline\hline
Species, column & Band & $T_\mathrm{ex}$ (K) & Lines: MHz, transition ($E_\mathrm{u}$, peak) \\
\hline
Glycolic acid, $N=1.8\times10^{12}$~cm$^{-2}$ (0.020$\times$GA) & 7~mm, 31--50~GHz & 10 & 45079.6, $7_{0,7}$--$6_{0,6}$ (9~K, 0.40); 44523.2, $7_{1,7}$--$6_{1,6}$ (9~K, 0.39); 39192.2, $6_{0,6}$--$5_{0,5}$ (7~K, 0.38); 44285.9, $6_{1,5}$--$5_{1,4}$ (8~K, 0.37) \\
 & 3~mm, 83--115~GHz & 10 & 99802.2, $5_{5,1}$--$4_{4,0}$ and $5_{5,0}$--$4_{4,1}$ (14~K, 0.21); 92640.8, $14_{1,13}$--$13_{1,12}$ and $6_{4,2}$--$5_{3,3}$ (36~K, 0.20); 87856.2, $12_{2,10}$--$11_{2,9}$ (29~K, 0.19); 86710.5, $14_{0,14}$--$13_{0,13}$ (32~K, 0.17) \\
 & 2~mm, 132--174~GHz & 10 & 142579.9, $7_{7,1}$--$6_{6,0}$ and $7_{7,0}$--$6_{6,1}$ (26~K, 0.13); 135340.8, $8_{6,3}$--$7_{5,2}$ and $8_{6,2}$--$7_{5,3}$ (25~K, 0.12); 149660.3, $8_{7,2}$--$7_{6,1}$ and $8_{7,1}$--$7_{6,2}$ (29~K, 0.10); 142391.1, $9_{6,4}$--$8_{5,3}$ and $9_{6,3}$--$8_{5,4}$ (28~K, 0.08) \\
 & 7~mm, 31--50~GHz & 20 & 45079.6, $7_{0,7}$--$6_{0,6}$ (9~K, 0.25); 44523.2, $7_{1,7}$--$6_{1,6}$ (9~K, 0.25); 48433.3, $7_{2,6}$--$6_{2,5}$ (11~K, 0.23); 44285.9, $6_{1,5}$--$5_{1,4}$ (8~K, 0.22) \\
 & 3~mm, 83--115~GHz & 20 & 104666.3, $17_{1,17}$--$16_{1,16}$ and $17_{0,17}$--$16_{0,16}$ (46~K, 0.53); 110653.9, $18_{1,18}$--$17_{1,17}$ and $18_{0,18}$--$17_{0,17}$ (52~K, 0.49); 98681.1, $16_{1,16}$--$15_{1,15}$ and $16_{0,16}$--$15_{0,15}$ (41~K, 0.48); 92695.2, $15_{0,15}$--$14_{0,14}$ and $15_{1,15}$--$14_{1,14}$ (37~K, 0.34) \\
 & 2~mm, 132--174~GHz & 20 & 134602.1, $22_{1,22}$--$21_{1,21}$ and $22_{0,22}$--$21_{0,21}$ (76~K, 0.23); 142579.9, $7_{7,1}$--$6_{6,0}$ and $7_{7,0}$--$6_{6,1}$ (26~K, 0.18); 140588.7, $23_{1,23}$--$22_{1,22}$ and $23_{0,23}$--$22_{0,22}$ (82~K, 0.18); 149660.3, $8_{7,2}$--$7_{6,1}$ and $8_{7,1}$--$7_{6,2}$ (29~K, 0.17) \\
DHA, $N=3.9\times10^{12}$~cm$^{-2}$ (0.04$\times$GA) & 7~mm, 31--50~GHz & 10 & 41953.2, $11_{1,11}$--$10_{0,10}$ (12~K, 0.32); 39003.6, $10_{1,10}$--$9_{0,9}$ (10~K, 0.31); 44970.4, $12_{1,12}$--$11_{0,11}$ (14~K, 0.31); 48059.5, $13_{1,13}$--$12_{0,12}$ (16~K, 0.30) \\
 & 3~mm, 83--115~GHz & 10 & 90104.2, $5_{5,1}$--$4_{4,0}$ and $5_{5,0}$--$4_{4,1}$ (12~K, 0.78); 93893.3, $6_{5,2}$--$5_{4,1}$ and $6_{5,1}$--$5_{4,2}$ (13~K, 0.74); 109704.9, $6_{6,0}$--$5_{5,1}$ and $6_{6,1}$--$5_{5,0}$ (17~K, 0.71); 97681.2, $7_{5,3}$--$6_{4,2}$ and $7_{5,2}$--$6_{4,3}$ (15~K, 0.70) \\
 & 2~mm, 132--174~GHz & 10 & 133094.1, $8_{7,1}$--$7_{6,2}$ and $8_{7,2}$--$7_{6,1}$ (25~K, 0.49); 136882.9, $9_{7,3}$--$8_{6,2}$ and $9_{7,2}$--$8_{6,3}$ (27~K, 0.43); 140670.7, $10_{7,4}$--$9_{6,3}$ and $10_{7,3}$--$9_{6,4}$ (29~K, 0.38); 148903.6, $8_{8,0}$--$7_{7,1}$ and $8_{8,1}$--$7_{7,0}$ (31~K, 0.36) \\
 & 7~mm, 31--50~GHz & 20 & 48059.5, $13_{1,13}$--$12_{0,12}$ (16~K, 0.27); 49177.6, $14_{0,14}$--$13_{1,13}$ (18~K, 0.26); 44970.4, $12_{1,12}$--$11_{0,11}$ (14~K, 0.25); 45391.0, $13_{0,13}$--$12_{1,12}$ (16~K, 0.25) \\
 & 3~mm, 83--115~GHz & 20 & 109704.9, $6_{6,0}$--$5_{5,1}$ and $6_{6,1}$--$5_{5,0}$ (17~K, 0.64); 113494.2, $7_{6,2}$--$6_{5,1}$ and $7_{6,1}$--$6_{5,2}$ (19~K, 0.63); 97681.2, $7_{5,3}$--$6_{4,2}$ and $7_{5,2}$--$6_{4,3}$ (15~K, 0.56); 93893.3, $6_{5,2}$--$5_{4,1}$ and $6_{5,1}$--$5_{4,2}$ (13~K, 0.56) \\
 & 2~mm, 132--174~GHz & 20 & 133094.1, $8_{7,1}$--$7_{6,2}$ and $8_{7,2}$--$7_{6,1}$ (25~K, 0.64); 148903.6, $8_{8,0}$--$7_{7,1}$ and $8_{8,1}$--$7_{7,0}$ (31~K, 0.62); 136882.9, $9_{7,3}$--$8_{6,2}$ and $9_{7,2}$--$8_{6,3}$ (27~K, 0.61); 152692.9, $9_{8,1}$--$8_{7,2}$ and $9_{8,2}$--$8_{7,1}$ (32~K, 0.59) \\
\hline

\end{tabular}
\end{table*}

\end{appendix}

\end{document}